# *In-situ* Optical Characterization of Noble Metal Thin Film Deposition and Development of a High-performance Plasmonic Sensor

by

David Joseph Mandia

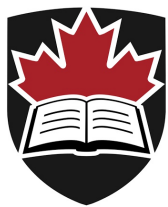

**A thesis submitted to the Faculty of Graduate and Postdoctoral Affairs in partial fulfillment for the degree of**

**Doctor of Philosophy**

**in**

**Chemistry**

**Department of Chemistry, Carleton University, Ottawa, Ontario**

**©2016-David J. Mandia**

# Abstract

With ever-growing industry demand for more uniform and conductive metal or metal oxide thin film coatings, an improvement to the techniques that produce these materials must be done in parallel. Two techniques that have shown great utility in this regard are referred to as chemical vapour deposition (CVD) and atomic layer deposition (ALD). While both techniques offer highly modular reactor designs and robust precursor-substrate chemistries (e.g., the benchmark $Al_2O_3$ process), many processes result in non-uniformity issues, ultimately leading to poor device performance. One such challenging process has been the fabrication of high-purity and uniform noble metal thin films such as gold. Equally as challenging is the ability to characterize metal CVD and ALD processes when the metallic film is just beginning to nucleate. This "nucleation delay", or induction process,is especially common in metal ALD processes and techniques such as ellipsometry lack sensitivity in the low-cycle regime of ALD (or pulsed CVD) processes. The present work addressed in this thesis introduces, for the first time, the use of tilted fiber Bragg grating (TFBG) sensors for accurate, real-time, and *in-situ* characterization of CVD and ALD processes for noble metals, but with a particular focus on gold due to its desirable optical and plasmonic properties. Through the use of orthogonally-polarized transverse electric (TE) and transverse magnetic (TM) resonance modes imposed by a boundary condition at the cladding-metal interface of the optical fiber, polarization-dependent resonances excited by the TFBG are easily decoupled. It was found that for ultrathin thicknesses of gold films from CVD (~6-65 nm), the anisotropic property of these films made it non-trivial to characterize their effective optical properties such as the real component of the permittivity. Nevertheless, the TFBG introduces a new sensing platform to the ALD and CVD community for extremely sensitive *in-situ* process monitoring. We later also demonstrate thin film growth at low (<10 cycle) numbers for the well-known $Al_2O_3$ thermal ALD process, as well as the plasma-enhanced gold ALD process. Finally, the use of ALD-grown gold coatings has been employed for the development of a plasmonic TFBG sensor with ultimate refractometric sensitivity (~550 nm/RIU).

# Acknowledgements

Capturing the name of every person that has made a positive impact on my research and in my life throughout the PhD is a tall order, and to fit them all within the margins of this page would simply be impossible. That said, I would like to start off by thanking my supervisor Prof. Seán T. Barry for all of his helpful mentoring and motivation throughout the years. Between all of the grant writing, scrounging funds so we can go to conferences, teaching courses, and, of course, finding out there's no coffee left, Seán has always displayed a steadfast patience that is second to none. Seán puts a lot of trust into his group to be independent and autonomous researchers, so that when "crazy" research ideas abound (which is an inevitability for all grad students) he can expertly get us back onto the path of logic and intuition.

I need to extend my appreciation to our collaborators in the department of electronics, Dr. Wenjun Zhou and Prof. Jacques Albert. It was very challenging at times to navigate through what must now be 1 TB of optical data. Your collective expertise in photonics made for many interesting and fruitful discussions. Wenjun is a fantastic engineer and I am very fortunate to have collaborated with him the past few years. I have learned a great deal about nanophotonics in our lengthy discussions with Jacques and they have certainly strengthened me as a scientist.

Thanks to Prof. Javier B. Giorgi (my former MSc supervisor) at the University of Ottawa for his continued mentorship and use of his UHV system for many XPS experiments. This has also allowed me to maintain my skills in a variety of surface science techniques going forward. Thanks to Dr. Yun Liu and Sander Mommers from the Centre for Catalysis and Research Innovation (CCRI) at the University of Ottawa for use of microscopy tools (SEM & AFM) and help with the odd vacuum pump repair, respectively. Thanks to many of the Barry lab members, past and present (Matthew Griffiths, Sara Koponen, Sydney Buttera, Goran Bačić, Peter Pallister, Peter Gordon, Jason Coyle, Paul Johnson, Awstin Chubb, Zack Dubrawski). Thanks to Matt Griffiths in particular; our initial work with the gold CVD monitoring experiments laid the groundwork for our eventual success in monitoring thermal and plasma-enhanced ALD processes. Thanks to professors Robert Crutchley, Maria DeRosa, Jeff Manthorpe, Anatoli Ianoul and Robert Burk for your wonderful insight throughout those very crucial undergraduate years in chemistry. I would also like to thank Howie Joress and Matthew J. Ward for their help during my beam-time at the Cornell High Energy Synchrotron Source.

Finally, I would like to thank my family (George, Mona, and Sarah) as well as my fiancée, Meaghan Morgan, for supporting me –and sometimes just plain "dealing" with me–through all of the peaks and valleys that graduate school entails. I will forego the hallmark "inspirational" quote that is supposed to terminate the acknowledgments section. Let the analysis roll forth.

# Table of Contents

## User-Defined

# Preface

This preface provides full bibliographical details for each article included in this thesis, as well as whether the article is reproduced in whole or in part. Use of copyrighted material is likewise acknowledged here. When citing material from this thesis, please cite the article relevant to the chapter, if the chapter is based on a publication.

Pursuant to the Integrated Thesis policy of Carleton University, the supervisor (Seán Barry) and the "student" (**David J. Mandia**) confirm that the student was fully involved in setting up and conducting the research, obtaining data and analyzing article(s) integrated in the thesis. Additionally, the supervisor confirms the information by the student in this preface.

The articles:

## Chapter 2

**Mandia, D.J.\***; Zhou, W.; Albert, J.; Barry, S.T. "CVD on Optical Fibers: Tilted Fiber Bragg Gratings as Real-time Sensing Platforms" *Chem. Vap. Depos.* **2015**; *21* (1); pp. 4-20.

This article is wholly reproduced and edited for formatting and clarity of presentation. David J. Mandia and Wenjun Zhou performed all experiments described in this invited review article with the exception of pulsed-CVD of copper. This publication was written by David J. Mandia.

## Chapter 3

**Mandia, D.J.\***; Zhou, W.; Wells, A.P.; Albert, J.; Barry, S.T. "Metallic Nanocoatings on Optical Fibers as a Sensor Platform" *ECS Trans.* **2015**; *69*(7); pp. 171-179.

This article is wholly reproduced and edited for formatting and clarity of presentation. David J. Mandia and Wenjun Zhou collaboratively performed all CVD experiments, *in-situ* spectral monitoring TFBG spectral data analysis. David J. Mandia performed the synthesis of CVD precursors , as well as microanalysis of the films (XPS/Ar$^+$-sputtering and SEM). Adam P. Wells performed any thermolysis kinetics experiments. This publication was written collaboratively with Seán T. Barry.

Coyle, J.P.*; Gordon, P.G.; Wells, A.P.; **Mandia, D.J.**; Sirianni, E.R.; Yap, G.P.A.; Barry, S.T.1 "Thermally Robust Gold and Silver Iminopyrrolidinates for Chemical Vapor Deposition of Metal Films" *Chem. Mater.* **2013**; *25*(22); pp. 4566-4573.

This article is partially reproduced and edited for formatting and clarity of precedence. David J. Mandia performed all X-ray photoelectron spectroscopy and Ar⁺-sputtering experiments. Jason P. Coyle wrote this publication.

# Chapter 4

Zhou, W.* ; **Mandia, D.J.**; Griffiths, M.B.E.; Bialiayeu, A.; Zhang, Y.; Gordon, P.G.; Barry, S.T.; Albert, J. "Polarization-dependent Properties of the Cladding Modes of a Single Mode Fiber Covered with Gold Nanoparticles" *Optics Express.* **2013**; *21*(1); pp. 245-255.

This article is wholly reproduced and edited for formatting and clarity of presentation. David J. Mandia and Wenjun Zhou collaboratively performed all CVD experiments, *in-situ* spectral monitoring, TFBG spectral data analysis. David J. Mandia, Matthew B.E. Griffiths, and Peter G. Gordon synthesized the precursor prior to CVD. David J. Mandia and Peter G. Gordon performed the SEM, Alex Bialiayeu performed the simulation of the polarization disperion in the gold coating and Yang Zhang assisted David. J Mandia and Wenjun Zhou in the IR scattering experiments. This publications was written colloabatively between David J. Mandia, Jacques Albert, and Wenjun Zhou.

# Chapter 5

Zhou, W. ; **Mandia, D.J.**; Griffiths, M.B.E.; Barry, S.T.; Albert, J.* "Effective Permittivity of Ultrathin Chemical Vapor Deposited Gold Films on Optical Fibers at Infrared Wavelengths" *J. Phys. Chem. C.* **2014**; *118*(1); pp. 670-678.

This article is wholly reproduced and edited for formatting and clarity of presentation. David J. Mandia and Wenjun Zhou collaboratively performed all CVD experiments, *in-situ* spectral monitoring, TFBG spectral data analysis. David J. Mandia, Matthew B.E. Griffiths synthesized the precursor prior to CVD. David J. Mandia performed all AFM/SEM measurements. Wenjun Zhou performed the FIMMWAVE simulations and "complex mode solver" simulations.

# Chapter 6

This article is wholly reproduced and edited for formatting and clarity of presentation. David J. Mandia aided with the SRI measurements and performed the AFM/SEM measurements. Wenjun Zhou performed the TFBG spectral measurements and calculations. David J. Mandia aided in manuscript preparation and interpretation of the spectral data. This publication was written mostly by Wenjun Zhou.

# Chapter 7

This article is wholly reproduced and edited for formatting and clarity of presentation. David J. Mandia performed the synthesis of precursor, *in-situ* TFBG spectral collection and analysis, XPS and $Ar^+$-sputtering, synchrotron GID, XANES, and XAFS work (done in collaboration with the Cornell High Energy Synchrotron Source), and AFM/SEM measurements. Wenjun Zhou aided with analysis of the TFBG spectral data and setup of the telecommunications equipment. Peter G. Gordon performed the ALD of the $Al_2O_3$ pre-coatings. Javier B. Giorgi and Jeff J. Sims aided with XPS collection. Howie Joress and Matthew J. Ward aided David J. Mandia with the the GID and XANES work at the synchrotron source).

# Chapter 8

**Mandia, D.J.\***; Zhou, W.; Ward, Matthew J.; Joress, H.; Sims, J.J.; Giorgi, J.B.; Albert, J.; Barry, S.T. "The Effect of ALD-grown $Al_2O_3$ on the Refractive Index Sensitivity of CVD Gold-coated Optical Fiber Sensors" *Nanotechnology*. **2015**; *26*(43) 43002; pp. 1-12

This article is wholly reproduced and edited for formatting and clarity of presentation. David J. Mandia performed the following: ALD of $Al_2O_3$ for pre-coatings, the *in-situ* CVD experiments, refractometry measurements (SRI tests), XPS and $Ar^+$-sputter depth profile of the CVD witness slides, AFM imaging, Jeff J. Sims and Javier B. Giorgi assisted with the XPS instrumentation. Howie Joress and Matthew J. Ward aided David J. Mandia with the the XRR and GID measurements at the Cornell synchrotron source (CHESS). Wenjun Zhou performed the simulation to generate power density profiles of the $Al_2O_3$-coated TFBGs. This publication was written jointly by David J. Mandia and Wenjun Zhou.

# Chapter 9

Griffiths, M.B.E.; Pallister, P.J.; **Mandia, D.J**.; Barry, S.T.\* "Atomic Layer Deposition of Gold Metal" *Chem. Mater.* **2016**; *28*(1); pp. 44-46.

This article is wholly reproduced and edited for formatting and clarity of presentation. Peter J. Pallister performed the nucleation studies and TEM characterization as well as several ALD experiments required for the saturation curve in Figure 9.2. Matthew B.E. Griffiths performed the precursor synthesis and characterization, film microanalysis, and process optimization. David J. Mandia performed the XPS experiments. This publication was written collaboratively between all authors.

# Chapter 10

**Mandia, D.J.\***; Zhou, W.; Pallister, P.J.; Albert, J.; Barry, S.T. "*In-Situ* Optical Monitoring of Gold and Alumina Atomic Layer Deposition: Development of an Ultra-High Sensitivity Plasmonic Optical Fiber-Based Sensor" *Manuscript in preparation.***2016.** ALD monitoring experiments were set up by David J. Mandia, Wenjun Zhou, and Peter J. Pallister. Refractometry measurements were performed by David J. Mandia and Wenjun Zhou. AFM and SEM imaging was performed by David J. Mandia. This chapter was wholly written by David J. Mandia. *As-yet unpublished.*

# List of Figures

X

XI

XII

XIII

# List of Tables

# List of Abbreviations

**AFM**        atomic force microscopy

**ALD**        atomic layer deposition

**BBS**        broadband source

**CDI**        carbodiimide

**CM**        Clausius-Mossotti

**CVD**        chemical vapour deposition

**DCM**        double-crystal monochromator

**EDX**        energy dispersive X-ray spectroscopy

**EMA**        effective medium approximation

**FBG**        fiber Bragg grating

**GID**        grazing incidence diffraction

**GPC**        growth per cycle

**IR**        infrared

**LSPR**        localized surface plasmon resonance

**MOCVD**    metal-organic chemical vapor deposition

**NC-AFM**   non-contact atomic force microscopy

**NIR**        near-infrared

**NMR**      nuclear magnetic resonance

**OSA**      optical spectrum analyzer

**p-CVD**    pulsed chemical vapor depsosition

**PEALD**    plasma-enhanced atomic layer deposition

**PID**      proportional-integral-derivative

**QCM**      quartz crystal microbalance

**RI**       refractive index

**RIU**      refractive index units

**SE**       spectroscopic ellipsometry

**SEM**      scanning electron microscopy

**SRI**      surrounding refractive index

**SPP**      surface plasmon polaritons

**SPR**      surface plasmon resonance

**TE**       transverse-electric

**TFBG**     tilted fiber Bragg grating

**TGA**      thermogravimetric analysis

**TM**       transverse-magnetic

**TMA**      trimethylaluminum

**UHV**      ultra-high vacuum

**XANES**    X-ray absorption near-edge spectroscopy

**XAFS**     X-ray absorption fine structure spectroscopy

**XAS**      X-ray absorption spectroscopy

**XPS**      X-ray photoelectron spectroscopy

**XRR**      X-ray reflectivity

# List of Publications (PhD: September '12 -May '16)

18.   **<u>Mandia, D.J.</u>**; Zhou, W.; Pallister, P.J.; Albert, J.; Barry, S.T. Real-time Optical Monitoring of Gold Metal Atomic Layer Deposition and Fabrication of an SPR-based Sensor. **2016**; *In preparation.*

17.   Buttera, S.C.; **<u>Mandia, D.J.</u>**; Barry, S.T. Tris(dimethylamido)aluminum(III): An overlooked ALD precursor. **2016**; *Journal of Vacuum Science and Technology A. Submitted.* **Manuscript ID:** *ALD16JVSTA-A-16-313*

16.   Zhang, W.; Dey, G.; **<u>Mandia, D.J.</u>**; Barry, S.T. Using a Vapor Phase Surfactant to Control Gold Metal Plate Growth. **2016**; *Advanced Materials Interfaces. Submitted.* **Manuscript ID:** *admi.201600864*

15.   Griffiths, M.B.E.; Pallister, P.J.; **<u>Mandia, D.J.</u>** Barry, S.T. Atomic Layer Deposition of Gold Metal. *Chemistry of Materials*; **2016**; *28* (1); pp. 44-46. **DOI:** 10.1021/acs.chemmater.5b04562

14.   **<u>Mandia, D.J.</u>**; Zhou, W.; Wells, A.P.; Albert, J.; Barry, S.T. Metallic Nanocoatings on Optical Fibers as a Sensing Platform. *ECS Transactions*; **2015**; *69* (7); 171-179. **DOI:** 10.1149/06907.0171ecst

13.   **<u>Mandia, D.J.</u>**; Zhou, W.; Ward, M.J.; Joress, H.; Sims, J.; Giorgi, J.B.; Albert, J.; Barry, S.T. The Effect of ALD-grown $Al_2O_3$ on the Refractive Index Sensitivity of CVD Gold-coated Optical Fiber Sensors. *Nanotechnology;* **2015**; *26* (43); pp. 434002. **DOI:** 10.1088/0957-4484/26/43/434002

12.   Griffiths, M.B.E.; Koponen, S.E.; **<u>Mandia, D.J.;</u>** McLeod, J.F.; Coyle, J.P.; Sims, J.J.; Giorgi, J.B.; Sirianni, E.R.; Yap, G.P.A.; Barry, S.T. Surfactant Directed Growth of Gold Metal Nanoplates by Chemical Vapour Deposition. *Chemistry of Materials*; **2015**; *2 7* (17); pp. 6116-6124. **D O I :** 10.1021/acs.chemmater.5b02712

11.   Zhou, W.; **<u>Mandia, D.J.</u>**; Barry, S.T.; Albert, J. Monitoring of the Insulator-to-Metal Transition of Ultrathin Gold Coatings on Optical Fibers. (NOMA 2015) *Advanced Photonics;* **2015**;*1*(1);pp.1-4. **DOI:** 10.1364/NOMA.2015.NM4C.4

10.   Zhou, W.; **<u>Mandia, D.J.</u>**; Barry, S.T.; Albert, J. Absolute Near Infrared Refractometry With A Calibrated Tilted Fiber Bragg Grating. *Optics Letters.* **2015**; *40* (8); pp.1713-1716. **DOI:** 10.1364/OL.40.001713

9.   **<u>Mandia, D.J.</u>**; Zhou, W.; Albert, J.; Barry, S.T. Chemical Vapor Deposition on Optical Fibers: Tilted Fiber Bragg Gratings as Real-Time Sensing P l a t f o r m s . *Chemical Vapor Deposition;* **2015**; *2 1* (1-3); pp. 4-20. Invited Review. **DOI:**

10.1002/cvde.201400059

# List of Conferences

**33.** Griffiths, M.B.E.; Pallister, P. J.; **Mandia, D.J.;** Barry, S.T. Conformal Gold Nanoparticle Deposition Using Plasma Enhanced Atomic Layer Deposition. The 16[th] International Conference on Atomic Layer Deposition (AVS ALD 2016). Dublin, Ireland. July 24-27, 2016. Oral Presentation by Matthew B. E. Griffiths

**32.** **Buttera, S.C.; Mandia, D.J.**; Pedersen, H.; Barry, S.T. Preliminary Studies of Aluminum Guanidinates as Precursors for Aluminum Nitride PEALD. The 16[th] International Conference on Atomic Layer Deposition (AVS ALD 2016). Dublin, Ireland. July 24-27, 2016. Poster Presentation by Sydney C. Buttera. (XPS work completed after abstract submission)

**31.** **Mandia, D.J.**; Zhou, W.; Pallister, P.J.; Albert, J.; Barry, S.T.; In-Situ Optical Monitoring Of Gold Metal ALD And Development f An Ultra-High Sensitivity Plasmonic Optical Fiber- Based Sensor. The 16th International Conference on Atomic Layer Deposition (AVS ALD (2016). Dublin, Ireland. July 24-27, 2016. Oral Presentation by **Mandia, D.J.**

**30.** Barry, S.T.; **Mandia, D.J.** Metallic Nanocoatings on Optical Fibers as a Sensor Platform. The 228th Electrochemical Society (ECS) Meeting. Phoenix, AZ, US. October 11-16, 2015. Oral Presentation by Barry, S.T.

**29.** Koponen, S.E.; Griffiths, M.B.E.; McLeod, J.F.; **Mandia, D.J.**; Giorgi, J.B.; Coyle, J.P.; Barry, S.T. Low-dimension Au Nanostructures by Chemical Vapour Deposition. The 98th Canadian Chemistry Conference and Exhibition (CSC 2015. Shaw Centre, Ottawa, ON, Canada. June 13-17, 2015. Poster Presentation by Koponen, S.E.

**28**. Barry, S.T.; Albert, J.; **Mandia, D.J**; Zhou, W. Metallic Nanocoatings on Optical Fibers as a Sensor Platform. The 98th Canadian Chemistry Conference and Exhibition (CSC 2015. Shaw Centre, Ottawa, ON, Canada. June 13-17, 2015. Invited Oral Presentation by Barry, S.T.

**27.** McLeod, J.F.; Griffiths, M.B.E.; Koponen, S.E.; **Mandia, D.J.**; Giorgi, J.B.; Coyle, J.P.; Barry, S.T. Novel CVD Gold Morphologies by Functional Ligand Precursors. The 98th Canadian Chemistry Conference and Exhibition (CSC 2015. Shaw Centre, Ottawa, ON, Canada. June 13-17, 2015. Oral Presentation by McLeod, J.F.

**26.** **Mandia, D.J.**; Zhou, W.; Albert, J.; Barry, S.T.; Use of Dielectrics for Refractometric Sensitivity Enhancement in Tilted Fiber Bragg Gratings. IC-Impacts Summer Institute: Optical Sensing Technologies. University of Toronto, Toronto, ON, Canada. June 14-19, 2015. Poster Presentation by

**Mandia, D.J.**

25. Griffiths, M.B.E.; **Mandia, D.J.**; Pallister, P.J.; Barry, S.T.; Towards Atomic Layer Deposition of Gold Metal from Volatile, Ambient-stable, Gold Phosphine Compounds. The 15th International Conference on Atomic Layer Deposition (AVS ALD 2015). Portland Hilton, Portland, Oregon, USA. June 28-July 1, 2015.Oral Presentation by Griffiths, M.B.E.

24. **Mandia, D.J**.; Zhou, W.; Sims, J.J.; Giorgi, J.B.; Ward, M.J.; Joress, H.; Albert, J.; Barry, S.T. The Effect of ALD-grown $Al_2O_3$ Coatings on the Refractive I n d e x Sensitivity of an Optical Fiber Sensor During Metal D e p o s i t i o n . T h e 1 5 t h International Conference on Atomic Layer Deposition (AVS ALD 2015). Portland Hilton, Portland, Oregon, USA. June 28-July 1, 2015.Oral Presentation by **Mandia, D.J.**

23. Zhou, W.; **Mandia, D.J.**; Barry, S.T., Albert, J.; Monitoring of the Insulator to Conductor Transition of Ultrathin Gold Coatings on Optical Fibers. Novel Optical Materials and Applications (NOMA). Omni Parker House, Boston, Massachusetts, USA. June 26-July 1, 2015. Oral Presentation by Zhou, W.

22. Griffiths, M.B.E.; Koponen, S.E.; **Mandia, D.J.**; McLeod, J.F.; Coyle, J.P.; Giorgi, J.B.; Barry, S.T. Chemical Vapor Deposition of Gold: Morphology Control from Ligand Design. *The 47th Inorganic Discussion Weekend (IDW 47)*. Université de Montréal, Montreal, Quebec, Canada. November 14-16, 2014. Poster Presentation by Griffiths, M.B.E.

21. Koponen, S.E; Griffiths, M.B.E.;., **Mandia, D.J.**; McLeod, J.F.; Coyle, J.P.; Giorgi, J.B.; Barry, S.T. Gold Chemical Vapor Deposition Using Gold (I) Amide Precursors. *The 47th Inorganic Discussion Weekend (IDW 47).* Université de Montréal, Montreal, Quebec, Canada. November 14-16, 2014. Oral Presentation by Koponen, S.E.

20. **Mandia, D.J.;** Zhou, W.; Albert, J.; Barry, S.T. XAS and Optical Study of an Interfacial $Al_2O_3$ Coating: Enhancement of Refractive Index Sensitivity in Optical Fibers for CVD Monitoring. *The 47th Inorganic Discussion Weekend (IDW 47).* Université de Montréal, Montreal, Quebec, Canada. November 14-16, 2014. Poster Presentation by **Mandia, D.J.**

19. Sims, J.; **Mandia, D.J.**; Giorgi, J.B. Adsorption and Decomposition of Formic Acid on Cobalt. *64th Canadian Society of Chemical Engineering Conference (CSChe 2014).* Niagara Falls, Ontario, Canada. October 19-22, 2014. Poster Presentation by Sims, J. (3rd place winner in Graduate StudentPoster Competition).

18. **<u>Mandia, D.J.</u>**; Zhou, W.; Albert, J.; Barry, S.T. Optical Fiber-Based Approach t o *In-situ* Monitoring of Group 11 Metal-Organic Chemical Vapor Deposition: Generation of Isotropic and Anisotropic Ultrathin Films. *Materials R e s e a r c h Society (MRS) Fall Meeting 2014.* Hynes Convention Center and Sheraton Boston, Boston, Massachusetts, USA. November 30- December 5 , 2 0 1 4 . Poster Presentation by **<u>Mandia, D.J.</u>**

17. **<u>Mandia, D.J.</u>**; Zhou, W.; Albert, J.; Barry, S.T. Optical Fiber-based Approach t o *In-situ* Monitoring of Group 11 Metal-organic Chemical Vapor Deposition. *9ᵗ ʰ International Conference on Surfaces, Coatings, and Nano-structured Materials (NANOSMAT 2014).* Trinity College Dublin, Dublin, Ireland. September 8-11, 2014. Oral Presentation by **<u>Mandia, D.J.</u>**

16. **<u>Mandia, D.J.</u>**; Zhou, W.; Albert, J.; Barry, S.T. Chemical Vapor Deposition of Ultrathin Gold Films on Optical Fibers: Real-time Optical Monitoring of Film Growth by a Tilted Fiber Bragg Grating. *The 16th Photonics North Conference (SPIE Photonics North 2014).* Montreal Convention Centre, Montreal, QC, Canada. May 28-30, 2014. Poster Presentation by **<u>Mandia, D.J.</u>**

15. **<u>Mandia, D.J.</u>**; Zhou, W.; Albert, J.; Barry, S.T. Chemical Vapor Deposition of Ultrathin Gold Films on Optical Fibers: Real-time Optical Monitoring of Film Growth by a Tilted Fiber Bragg Grating. *The 16th Photonics North Conference (SPIE Photonics North 2014).* 2014. Montreal Convention Centre, Montreal, QC, Canada. May 28-30, 2014. Oral Presentation by **<u>Mandia, D.J.</u>**

14. **<u>Mandia, D.J.</u>**; Griffiths, M.B.E.; Pallister, P.J.; Coyle, J.P.; Zhou, W.; Albert, J.; Barry, S.T. In situ Monitoring of the Chemical Vapour Deposition of Gold Films onto TFBG-Inscribed Optical Fibers. *The 46th Inorganic Discussion Weekend (IDW 46).* York University, Toronto, ON, Canada. November 8-10, 2013. Poster Presentation by Griffiths, M.B.E.

13. **<u>Mandia, D.J.</u>**; Griffiths, M.B.E.; Pallister, P.J.; Coyle, J.P.; Zhou, W.; Albert, J.; Barry, S.T. Preliminary Surface Mechanisms and In situ Optical Monitoring for Metal Deposition from Group 11 Iminopyrrolidinates by CVD. *The 46th Inorganic Discussion Weekend (IDW 46).* York University, Toronto, ON, Canada. November 8-10, 2013. Oral Presentation by **<u>Mandia, D.J.</u>**

12. Barry, S.T.; Griffiths, M. B.; <u>Mandia, D. J.</u>; Coyle, J. P.; Gordon, P. G.; Zhou, W. T.; Shao, L.-Y. T.; Albert, J. Chemical Vapour Deposition and Atomic Layer Deposition: Metals for Optical Fibers. In Workshop on Specialty Optical Fibers and their Applications; (WSOF 2013) *Optical Society of America: Sigtuna* 2013; p. W4.3. August 28-30, 2013, Sigtuna, Sweden. Invited Oral Presentation by Barry, S.T.

11. **Mandia, D.J.**; Griffiths, M.B.E.; Zhou, W.; Gordon, P.G.; Albert, J.; Barry, S.T. *In-situ* Deposition Monitoring by a Tilted Fiber Bragg Grating Optical Probe: Probing Nucleation in Chemical Vapour Deposition of Gold. 19th International Conference on Chemical Vapor Deposition (EuroCVD 19). Varna, Bulgaria. Invited Oral Presentation by Seán T. Barry.

10. **Mandia, D.J.**; Griffiths, M.B.E.; Gordon, P.G.; Zhou, W.; Albert, J.; Barry, S.T. Optical Monitoring of the Chemical Vapor Deposition of Gold and Silver Metal Films from Gold (I) and Silver (I) Guanidinates. *Plasmonics Summer School 2013.* University of Toronto, ON, Canada. Invited Poster Presentation by **Mandia, D.J.**

9. **Mandia, D.J.**; Griffiths, M.B.E.; Gordon, P.G.; Zhou, W.; Albert, J.; Barry, S.T. Coinage Metal Chemical Vapour Deposition on Tilted Fiber Bragg Gratings: Novel Processes and Fiber-optic Based Sensors for Thin Film Growth. The *96th Canadian Chemistry Conference (CSC 2013).* June 1-5, 2013, Quebec City, QC, Canada. Oral Presentation by Griffiths, M.B.E. (1st prize, oral presentation competition)

8. **Mandia, D.J.**; Griffiths, M.B.E.; Gordon, P.G.; Zhou, W.; Albert, J.; Barry, S.T. *In situ* Deposition Monitoring by a Tilted Fiber Bragg Grating Optical Probe: Probing Nucleation and Polarization-dependent Properties of Gold Grown Via Chemical Vapour Deposition. The *96th Canadian Chemistry Conference (CSC 2013).* June 1-5,2013, Quebec City, QC,Canada.Poster Presentation by **Mandia, D.J.** (1st prize poster competition, surface science division, > 150 participants)

7. **Mandia, D.J.**; Griffiths, M.B.E.; Gordon, P.G.; Zhou, W.; Albert, J.; Barry, S.T. Coinage Metal Chemical Vapour Deposition on Tilted Fiber Bragg Gratings: Novel Processes and Fiber-optic Based Sensors for Thin Film Growth. *20th Annual Ottawa-Carleton Chemistry Institute (OCCI) Day.* May 31st, 2013, University of Ottawa, Ottawa, ON, Canada.Oral Presentation by Griffiths, M.B.E.

6. **Mandia, D.J.**; Griffiths, M.B.E.; Gordon, P.G.; Zhou, W.; Albert, J.; Barry, S.T. *In-situ* Deposition Monitoring by a Tilted Fiber Bragg Grating Optical Probe: Probing Nucleation and Polarization-dependent Properties of Gold Grown Via Chemical Vapour Deposition. *20Th Annual Ottawa-Carleton Chemistry Institute (OCCI) Day.* May 31st, 2013, University of Ottawa, Ottawa, ON, Canada. Poster Presentation by **Mandia, D.J.**

5. **Mandia, D.J.**; Barry, S.T.; Griffiths, M.B.E.; Zhou, W.; Gordon, P.G. Albert, J. *In-situ* Deposition Monitoring by a Tilted Fiber Bragg Grating Optical Probe: Probing Nucleation and Polarization-dependent Properties of Gold Grown

Via Chemical Vapour Deposition. *Surface Canada 2013*. May 7-10, 2013, Western University, London, ON, Canada. Poster Presentation by  **Mandia, D.J.**

4. Griffiths, M.B.E.; **Mandia, D.J.**; Coyle, J.P.; Zhou, W.; Gordon, P.G.; Albert, J.; Barry, S.T. *In-situ* Monitoring of the Chemical Vapour Deposition of Copper and Gold Films onto TFBG-Inscribed Optical Fibers. *The 45th Inorganic Discussion Weekend (IDW 45)*. November 2-4, 2012, University of Ottawa, Ottawa, ON, Canada. Poster Presentation by Griffiths, M.B.E.

3. **Mandia, D.J.**; Zhou, W.; Griffiths, M.B.E.; Gordon, P.G.; Albert, J.; Barry, S.T. *In-Situ* Deposition Monitoring of a Novel Gold Metal CVD Process by a Tilted Fiber Bragg Grating Optical Filament. *The 45th Inorganic Discussion Weekend (IDW 45)*. November 2-4, 2012, University of Ottawa, Ottawa, ON, Canada. Oral Presentation by **Mandia, D.J.**

2. **Mandia, D.J.**; Wee, T.L.; Scaiano, J.C.; Giorgi, J.B. Single-crystal Supported Growth of  Cobalt Oxide by a Novel Photochemical Method: A NC-AFM and XPS Investigation. 1*9th Annual Ottawa-Carleton Chemistry Institute (OCCI) Day*. June 3rd, 2012, Carleton University, Ottawa, ON, Canada.  Poster Presentation by **Mandia, D.J.**

1. **Mandia, D.J.**; Wee, T.L.; Scaiano, J.C.; Giorgi, J.B. Noncontact Atomic Force Microscopy Study of Photochemically Induced Cobalt (III) Oxide Nanostructure Growth Supported on Single-crystal Substrates: YSZ (100), YSZ(111), MgO(100), and HOPG. *18th Annual Ottawa-Carleton Chemistry Institute (OCCI) Day*. June 1st, 2011, University of Ottawa, Ottawa, ON, Canada. Poster Presentation by **Mandia, D.J.**

*…this thesis is dedicated to my parents (George and Mona) and my uncle Burke Blanchfield...*

# Chapter 1

Introduction

## 1.1 Chemical Vapor Deposition and Atomic Layer Deposition

Chemical vapour deposition (CVD) is a technique that comprises a broad range of vapour-phase deposition techniques that also includes atomic layer deposition (ALD). The applications of the CVD technique are quite diverse and varied, spanning a wide array of applications from protective coatings[1,2] to microelectronics[3] and thin film optics.[4,5] CVD and ALD were introduced as solutions to the uniformity and conformality issues that plagued the line-of-sight techniques such as sputtering (physical vapour deposition or "PVD"),[6] pulsed laser deposition,[7] and many other adaptations involving electron-beam evaporation of a target metal source,[8,9] which is effectively the "precursor" in these cases. CVD involves a more careful, surface-chemical approach wherein one (or many) precursors are introduced to a heated surface (substrate) and undergo deposition of a target metal (or metal oxide) film with desirable properties and often on topographically challenging substrates. Unlike PVD where the precursor is the same composition as the target metal film, CVD and ALD involve the use of robust and highly-reactive inorganic or organometallic precursors. In order to obtain high-quality target films, a rational precursor design approach must be taken in order to generate precursors that are chemically reactive (low melting point), volatile (high vapour pressure, low molecular weight), and thermally stable enough to allow for ease-of-transport in the gas-phase with no decomposition. The most typical generic growth mode observed for CVD and some PVD cases is Vollmer-Weber growth, which involves nucleation of islands that coalesce into larger islands if it is more energetically favourable than surface wetting.[10] A similar growth mode that is more common in layer-by-layer processes like ALD is that of Stranski-Krastanov growth wherein 2D growth is initiated and then transitions into 3D growth, resulting in islands that coalesce to form a continuous film.[11] Looking at Figure 1.1, CVD (and ALD) can occur in a variety of ways depending on the nature of the desired target film and the precursors involved. In general, the process involves entraining the reaction chamber with the precursor(s) (**A, B**) by heating a bubbler source and then allowing transport of the molecules to the surface (processes **(i)** and **(ii)**). In the case of ALD and some CVD, the target film can be a metal oxide (e.g. $Al_2O_3$) which means a secondary precursor (**B**) (co-reagent) such as water (or oxygen, ozone, peroxide, etc) is required for the heteroatom (oxygen source). CVD, particularly for metal films, often involves the use of a single-source precursor, making it unnecessary for the use of a secondary precursor (**B**) or a purge gas (**C**). In some cases, the use of a heated carrier gas (**C**) (e.g. $N_2$) is necessary to aid the transport of the precursors to the reaction chamber. In ALD, intermittent purge cycles are often required between exposure of precursors **A** and **B** for efficient removal of unreacted precursor or surface reaction byproducts (process **(v)**). After transport to the reactive surface (e.g. hydroxyl-terminated silicon), a physisorption process occurs followed by a diffusion on the surface (**(iv)**) and subsequent chemisorption. The latter step begins the nucleation and growth process for the target film, which can be interrogated *in situ* by impinging the reaction surface with a light source (**(vii)**) and collecting the attenuated light signal with a detector (**(vi)**), which is an optical spectrum analyzer (OSA) in the case of this work. The most common optical technique for real-time CVD and ALD monitoring is called spectroscopic ellipsometry (SE), however, in this thesis we introduce a new optical fiber-based technique relying on fiber Bragg grating (FBG) technology.

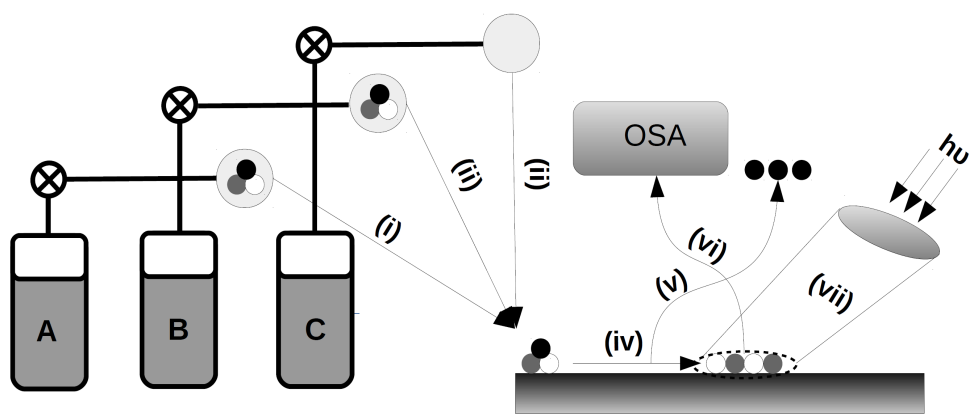

**Figure 1.1**: General schematic of the thermal vapour-phase deposition processes of CVD and ALD. Note that processes **(vi)** and **(vii)** represent the *in-situ* interrogation of thin film growth by CVD or ALD through the use of a tilted fiber Bragg grating (TFBG) and transmission detection via an optical spectrum analyzer (OSA).

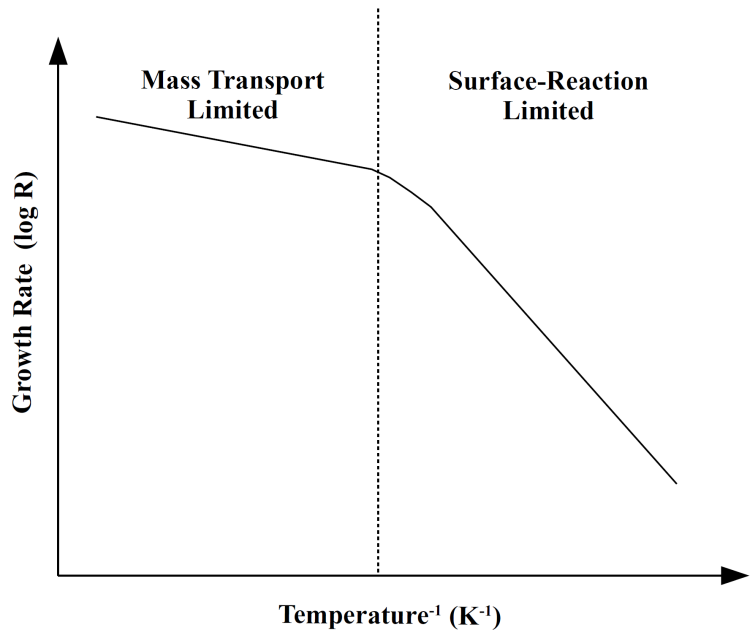

**Figure 1.2**: General schematic showing the growth rate of a generic CVD process with respect to temperature. The two regimes separated by the dashed line indicate that the rate-determining step for CVD processes is mass-transport limited and surface-reaction limited at high and low reactor temperatures, respectively. Adapted from Pierson.[209]

As mentioned, CVD must occur through the following processes: impingement, adsorption (physisorption), reaction with surface (chemisorption), surface mobilization (diffusion), nucleation, byproduct desorption. However, the hundreds of CVD reactor designs, which typically range from tubular cold-walled to low-pressure hot-walled chamber designs, make the surface reaction kinetics of even the simplest CVD processes difficult to probe. Figure 1.2 is a simplified and generic model for CVD growth and it highlights two salient points: (1) At high temperatures the mass transport to the surface is the effective rate-determining step.; (2) In the low-temperature regime, the growth rate of CVD is highly-sensitive to temperature gradients and follows exponential, Arrhenius-type growth kinetics. Uniformity of the temperature and flow parameters across the deposition zone are typically assumed in most mass-transport models,[12–14] but the quality of the resultant films will always depend on a delicate balance of mass-transport dynamics imposed by the reactor design and the surface reaction kinetics that are dominated by the reactivity between the precursor molecule(s) and the substrate.

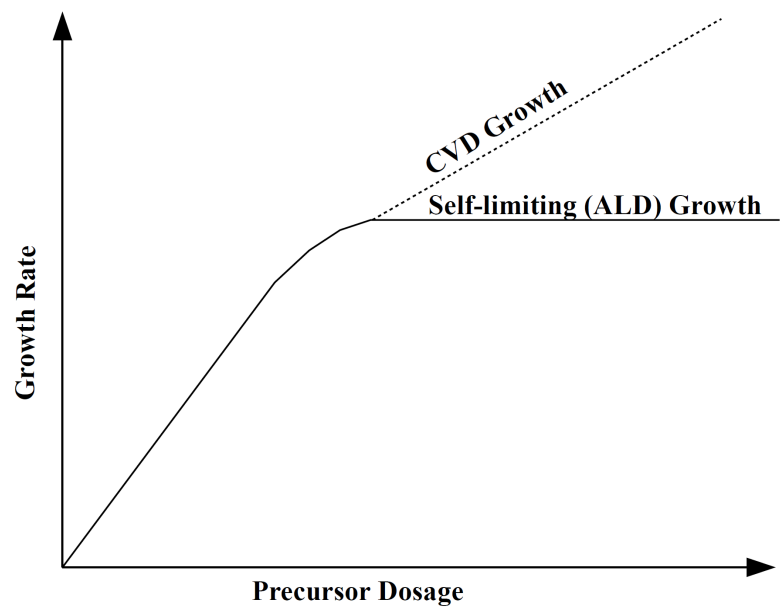

**Figure 1.3**: General schematic comparing the typical growth rates of CVD and ALD processes.

ALD is a layer-by-layer CVD process involving alternating exposures of two or more precursors to a surface and resulting in self-limiting growth. In a typical AB-type process, the first precursor (**A**) chemisorbs to the surface and continues to do so until the gaseous precursor no longer reacts with the surface; this results in the formation of a stable monolayer and effectively completes the first ALD half-cycle. Similarly, after purging the reactor with an inert gas, the second precursor (**B**) is introduced and reacts with the monolayer that was formed in the first ALD half-cycle until no further uptake of the secondary precursor occurs; this effectively completes the second ALD half-cycle, as well as completes the entire first ALD cycle for an AB-type process. The secondary precursor (**B**; "co-reactant") in any ALD process is typically the heteroatom of the target film and is often less molecularly complex than the primary, metal-containing precursor (**A**). For example, the benchmark AB-type thermal ALD process that produces $Al_2O_3$ films is $Al(CH_3)_3$ and $H_2O$, as precursors **A** and **B**, respectively.[15,16] Depending on the nature of the target film, it is desirable to use $O_2$ plasma, $O_3$, or $H_2O$ for metal-oxide films; $NH_3$ for metal-nitride films; $H_2$ for metallic films; and $H_2S$ or $H_2Se$ for metal-chalcogenide films. As will be discussed in the context of gold ALD in Chapter 9, the nature of the co-reactant in ALD sometimes plays crucial role in removing reactive reaction byproducts from the first ALD half-cycle. Similar to CVD, ALD processes have a growth per cycle (GPC) that varies with temperature, but only ever in a small temperature range that is often termed the "ALD window".[15] The so-called ALD window means that the self-limiting growth rate shown in Figure 1.3 can only exist for a process temperature that falls within this window, otherwise the growth is more sparse and "CVD-like" with the one main exception being the $Al_2O_3$ process, which undergoes a 20% decrease of its average growth rate in thermal ALD (~1.1 Å/cycle) at temperatures 100 ºC above the typical ALD window of 180-300 ºC.[16]

Generally, the typical ALD growth rate will be less than or equal to the thickness of a molecular monolayer of the target film (e.g. ~0.5 Å/cycle for gold ALD[17] with average bulk monolayer thickness of ~2.4 Å). Ideal ALD occurs if the precursor is cleanly volatilized and subsequently undergoes a clean de-coordination of one or more ligands at the surface. The signature self-limiting growth "plateau" presented in Figure 1.3 can only occur if the ALD window sufficiently separates the decomposition and onset-of-volatility temperatures for a given precursor. Both of those metrics are easily determined through the use of thermogravimetric analysis (TGA), which is a typical pre-screening process for precursor viability in vapour-phase deposition processes such as CVD or ALD; it has been employed for all CVD and ALD precursors presented in this work.

## 1.2    Chemical Vapor Deposition and Atomic Layer Deposition of Gold Films

Noble metal thin films have sustained a broad interest for a variety of applications that include–but are not limited to–catalysis,[18] optics and photonics,[19] microelectronics,[13] and plasmonics.[20,21] Surprisingly, there is a sparse literature precedent for gold CVD processes, but nevertheless some interesting gold(III) compounds such as dimethylgold(III) acetate,[22,23]dimethylgold(III) dithiophosphinates,[24] dimethylgold(III) pivalates,[25] and dimethylgold(III) trifluoroacetylacetonate[26] have all proven to be viable CVD precursors. In the present work we explore the use of gold(I) guanidinate and iminopyrrolidinate precursors (as well as one silver(I) guanidinate example) in CVD and a gold(I) phosphine for ALD. Guanidinates are a subset of amidinate ligands that have recently seen a lot of use in CVD and ALD,[27] and they exhibit both a compatibility and robustness that makes them more stable in CVD and ALD process conditions than other ligand systems such as cyclopentadienyl.[28,29] Figure 1.4 shows the typical resonance structures of the monoanionic amidinate, guanidinate, and iminopyrrolidinate ligands employed in this work. It is worth noting that the noble metal *N,N′*-dialkylamidinates typically favour dimeric structures with one notable exception being the silver amidinates, which tend to form dimeric/trimeric mixtures in solution.[30]

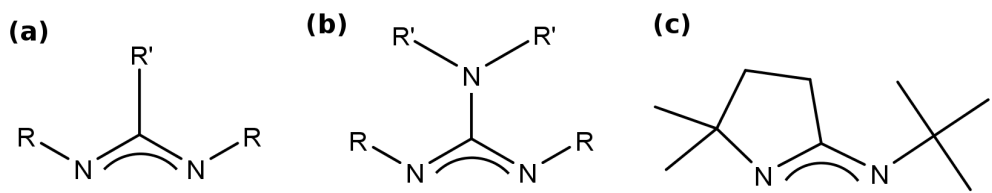

**Figure 1.4:** (a) Amidinates, (b) guanidinates, and (c) iminopyrrolidinates used as ligands for CVD precursors. R, R'=alkyls

The homoleptic (dimeric or trimeric structures bearing identical ligands) gold(I) and silver(I) guanidinates used in this particular work bear isopropyl and methyl moieties as the R and R' substituents, respectively, and the iminopyrrolidinates were used as shown above (Figure 1.4). The metal complexes containing guanidinates are typically synthesized through a salt metathesis route which involves reaction between the lithiated guanidinate and the metal-halide species containing the desired group 11 metal centre. These reactions are typically performed at low temperatures (e.g. -30 ºC) due to the thermodynamic driving force (exothermicity) of producing an insoluble (LiCl) salt. Another common synthetic route for homoleptic guanidinate compounds involves insertion of the alkylated metal centre over carbodiimide (CDI) to form the M-N-C-N metallocycle.[30] The gold(I) and silver(I) guanidinates ([Au(N$^i$Pr)$_2$CNMe$_2$]$_2$ and Ag$_3$[($^i$PrN)$_2$CN(Me)$_2$]$_3$, respectively) synthesized in our group tend to undergo *β*-hydride elimination in the gas-phase during CVD, which has been attributed to undesirable thermolysis (i.e. decomposition) prior to reaching the substrate. To address this issue, the guanidinate framework was re-designed by linking the exocyclic carbon to one of the carbon moieties on the chelating nitrogen group in the metallocycle and forming a *tert*-butyl substituent on the other chelating nitrogen, thereby eliminating any hydrogens beta to the metal centre.[31] This not only prevents decomposition in the gas phase, but it also eliminates the liquid-phase decomposition pathway of the guanidinate compounds, which occurs through CDI deinsertion.[31] Overall, this slight modification of the guanidinate precursor imparts much more volatility and thermal stability to the precursor. More information on this decomposition pathway and a contrasting study of the gold(I) guanidinate and iminopyrrolidinate CVD precursors is found in Chapter 3. Briefly, the synthesis of the silver(I) and gold(I) *tert*-butyl-imino-2,2-dimethylpyrrolidinates (Figure 1.5) was made through a simple salt metathesis of the lithiated *tert*-butyl-imino-2,2-dimethylpyrrolidinate and a metal chloride salt, affording reasonable (> 60%) yields.[31] The *tert*-butyl-imino-2,2-dimethylpyrrolidinate ([Au(N$^t$Bu)(Me$_2$ip)]$_2$) ligand itself involves a much less trivial synthetic route and is described thoroughly elsewhere.[32]

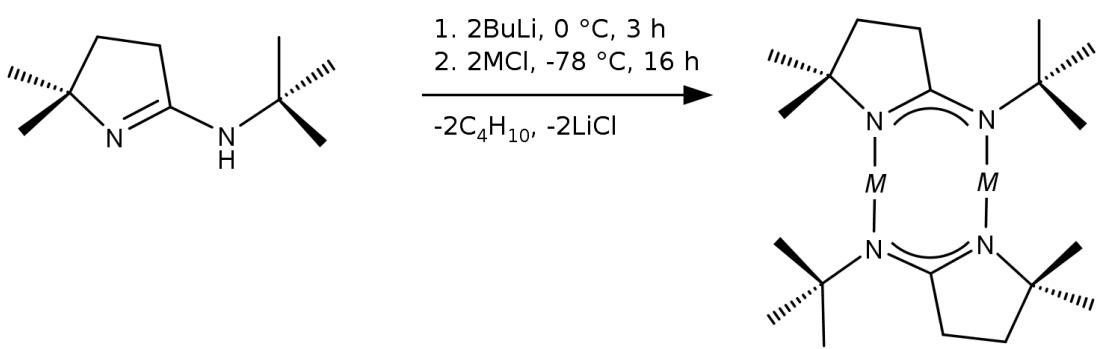

**Figure 1.5:** Synthesis of metal(I) *tert*-butyl-imino-2,2-dimethylpyrrolidinate (M=Cu, Ag, or Au)

To summarize, the use of [Au(N$^i$Pr)$_2$CNMe$_2$]$_2$ and [Au(N$^t$Bu)(Me$_2$ip)]$_2$ as single-source precursors for gold CVD are reported throughout this document (Chapters 2-5, 7, and 8), as well as an interesting silver CVD example with a trimeric silver(I) guanidinate (Ag$_3$[($^i$PrN)$_2$CN(Me)$_2$]$_3$) in chapter 2. Additionally, the first report of an ALD process for gold using (Me)$_3$AuP(Me)$_3$ with O$_2$ plasma and H$_2$O co-reactants is described in Chapter 9. Attempts to make 1,3 diisopropyl-imidazolin-2-ylidene and triphenylphosphine gold(I) pivalate compounds were successful by [1]H-NMR and [13]C-NMR, but resulted in very unstable precursors not viable for CVD (data not included in this thesis). A thermally stable 1,3  diisopropyl-imidazolin-2-ylidene gold(I) chloride was also successfully synthesized and recrystallized in toluene to produce a single rectangular crystal, but also did not lead to any film deposition during attempted CVD experiments (published in supporting information section ref. [12][33] of the List of Publications section and is not included as a chapter of this thesis).

### 1.3    Tilted Fiber Bragg Grating Sensors: A New *In-situ* Monitoring Tool for CVD and ALD

Over the past few decades, numerous advancements have been made with respect to *in-situ* characterization of CVD and ALD processes. Techniques such as quartz crystal microbalance (QCM),[34] mass spectrometry (MS),[15,35] spectroscopic ellipsometry (SE),[36,37] infrared spectroscopy (IR),[38] and, more recently, synchrotron radiation studies[39] have all been successful in proving certain surface-mechanistic aspects of CVD and ALD processes. Now adding to this surface science toolkit is the use of optical fiber-based tilted fiber Bragg grating (TFBG) technology. First introduced by Meltz *et al* in the late '80s,[40] FBGs have demonstrated remarkable utility as temperature, strain, pressure, and bend sensors, but more recently there has been an emerging interest in TFBGs for chemical sensing.[4] These optical fiber devices have numerous advantages including portability, inexpensiveness, and robustness. Moreover, there are now more ways than ever to fabricate the TFBG structures (see Chapter 2 for fabrication methods) such as the UV excimer[41] and femtosecond laser direct-write[42] methods for ultra-low loss fibers. Unlike the previous fabrication methods, which often involved an etching step with HF in order to expose the fiber core,[43,44] these direct-write techniques (Figure 1.6) preserve the structural integrity of the fiber while still yielding a sensor platform with extreme sensitivity to the surrounding refractive index (SRI). Since the TFBG sensors used in this work had to be exposed to CVD and ALD process conditions (e.g. T > 120 ºC), it was crucial to use fibers with sufficient mechanical strength.

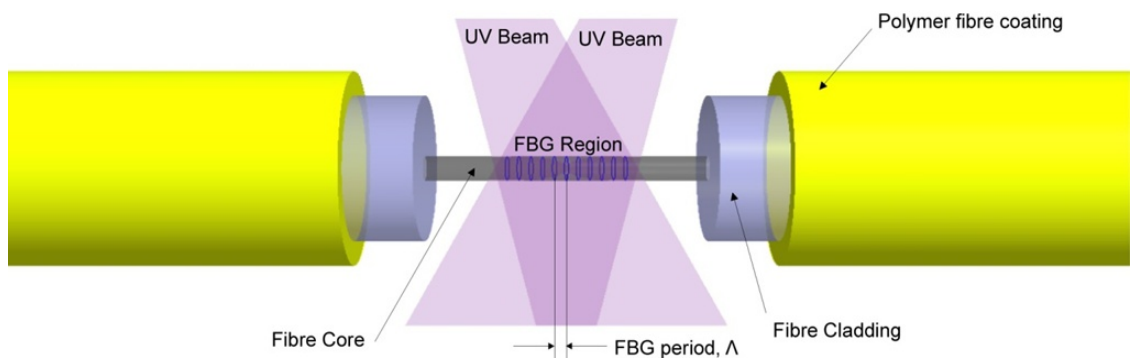

**Figure 1.6:** UV direct-write method used for fabrication of long or short-period (Λ) FBGs. Prior to fabrication, the polymer jacket must be stripped away. (note: the gratings are written within the core, as the cladding is SiO$_2$ and therefore UV-transparent).

The standard single-mode fiber (SMF-28) core is doped with GeO$_2$, which is UV-absorbent, a densification of the surrounding SiO$_2$ results in the TFBG pattern with an associated grating period spacing (Λ) of ~550-556 nm, which is the grating-to-distance in the FBG region (Figure 1.6).The resulting modulated index yields a family of a cladding mode resonances with varied sensitivities to the SRI. The wide spectral comb of cladding modes in the 1520-1610 nm telecommunications window (NIR range) shown in Figure 1.7 have evanescent fields that are excited by the TFBG structure and when this field is perturbed by a change in the SRI, the cladding modes can remain "leaky" within the core or guided. In the latter case, the SRI must be greater than the effective index of the cladding mode, effectively "pulling" more light from the core than in the guided case. When the effective index of the cladding mode is phase-matched with the SRI, this is known as the "cut-off" mode and due to the high density of cladding modes for a 10º-tilted FBG in the NIR region, this mode is usually observed. Although it is mentioned thoroughly in the following chapters, the phase-matching condition described below in equation [2] is the most important relationship for describing interaction between the core and the cladding mode resonances:[41]

$$\lambda_{co} = 2\, n^{co}_{eff}(\lambda_{co})\Lambda \qquad\qquad [1]$$

$$\lambda_{cl} = [2\, n^{co}_{eff}(\lambda_{co}) + n^{cl}_{eff}(\lambda_{cl})]\Lambda \qquad\qquad [2]$$

where $\lambda_{co}$ and $\lambda_{cl}$ are the wavelengths of the core and cladding modes, respectively. Λ is the grating period spacing. As mentioned, the effective index for any cladding mode $\left(n^{cl}_{eff}\right)$ can be easily calculated from the TFBG transmission spectrum and, through substitution of equation [1] into [2], we find:[45]

$$n_{eff}^{cl}(\lambda_{cl}) = \frac{2\lambda_{cl}}{\lambda_{co}} n_{eff}^{co}(\lambda_{co}) - n_{eff}^{co}(\lambda_{cl})$$ [3]

The core effective index $\left(n_{eff}^{co}\right)$ is less trivial and relies solely on the mechanical and physical properties of the optical fiber core (e.g. $GeO_2$ dopant concentration, etc), but the TFBG refractometers used throughout this work have been calibrated for absolute refractometry measurements with a high degree of accuracy (RI measurement error $<10^{-3}$ RIU) and comparable to the analogous, prism-based Abbe refractometers.[46] Finally, Figure 1.7 (right) shows, with added clarity, that the light coupling between the core, cladding, and SRI in the TFBG configuration is analogous to the waveguide optics mechanism for bulk refractometry.[45] The previous discussion regarding the careful calibration of the TFBG refractometer has omitted one detail until this point, which is that when the circular symmetry of the core-guided light is broken up by the TFBG, a boundary condition is imposed whereby the core mode is decomposed into two separate polarization states. These two modes, referred to as transverse-electric (TE or "S-") and transverse-magnetic (TM or "P-"), represent in-plane and out-of-plane electric fields, respectively, and are described in much more detail throughout the chapters that follow. However, it needs mentioning that the introduction of an additional interface between the TFBG cladding modes and a thin gold coating deposited by CVD or ALD highlights the necessity of the orthogonally polarized TE and TM cladding modes.

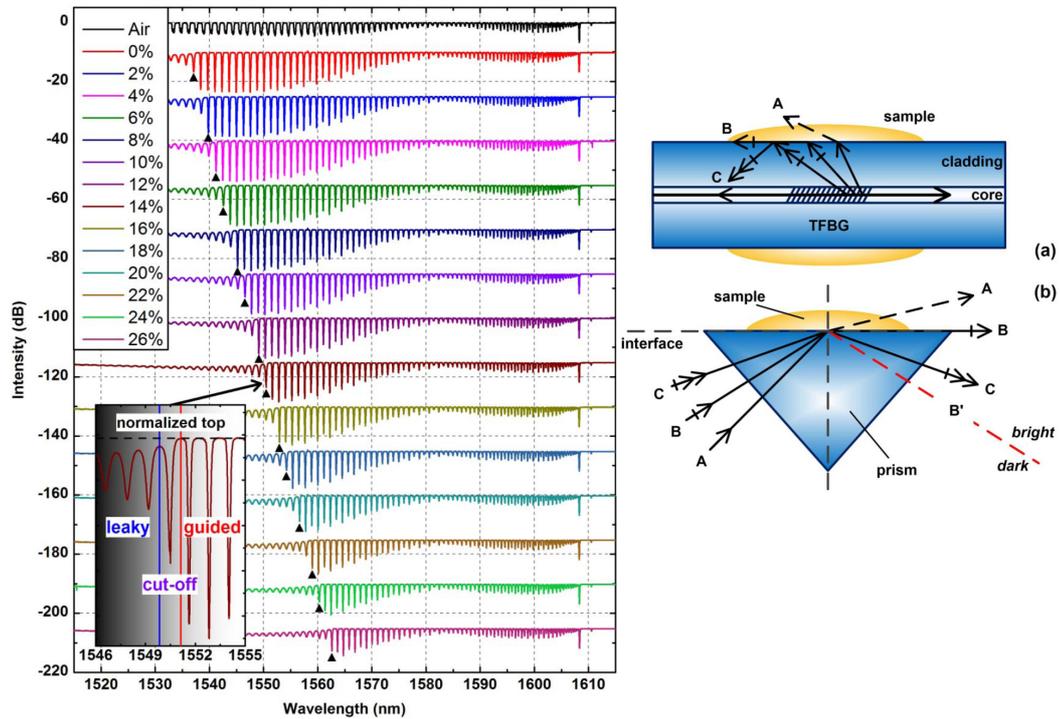

**Figure 1.7:** (left) TFBG spectra (non-polarized) under various surrounding media (air, water, and NaCl solutions of varying mass concentrations). In each case the corresponding cut-off mode resonance is indicated with a black arrow. Inset: detailed snapshot of a cut-off mode, guided mode, and leaky mode for 14 % NaCl solution. (right) Schematic diagram of the leaky, cut-off, and guided modes traced by rays A, B, and C, respectively, for a TFBG-based refractometer (a) and standard Abbe refractometer (b).

The spectral response of the cladding modes, as will be shown throughout this work, varies drastically depending on the polarization state of the light interrogating the growth of ultrathin gold coatings by CVD or ALD. Through an effective medium approximation (EMA) treatment of the separate perturbations imposed by the gold coatings under the TE- or TM-polarized light, the observed spectral evolution in the form of wavelength shifts and amplitude attenuations of cladding mode resonances near the cut-off wavelength can be accurately modeled. This ultimately leads to the development of the TFBG-based "ellipsometer" that can be incorporated into any CVD or ALD system and yield the same thin film metrics (e.g. growth rate or thickness) in real-time.

In addition to the development of an *in-situ* characterization technique, the particular focus on gold thin films (by CVD and ALD) in this work is towards the development of a TFBG surface plasmon resonance (SPR-TFBG) sensor that performs better than the current SPR-TFBG sensor devices using PVD-grown gold films.[47,48] Current SPR-TFBG sensors rely upon the propagation of surface plasmon polaritons (SPPs) in the TM-polarized cladding modes at the metal-SRI interface in the NIR range; a phenomenon that has been compared to the classic prism-based Kretschmann configuration for SPR on planar surfaces.[47] This is a reasonable assumption in this work as well since the optical sensing window of 1510-1620 nm is much larger than the average penetration depth of the TM-polarized evanescent field across the cladding-gold-SRI interface, which is typically between $\lambda/2$ and $\lambda/5$.[47] The SPR phenomenon in this work occurs if and only if the real part of the permittivity (dielectric constant) of the gold coating has a negative value in the telecommunications wavelength (NIR) range. As discussed in the following chapters, SPP propagation and the SPR property was not retained for any of the CVD-grown films directly on TFBGs or on a dielectric spacer layer such as $Al_2O_3$, however, it was readily observed for sputtered gold coatings and ALD-grown gold (Chapter 10). The dispersion relation for an SP resonance at the gold-cladding interface can be expressed generally as:[49]

$$k_{SP} = \frac{\omega}{c}\sqrt{\frac{\varepsilon_m \varepsilon_d}{\varepsilon_m + \varepsilon_d}} \qquad [4]$$

where $\varepsilon_m$, and $\varepsilon_d$, are the dielectrics constants of gold metal and cladding, respectively. $\omega$, and $k_{SP}$ represent the SPR frequency and wave vector, respectively, and $c$ is the speed of light constant Alternatively, when measurements yield the SPR wavelength such as in this work, $2\pi/\lambda = \omega$ is easily invoked. In fact, recalling the phase-matching condition in equation [2], we now introduce a phase-matching condition for a three-layer slab waveguide (prism-based, Kretschmann SPR is assumed here):[49]

$$\sqrt{\varepsilon_p}\sin(\theta) = \Re\left\{\sqrt{\frac{\varepsilon_m \varepsilon_d}{\varepsilon_m + \varepsilon_d}}\right\}$$  [5]

where $\varepsilon_p$, $\varepsilon_m$, $\varepsilon_d$ are the dielectric constants for the bottom of the Kretschmann prism (of TFBG SiO$_2$ cladding layer), thin metal coating (ultrathin gold film), the surrounding medium (SRI), respectively. Interestingly, because the SP wave (evanescent field) propagation through the fiber cladding would be geometrically similar if not identical to the prism-based configuration, the following equation:

$$\sqrt{\varepsilon_p} = n_{eff}^{cl}(\lambda_{cl})$$  [6]

along with equation [2], can be substituted into equation [5] to generate:

$$SRI = \sqrt{\frac{\varepsilon_p \varepsilon_m \sin\left(\arcsin\left(\frac{\frac{\lambda_{cl}}{\Lambda/\cos(\theta)} - n_{eff}^{co}}{n_{eff}^{cl}(\lambda_{cl})}\right)\right)}{\varepsilon_p - \varepsilon_m \sin\left(\arcsin\left(\frac{\frac{\lambda_{cl}}{\Lambda/\cos(\theta)} - n_{eff}^{co}}{n_{eff}^{cl}(\lambda_{cl})}\right)\right)}}$$  [7]

which is an expression that directly relates the TM-polarized cladding mode wavelength that supports SPPs (lossy SPR cladding mode) to the refractive index of the surrounding medium or environment (SRI). $n_{eff}^{co}$ , $n_{eff}^{cl}$ , $\Lambda$, $\lambda_{cl}$, $\varepsilon_p$, and $\varepsilon_m$ are described previously (equations [1]-[6]). In the case where the gold coating does not support SPPs such as the gold CVD work presented in this thesis, the dielectric and permittivity constants of the gold coatings must be back-calculated from the wavelength shift and attenuation profiles of the cladding modes. This process incorporated the use of multiple EMA models in order to properly characterize the anomalous permittivity of the CVD-grown gold coatings (a fully-detailed procedure of our approach is presented in Chapter 5).

Although the TFBG-polarized transmission spectra obtained from *in-situ* CVD and ALD monitoring contain a great deal of information regarding the optical properties of the resultant gold coatings, the use of polarization-dependent loss (PDL) spectra can further characterize the often dichroic (polarization-dependent) nature of the gold films. It is used primarily in this work (Chapter 10) to unambiguously characterize the SPR signature of highly-continuous and uniform films generated by PVD and ALD methods. PDL is essentially the relative variation of all cladding modes across both polarization states,[50] yielding an intensity ratio:

$$PDL_{dB} = 10 \left| \log\left( \frac{T_{TM}(\lambda_{cl})}{T_{TE}(\lambda_{cl})} \right) \right| \qquad [8]$$

where $T_{TM}(\lambda_{cl})$ and $T_{TE}(\lambda_{cl})$ are the transmission amplitudes (intensities in dB) of the TM-polarized and TE-polarized resonances, respectively. The PDL technique is applied herein to quickly and efficiently quantify the refractive index sensitivity of our gold-coated SPR-TFBG sensors by using single-resonance tracking of the SPR signature in a variety of SRIs. Naturally, the development of these improved sensors can lead to broader applications in (bio)chemical sensing and applications where extreme refractometric sensitivity is required.

## 1.4      Structure of Thesis

This 11-chapter thesis comprises a comprehensive study of noble metal thin film growth on an optical fiber sensor, with a particular emphasis on gold due to its desirable optical properties. Orthogonally polarized cladding modes (TM and TE) generated in the optical fiber core by the TFBG structure are used to selectively interrogate the growth of noble metal (Au, Ag, Cu) films during the very early stages of nucleation (sub-nanometer) all the way to film closure (100-200 nm thickness) in real-time during CVD and ALD processes. Depending on the polarization state of the input light, the light that is out-coupled from the core through the optical fiber cladding is either absorbed (TE; in-plane with respect to grating axis) or scattered (TM; out-of-plane with respect to grating axis) through the metal coating, providing very clear, polarization-resolved transmission data that is used to calculate parameters such as thickness or growth rate. In this optical fiber-based ellipsometric approach, attenuation of the TE- and TM- polarized cladding-guided light modes at the metal-cladding interface reveals how optically uniform and continuous the resultant film will be. Effective optical properties of the gold films generated from CVD and ALD processes are compared to the bulk optical properties of gold films obtained from conventional sputtering or PVD methods throughout this thesis. While being a highly effective (and novel) *in-situ* diagnostic tool for any CVD or ALD process, gold coatings, for example, that are optically conductive (i.e. support SPR) can enhance the overall SRI sensitivity of the TFBG by orders of magnitude compared to the bare TFBG.

Chapter 1 introduces the precursor molecules used in the CVD experiments and introduces the fundamental aspects of the TFBG sensor in the context of optical CVD and ALD monitoring, as well as in bulk refractometry measurements to assess the sensor performance (SRI sensitivity, polarization-dependent loss, etc) post-deposition.

Chapter 2 is an invited contribution (review) that reviews the use of TFBGs in CVD monitoring against a variety of other closely related optical fiber-based monitoring techniques for deposition processes of all sorts. It also presents a thorough work-up of the TFBG optics as well as a perspective on the use of TFBGs in CVD and eventually in ALD processes. The use of the TFBG as way to assess growth modes (eg. Vollmer-Weber vs. Stranski-Krastanov growth) using a particular silver(I) guanidinate as an example is also presented.

Chapter 3 is a contrasting study that compares (2 published works, adapted) the thermal stability of a gold(I) guanidinate and a gold(I) iminopyrrolidinate compound. By eliminating a common decomposition pathway ($\beta$-hydrogen elimination) through rational ligand design, the latter compound shows enhanced thermal stability in a variety of thermolysis studies. Also presented is the corroborating evidence of a slower growth rate of 1.1 nm/min observed by optical monitoring during CVD (compare to 37 nm/min for the gold(I) guanidinate).

Chapter 4 describes the first published work of TFBG-based monitoring of gold nanoparticle layers as they decorate the TFBG cladding, as well as the evolution of a polarization-dependence property as the gold nanoparticle islands coalesce into more continuous structures. Unlike all other cases for the gold CVD and ALD work, the attenuation profiles for both the radially (TM) or azimuthally (TE)-excited cladding modes originates from scattered (not absorbed) light at the metal-cladding boundary.

Chapter 5 assesses the size-dependent effective optical properties of ultrathin (6-10 nm) and thicker (15-200 nm) gold films from optical CVD monitoring using the gold(I) iminopyrrolidinate precursor. The positive real effective permittivity of the resulting gold nanoparticle coatings indicated that the films consist of discontinuous aggregates of nanospheroidal islands that do not efficiently couple the evanescent field that extends from the cladding into the metal inclusion. The slightly ellipsoidal shape of the nanoparticles (determined by calculation of aspect ratio based on lateral size and thickness determinations from AFM) at thicknesses higher than 25 nm allow the coatings to discriminate between the TE and TM-polarization states, but not strongly enough to support SPP propagation in the NIR range. Interestingly, the lateral size of the nanoparticles remains relatively constant for film thicknesses ranging from 25-200 nm.

Chapter 6 introduces the use of the TFBG probe as an SRI sensor for bulk refactometric measurements in various media (SRIs). The TFBG sensor has a 5-5.5 nm layer of gold grown by the non-chemical PVD technique in order to model the effective permittivity of gold while it is effectively still an "insulator" (i.e. close to the insulator-to-metal transition for gold). This provides a basis for the following chapters that use an $Al_2O_3$ interface to cause better wetting and adhesion of the CVD-grown gold coatings on top. This 5 nm gold coating (by PVD) has strikingly similar effective optical properties to gold films with thicknesses as high as 200 nm grown by CVD (chapter 4 and 5). Moreover, this chapter serves to characterize the hitherto CVD-grown gold films as effectively "dielectric". Also presented is the glaring inconsistency between the various EMAs when attempting to model the effective permittivity of insulating gold on an optical fiber.

Chapters 7 and 8 explore the use of a dielectric interfacial layer (50 and 100 nm ALD-grown $Al_2O_3$) in order to improve the TFBG sensor performance during gold CVD monitoring and also the uniformity of the resulting gold coating. While somewhat negligible, the addition of a uniform $Al_2O_3$ "pre-coating" increases the overall SRI sensitivity of the TFBG sensor after the gold layer is deposited (from gold (I) iminopyrrolidinate by CVD). Microstructural (SEM, AFM) analysis and a variety of X-ray analysis (synchrotron XAS methods and XPS) confirms that the gold coatings grown directly on the optical fiber (i.e. no $Al_2O_3$ pre-coating) and witness slides has higher grain size and RMS roughness than those grown on the $Al_2O_3$ pre-coating.

Chapter 9 introduces a major breakthrough with respect to achieving ultimately uniform gold thin films with the discovery of the first successful ALD process for gold metal. This chapter presents the recently published plasma-enhanced ALD process (ABC-type) that uses a methylated gold(III) phosphine precursor with oxygen plasma (O*) and water as co-reagents. A detailed synthesis along with XPS characterization of the films after 1000 cycles (~50 nm thickness) with and without water (AB vs. ABC-type) as a co-reagent is presented. A detailed nucleation study using TEM grids reveals that this ALD process can also yield very monodisperse gold nanoparticles after only a few ALD cycles.

Chapters 10 is the final research chapter of this thesis and a continuation of the work with the gold ALD process. This chapter comprises the preliminary work using the TFBG, for the first time, to monitor thermal ALD of $Al_2O_3$ (as a calibration experiment) and also the plasma-enhanced ALD process for gold metal described in chapter 9. Cycle-for-cycle resolution is obtained for both thermal ($Al_2O_3$) and plasma-enhanced ALD (Au metal) processes, highlighting the versatility of the TFBG sensor platform. Amplitude attenuation of the TE-polarized cladding modes occurs after only a few cycles (< 10 cycles) of the gold ALD process, indicating that this sensor can be used to study ALD processes for various metals that have a perceived nucleation "delay" by other characterization techniques such as QCM. In addition to confirming the growth rate determined in chapter 9 for the gold ALD process (from *ex-situ* thickness measurements), the gold-coated TFBG after 900 cycles (~45 nm thickness by AFM) shows refractometric sensitivity at the theoretical limit and nearly 50 nm/RIU higher than a gold-coated TFBG sensor from PVD at a similar thickness. That is, the TM-polarized cladding modes easily excite SPR in this case.

Chapter 11 is a summary of the thesis and future outlook.

# Chapter 2

## CVD on Optical Fibers: Tilted Fiber Bragg Gratings as Real-time Sensing Platforms

[1] Department of Chemistry, Carleton University, 1125 Colonel By Drive, Ottawa, Ontario, Canada, K1S 5B6

[2] Department of Electronics, Carleton University, 1125 Colonel By Drive, Ottawa, Ontario, Canada, K1S 5B6

*Corresponding author

Invited review article.

## 2.1 Abstract

The tilted fiber Bragg grating (TFBG), as a versatile and robust tool for many sensing applications with a particular focus on vapor deposition processes, is reviewed. Recent work employing the TFBG as an optical probe for monitoring the metal-organic chemical vapor deposition (MOCVD) of noble-metal, single-source precursors is discussed extensively. This work also establishes the broad scope that utilizes TFBGs and other optical fiber configurations to interrogate thin film growth and the associated optical properties. While it cannot possibly cover the full scope of applications with respect to the TFBG device, this review highlights the recent advances of TFBG-based sensing for CVD, and progress towards the applicability of TFBGs as evanescent field-based sensors.

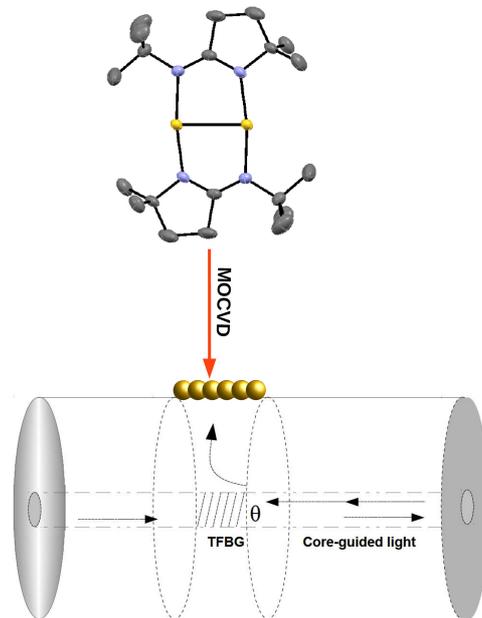

**Figure 2.1:** Table of contents image used for publication described in Chapter 2.

## 2.2    Introduction

### Tilted Fiber Bragg Gratings

Tilted fiber Bragg gratings (TFBGs) are a type of specialized optical fiber device whose fabrication relies upon a photochemical process,[40] which involves writing the periodic pattern (grating) into the core of a fiber using coherent UV radiation (Figure 2.2).[41] The resulting periodic array couples light modes from the core to the cladding, where they become highly sensitive to refractive index changes that occur around the fiber-air interface. Inducing a "tilt" in the otherwise vertical gratings enhances the controllable excitation of cladding modes from the core to the cladding, and facilitates the measurement of the surrounding refractive index (SRI) using an optical spectrum analyzer (OSA). The coupling to the cladding modes is a process that leads to the formation of a discrete "comb" of narrowband features in the spectral transmission of the devices, as measured over a broadband spectrum (BBS), where wavelength ($\lambda$) is in the range 1520–1620 nm. These mode resonances in the core transmission result from light out-coupled into the cladding at very specific wavelengths as it travels through the interference pattern in the core (the vertical or tilted grating planes). It is interesting to note that for many applications of fiber optics these lossy resonance (cladding) modes are undesirable due to the loss of spectral efficiency and purity.[51] The highly lossy and resonant cladding modes of the TFBG, however, effectively make the BBS a spectral window for many sensing applications and, in the context of this review, direct, *in-situ* monitoring of CVD.[52–54]

**Effect of Tilting FBGs Relative to the Fiber Axis**

The utility of the TFBGs, particularly towards sensing, hinges upon the cladding modes that result from the "tilt" of the grating. In un-tilted gratings, transmitted core-guided light will reflect directly back through the core, so the light does not effectively scatter or propagate to any measurable degree outwards. But as the grating is tilted with respect to the fiber axis, the light is then excited outwards from the core as it couples through the cladding. Figure 2.3 depicts this angle dependence of the cladding modes for a TFBG; as the grating is tilted from 2º to 10º, more and more cladding resonances appear at increasingly smaller wavelengths. The transmission dip occurring at the longest wavelength on these graphs is due to core-mode back-reflection coupling, and its intensity changes with tilt angle. This resonance, hereafter named the "Bragg" mode resonance, is insensitive to the SRI and can be used as a reference "channel". Beyond approximately 1530nm in Figure 2.3d, there is a discontinuity in the amplitude of the cladding-mode resonances, which

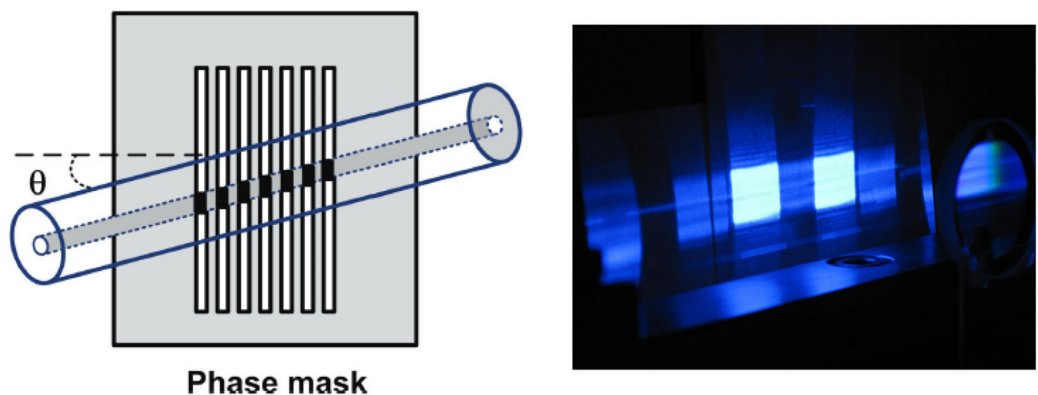

**Phase mask**

**Figure 2.2:** (Left) Fabrication of the TFBG with UV light (in the plane of the page) through the phase mask as it is tilted ($\theta$=10º) off-axis with respect to the single-mode fiber. (Right) Photograph of the diffraction pattern generated on the phase mask during fabrication. Adapted with permission.[41] Copyright 2012, Wiley-VCH.

indicates that, for shorter wavelengths, light is no longer guided by the cladding but is instead radiating in the external medium. In between these two extremes are the cladding modes that are used for sensing studies.[41,55]

## TFBGs: Temperature, Strain, and Polarization-dependence

As well as the tilt, there are many other intrinsic properties of the TFBG that need to be addressed if they are to be discussed in the context of a sensing platform. Three such properties that are important in terms of TFBG performance are temperature, axial strain, and polarization. Temperature changes imposed on the optical fiber during high-temperature (>200°C) processes, such as CVD, affect the local geometry of the grating of the TFBG, and therefore also induce a reversible wavelength shift of the cladding modes. If the device is fixed on a support, strain on the fiber section that includes the TFBG has a similar effect. Discrimination between the effect of the strain or elevated temperature and the quantity to be measured in the overall transmission spectra is difficult, and has been the subject of numerous studies.[56–58] The overall wavelength shift in the cladding mode resonances ($\Delta\lambda_r$) as well as the Bragg mode ($\Delta\lambda_B$) of the TFBG as a function of both the axial strain ($\varepsilon$) and temperature ($T$) is modeled in Equations 1 and 2.[41]

$$\Delta\lambda_B = \left( \frac{2N_{eff}^B}{cos(\theta)} \frac{d\Lambda}{d\varepsilon} + \frac{2\Lambda}{cos(\theta)} \frac{dN_{eff}^B}{d\varepsilon} \right) \Delta\varepsilon + \left( \frac{2N_{eff}^B}{cos(\theta)} \frac{d\Lambda}{dT} + \frac{2\Lambda}{cos(\theta)} \frac{dN_{eff}^B}{dT} \right) \Delta T \qquad [1]$$

$$\Delta\lambda_r = \left(\frac{\left(N_{eff}^B + N_{eff}^r\right)}{\cos(\theta)}\frac{d\,\Lambda}{d\,\varepsilon} + \frac{\Lambda}{\cos(\theta)}\frac{d\left(N_{eff}^B + N_{eff}^r\right)}{d\,\varepsilon}\right)\Delta\varepsilon$$

$$+ \left(\frac{\left(N_{eff}^B + N_{eff}^r\right)}{\cos(\theta)}\frac{d\,\Lambda}{dT} + \frac{\Lambda}{\cos(\theta)}\frac{d\left(N_{eff}^B + N_{eff}^r\right)}{dT}\right)\Delta T$$

$$[2]$$

The factors most affected by the axial strain and high temperatures of processes such as CVD are the grating period (Λ) of the Bragg grating, which describes the spacing between the grating fringes (557 nm with Bragg wavelength of 1612 nm), and the refractive index of the core and cladding glass of the fiber. The wavelength sensitivities implied by Equations 1 and 2 are also a function of the effective refractive indices of the cladding mode resonances, $N_{eff}^r$ , and the Bragg mode resonance, $N_{eff}^B$ :

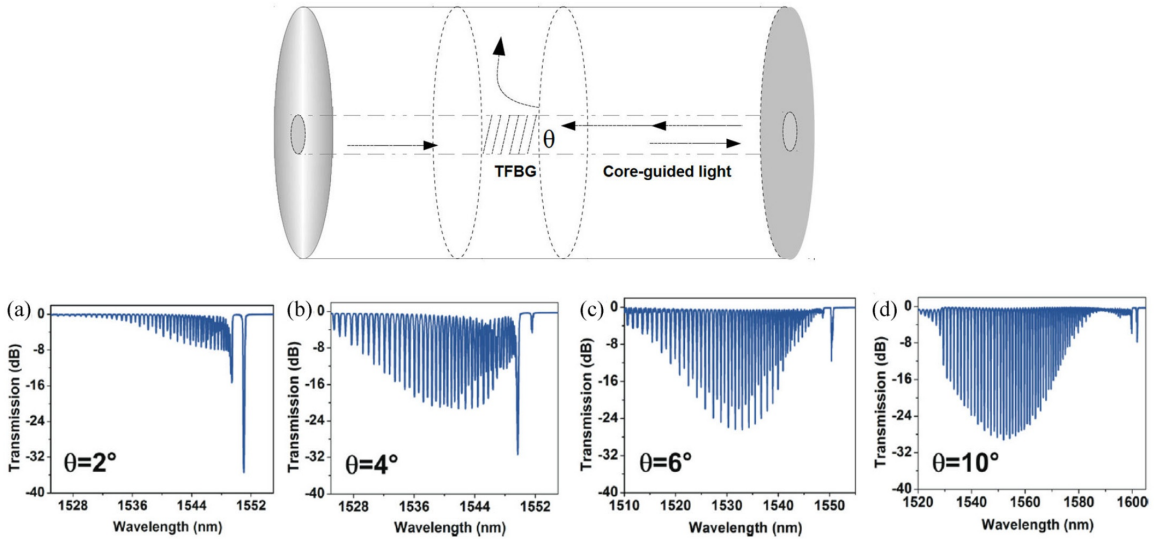

**Figure 2.3**: Transmission spectra of the TFBG as a function of tilt angle, θ a) 2°, b) 4°, c) 6°, and d) 10°). The sharpest cladding modes appear under both polarization states at 10°. Adapted with permission.[41] Copyright 2012, Wiley-VCH.

in addition to the thermal expansion, temperature effects change the values $N_{eff}^{r}$ and $N_{eff}^{B}$. The general wavelength functions for the Bragg and cladding-mode resonances ($\lambda_B$ and $\lambda_r$, respectively) are defined using Equations 3 and 4.[41,54]

$$\lambda_B = \frac{2N_{eff}^{B}\Lambda}{\cos(\theta)} \qquad [3]$$

$$\lambda_r = \frac{(N_{eff}^{B} + N_{eff}^{r})\Lambda}{\cos(\theta)} \qquad [4]$$

It is interesting to note that $\lambda_r$ contains the refractive index-dependence of both the cladding modes and the general Bragg mode, whereas Equation 3 depends only on the Bragg mode. Since Equations 3 and 4 are the basis for many of our calculations and simulations of a nucleating thin film in CVD, the effective refractive index ($N_{eff}$) is now explained in more detail. The effective refractive index of a mode (core or cladding-guided) represents a measure of the optical propagation properties of light for this mode; the real part of $N_{eff}$ reflects the speed of light along the fiber longitudinal axis, while its imaginary part corresponds to the loss per unit length. The effective indices of the guided electromagnetic modes depend strongly on the TFBG (waveguide) geometry, material refractive index, and wavelength. This will be of critical importance when later discussing the wavelength-shift-dependence on the polarization mode of the light that is interrogating a growing CVD film. Ultimately, it is crucial that the effects of temperature and axial strain are minimal and properly considered when interpreting the *in-situ* optical data from film growth. By tracking a number of cladding-mode resonances of varying line-widths and sensitivities during heating of a bare TFBG, Albert *et al.*[41] reported an average wavelength shift of 10.61

pm/°C induced in a temperature range 0–120°C. In temperature ranges higher than 120°C (typical of CVD), these effects are not considerably exacerbated and result in similar wavelength shifts of 10–11 pm/°C.[41,50,51,59,60] Furthermore, since all resonances shift by similar amounts with temperature when the TFBG is not fixed to a substrate (which is the case for the results reported herein), this means measurements of the SRI at the air-fiber interface of the TFBG can be made accurately, without temperature cross-sensitivity, by measuring all wavelength and power level shifts relative to the Bragg mode resonance, which is by definition insensitive to the SRI.[41]

As mentioned, another important property affecting the SRI sensitivity of the excited TFBG is the polarization state of the core-guided light. When the input light is linearly polarized in the tilt plane (TM or "P") or perpendicular (TE or "S"), there are two peak families associated with the transmission amplitude spectrum corresponding to cladding modes with azimuthally (S) or radially (P) polarized light at the cladding-SRI boundary.[61] The SRI sensitivity of the TFBG can be either isotropically or anisotropically dependent on the polarization state of the light interrogating the TFBG surface where both peak families (S and P) respond the same way (i.e. isotropically) or only one peak family shows a sensitivity (S or P in an anisotropic manner). This is a well-defined property described extensively in the literature.[41,50,52–54,61–65] Allsop *et al.* reported that, in addition to a polarization dependence, the SRI sensitivity and penetration depth of an evanescent field induced by a 35 nm silver thin film on a TFBG is orders of magnitude higher in the NIR range (1500 nm) than the conventional reports with FBG sensors up to that point.[66] Also, because the tilt angle was tuned from 1°to 9°in that study, the deeper (i.e., higher amplitude) resonances in both peak families (Fig. 2.3) permitted a more sensitive response to changes in the transmission amplitude of the TFBG, and enhanced the coupling of the evanescent field of certain cladding modes to a surface plasmon resonance (SPR) of the film when exposed to aqueous media. In fact, the excitation of SPR by cladding modes is frequently used to monitor the refractive index near the fiber surface, as long as the light polarization is radial. When this occurs, light is absorbed by the metal and leads to additional transmission loss which, in gratings, results in a diminishing of the resonance amplitude. Furthermore, in the case of nanostructured metal films, the localized SPR (LSPR) modes of individual nanoparticles can be detected, as well as the red shift of the SPR resonances that occurs when the films evolve into a situation where LSPR delocalizes across several particles.[41,52] By observing the spectral evolution of the transmission amplitude of the TFBG during the nucleation of a film, the peak families under either polarization can be used to probe features about the film, such as the connectivity, growth mode (e.g., Vollmer-Weber vs. Stranski-Krastanov), thickness, and surface roughness, all in real-time. Of course, the optical properties (complex refractive index of the film layer) can also be inferred from the TFBG spectra measured *in-situ* as the films grow through nucleation and coalescence (Fig. 2.4), a process sometimes referred to as evanescent wave absorption spectroscopy.[67] This underlying mechanism of TFBGs as sensing platforms for CVD is the main focus of this chapter

.

## 2.3 Current State of Fiber-based Characterization of Deposition or Growth Processes

It is clear from even a brief survey of the literature that there is no dearth of fiber optic-based processes applied to *in-situ* monitoring or *ex-situ* assessment of a variety of deposition or growth processes. This consequently leads to a variety of fiber configurations (Table 2.1) that, depending on the metric, can deliver extremely sensitive and accurate thin film data.[48,52–54,63,65,66,68–77] Table 2.1 outlines some interesting literature-reported examples of optical monitoring for processes ranging from CVD, and adaptations thereof, to crystallization and protein binding. Sputtering (or physical vapor deposition (PVD)) has also been used to deposit metal or metal oxide films on TFBGs. In one case, Shevchenko and Albert[48] show that gold films, deposited on a TFBG by sputtering at thicknesses of 20–30 nm, caused the wavelength of the deepest cladding modes of the Au-TFBG to shift drastically when subjected to higher refractive index media (e.g., sucrose solutions). Bare TFBG resonances also shift when exposed to SRI changes, but less significantly than metal-coated ones that benefit from plasmonic enhancement. The maximum sensitivity for non-plasmonic sensors is obtained by following the cladding mode resonance closest to the cut-off point, beyond which the light extracted from the core by the grating is no longer guided by the cladding and leaks out.[41] Thus, the utility of TFBGs for gas or liquid sensing exists in this case due to ease of FBG fabrication and measurement, and also in the ability to change the interrogation wavelength window from the near-infrared to the visible range (more common for SPR-based sensors),[48] just by changing the period, grating length, or tilt angle (see Eq. 4). More recently, sputtering and magnetron sputtering (MS) have been employed to coat TFBGs and long-period gratings (LPGs) with ZnO[65] and Al$_2$O$_3$,[70] respectively. In the former study, Renoirt *et al*[78] used radio-frequency sputtering to coat a TFBG with ZnO at thicknesses of 200–800 nm, and studied the well-established effect of metal oxide-induced polarization-dependence of the TFBG.[69] An SRI sensitivity of both polarization states occurred in the presence of the ZnO films, but they report a 3.5 times larger SRI sensitivity in the P-polarized mode after monitoring a single resonance at $\lambda_r$=1565 nm.[65] Smietana *et al.*[70] did a comparative study of Al$_2$O$_3$ nano-overlays deposited onto LPGs by MS and atomic layer deposition (ALD). It was found that the optical properties, particularly the refractive index (n), of the ALD Al$_2$O$_3$ overlays were identical to those of the thicker films generated by the MS method, while benefiting from the better uniformity and sub-nanometric control over film thickness offered by ALD. Resonances of the LPGs at thicker overlays from both deposition techniques deepen and show a higher sensitivity to SRI, but it was found that this sensitivity tuning only begins to occur at overlay thicknesses > 100 nm. A similar study by Smietana et al.[71] reports ZnO coatings on LPGs by ALD at a variety of thicknesses to tune the distribution and sensitivity of the fiber modes. At thicknesses of 180 nm, wavelength shifts of the mode resonances of up to 50 nm and enhancements in the refractive index and extinction coefficient (k) sensitivities are reported. It is clear that the use of dielectric coatings on LPGs, FBGs, and a number of other fiber configurations make the resulting SPR-based sensors promising candidates in applications such as gas detection,[67,75] or as sensors in further vapor deposition processes such as CVD and ALD. In one unusual example of fiber-optic based sensing, Boerkamp *et al.*[74] make use of an intrinsic exposed core optical fiber sensor (IECOFS) to explore the crystallization kinetics of CaCO$_3$ crystal formation. Somewhat analogous to the use of TFBG or other fiber arrangements, the attenuation profile of the IECOFS was tracked as the crystallization occurred in real-time. As a second experiment, they assessed the ability of the IECOFS as a scale monitoring system for crystal growth and found that as the surface area of crystal faces in contact with a custom stainless steel wire (to serve as the optical fiber reference) increased, the amount of radiation drawn out of the IECOFS also increased, which generated highly accurate attenuation profiles. Very interesting examples of recent *in-situ* optical monitoring that employ single-mode fibers (SMFs) includes measurement of SnO$_2$ growth from a single-source precursor, (CH$_3$)SnCl$_2$, by NIR-diode laser spectroscopy (NIR-DLS),[73] nucleation and growth of diamond on fused silica SMFs by plasma-enhanced (PE)-CVD,[72] and optical emission spectroscopy (OES) of diamond growth by PE-CVD and direct measurement of a $C_2$ radical.[76] Other notable examples on a TFBG include the classic streptavidin-biotin protein binding system, typically on a gold-functionalized SPR-based fiber.[64,79] Recently, in our group, the TFBG has been applied to monitoring the CVD process of gold films from two different single-source precursors (a gold(I) guanidinate and gold(I) iminopyrrolidinate), as well as silver films from a silver (I) guanidinate compound analogous to the gold (I) guanidinate.[30,53,54,63] By employing the TFBG as a refractometer, deposition processes could be quenched during attenuation of the transmission amplitude spectrum at various times by flowing room-temperature $N_2$ through the reactor or halting the vacuum. The TFBG-extracted growth rates were on the order of 1 nm/min for gold films from the gold (I) iminopyrrolidinate,[63] whilst for the gold and silver films from gold (I) and silver (I) guanidinates much higher growth rates of 222 nm/min (during attenuation) and 39 nm/min (overall) were reported.[53] The origin of this work began in 2011 when Shao *et al.*[52] used a TFBG to monitor the pulsed-CVD (p-CVD) of copper films from a copper (I) guanidinate, and generated the desirable SPR response in the transmission spectrum of the TFBG for films in the thickness regime of 40–50 nm. Reported growth rates of those films was 1.52 nm per cycle (where the pulse cycle duration was 15 s). These results, as well as those for the mentioned gold and silver films, will be discussed further in the next section, but it is important to note that the fabrication of this Cu-TFBG waveguide was a strong motivation for pursuing other SPR-based sensors, particularly through group 11 (Cu, Ag, Au) MOCVD.

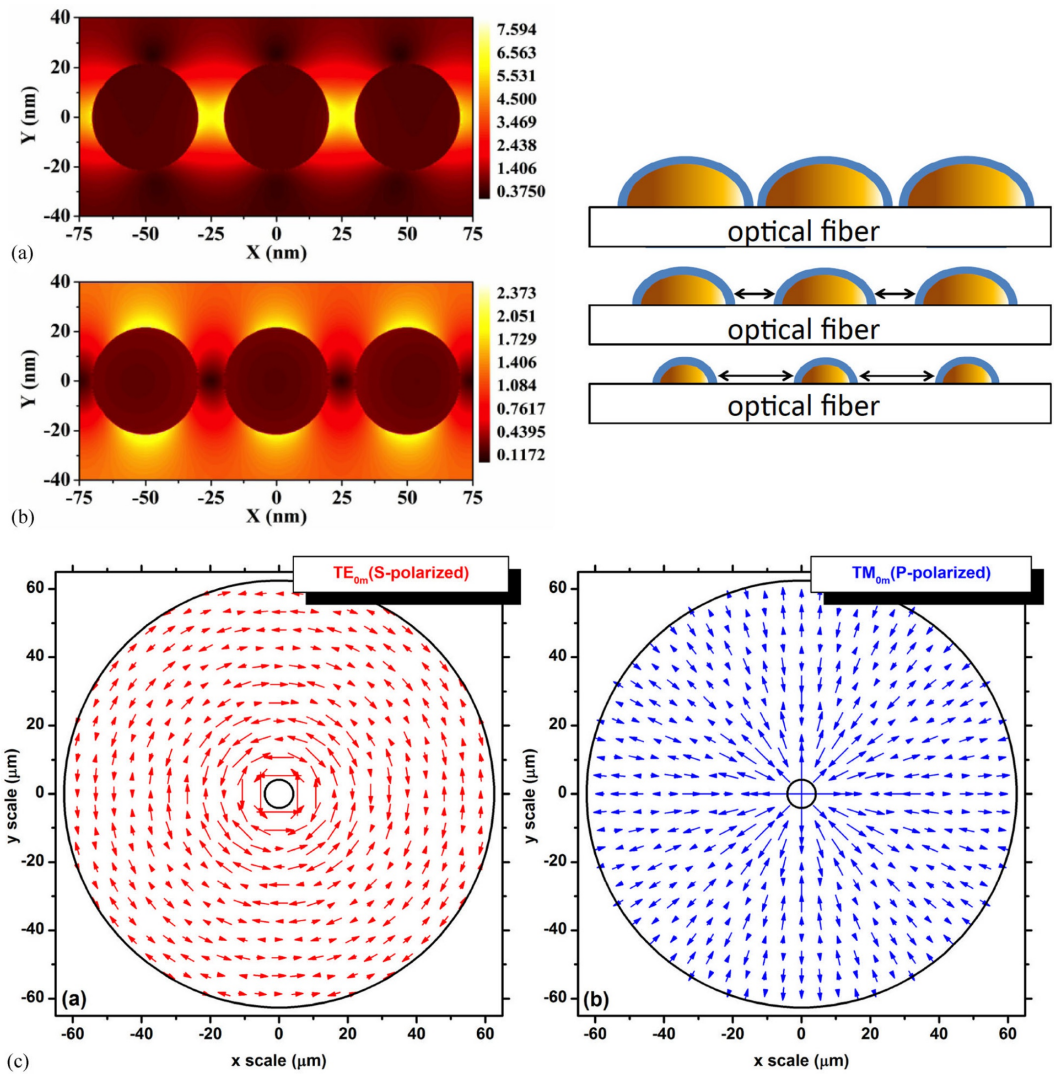

**Figure 2.4:** 2D, finite-difference, time-domain simulation of the electric field distribution in a layer of gold NPs (in air). Depending on the polarization mode, the core-guided light may a) couple the evanescent fields induced by the nanostructures as they get closer, S-polarization (transverse or azimuthal scattering with respect to the electric field vector which is coming in horizontally from the left), or b) they may be isolated and scattering outward, P-polarization (perpendicular or radial scattering with respect to electric field vector). c) Electric field distribution of the two typical polarized cladding modes (note, tilt planes are in the y-direction) for a TFBG. a) Reproduced with permission.[54] Copyright 2013, Optical Society of America.

**Table 2.1:** Selected examples of fiber-optic based characterization of deposition processes

| Fiber/Optical Configuration | Reactant(s) | Substrate[b] | Coating Material | Deposition Process | Interrogation Wavelength (nm) | Process Temperature [c] (°C) | Ref. |
|---|---|---|---|---|---|---|---|
| TFBG | [Cu(NMe₂(N¹P r)₂]₂ | SiO₂(fiber) Si(111) | Cu | p-CVD | 1520-1600 | ~160 | 52 |
| TFBG | [Au(NMe₂(N¹P r)₂]₂ | SiO₂(fiber) Si(111) | Au | CVD | 1515-1615 | 200-225 | 53,54 |
| TFBG | [Au(Me¹Bu-¹p)]₂ | SiO₂ (fiber) Si(111) SiO₂ | Au | CVD | 1520-1610 | 350 | 63 |
| TFBG | O₂-plasma, Zn source | SiO₂ (fiber) | ZnO | Sputtered | 1540-1580 | RT[a] | 65 |
| TFBG | Au source | SiO₂ (fiber) | Au | Sputtered | 1520-1550 | RT[a] | 48 |
| Intrinsic Exposed Core Optical Fiber Sensor (IECOFS) | CaCl₂, Na₂CO₃ | SiO₂(fiber) | CaCO₃ | Crystallization | 632.9 | 25-65 | 74 |
| SPR-TFBG | Biotin-Streptavidin | Au-SiO₂ (fiber) | Biotin-Streptavidin complex | Protein binding | 1525-1625 | RT | 64 |
| TFBG | Cysteamine, AuNP colloid | SiO₂(fiber) | Biotin-Avidin | Protein binding | 1530-1540 | RT | 79 |
| TFBG | AgNO₃, HCl, Na₂S | Ag-SiO₂(fiber) | Ag | Langmuir-Blodgettry, SAM, or LbL | 458-647 | RT | 73,80 |
| NIR-DLS/FTIR Spectroscopy | (CH₃)SnCl₂ | SiO₂ | SnO₂ | CVD | 1667-30000 | 400-500 | 75 |
| Au/SiO₂-coated fiber | H₂, CO, O₂ | Au/SiO₂-SiO₂(fiber) | H₂, CO, O₂ | Gas sensing | 500-650 | 20-950 | 67 |
| Long-period Grating (LPG) | Al(CH₃)₃, H₂O | SiO₂(fiber) | Al₂O₃ | Magnetron sputtering (MS) and ALD | 1200-1600 | 150 (ALD) RT (MS)[a],[d] | 70 |
| Single-mode fiber (SMF) | CH₄, H₂ | SiO₂(fiber) SiO₂ Si | Diamond | PE-CVD | 200-900 | 500 | 72 |
| SMF | CH₄, H₂ | Diamond | C₂ radical, Diamond | PE-CVD | 516.5 | 910 | 76 |
| LPG | Zn(CH₂CH₃)₂, H₂O | SiO₂(fiber) Si | ZnO | ALD[e] | 200-800 | 140 | 71 |

[a] Sputtering process, therefore a voltage is applied to the metal source with the substrate typically held at room-temperature (RT).[b] Substrates not defined with "fiber" or otherwise are planar substrates (slides/wafers). [c] Process temperatures correspond to the substrate temperature at the time of deposition and do not necessarily reflect temperature of the reactant(s). [d] 500-1300 W of power used for magnetron sputtering processes.[e] LPG measurements were performed *ex-situ*.

## 2.4    Optical Monitoring of Noble Metal (Cu, Ag, Au) MOCVD by TFBGs

A number of years ago, our group began rational precursor design towards noble/coinage metal CVD and ALD applications. In 2008, a number of copper (I) guanidinates were synthesized[81] and soon after, in 2011, the syntheses of the gold (I) and silver (I) analogues were reported,[30] (Fig. 2.5, top scheme) along with the first-ever TFBG optical monitoring of a group 11 metal film deposited by CVD;[52] in that case the p-CVD of a copper (I) guanidinate. To rule out an undesirable gas-phase decomposition pathway common with these precursors,[27] the hydrogens in a position "beta" to the metal center on the exo-cyclic isopropyl moiety were eliminated by tethering this position to the bridgehead carbon attached to the same carbodiimide (CDI).[27,31] Moreover, the isopropyl moiety on the other chelating nitrogen was replaced with a *tert*-butyl substituent to remove those *β*-hydrogens. The resulting ligand redesign into the iminopyrrolidinate framework (example for gold in Fig. 2.5, bottom scheme) has been shown to improve the thermal stability of the precursor, making it a better candidate for CVD and ALD.[27,31,82] Using a TFBG *in situ* for monitoring, the optical response could be correlated with film growth. By correlating the *in-situ* refractometric and attenuation data obtained from the transmission spectra of the TFBG with thickness, growth rate, and coverage data obtained *ex-situ* by techniques such as atomic force microscopy (AFM) and scanning electron microscopy (SEM), the TFBG makes an ideal alternative for similar *in-situ* measurements usually obtained by variable-angle ellipsometry[83] or quartz crystal microbalance (QCM) studies.[84] In this section we will highlight the flexibility of the TFBG as a sensing platform for CVD by presenting the results for the optical monitoring of copper, gold, and (in a preliminary manner) silver nucleation and growth from single-source precursors during CVD. Additionally, we will discuss how the refractive index data of the ultrathin films are correlated with attenuation profiles of TFBG-polarized transmission spectra, and how the growing films tend to show a strong discrimination between S- and P-polarized evanescent fields (i.e., polarization dependence). As will be highlighted, a lot of very useful thin film metrics are encoded in each TFBG attenuation profile that is obtained from CVD monitoring experiments.

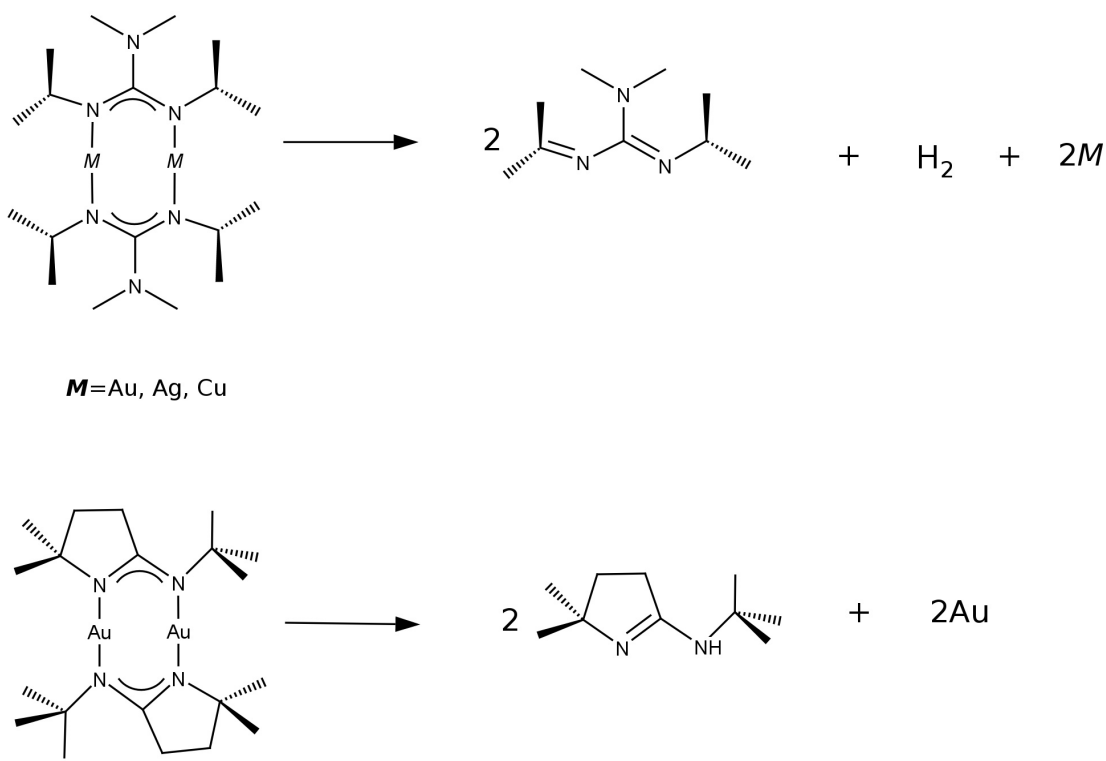

**Figure 2.5:** General gas-phase thermolysis of metal (I) guanidinate (top) and gold (I) *tert*-butyl-imino-2,2-dimethylpyrrolidinate (bottom) single-source CVD precursors. Note that for silver (I) guanidinates (M=Ag) the silver prefers a trimeric coordination geometry with the guanidinato ligands and therefore thermolyzes to 3 stoichiometric equivalents of silver (not 2 as in gold and copper CVD).

## 2.5    In-situ Optical Monitoring Studies of Copper MOCVD

As mentioned, the growth of copper nano-layers by p-CVD on a reflective TFBG sensor has been investigated in the NIR spectral window of 1520–1610 nm.[52] In order to facilitate the insertion of the TFBG in growth chambers, the fiber is cut about 2 cm downstream from the grating and a gold mirror is deposited on the end by sputtering. This way, the transmission spectrum is reflected towards the input where it is extracted (using a simple fiber coupler) and measured. The optical spectrum analyzer (OSA) and polarization controller (PC) were incorporated into the p-CVD tool (Fig. 2.6a), and the copper (I) guanidinate precursor (solid powder at RT) was pulsed into the hot-walled furnace from a separate evaporation zone (bubbler). Careful control is critical for obtaining uniform films, and therefore a specific 15 s pulse cycle was adopted; (i) purge with $N_2$ carrier gas (5.0 s), (ii) evacuate the furnace, fill bubbler with $N_2$ (3.0 s), (iii) entrain the contents of the bubbler into the furnace (2.0 s), (iv) purge furnace with $N_2$ (3.0 s), (v) evacuate the furnace (2.0 s).[52] At furnace temperatures of 160 °C and using 10–40 pulse cycles, smooth and regular plasmonic copper thin films were obtained on witness slides (Fig. 2.6b) and along the surface of the fiber (Fig. 2.6c). Connectivity of the films improved at higher cycles, rendering larger particles and smaller air-gaps. We speculate that as the particles agglomerate at increasing number of cycles (e.g., after 20 cycles), the films go from sparse and discontinuous to more continuous with a lower density of air-gaps (Fig. 2.7d). This observation corroborates the plasmonic properties of the films since in the sparse films (<20 cycles) the LSPR is still stable, but as the nanoparticles (NPs) agglomerate, their surface plasmons are excited at the TFBG surface and couple together, effectively delocalizing the LSPR modes of the individual NPs. Looking at Figure 2.7 and 2.8, an overall wavelength and peak-to-peak amplitude ($A_{\text{p-p}}$) shift of 2.0 nm and 14 dB, respectively, are observed for the resonance mode at 1526 nm after 40 cycles (growth rate=1.52 nm per cycle, 50 nm film). The $A_{\text{p-p}}$ attenuation is typically the metric used for probing the extent of the deposition, and it can also be useful in extracting growth rate data when the process is direct MOCVD of a single-source precursor.[53,54,63] It should be noted that the reflective spectral profile of the TFBG shown in Figure 2.8a was from the growth and nucleation of the film while considering the P-polarized mode. Interestingly, as has been shown with sputtered gold films on TFBGs,[48] the sensor response is strongly dependent on the polarization state of the light interrogating the TFBG surface. Thus, as the light is scattered radially out of the TFBG (at the fiber-air interface), it is also exciting the SPR modes of the copper NPs as they nucleate and coalesce into larger particles. This is not observed for the S-polarization peak modes, implying that the polarization-dependence of the TFBG sensor is the dominant mechanism for its application as an SPR-based sensor. Application of the 50 nm copper NP-coated TFBG as an SPR-based refractive index sensor was assessed by immersing the coated fiber in a variety of sugar solutions, with various refractive indices, and monitoring the transmission spectrum. Figures 2.7a–c feature the spectral response of the bare TFBG versus the copper NP-coated TFBG, as well as the formation of copper oxides on the NPs after constant exposure to air. The characteristic "notch" in the resonances between 1545 nm and 1560 nm of the copper NP-coated TFBG is the SPR signature of the thin film, and this portion of the spectrum is crucial for further calculation of the RI sensitivity of the coated TFBG. Almost serendipitously, the formation of the copper oxides and associated attenuation of the transmission amplitude potentially makes the copper NP-coated TFBG an effective SPR-based gas sensor. The oxide coatings could be easily removed in acetic acid to regenerate the original copper NP-coated TFBG transmission spectrum. Finally, a refractive index sensitivity of 583 nm per refractive index unit (RIU) was reported, which is comparable to gold-clad and other noble metal-decorated SPR-TFBG sensitivities.[41,48] Briefly, the RI sensitivity of TFBGs is typically carried out by modeling the SPR-induced wavelength shift (central wavelength) as a function of the associated change in SRI and extracting the slope data of the resulting linear fit (e.g., 2–3 pm shifts observed for SRI changes of the order of $10^{-5}$ RIU).[59]

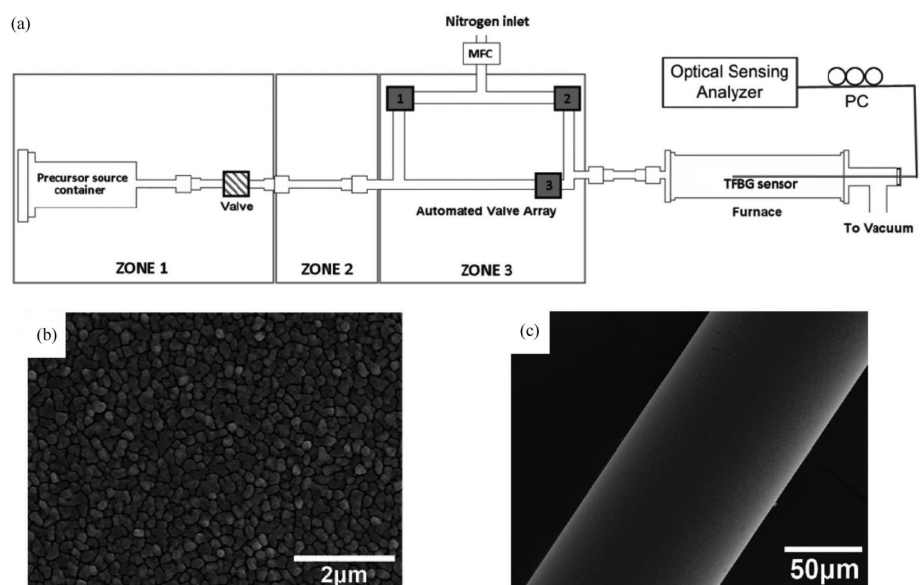

**Figure 2.6:** **(a)** Schematic of the p-CVD tool with optical sensing analyzer (OSA), polarization controller (PC) and TFBG sensor fitted in the deposition zone. **(b)** SEM image of the Cu nanolayer on the fiber after 40 growth cycles. **(c)** SEM image of the whole coated fiber after 40 growth cycles. Adapted with permission.[52] Copyright 2011, Optical Society of America.

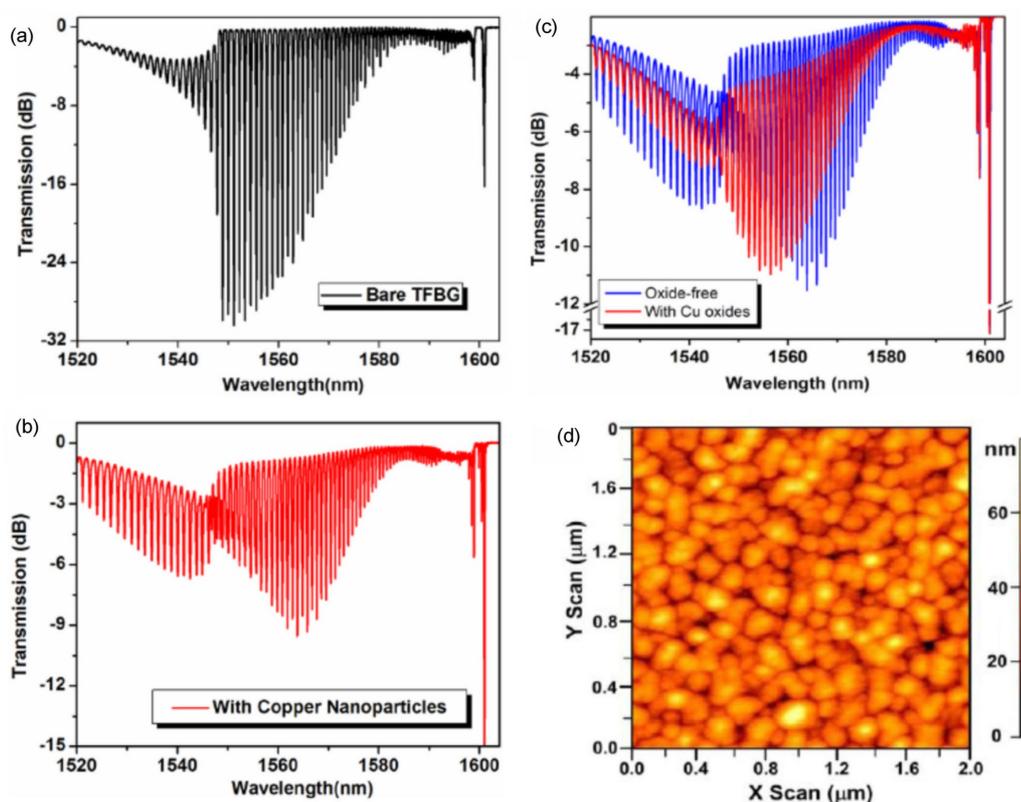

**Figure 2.7**: **(a)** Transmission spectrum of the bare TFBG prior to deposition. **(b)** Transmission spectrum of copper NP-coated TFBG in a refractive index solution (RI=1.3597), showing the characteristic SPR notch around 1545-1550 nm. **(c)** Copper NP-coated TFBG from **(b)** undergoing further oxidation to copper oxides (~0.031 nm/day, determined from *ex-situ* AFM measurements). **( d )** AFM topographic image of copper NP coating from the same experiment on a silicon witness slide (thickness of ~50 nm). Adapted with permission.[52] Copyright 2011, Optical Society of America.

More recently, studies employing the TFBG as a sensor for the monitoring of gold CVD in real-time were done using two different single-source MOCVD precursors also developed recently (Fig. 2.5).[53,54,63] Since the thermolysis of the gold (I) guanidinate ([Au(N$^i$Pr)$_2$CNMe$_2$]$_2$) is different from the gold (I) iminopyrrolidinate ([Au(Me$_2$-$^t$Bu-ip)]$_2$), due to the previously mentioned β-hydrogen decomposition pathway commonly found in the former compounds but not the latter, variability in the physical and optical thin film properties were both expected and investigated thoroughly. Figure 2.9 illustrates the experimental set-up of the CVD system and also the *in-situ* spectral measurement platform. Not shown is the fiber-coupled spectral interrogation system which is always composed of a broadband source, OSA, and PC, but they were incorporated into the CVD system analogously to the copper p-CVD system[52] (Fig. 2.6a). Multiple thermocouples ("T" in Fig. 2.9) and a pressure gauge ("P" in Fig. 2.9) were placed both upstream and downstream of the precursor vial, which was housed directly in the furnace for deposition of gold films from the gold(I) guanidinates,[53,54] and located in a separate evaporator tube (bubbler) in the case of the gold(I) iminopyrrolidinate precursor. The stainless-steel furnace was typically evacuated to a base pressure of 30 mTorr prior to deposition, and temperature ramp programs of 11°C/min were used to heat the furnace. The custom-built fiber guide consisted of a stainless-steel boat (passivated with ALD-grown $Al_2O_3$) fitted with a "w"-shaped holder, which was affixed to a KF-25 flange using UHV-grade epoxy, and fed out through fiber feed-throughs that were connected externally to the coupler/PC/OSA system. For use in *ex-situ* characterization, witness slides of Si(100) or Si(111) were also placed in the boat attached to the fiber guide during all depositions. While the planar witness slides serve as a much less topographically complex surface than the optical fiber, AFM measurements directly on the fiber are extremely non-trivial. However, thicknesses from cross-sectional SEM on the fiber could be directly compared to the AFM thickness data from the witness slides. Further experimental details pertaining to the gold CVD monitoring processes are reported thoroughly elsewhere.[52,53,63,85] Figures 2.10a-b show the attenuation profiles of the gold films under the S- and P-polarization modes for a pair of cladding modes around 1559 nm. The S- and P-polarized spectra undergo similar spectral evolution as found in the preceding copper films, however the nature of this CVD process is not pulsed and involves evaporation of the precursor directly adjacent to the TFBG zone. The onset of volatility of the precursor had an associated pressure spike in the reactor, and this coincided exactly with the initial attenuation in the transmission amplitude spectrum of the TFBG; this is effectively the time-scale of the attenuation profiles. As the film grew, a maximum attenuation in both polarization modes followed by the regeneration of the S- and P-polarized resonances back to roughly the same amplitude as they were measured in air (i.e., prior to CVD) was noticed. Subtle differences in the spectral evolution in the S- and P-polarization modes were, however, observed. (i) Maximum attenuation of the S-polarized resonances occurred after roughly 25 s (complete spectral recovery after 60 s) whereas this occurred only after 39 s for the P-polarized resonances (complete spectral recovery after 80s). (ii) The resonances of the S-polarized TFBG spectrum all red-shifted roughly by 0.08 nm during the first 20 s of attenuation, and then they discontinuously blue-shifted by 0.4 nm, whereas the P-polarized resonances red-shifted by 0.06 nm after 30 s followed by a 0.05 nm blue-shift after an additional 5 s, and then finally followed by a discontinuous red-shift by 0.14 nm where it remained. Interestingly, the net result was a direct interchange between the S- and P-polarized peak families and this had also been observed for sputtered gold films of a similar thickness (50 nm), wherein the SPR functionality of the coated TFBG sensor could effectively be "switched" on and off by changing the polarization of the core-guided light by 90°.[50] This rather interesting result was probed further since this polarization-dependence particularly in the S-polarization modes was directly opposite to what was found in the previously discussed 50 nm copper films generated by p-CVD.[52] Since the time-scale of the attenuation was slower for the P-polarized peak family, the deposition process could be quenched (i.e., halted) at various points during film growth by flowing room-temperature $N_2$ carrier gas into the reactor, hindering further precursor evaporation. This was done at the three points (onset of attenuation, maximum attenuation, regeneration of resonances) shown in Figure 2.10b, and the associated SEM images (taken *ex-situ*) in Figure 2.10c indicate that the film thickness and connectivity did increase along the attenuation profile from 29–50 s, where thickness increased from 50nm to 200 nm. The remarkably high growth rate of 222 nm/min seen within this attenuation window was calculated based on thicknesses obtained from cross-sectional SEM imaging of the coated fibers and witness slides. Thickness values were correlated with energy dispersive X-ray (EDX) spectroscopy data using the GMR Film software package.[86] To verify the purity of the gold films from the gold (I) guanidinate precursor, high-resolution (HR) X-ray photoelectron spectroscopy (XPS) was employed. The carbon, nitrogen, and oxygen 1s peak envelopes were investigated and, as shown in the survey spectrum in Figure 2.11, no persistent impurities, particularly decomposition species from the ligand system, existed after multiple $Ar^+$-sputtering cycles (0.7 keV@2 min per cycle). A base value of 13 at.% of carbon was obtained after sputtering (pre-sputtering composition of 26.9 at.% C), which is attributed to adventitious carbon.[87] No oxygen or nitrogen was present after only one sputtering cycle and, expectedly, no signal from the underlying silicon substrate was found. Because of the single-source nature of the precursor and the actual deposition process itself, overall growth rates of 39 nm/min of these gold films were considered to be quite high when compared to other gold CVD processes.[22–24,87]

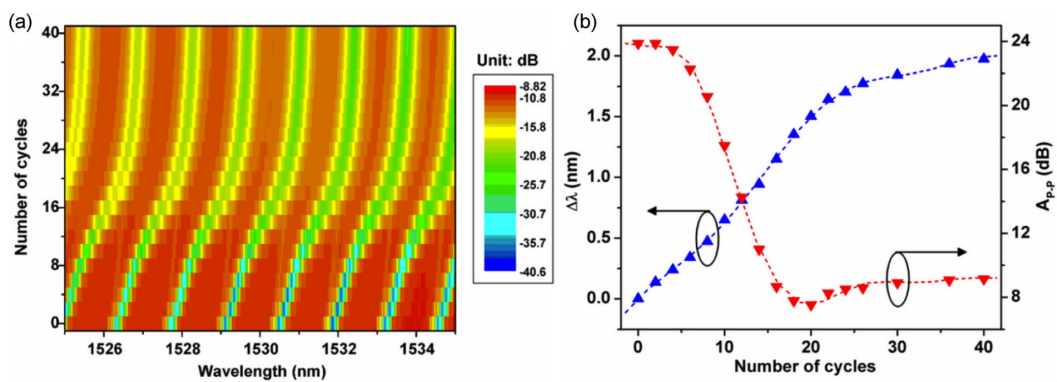

**Figure 2.8: (a)** Spectral evolution of the TFBG during p-CVD of copper NPs under the P-polarization light mode (colour indicates amplitude of the resonances during attenuation). **(b)** Associated wavelength shifts and peak-to-peak amplitude change of the resonance mode at 1526 nm as function of the thickness (shown as number of cycles). Reproduced with permission.[52] Copyright 2011, Optical Society of America.

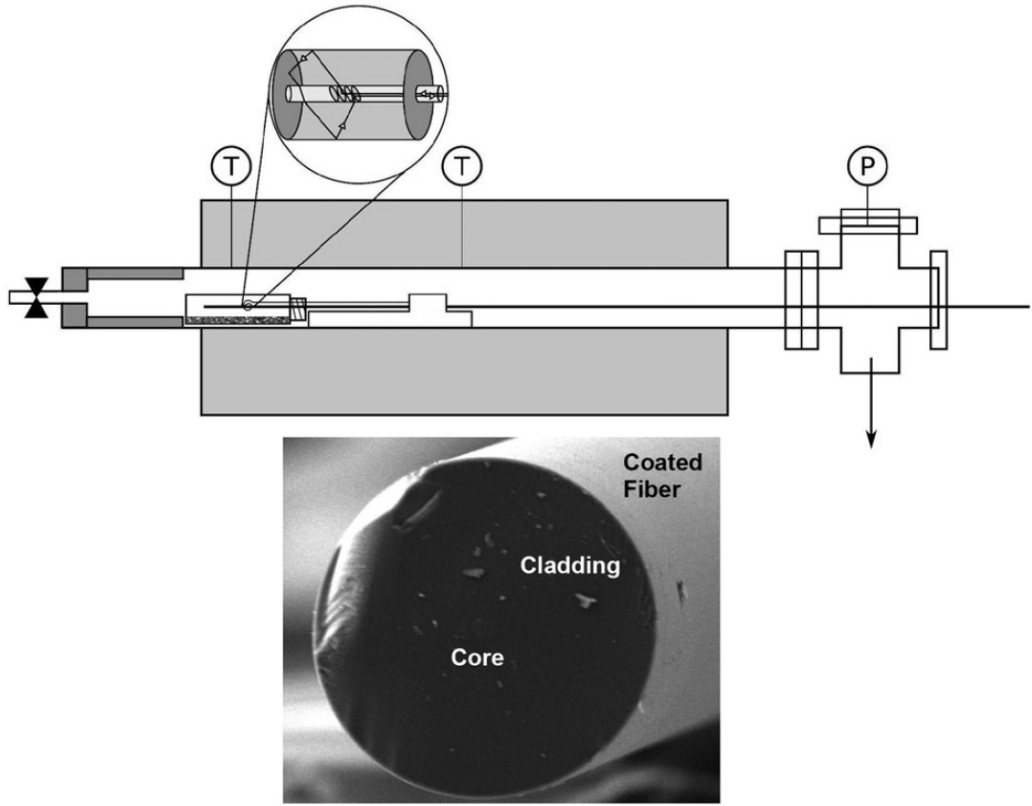

**Figure 2.9:** Schematic of the CVD tool with the TFBG sensor inserted directly into the furnace. Two thermocouples (labelled "T") are used both upstream and downstream of the precursor to obtain accurate reaction profiles and compensate for any thermal lag. The system is evacuated to <30 mTorr ("P" is pressure gauge) prior to deposition with $N_2$ as a carrier gas. The low-magnification SEM image below shows the cladding of a typical fiber after deposition coated with a gold NP film. Adapted with permission.[53] Copyright 2013, Elsevier.

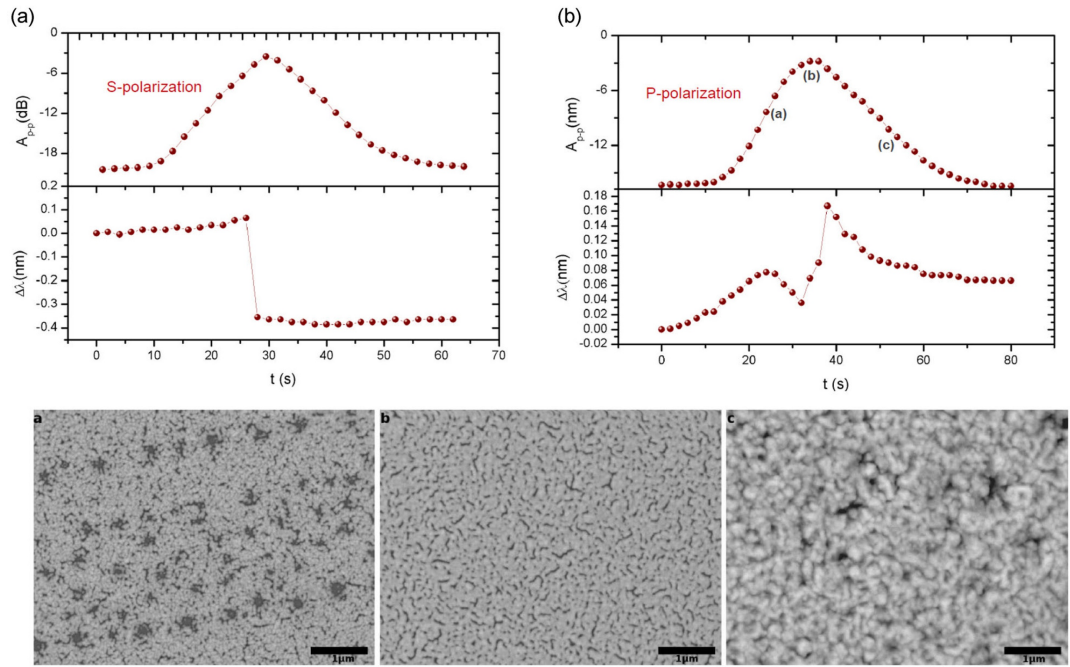

**Figure 2.10:** Attenuation profiles of the TFBG transmission amplitude spectra for the growth of gold thin films from a gold (I) guanidinate. Film growth was interrogated by the TFBG in both the **(a)** S-polarization mode and **(A)** P-polarization mode. The SEM images in the lower panel correspond to the 3 labeled points in the P-polarized attenuation profile in **(B)**. Film growth was quenched at 29, 39, and 50 seconds (corresponding to points (a), (b), and (c), respectively) in order to study the films *ex-situ* for characterization and growth rate determination. Adapted with permission.[53] Copyright 2013, Elsevier.

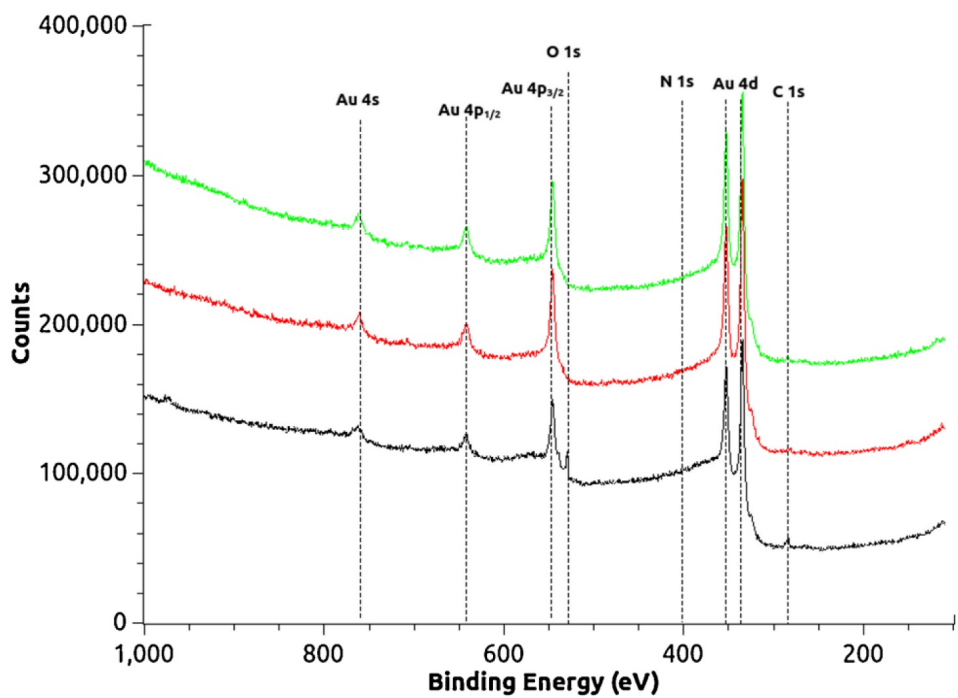

**Figure 2.11:** XPS survey spectrum of gold thin film from the gold (I) guanidinate precursor. All survey spectra were fitted to a Shirley background. Bottom trace (black) is as-deposited film, middle trace (red) is the film after 1 Ar$^+$-sputtering cycle (cycle: 2 min @ 0.7 keV), top trace (green) is the film after two Ar$^+$-sputtering cycles.

## 2.7 In-situ Optical Monitoring Studies of Silver MOCVD: Preliminary Results

A trimeric silver (I) guanidinate precursor,[30] analogous to the dimeric gold(I) guanidinate previously discussed (Fig. 2.5, top scheme), was recently employed in the fabrication of silver thin films. Deposition parameters were identical to those reported for the gold (I) guanidinate,[53,85] however the silver (I) guanidinate precursor requires more protection from light prior to deposition. The transmission amplitude behavior of these films was remarkably similar to the copper films from the analogous precursor, with the exception of a double attenuation event in the P-polarized spectra (Fig. 2.12, blue trace). That is, the S-polarized resonances underwent an attenuation of 7 dB with an associated red-shift of 0.5 nm, while the P-polarized resonances underwent an initial attenuation of 10 dB followed by a regeneration of the resonance to half its initial amplitude for 30 s, followed by re-attenuation to 10 dB with an associated blue-shift of 0.2 nm.

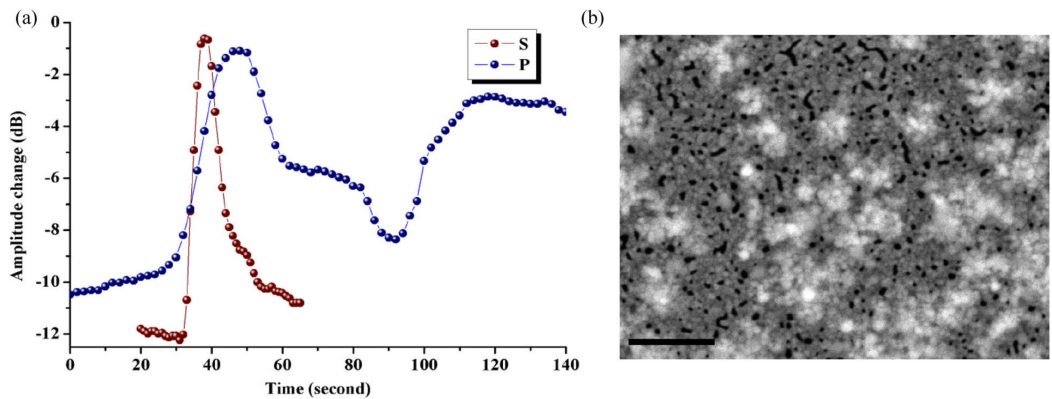

**Figure 2.12:** **(a)** S- and P-polarized TFBG attenuation profile of silver NP film after CVD employing a silver (I) guanidinate precursor. **(b)** associated SEM image (note: scale bar is 2 μm).

We speculate that the second attenuation may be due to an additional scattering component by the islands that are forming on top of the underlying film. We suggest that this is spectral evidence of a Stranski-Krastanov-type[88] of growth mode, wherein cycles of silver NP island nucleation and growth followed by coalescence into a relatively uniform film occurred. As mentioned, these particular silver films exhibit a strong polarization-dependence not dissimilar to the reported copper films, but this reproducible double attenuation of the P-polarized cladding modes will require extensive theoretical modeling. Since the underlying silver layer is more uniform than a silver NP film generated by a Vollmer-Weber growth mode (very common in CVD), the waveguide properties and scattering efficiency are also drastically different. A strong coupling constant calculated for these silver films would imply that they are far more lossy than their gold counterparts, making them more suitable for a TFBG sensor that utilizes LSPR[80] (instead of SPR) and also plasmon-enhanced refractometry. Effective medium approximations such as the ones presented herein are, however, required to confirm that these films will support SPPs in the NIR range.

## 2.8    CVD on Optical Fibers: Summary and Future Scope

This review is aimed at presenting the substantial body of work associated with the application of optical fibers to a variety of vapor deposition techniques, with a primary focus on optical monitoring of CVD. It should also serve to stimulate further research employing the TFBG as a robust sensing device for many applications. The TFBG probe is a highly efficient and convenient device for the in-situ monitoring of CVD and adaptations thereof. Perturbations in the cladding modes of the TFBG-polarized spectra during deposition of a thin metal film correspond to many physical changes of the TFBG, including the real and complex refractive indices/permittivities, wavelength, and polarization-dependent scattering. Because of the high signal-to-noise ratio of the uncoated TFBG transmission spectra, extremely accurate thin film metrics can be obtained in real-time. When real-time data are coupled with relatively straightforward effective medium modeling, the TFBG proves to be an invaluable tool for optical thin film characterization and a potential alternative to spectroscopic ellipsometry, QCM, and other *in-situ* tools adapted to CVD and ALD processes. Our ongoing research with TFBGs includes the fabrication of an effective TFBG ellipsometer and an SPR-TFBG sensor with noble metal (gold, silver, copper, etc) coatings from CVD and eventually ALD. The effect of adding a dielectric coating (e.g., $Al_2O_3$) to the TFBG enhances the polarization dependence of the TFBG and our current work is focused on exploiting this effect in order to enhance the plasmonic properties of CVD gold coatings in the NIR spectral window. Finally, we intend to incorporate the TFBG sensors into PEALD and thermal ALD processes where finer layer-by-layer thickness control is a promising advantage that advances progress towards simpler TFBG-based sensor fabrication.

# Chapter 3

## Metallic Nanocoatings from CVD of Gold(I) Guanidinates and Iminopyrrolidinates

*Modified from the original manuscripts published as:*

[1] Department of Chemistry, Carleton University, 1125 Colonel By Drive, Ottawa, Ontario, Canada, K1S 5B6

[2] Department of Electronics, Carleton University, 1125 Colonel By Drive, Ottawa, Ontario, Canada, K1S 5B6

[3] Department of Chemistry and Biochemistry, University of Delaware, Newark, Delaware, United States, 19716

*Corresponding author

## 3.1 Abstract

Deposition of gold metal on a tilted fiber Bragg grating (TFBG) demonstrates the utility of this optical fiber as a sensor for the conditions at the fiber's surface, including the continuity of a deposited metal film. Two precursors for the deposition of metallic gold are contrasted: gold(I) N'N'-dimethyl N,N''-diisopropylguanidinate (**1**) and gold(I) tert-butyl-imino-2,2-dimethylpyrrolidinate (**2**). Characterization of the deposition by optical fiber spectroscopy highlights the difference in thermal stability and growth rate of these two compounds. Compound **1** has a tremendous growth rate (222 nm/min) but is very thermally unstable, while **2** is much more thermally stable, and shows a lower growth rate of 1 nm/min.

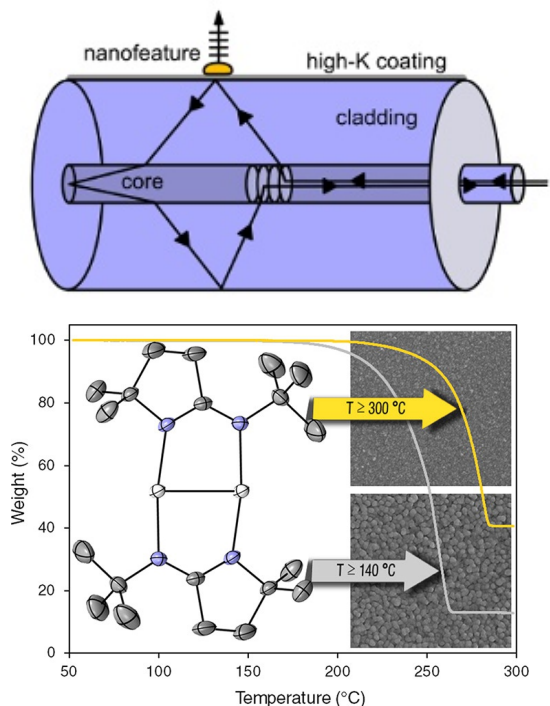

**Figure 3.1:** Table of contents images used for the publications discussed in Chapter 3. (top) for *ECS Trans.*, (bottom) for *Chem. Mater.*

## 3.2    Introduction

Since the recent employment of optical fibers as an ellipsometric witness probe during CVD, the deposition of gold nanoparticulate coatings is an active area of research in our laboratory.[4] We have employed a tilted fiber Bragg grating (TFBG) fiber to diffract light launched into the fiber's core such that it encounters the surface of the fiber, and can interact with the metal nanoparticle found there (Figure 3.1).

### Tilted Fiber Bragg Gratings

The interaction of the diffracted light mode with a metal nanoparticle at the fiber's surface is dependent on the polarization of the light.[41] If the light has a transverse magnetic moment (TM, the magnetic moment is perpendicular to the plane of incidence), the light can interact with the electron density in the particle and allow an evanescent wave to form. If the light has a transverse electrical moment (TE, the magnetic moment is parallel to the plane of incidence), the light cannot interact with the particle, but can escape through gaps in the growing film. Because the grating is not circular, the polarization of the core light mode is preserved when it is diffracted. This allows the polarity of the light to be defined at the source and so the different effects of polarization can be tested. This difference in interaction due to the light's polarization allows a coated fiber to act as a sensor by two different mechanisms. Using TM modes, the evanescent field propagated through the metal nano-features can sense the immediate surrounding material. This has been demonstrated using this arrangement by sensing the surrounding refractive index when a coated fiber is submersed in a calibrated saline solution.[45] Using TE modes, the deposition process itself can be monitored. When gold was deposited on the fiber surface using gold(I) N'N'-dimethyl N,N"-diisopropylguanidinate by CVD, the nanoparticulate film showed loss of the TE polarized light modes until a deposition time of 50 s, where the 220 nm thick film was continuous and no longer allowed light to propagate out of the film.[53] One of the surprising aspects of this study was the incredibly high growth rate. Because we could sense the deposition very accurately with the TFBG, the initial growth rate was found to be 37 Å/s (220 nm/min) for the first 50 s, before the growth rate plateaued significantly. This led us to undertake the present study, where the deposition of gold metal from two different chemical precursors were compared.

**Gold(I) Precursors**

The two precursors in this study are gold(I) N'N'-dimethyl N,N"-diisopropylguanidinate (**1**) and gold(I) tert-butyl-imino-2,2-dimethylpyrrolidinate (**2**) (Figure 3.2). There are some important design criteria that will inform thermal stability of these two compounds.

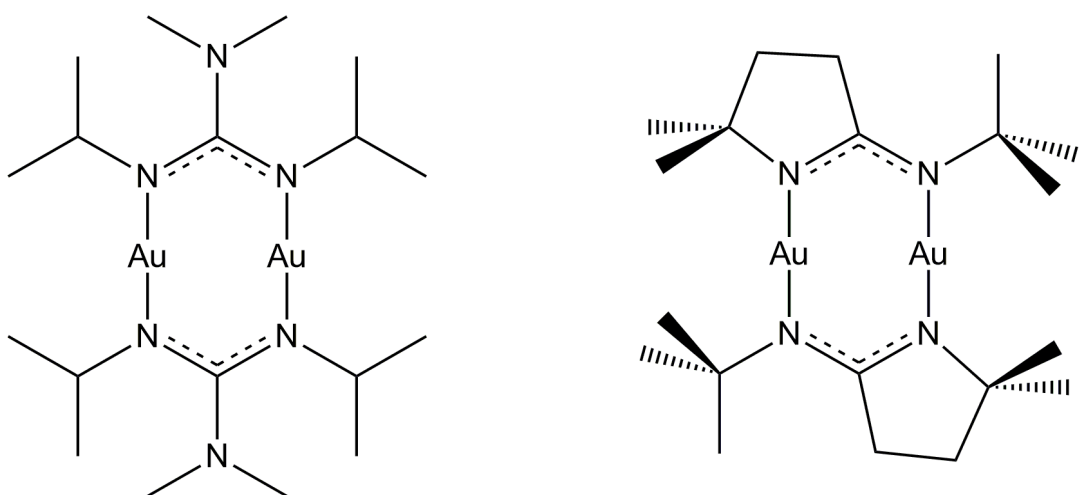

**Figure 3.2:** The structures of gold(I) N'N'-dimethyl N,N"-diisopropylguanidinate (**1**) and gold(I) tert-butyl-imino-2,2-dimethylpyrrolidinate (**2**).

Compound **1** has β-hydrogen atoms (i.e., two bonds away from the metal center). These guanidinate ligands have been shown to be thermally labile, causing the compound to undergo a redox reaction to produce metal (Figure 3.3).[81] In addition, **1** can undergo a second thermally accessible decomposition where a carbodiimide can be produced by transferring the exocyclic amido moiety to the metal center (Figure 3.3).[89] These two mechanisms have previously been compared through modeling and experimental study.[90] While the pathways are energetically similar, it was found – in the particular case of copper – that *β*-hydrogen elimination is the more likely thermal decomposition pathway in the gas phase and the predominant one under conditions that emulated CVD deposition of copper metal. The synthesis for compounds **1** and **2** both follow straightforward salt metathesis routes between a gold (I) chloride salt and the lithiated ligand system, and are reported elsewhere.[30,31]

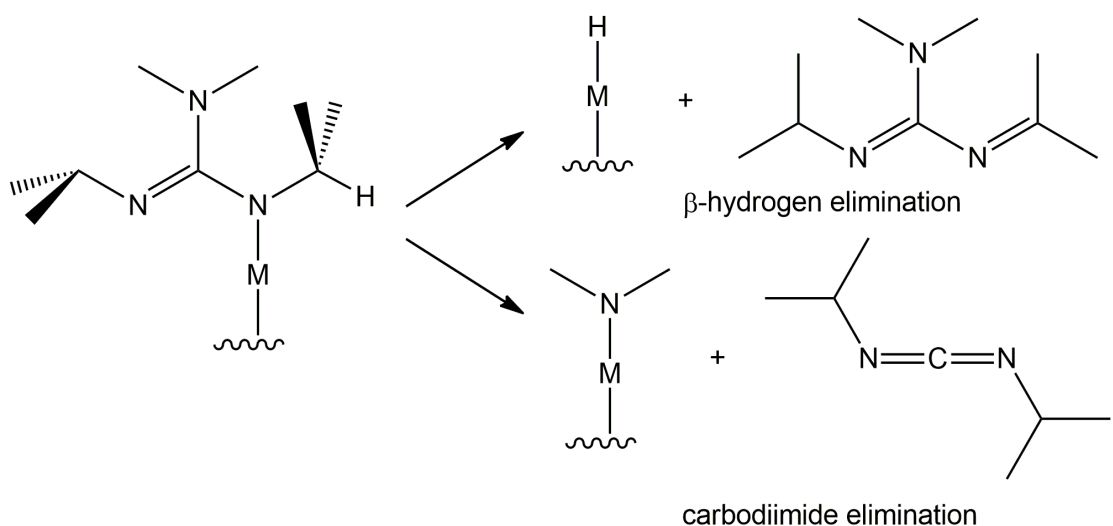

**Figure 3.3:** The two predominant thermal decomposition pathways known for guanidinate ligands.

Compound **2** was specifically designed to eliminate these two pathways. The iminopyrrolidinate ligand is an amidinate, with no third nitrogen center on the central carbon. In this ligand, one of the chelating nitrogens is tethered through a five-membered ring to the central carbon, making carbodiimide elimination extremely energetically unfavorable. Additionally, the thermally reactive $\beta$-hydrogen atoms have been replaced by methyl groups. In the case of copper, this redesign resulted in a precursor compound the that had a thermal stability ~200°C higher than the comparable guanidinate.[27]

### 3.3        Results and Discussion

Both **1** and **2** were easy to synthesize by a salt metathesis reaction between the lithiated ligand salt and gold(I) chloride adducted with tetrahydrothiophene. It should be noted that **1** is a better precursor from this point of view due to the fact that it can be made in a higher yield[31] (87.4%, compared to 65.3% for **2**[30]), but also because the ligand is far simpler to make: the guanidiante ligand in **1** can be made using a commercially available carbodiimide and lithium reagent. The reaction occurs immediately and in quantitative yield. The iminopyrrolidinate ligand, on the other hand, was made in 78.2% yield. The synthesis is more elaborate and time-consuming, making this ligand slightly less favored. Proof of chemical structure and purity for **1** and **2** are reported in thorough detail elsewhere[30,31] in the original synthetic papers.

**Thermolysis**

The thermal chemistry of these two compounds was very different. Simple thermogravimetric analysis showed that compound **1** had a much lower onset of mass loss but almost quantitatively decomposed to gold metal, undergoing very little volatilization (Table 3.1). Note that the onset of volatility was taken to be 99.5% mass loss in the thermogravimetric analysis. Compound **2** had a residual mass of 40%, indicating that its very high onset of volatilization (Figure 3.4) occurred concurrently with a thermal decomposition. During synthesis, **1** was purified by sublimation, indicating that there was a low vapor pressure of the compound (20 mTorr at 80°C). Compound **2** could not be sublimed from the reaction mixture, and was purified by recrystallization.

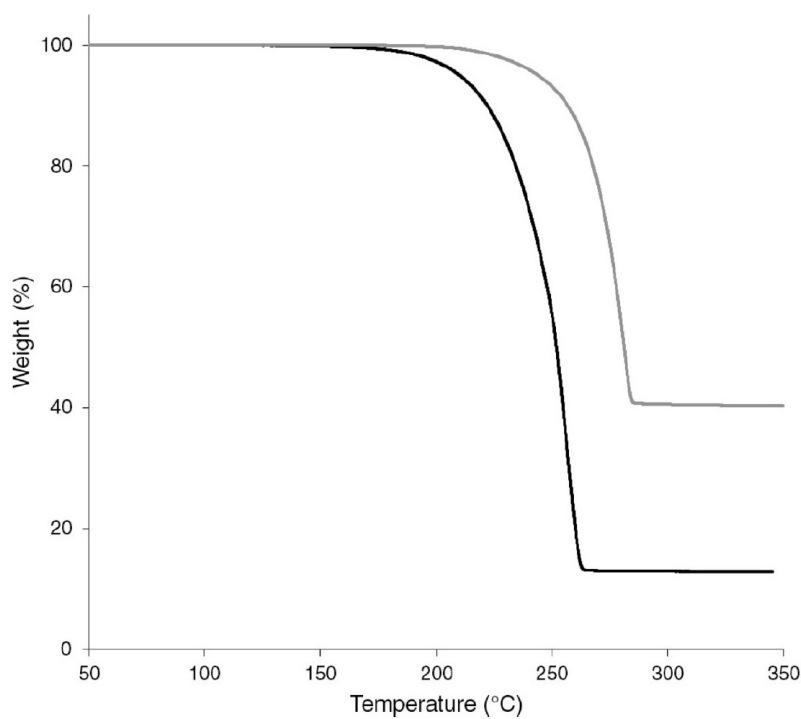

**Figure 3.4:** Thermogravimetric analysis of **2** (gray). Experiments were conducted under 1 atm of nitrogen with a ramp rate of 10 °C/min and sample size for **2** was 13.65 mg. For comparison, the thermolysis of the silver(I) analogue of **2** is also shown (black).

| | Compound 1 | Compound 2 |
|---|---|---|
| Onset of mass loss (°C) | 52.0 | 209 |
| Residual mass (compositional mass of gold) (%) | 53.0 (53.6) | 40 (53.6) |
| Synthetic sublimation temperature @ pressure (°C/mTorr) | 85 @ 20 | N/A |

Solution decomposition (solvent: $C_6D_6$) experiments further highlighted the difference in thermal stability of these two compounds. Compound **1** completely decomposed at 100 °C over four hours leaving a gold mirror in the reaction vessel and showing chemical byproducts by $^1$H NMR that suggested that the compound decomposed by both carbodimide and β-hydrogen elimination,[31] as expected. Compound **2** showed different decomposition kinetics at this temperature. The chemical byproducts suggested that the iminopyrrolidinate ligand in **2** was providing a hydrogen atom for the decomposition, suggesting that the decomposition was similar to the analogous copper compound.[82]

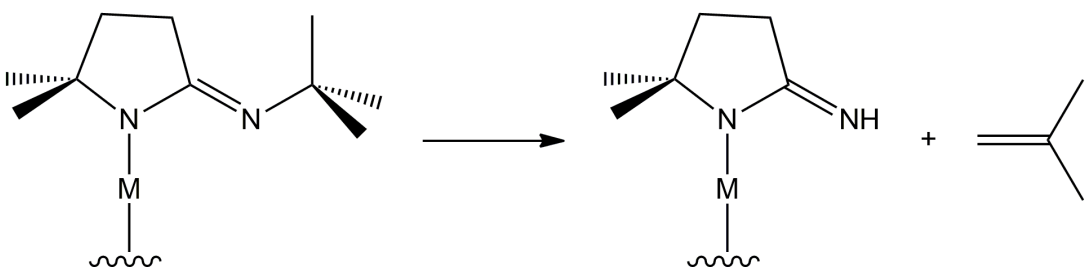

**Figure 3.5:** Thermal decomposition of the iminopyrrolidinate ligand. Hydrogen abstraction from tert-butyl group produces a metal imine and isobutene, which undergoes subsequent thermolysis to produce a metallic film.

Here, the iminopyrrolidinate undergoes a deprotonation of the *tert*-butyl group to produce *iso*-butene and a metal imine (Figure 3.5). This imine is suspected to be the source of the proton at elevated temperature, although this mechanistic step has yet to be observed. This reaction occurs over a significantly longer time frame, with a half-life of 65 ± 2 h (assuming second order rate kinetics). With respect to decomposition experiments, **2** is the favored precursor due to its significantly slower decomposition kinetics. Compound **2** decomposed by a second-order reaction with a rate constant of 0.0155 mL·mmol$^{-1}$s$^{-1}$ ± 0.0005 mL·mmol$^{-1}$s$^{-1}$ at 100°C.

## Deposition

Deposition experiments on silicon substrates as well as on TFBGs were undertaken using both **1** and **2** (Table 3.2). The silicon and fibers were assembled in the CVD apparatus such that both substrates were deposited upon at the same time. The optical fibers were monitored *in situ* while the silicon slides were used in post-deposition characterizations.

Compound **1** showed a very fast deposition such that the growth rate over the entire deposition experiment was 6.5 Å/s over 7 minutes. Deposition on the fiber was measured using TE polarized light, since this light mode will exit the fiber at a silica/air interface in the CVD reactor, but undergoes reflection when it encounters a gold feature. Thus, the attenuation of the TE polarized cladding mode showed that deposition was mostly complete in the first 80 s. The TE mode underwent attenuation starting at 10 s and showed maximum attenuation at 29 s. At 50 s, the attenuation had abated, giving a signal strength similar to pre-deposition levels. The growth rate over this period was measured as 37 Å/s (220 nm/min). In this deposition experiment, the fiber was held directly over the evaporating precursor, since the thermal stability of **1** was too poor for the vapor to survive for a significant period of time at the deposition temperatures.

Compound **2** again showed very different behavior. The precursor was evaporated at 160 °C, which produced a sufficient partial pressure of **2** for deposition to occur. The precursor was very stable at this temperature, and showed very little decomposition during the deposition experiment. The deposition temperature (heated resistively from a separate bubbler) was higher than with **1** again due to the thermal stability of **2**. The growth rate with **2** was extremely small, on the order of 200 times slower than with **1**. In this case, **2** again appears to be the superior precursor. Although the growth rate is significantly lower, the precursor is easier to handle in the gas phase and more amenable to the CVD process.

**Table 3.2:** The deposition characteristics for **1** and **2** using an optical fiber as a substrate

| Process Parameter | Compound 1 | Compound 2 |
|---|---|---|
| Precursor volatilization temperature (°C) | 80 | 160 |
| Deposition temperature (°C) | 200 | 300 |
| Growth rate (Å/s) | 37 | 0.17 |

A scanning electron micrograph (SEM) reveals that the deposited films are quite different. A cross-sectional SEM shows the metallic gold film deposited from **1** (Figure 3.6a). On the right side of the image, it is apparent that the gold film ripped away from the surface when the fiber was cleaved prior to imaging. In the same micrograph, the core of the fiber is visible as a slightly lighter dot in the center of the cladding. Figure 3.6b shows the typical film morphology of gold metal deposited by **1**. Note that the grains are not very well defined and somewhat amorphous. This could be a function of the extremely high deposition rate, but is more likely due to the fact that the deposition temperature is on the order of 200°C. This contrasts with Figure 3.6c, which is a gold film deposited from **2** at 300°C. The grains are much more well defined, likely caused by the higher mobility of gold adatoms at this temperature, as well as the fact that the very low deposition rate allowed reorganization of the film to a greater degree.

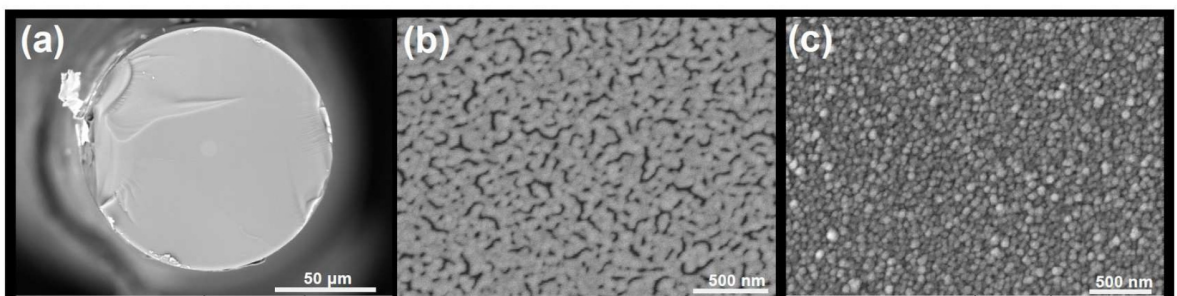

**Figure 3.6:** (a) Cross-sectional field-emission SEM image of a gold coating on the optical fiber deposited from **1**. (b) SEM image of a gold coating on a Si(111) witness slide (thickness ~65 nm) from **1**. (c) SEM of a gold coating on a Si(111) witness slide (thickness ~65 nm by AFM) from **2**

### X-ray Photoelectron Spectroscopy (XPS)

XPS analysis of films from the two precursors reveals some other slight differences in deposition (Table 3.3). The films show comparable impurities when characterized after deposition, with **2** giving a nitrogen impurity that is not found in **1**. We attribute this again to the superior stability of **2**. In the as-deposited state, it is possible that the nitrogen impurity comes from a surface coating of intact precursor, which appears in the film deposited by **2** but not in **1** due to the better thermal stability of **2**. This surface contaminant is removed under sputtering. After sputtering with an Ar$^+$ source, **2** again showed superior film deposition. It had a smaller carbon impurity compared to **1**, and the nitrogen impurity was removed as a surface contamination.

| Surface Species | Compound 1 (as-deposited) | Compound 2 (as-deposited) | Compound 1 (post-sputtering) | Compound 2 (post-sputtering) |
|---|---|---|---|---|
| Au (at. %) | 83.05 | 84.90 | 91.20 | 96.80 |
| C (at. %) | 12.36 | 10.90 | 8.80 | 3.20 |
| O (at. %) | 4.60 | 0.20 | -- | -- |
| N (at. %) | -- | 4.00 | -- | -- |

An example of the removal of the impurities by plasma treatment can be seen by monitoring the carbon signal as a function of sputter time. In both cases, the as-deposited films (top traces in Figure 3.7a) and b) show significant carbon impurities. Even after one sputtering cycle of two to three minutes, the carbon content dropped appreciably, demonstrating that the originally high carbon content was likely adventitious or from condensed, partially decomposed precursor that adsorbed as the deposition furnace cooled.

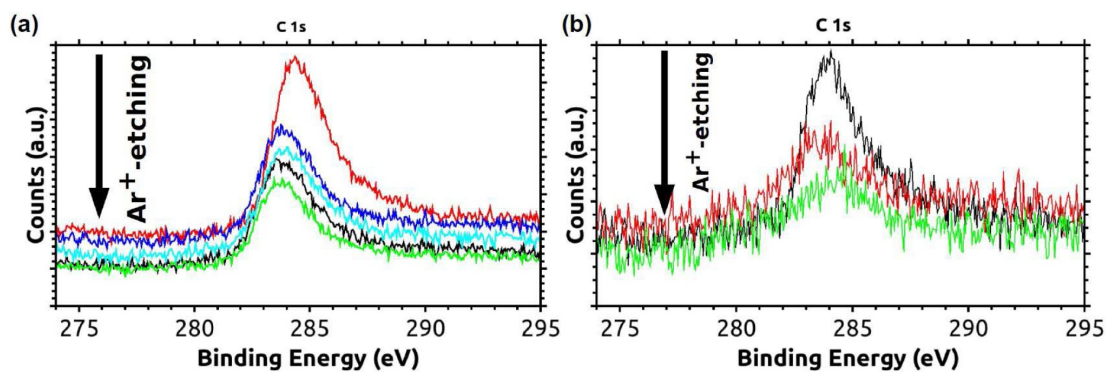

**Figure 3.7:** High-resolution XPS C 1s peak envelopes for (a) Au on Si(111) from **1** and (b) Au on Si(111) from **2**, both as a function of Ar$^+$-sputtering. The top traces correspond to the as-deposited films on the witness slides. A typical Ar$^+$-sputter cycle was 2-3 min of Ar$^+$ at ~1.7-2 keV acceleration voltage.

### 3.4 Conclusions

Gold metal films were deposited from gold(I) N'N'-dimethyl N,N"-diisopropylguanidinate (**1**) and gold(I) tert-butyl-imino-2,2-dimethylpyrrolidinate (**2**) at 200 °C and 300 °C respectively. Although **1** has a lower apparent volatilization temperature and extremely high growth rate, **2** showed itself to be the superior precursor. Under process conditions, the slower decomposition kinetics of **2** meant that a more controlled growth of film with lower bulk impurities resulted. From the point of view of depositing nanoparticles for sensor applications, the more faceted film deposited by **2** would be more appropriate: in order to get this type of film, the deposition temperature needs to be high, and this is unachievable by **1**.

### 3.5 Experimental

Compounds **1**[30] and **2**[31] were prepared using literature methods. Thermogravmetric analysis was carried out on a TA Instruments Q500 apparatus housed in an MBraun Labmaster dry-box under a nitrogen atmosphere. Thermal decomposition analysis was undertaken in vacuum sealed, heavy walled NMR tubes that were kept in a dark oven at 120 °C. Periodically, the NMR tubes were removed from the oven, allowed to reach room temperature, and then a $^1$H-NMR was measured on a Bruker 300 MHz Avance spectrometer using $C_6D_6$ as a solvent and tetramethylsilane as an internal standard. Selected peaks were integrated against the internal standard and the change in area of two different peaks for **2** were plotted over time to collect the kinetic data. These data were graphed as the inverse of the normalized peak area with respect to time, and the rate constant and kinetic half-life was extracted from a linear fit of these data. Deposition was undertaken in a homemade reactor. The design of the reactor was slightly different for each case to accommodate the specific thermal behavior of the compound. For **1**, the fiber was suspended directly over the precursor to permit the highly thermally unstable precursor reach the fiber intact. In the case of **2**, the precursor was independently heated using a heating tape, and the precursor was entrained to the hot deposition zone under vacuum. Growth rate was measured *ex situ* from a micrograph collected using a Tescan Vega-II XMU SEM systems and corroborated using a program (GMRFilm) to calculate film thickness from the k-alpha ratio of Au(0) measured using Oxford Instruments Inca energy dispersive spectrometer. XPS and $Ar^+$-sputtering experiments were performed in a custom-built multi-technique ultra-high vacuum system (Specs, Gmbh). The samples were mounted to a copper and iron sample holder (puck) on the main manipulator arm and held in place using molybdenum clips. The XPS in this study used a 14.26 keV Al Kα source and performed at a background pressure between 8 x $10^{-10}$ and 1 x $10^{-9}$ Torr. A fixed holder angle of 0.2 ° was used during spectral acquisition and intermittent $Ar^+$-sputtering cycles at a filament potential between 1.71 – 2 keV were performed in 2 – 3 minute periods. $Ar^+$-sputtering was performed with a background chamber pressure of 1 x $10^{-5}$ Torr and was halted once the base concentration (at.%) of carbon (i.e. the C 1s signal) was constant after 3 $Ar^+$-sputtering cycles. Spectra were collected in SpecsLab2 software and post-processing was performed using CasaXPS.

## 3.6 Supporting Information

**Table 3.4:** Typical XPS hemi-spherical analyzer values used for ex-situ film analysis

| HSA3500 Parameters | | Scan Mode (SpecsLab2) | | |
|---|---|---|---|---|
| **Entrance Slit Aperture** | 7 mm x 20 mm (medium) | **Scan Parameter** | **Survey** | **Component** |
| **Exit Slit** | Open | **Energy Step (eV)** | 0.2-0.4 | 0.05-0.1 |
| **Iris Aperture** | 32 mm | **Dwell Time (s)** | 0.3 | 0.1-0.3 |
| **Lens Mode** | Medium | **Number of Scans** | 1 | 2-5 |
| **Analysis Mode** | Fixed Analyzer Transmission | **Pass Energy (eV)** | 10 | 30 |

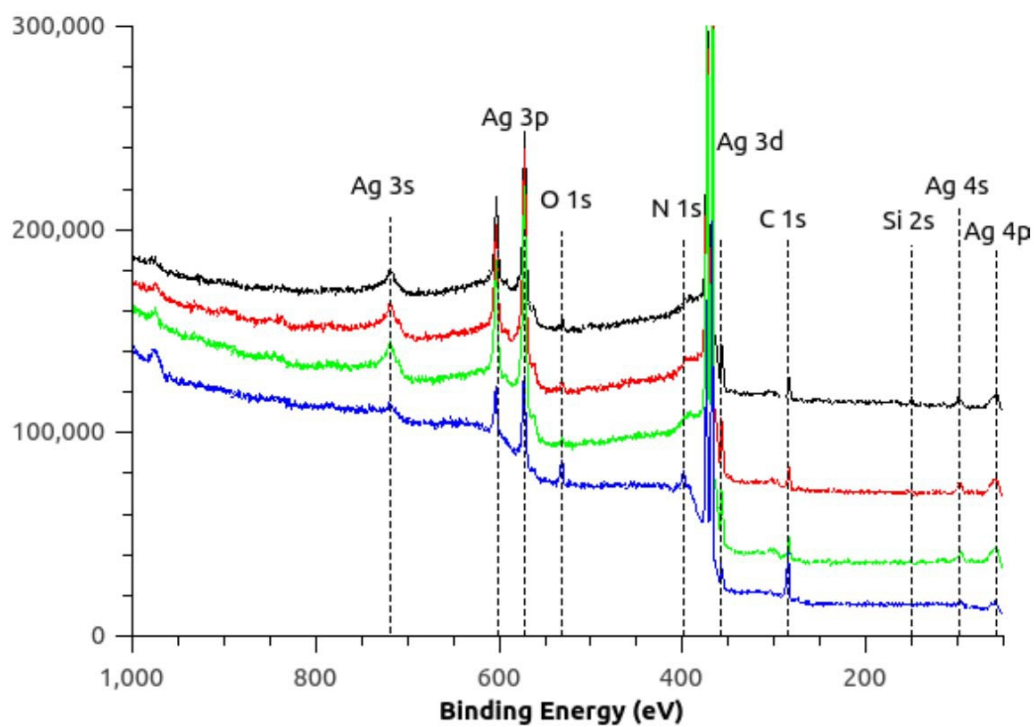

**Figure 3.8:** XPS survey spectrum of Ag on Si(100) substrate as-deposited from silver(I)-*tert*-butyl-2,2-dimethyliminopyrrolidinate (blue), after 1 sputtering cycle (green), after 2 sputtering cycles (red), and after 3 sputtering cycles (black). Ar+ sputtering cycles lasted 2 min and were done under base Ar pressures of 1 x 10^-5 Torr.

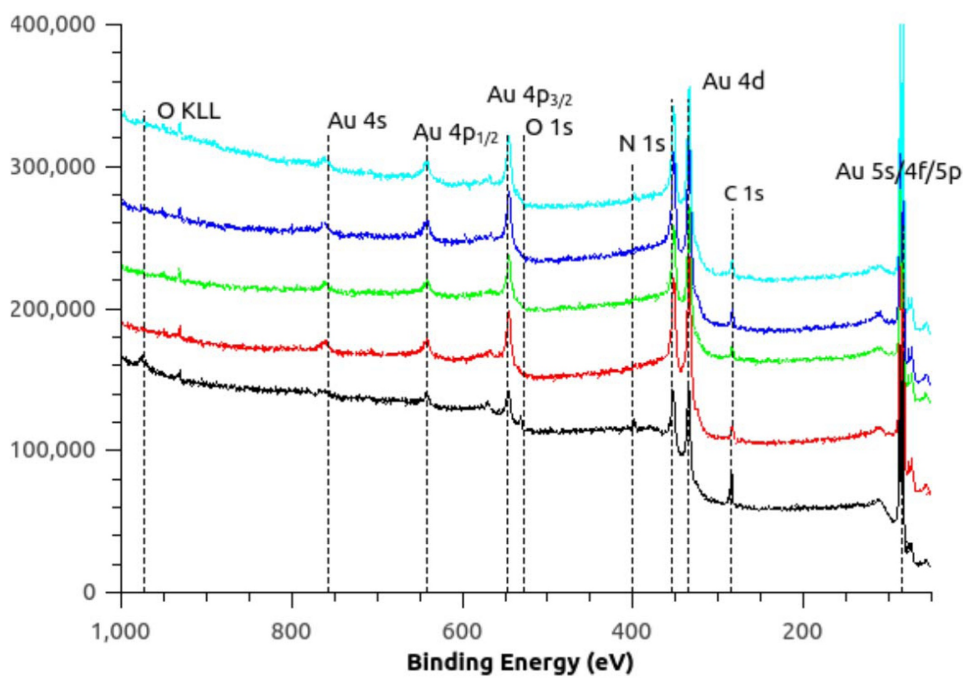

**Figure 3.9:** XPS survey spectrum of Au on Si (100) substrate as deposited from gold(I)-*tert*-butyl-2,2-dimethyliminopyrrolidinate (black), after 1 sputtering cycle (red), after 2 sputtering cycles (green), after 3 sputtering cycles (blue), and after 4 sputtering cycles (cyan). Ar[+] sputtering cycles lasted 2 min and were done under Ar pressures of 1 x 10[-5] Torr.

# Chapter 4

## Polarization-dependent Properties of the Cladding Modes of a Single-mode Fiber Covered with Gold Nanoparticles

Zhou, W.*[2]; **Mandia, D.J.**[1]; Griffiths, M.B.E.[1]; Bialiayeu, A.[2]; Zhang, Y.[2]; Gordon, P.G.[1]; Barry, S.T.[1]; Albert, J.[2] "Polarization-dependent Properties of the Cladding Modes of a Single Mode Fiber Covered with Gold Nanoparticles" *Optics Express.* **2013**; *21*(1); pp. 245-255.

[1] Department of Chemistry, Carleton University, 1125 Colonel By Drive, Ottawa, Ontario, Canada, K1S 5B6
[2] Department of Electronics, Carleton University, 1125 Colonel By Drive, Ottawa, Ontario,
Canada, K1S 5B6
*Corresponding author

## 4.1 Abstract

The properties of the high order cladding modes of standard optical fibers are measured in real-time during the deposition of gold nanoparticle layers by chemical vapor deposition (CVD). Using a tilted fiber Bragg grating (TFBG), the resonance wavelength and peak-to-peak amplitude of a radially polarized cladding mode resonance located 51 nm away from the core mode reflection resonance shift by 0.17 nm and 13.54 dB respectively during the formation of a ~200 nm thick layer. For the spectrally adjacent azimuthally polarized resonance, the corresponding shifts are 0.45 nm and 16.34 dB. In both cases, the amplitudes of the resonance go through a pronounced minimum of about 5 dB for thickness between 80 and 100 nm and at the same time the wavelengths shift discontinuously. These effects are discussed in terms of the evolving metallic boundary conditions perceived by the cladding modes as the nanoparticles grow. Scanning Electron Micrographs and observations of cladding mode light scattering by nanoparticle layers of various thicknesses reveal a strong correlation between the TFBG polarized transmission spectra, the grain size and fill factor of the nanoparticles, and the scattering efficiency. This allows the preparation of gold nanoparticle layers that strongly discriminate between radially and azimuthally polarized cladding mode evanescent fields, with important consequences in the plasmonic properties of these layers.

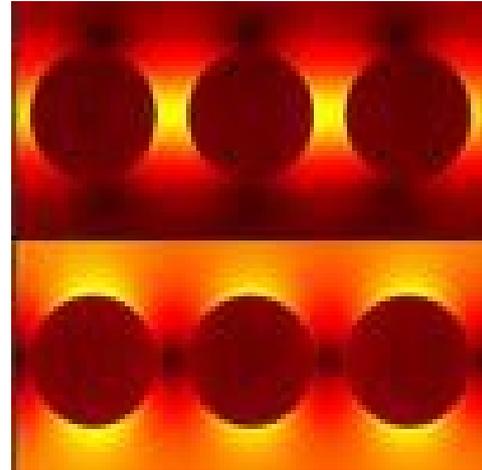

**Figure 4.1:** Table of contents image used in the publication for Chapter 4.

## 4.2 Introduction

The interaction between propagating light waves and nanometer-sized gold particles presents many interesting features that have been studied in several contexts.[91,92] It is therefore expected that depositing gold nanoparticles on the cladding of an optical fiber will have an impact on the cladding-guided modes of such fibers and may lead to new applications in sensing, all optical switching, and nonlinear optics. Tilted fiber Bragg gratings (TFBGs) are ideally suited for such studies because of their capability to excite a number of cladding modes whose properties can be probed very precisely by measuring the wideband spectral response of the grating transmission.[41,93] Furthermore, due to the breakup of the circular symmetry caused by the tilted grating planes, the polarization state of the incident core mode controls the orientation and polarization of the excited cladding modes at the cladding boundary.[41] This is of utmost importance for metal coatings as the boundary conditions for metal-dielectric interfaces depend very strongly on the polarization state of the light. For example, cladding modes with radial polarization at the cladding boundary can be excited selectively and be used to couple light into surface plasmon waves on metal coated fibers. A gold-coated TFBG refractometer based on Surface Plasmon Resonance (SPR) was demonstrated using this technique,[94] and shown to yield a great enhancement in the minimum surrounding refractive index (SRI) detection level,[50] compared to non-TFBG fiber SPR devices.[95,96] In all these fiber-based SPR configurations the quality and uniformity of the (typically) 50 nm thick metal layer is critical but also difficult to control; especially to make it uniform around the circumference of the fiber. The gold coatings in the papers mentioned above were fabricated by sputtering or evaporation techniques, which require fiber rotations to obtain approximately uniform films. Gold films with better uniformity are expected from conformal coating approaches such as electroless plating and chemical vapor deposition (CVD), which have been demonstrated for gold and copper, respectively.[52,97] Recently, a new precursor for gold CVD was developed[30] and the present work deals with an investigation of the effect of this new CVD coating process on the properties of the cladding modes of such coated fibers. As shown in Fig. 4.2, we used single-ended 10° tilt TFBGs inserted in the CVD process chamber and monitored their transmission spectrum from the input side by using the reflection from a gold mirror deposited on the fiber end. This avoids having to loop back the fiber inside the chamber. The wavelengths and amplitudes of resonances associated with cladding modes of different polarization states were obtained throughout the CVD process. The observations obtained from inside the fiber using the TFBG resonances were compared with scanning electron microscope images of the films at various stages of deposition, and with infrared camera images of the light scattered out of the fiber by the nanoparticles. The results indicate three different regimes according to the density and thickness of the CVD gold nanoparticle coatings. In particular, during the coating process we observed a well-defined transition between isolated nanoparticles and a semi-continuous film that has a strong impact on the optical properties of the optical fiber cladding modes.

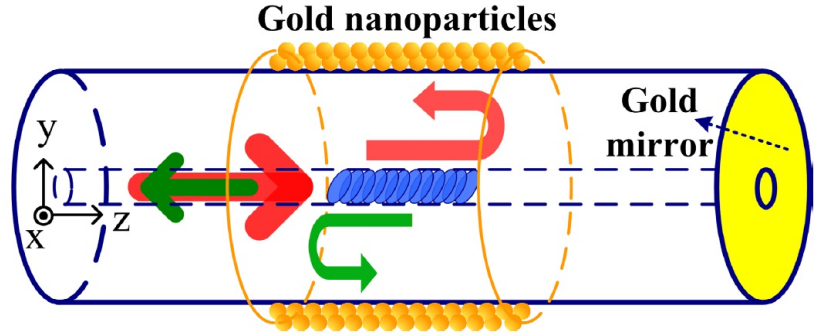

**Figure 4.2:** Schematic diagram of TFBG coated by gold nanoparticles (the arrows show how the incident core guided light (red) goes through the grating twice, each time coupling light to cladding modes). The light remaining in the core (straight green arrow) returns to the interrogation system.

## 4.3 Principle and Sensor Fabrication

In order to clarify how nanoscale metal coatings influence cladding mode resonances, a brief review of TFBG properties is required before the description of the experiments. A periodical refractive index perturbation with tilted grating planes along the core axis couples the forward-propagating core mode to a number of backward-propagating cladding modes in addition to a backward-propagating core mode (the "Bragg" mode). The wavelength of the resonance of the Bragg mode, $\lambda_{Bragg}$, and of the $i$th-order cladding mode resonance, $\lambda_{cladding}^{i}$, in the transmission spectrum of TFBG can be expressed as [60]

$$\lambda_{Bragg} = \frac{2 N_{eff(Bragg)} \Lambda}{\cos(\theta)} \qquad [1]$$

$$\lambda_{Cladding}^{i} = \frac{\left( N_{eff(Bragg)}^{i} + N_{eff(Cladding)}^{i} \right) \Lambda}{\cos(\theta)} \qquad [2]$$

where $N_{eff(Bragg)}$ is the effective index of core mode at the Bragg wavelength, $\Lambda$ is the grating period, $\theta$ is the tilt angle of the grating planes, $N_{eff(Bragg)}^{i}$, and $N_{eff(Cladding)}^{i}$ are the effective indices of the core mode and $i$th cladding mode at the wavelength of the $i$th cladding resonance. The guiding properties of these cladding modes and hence of $N_{eff(Cladding)}^{i}$ depend on the permittivity of the medium in which the cladding is located. Therefore, we expect the wavelengths of the cladding mode resonances to shift during the deposition of gold nanoparticles on the fiber. The amplitudes of these cladding resonances would also change under the influence of the imaginary part of the complex refractive index of gold (as shown for LPGs in [98]) and of the loss due to scattering by the nonuniform layer of nanoparticles on the surface of the TFBG.

We further expect these effects to depend strongly on the polarization state of the cladding modes at the cladding boundary. The tilted grating planes break the azimuthal symmetry of the fiber and two orthogonal polarization states of the electrical field input light can be defined relative to the tilt plane: S-polarized light with the electrical field perpendicular to y-z plane and P-polarized light with the electrical field parallel to y-z plane (as shown in Fig. 4.3). In previous reports, we showed that when the polarization state of the light at the TFBG contains only S- or P-polarized light (two extreme cases), the electrical field of the excited high order cladding modes also has radically different polarization properties.[62,94] Simulations obtained with a finite difference mode solver,[99] and shown in Fig. 4.3, indicate that S-polarized light can only couple into high-order cladding modes that have their electrical field tangential to the cladding boundary, while P-polarized light excites cladding modes with predominantly radial electrical fields. This difference is highlighted in Fig. 4.4, where two transmission spectra of the same grating under S- and P-polarized light interrogation (measured in air without coating) show clearly different sets of resonances that occur in closely spaced pairs. Similarly to TFBGs with uniform thin metal layers, we use this polarization selectivity to obtain additional information about the gold nanoparticle coating as it grows.

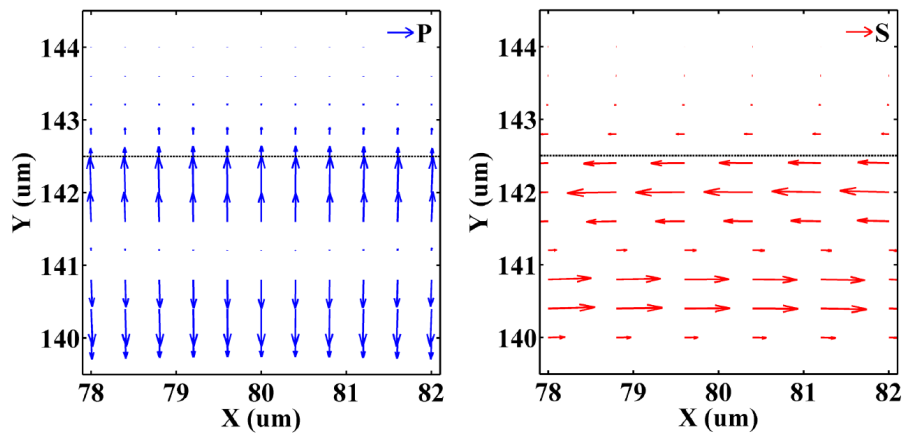

**Figure 4.3:** Simulated electrical fields of a high-order cladding mode excited by P- (left) and S-(right) polarized light. The figures represent a small area of 4 by 4 μm² close to cladding boundary (indicated by a horizontal line at Y = 142.5 μm). All results were calculated with FIMMWAVE.

The TFBGs were written in hydrogen-loaded photosensitive CORNING SMF-28 fiber with a pulsed KrF excimer laser using the phase-mask method. The Bragg wavelength is around 1610 nm so that important cladding modes are observed in the middle of the C-band (near 1550 nm). The length of TFBG is only 4 mm long in order to minimize the impact of eventual coating thickness non-uniformities on the grating response. A tilt angle of 10° leads to a large number of strong, high order cladding mode resonances that have larger evanescent field penetration outside the cladding. In order to have a single entry port in the furnace used for the CVD process and to avoid bending the fiber, a reflective sensing configuration of TFBG was implemented by cleaving the fiber 1.5 cm downstream from the TFBG and coating the end with a sputtered gold mirror (Fig. 4.2). This configuration further enhances the response of the TFBG since core-guided light goes through the grating twice. The gold mirror at the downstream end was prepared separately by conventional sputtering, and its thickness (several microns) is such that the addition of more gold on the fiber end (during the nanoparticle deposition process) has no effect on its reflectance. There is unavoidable growth on the fiber end but it has no influence on the properties of the cladding modes at the location of the grating (a few cm away).

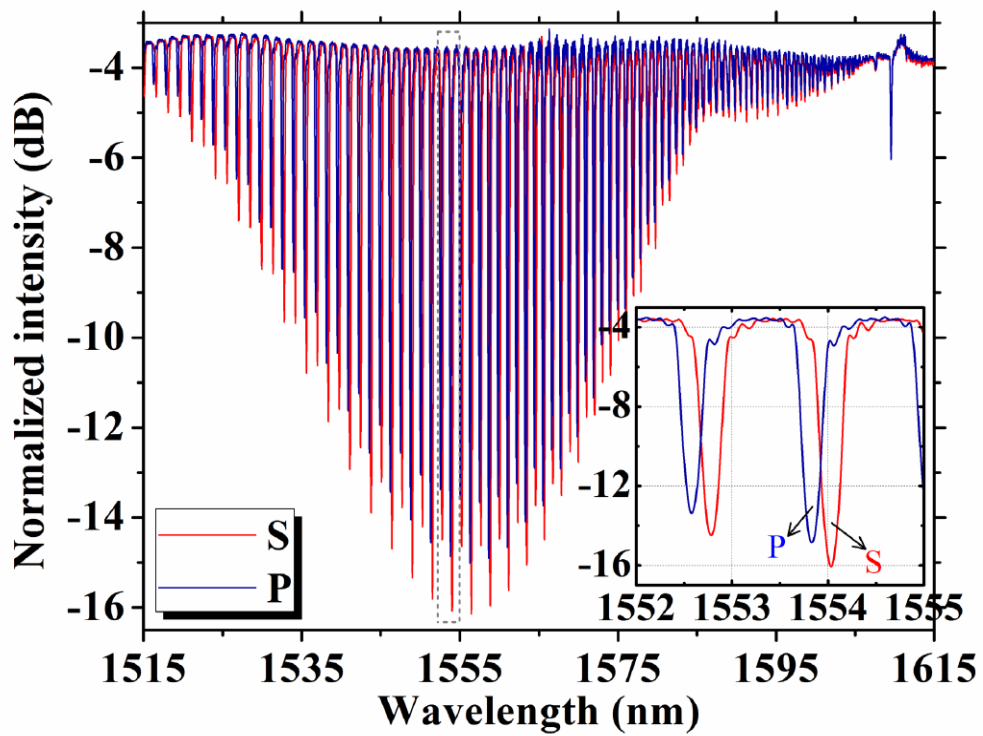

**Figure 4.4:** Experimental transmission spectra of 10° TFBG with S- and P-polarized input light in air, inset: detailed spectra from 1550 to 1556 nm.

## 4.4    Experiment and Results

Figure 4.5 illustrates the experimental setup of the gold CVD system and the *in situ* spectral measurement system. The fiber was fixed on a metal boat and the end containing the TFBG was inserted into a vial filled with a single-source gold precursor ([Au(N$^i$Pr)$_2$CNMe$_2$]$_2$ ).[30]

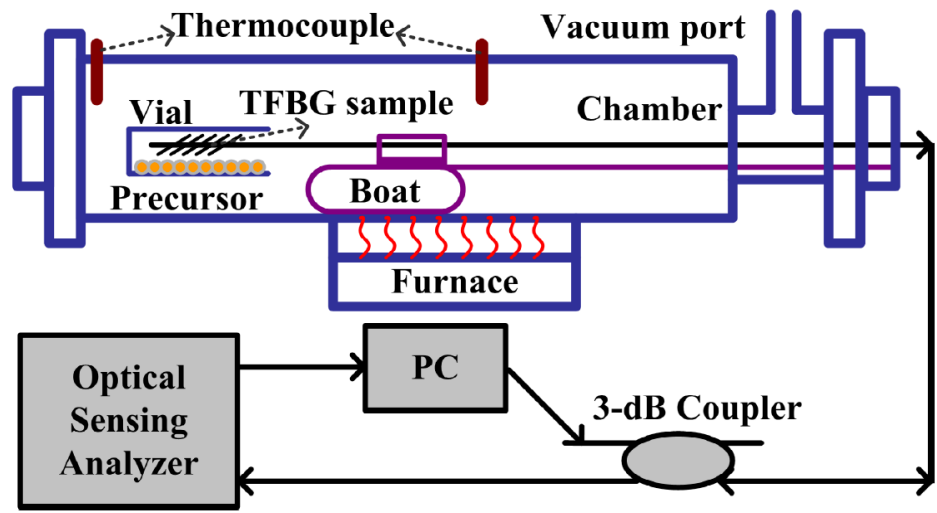

**Figure 4.5:** Schematic diagram of gold CVD system and spectral monitoring setup. The optical sensing analyzer includes a scanning laser source with a wavelength range from 1520 to 1570 nm and a synchronized photodetector. The precursor is heated within the CVD reactor and not in a separate bubbler due to the thermal instability of the precursor used.

The fiber and vial were located in a deposition chamber made of stainless steel chamber itself being located within a heating furnace. At the beginning of the experiment the chamber was pumped down to a base pressure of 30 mTorr and then the furnace temperature ramp (11 °C/min) program was set to 255 °C. Since the reaction chamber is situated far from the main thermocouple, temperature measurements were collected on a second thermocouple fitted further down the furnace tube to account for any thermal lag. As the temperature increases, it reaches the vaporization and decomposition point of the highly volatile precursor (~220 °C) and the gold nanoparticles begin to nucleate on all exposed surfaces including the cladding of the TFBG. EDS analysis performed on the fibers and planar samples with this gold compound showed no significant or detectable impurities (only signals for Si, O, and Au). The process self-terminates when the precursor has fully reacted and the furnace is then allowed to cool down. The whole process takes about 10 minutes and the onset of the deposition itself occurs after approximately 8 minutes, as corroborated by changes in the TFBG response. With this methodology, the final thickness of the gold nanoparticle coating on the fiber is determined by the amount of precursor in the vial. For instance, the (mass-equivalent) film thickness from a 30 mg sample of precursor was typically ~200 nm. In other experiments wherein the mass was varied, a linear relationship was found with respect to the film thickness, and a nominal growth rate of 5 nm/s (starting from onset of deposition) was calculated from *ex situ* thickness calculations using WSXM 5.0 image processing software.[100] Finally, the impact of the strong temperature changes on the spectral positions of the resonances was removed from the data analysis by recording the wavelength shift of the Bragg mode and subtracting it from the shifts of the cladding modes. The Bragg mode is well isolated from the cladding boundary (therefore from any effect resulting from the deposition) and it has been verified that all cladding mode resonances have the same intrinsic temperature dependence as the Bragg mode.[60] It must be noted that this deposition configuration yields coatings that vary in thickness as a function of distance from the precursor source. We usually work with longer TFBGs (between 10 and 20 mm long) in order to have narrower resonances to enhance the Q-factor and thus the minimum detectable level of sensors, but if the coating thickness varies over this length, the net effect is a chirp of the grating properties and widening of resonances. Also, we need only to deposit the particles on the fiber sections where the grating is located, because these are the only locations where light is emitted into the cladding.

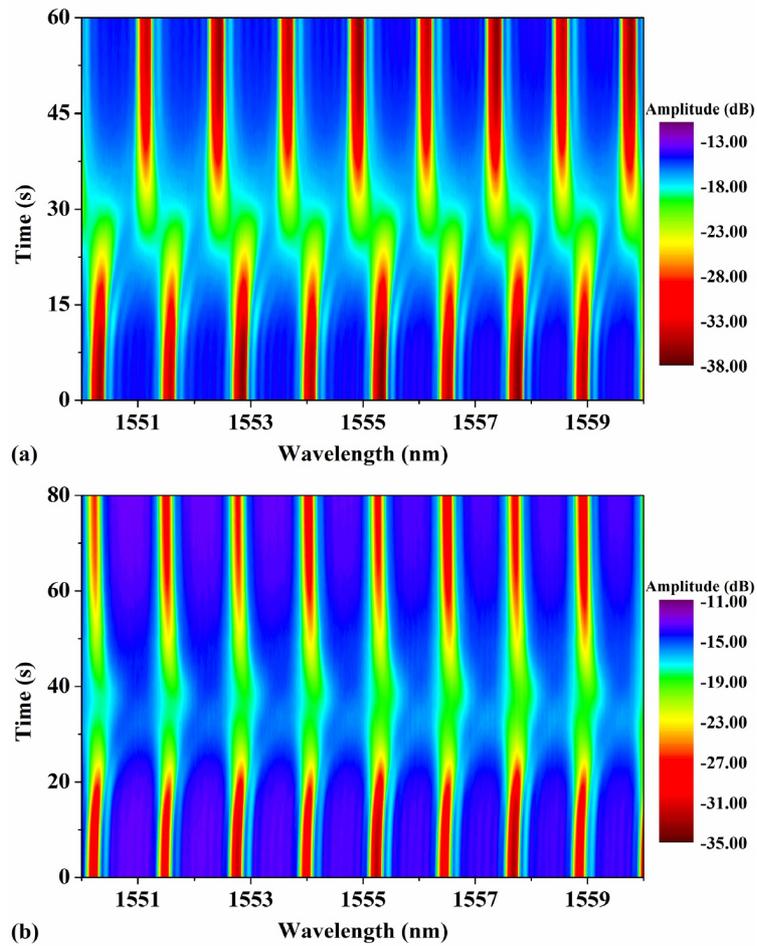

**Figure 4.6:** TFBG spectral evolutions during gold nanoparticles deposition under S- (a) and P-polarized (b) light (the color scale represent the amplitude of the resonances). Spectra are acquired every second and temperature-corrected with reference to the Bragg mode shifts.

Before the deposition, the polarization of the incident light was adjusted to the exact S- or P-polarization state using an electronic polarization controller (PC) (JDS

Uniphase) that contains one polarizer, a half-wave plate and a quarter-wave plate. This combination allows the preparation of arbitrary polarization states at the fiber input, which can compensate for any change of polarization state induced by fiber loops and twists in the optical path leading to the TFBG. The optimum launch polarization is simply determined by observing the transmission spectrum and maximizing either one of the two sets of transmission resonances. The spectral evolution was recorded continually during the deposition by an optical sensing analyzer (Micron Optics Si720) with a measurement frequency of 5 Hz. Even though S- and P-polarized spectra were obtained during separate experiments, it was verified that the results were reproducible for identical process parameters.

Figure 4.6 shows the evolution of the TFBG reflective transmission ranging from 1550 to 1560 nm (where the cladding modes have the largest amplitudes for 10° TFBG) with S- and P-polarized light. The origin of the time scale in Fig. 4.6 is determined by the moment when the precursor begins to evaporate, as indicated by a small pressure rise in the chamber (near 8 minutes into the process, as mentioned above). Basically, both the S- and P-polarized spectra go through a similar evolution in that all cladding modes become much attenuated at first, and then re-grow back to a shape very similar to the original one measured in air. There are subtle differences however: the initial attenuation occurs faster and lasts longer for P-polarized light, but most importantly the re-appearance of the resonances as the thickness increases is accompanied by a strong wavelength shift for S-polarized light but hardly any at all for the P-polarized case. This is further clarified in quantitative terms by plotting the amplitude and wavelength of one pair of resonances near 1559 nm as shown in Fig. 4.7.

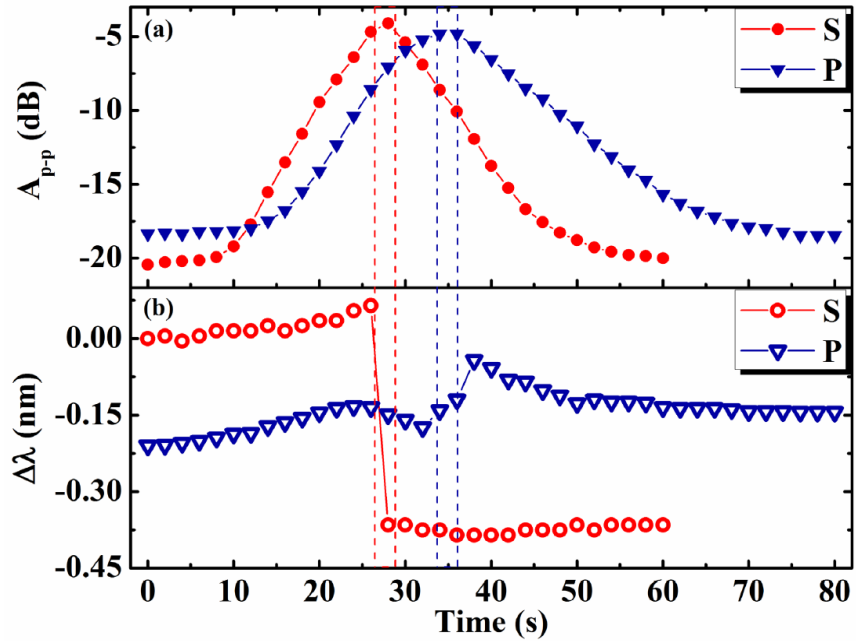

**Figure 4.7:** Normalized amplitude (a) and wavelength (b) evolutions versus time of gold deposition under S- and P-polarized light. The vertical dashed lines bound the minimum amplitudes for each polarization, highlighting the simultaneous occurrence of the corresponding wavelength change discontinuities.

The results indicate that the amplitudes of S-resonances decrease by 15 dB in the first 25 s of deposition and recover completely after 60 s. Still in the first 25 s, the resonance wavelength red-shifts by about ~0.08 nm. Most importantly, near the minimum amplitude point the wavelength discontinuously blue-shifts by ~0.4 nm and stays relatively constant for the remainder of the deposition, as the resonance gradually regains its full amplitude. The P-resonances also decrease in amplitude but by a slightly smaller amount and about 10 s later than the minimum of the S-resonance. The most significant difference however is that the wavelength shift of P-resonances is completely antagonistic to that of the S-resonance: it first increases slowly by about 0.06 nm, then decreases by 0.05 nm and when the amplitude minimum is reached near the 35 second point, jumps discontinuously upward by 0.14 nm. The net result is that there is an interchange of the S- and P-resonances during the deposition. This confirms our previous finding, obtained for continuous, 50 nm-thick sputtered gold films[50] that the S- and P- resonances were interchanged between air-clad and metal-clad fibers.

In order to further investigate the origin of the results obtained, we obtained scanning electron microscopy (SEM) images of the surface morphology of the gold nanoparticles deposited on three TFBG samples with different amounts of precursor in the vial (yielding different thicknesses and shorter process durations). Figure 4.8 shows the surface morphology of the growing gold film at beginning (a), middle (b), and end (c) of the whole TFBG spectral evolution, as determined by comparing the final TFBG spectra with those of the "full" process shown in Figs. 4.6 and 4.7. Also included (Fig. 4.8(d)) is an image of one of the fiber cross sections that were obtained to estimate the average coating thickness for each case. This correlation yields average thicknesses of 50, 100, and 200 nm for films equivalent to process durations of 20, 35, and 80 seconds on Fig. 4.8 (i.e. before the point of maximum attenuation, near the maximum and, finally later in the deposition). Average grain sizes were measured from SEM images to grow linearly with time from $55 \pm 3$ nm at 21 s, to $122 \pm 10$ nm at 50 s. Individual grains become difficult to identify beyond that time. The films are all quite rough but the metal coverage increases from very sparse (46.41%), to relatively dense (72.92%) (but with gaps remaining), and finally to complete, although with a rough top surface. Finally, two further TFBGs were prepared with thicknesses corresponding to deposition times of 20 s and 40 s, i.e. when the coated TFBG have the largest amplitude differences between S- and P-resonances according to Fig. 4.7.

The last two gratings were used to investigate how much of the cladding-guided light was scattered by the coatings under these circumstances. We used a broad band source (BBS), polarization controller, and optical spectrum analyzer (OSA) to achieve spectra with precise S- and P-polarized input light, and then imaged the IR-scattering emitted at right angle from the fiber axis with an infrared camera. The measured spectra and associated IR-scattering images with S- and P-polarized light are shown in Fig. 4.9. By comparing the spectra and the images of the two TFBGs, it is very clear that there is a strong correlation between the scattered light intensity and attenuation of the TFBG resonances (the brightest scattering occurring for Fig. 4.9(c), P-polarization, that also shows the strongest resonance amplitude attenuation). In comparison, uncoated TFBGs with similar grating periods look totally dark at these wavelengths because the evanescent field of cladding modes is non-radiative in the direction away from the fiber surface.

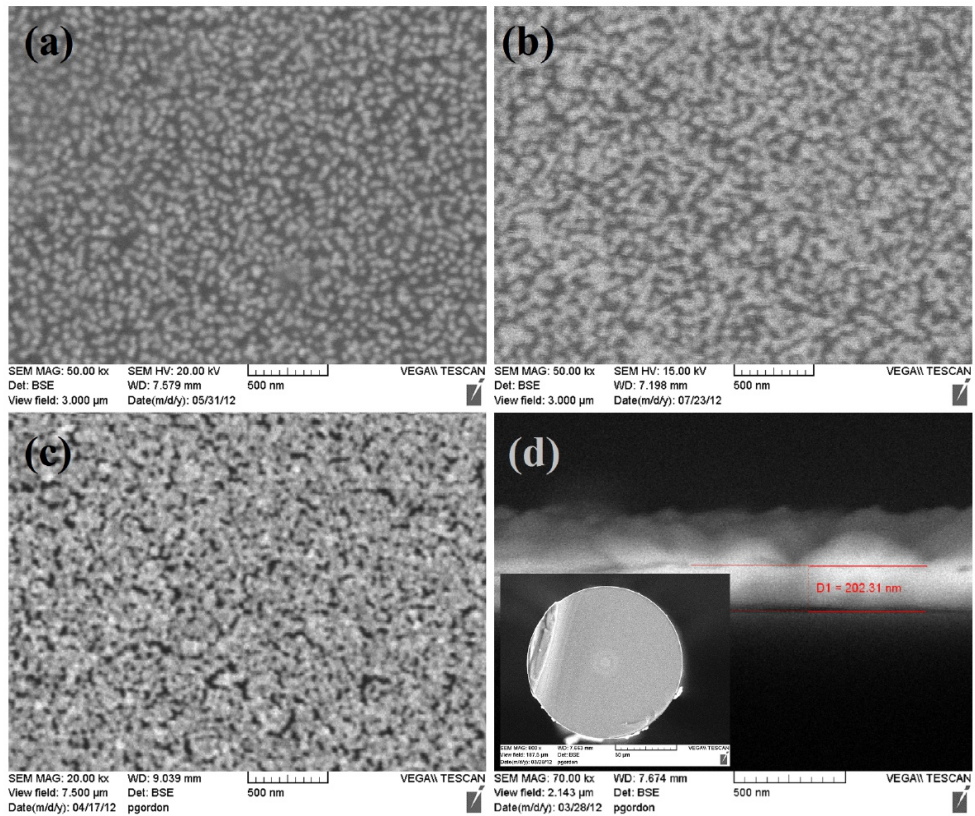

**Figure 4.8:** SEM images of gold nanoparticles deposited on the TFBG surface with different deposition times (with a scale of 500 nm). (a) Beginning, (b) middle, and (c) end of TFBG spectral evolution. (d) Sectional image of gold film in a small area with an inset of coated TFBG sectional view.

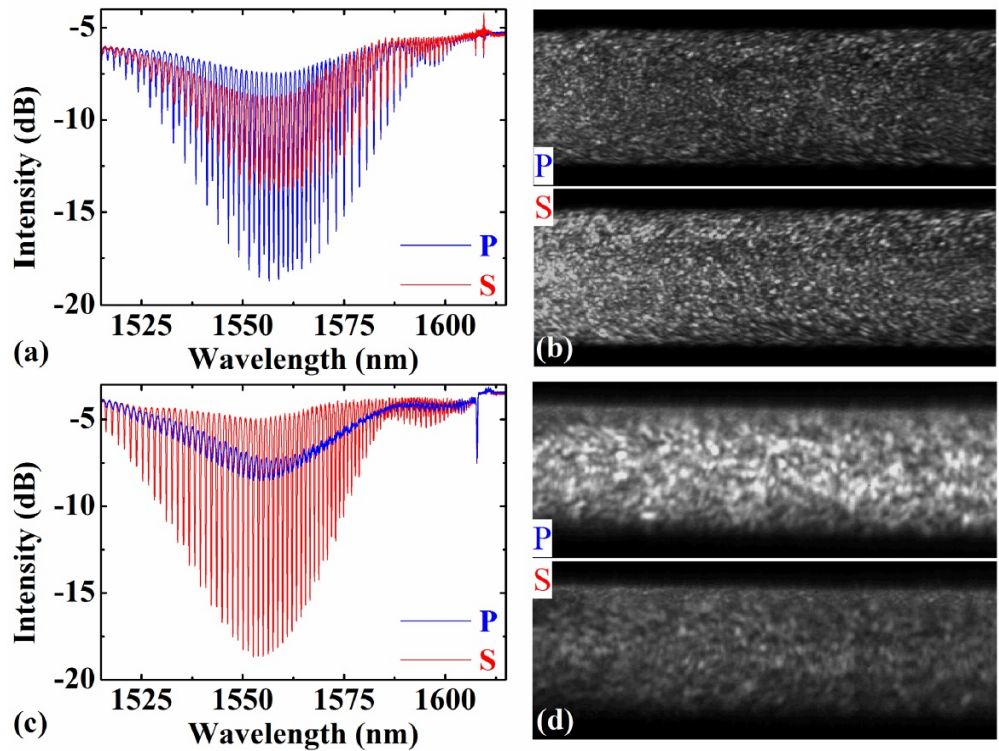

**Figure 4.9:** Reflective transmission spectra and IR-scattering images under S- and P-polarized lights of the coated TFBG with spectral response times of ~20 s (a), (b) and 40 s (c), (d).

## 4.5    Discussion

It is now clear that the attenuation of the amplitudes is mostly due to scattering by the metal nanoparticles rather than absorption of light by the metal. Furthermore we can extrapolate this finding by claiming that the full recovery of the resonance amplitudes (both S- and P-polarized) arises when the coatings "fill up" and no longer allow the light to leak out (this spectral recovery was also observed for thick electro-plated gratings in.[97] The almost full recovery of the resonance amplitudes also confirm that very little energy is dissipated in the metal itself, otherwise the resonances would remain lossy, losing amplitude and increasing in spectral width as a result.

The peculiar wavelength shift differences observed between the two polarization states can be explained by the granular nature of the films grown. Based on previous reports about the anomalous permittivity of thin metallic film,[49,52,83] the real part of the complex refractive index of a thin film composed of gold nanoparticles increases significantly above the value for bulk gold (a value near 0.55 at these wavelengths) when the particle densities are sparse and the sizes are small. So during the initial growth of the nanoparticles, a thin material layer with a relatively large average refractive index is formed on the cladding, resulting in a red-shift of both kinds of cladding mode resonances. Recalling from elementary electromagnetic theory that tangential fields cannot penetrate high-conductivity metals, S-polarized light (tangential to the cladding surface) can only exit the cladding through the coating material, while P-polarized light (normal to the interface) can tunnel across the gold particles and also exits from the air gaps. In order to confirm this idea and get a deeper insight into the experimentally observed process, the solution of the full problem of the interaction of the evanescent field of various cladding modes with the films shown in Fig. 4.8 will require exhaustive additional investigations. In the meantime, in order to validate some of the hypotheses brought forth to explain our experimental results, 2D-finite difference time domain (FDTD) numerical simulations of a much simplified configuration were carried out (with software from Lumerical Solution, Inc.). We modeled a linear array of conducting disks with a diameter of 40 nm and a center-to-center spacing of 50 nm (Fig. 4.10). The disks were surrounded with air and had a complex refractive index (n + ik) of 0.51 + i10.8. The simulation domain was meshed with cells on a 0.2 nm grid. The incident electromagnetic field was a plane wave pulse centered at wavelength of 1.5 μ m with its electric field vector **E** lying in the plane of the simulation domain. Two orthogonal cases were investigated: a) wave incident from the bottom of the image with the vector **E** oriented horizontally (corresponding to azimuthally polarized modes, from S-polarization coupling); b) wave incident from the left with the vector E oriented vertically (approximating a radially polarized evanescent field, i.e. a mode generated by P-polarized light).

The optical intensity distributions around the metal disks under these two excitation conditions are shown in Fig. 4.10. It is very clear that the light with the electric field tangential to the nanoparticle layer (S-polarization) is strongly localized between the metal particles but is forbidden to go through the particles themselves. On the other hand, light associated with P-polarized excitation of the TFBG (case b) has strong electric field maxima on the top and bottom of the nanoparticles while being "forbidden" in between because it is polarized tangentially to the metal there (and the gap is too narrow). Given these simulated results, we can infer that as the particles grow and the air gaps decrease, light from S-polarized resonances reaches a point, when most of the gaps between particles close, where it can no longer penetrate the nanoparticle coating and becomes bounded by what appears from the inside of the fiber as a continuous (i.e. bulk) metal film with a low refractive index of 0.5, hence the large sudden blue shift of the resonances. On the other hand, when the air gaps close the P-resonances continue to probe deeply into the metal and to continue to red shift slowly as long as their evanescent field can tunnel across the whole layer thickness (apart from some irregular behavior in the transition zone where the gaps become small before closing). The net result is that the presence of the nanoparticles introduces loss for the cladding modes propagating underneath. This loss is reflected in the attenuation of the cladding mode resonances (they become less pronounced, as seen on Figs. 4.6, 4.7, and especially Fig. 4.9(c)). Figure 4.9 clearly shows that most of the attenuation arises because of scattering, by opposition to absorption by the metal. If absorption was involved, the resonances would not recover their full amplitudes (see Fig. 4.7(a)) as the films become continuous and thicker. Finally, for the most strongly attenuated cladding modes (and strongest scattered light), the total net loss of the core guided mode for P-polarized light (as determined by its transmission spectrum) is only about 5 dB: this means that about 70% of the light is extracted from the core and transferred to the nanoparticles and their surroundings. Regardless of the validity of these hypotheses, the main conclusion remains that we have identified, from the TFBG spectral signature, the conditions where the HE (and TE) cladding modes become isolated from the fiber surroundings while allowing EH (and TM) modes to tunnel across and scatter efficiently off the nanoparticles.

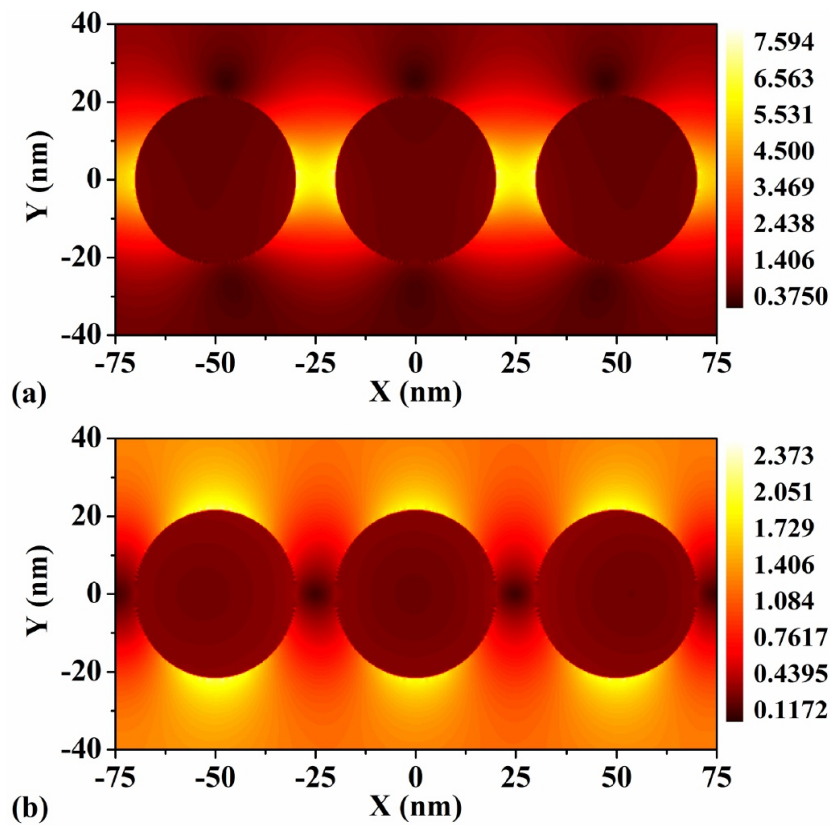

**Figure 4.10:** Simulated intensity distributions of electric fields on single layer gold nanoparticles interacted by S- (a) and P-polarized (b) light in free space. The incident light propagates along Y-axis in S case, while along X-axis in P case.

## 4.6        Conclusion

We have demonstrated that there is a significant correlation between the near infrared polarized transmission spectra of 10º TFBG inscribed in standard single mode fibers and the optical properties gold particle coatings with thicknesses ranging from zero to ~200 nm, and formed using a recently developed chemical vapor deposition process. In particular, we identified spectral signatures corresponding to coatings yielding different states of light confinement, scattering, and polarization. For instance, for films with particle sizes near 50 nm and 100 nm respectively we can selectively scatter out of the fiber light polarized predominantly azimuthally or radially and thus control the plasmonic enhancement properties due to these disordered nanostructured metal films.[101,102] However, the optimum film thickness and uniformity will vary from application to application and the scattering properties further depend on the size of individual nanoparticles and the porosity of the films, parameters that cannot be controlled independently at the present time with this method. What we show here is that the film thickness can be controlled by stopping the deposition before the precursor is exhausted and that several film conditions can be obtained as a result. Further experiments with different process temperatures, residual pressure in the chamber, or additional reactants would provide a richer parameter space but the results obtained so far are sufficient for developing interesting devices. It is to be pointed out that the purpose here was not to control the polarization of the guided light, as can be done very efficiently using uncoated TFBGs with large tilt angles,[103] or by using graphene coatings on side polished fibers for instance,[104] but rather to excite coatings on fibers with radially or azimuthally polarized light at infrared wavelengths. As a side benefit, we have also confirmed the usefulness of the TFBG as a process monitor for thin film coatings of CVD gold on other substrates in the same process chamber, *in situ* and in *real time*.

# Chapter 5

## Effective Permittivity of Ultrathin Chemical Vapor Deposited Gold Films on Optical Fibers at Infrared Wavelengths

*Modified from the original manuscript published as:*

Zhou, W.[2] ; **Mandia, D.J.**[1]; Griffiths, M.B.E.[1]; Barry, S.T.[1]; Albert, J.*[2] "Effective Permittivity of Ultrathin Chemical Vapor Deposited Gold Films on Optical Fibers at Infrared Wavelengths" *J. Phys. Chem. C.* **2014**; *118*(1); pp. 670-678.

[1] Department of Chemistry, Carleton University, 1125 Colonel By Drive, Ottawa, Ontario, Canada, K1S 5B6

[2] Department of Electronics, Carleton University, 1125 Colonel By Drive, Ottawa, Ontario,

Canada, K1S 5B6

*Corresponding author

## 5.1       Abstract

The geometry- and size-dependent effective medium properties of ultrathin gold films deposited on the bare cladding of single mode optical fibers by chemical vapor deposition are characterized by measuring the polarized transmission spectra of in-fiber gratings at wavelengths near 1550 nm. The real part of the complex refractive indices of films with average thicknesses ranging from 6 to 65 nm are about 10 times higher than that of bulk gold at these wavelengths, while the imaginary part values are 2 orders of magnitude lower. The films are essentially isotropic, apart from a small increasing dichroism between the in-plane and out-of-plane component of the imaginary part of the refractive index at thicknesses larger than 25 nm. Unlike gold films prepared by other means, the optical properties of the coatings do not converge rapidly toward bulk values at thicknesses larger than 10 nm but remain characteristic of gold films prepared by very slow physical deposition processes. The modified Clausius−Mossotti theory for anisotropic structures was used to confirm that the observed properties arise from a persistent granularity of the film at larger thicknesses, with metal filling fractions increasing from 30% to 68% and particle aspect ratios from 0.8 to 1.0 (spherical). These conclusions are supported by nanoparticle shape measurements obtained by atomic force microscopy and scanning electron microscope images.

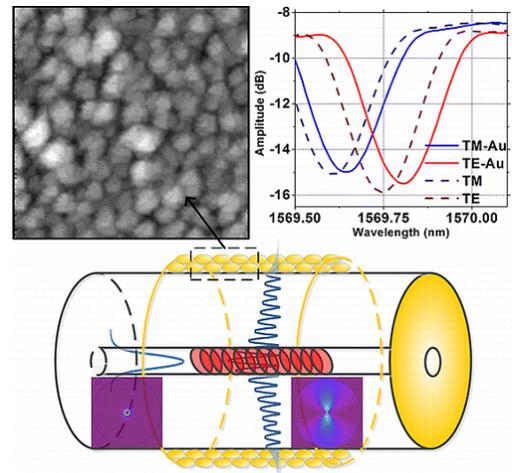

**Figure 5.1:** Table of contents image used for the publication in Chapter 5.

## 5.2 Introduction

The electromagnetic field intensity in the immediate vicinity of various ultrathin metal films and nanostructures can be enhanced by several orders of magnitude.[105–109] This phenomenon has led to a number of applications in plasmonic sensing,[91,110] surface-enhanced Raman scattering,[111] and thermoplasmonics.[112] There are many of those applications that can be implemented using coated optical fibers, and it is important in such cases that the thin metals films be coated uniformly around the fiber circumference. This is most easily done using liquid phase or gas phase chemical deposition techniques, because the more widely used thin film coating techniques such as evaporation and sputtering cannot reliably produce uniform thicknesses around curved surfaces. Once fabricated, the intensity and wavelength dependence of the plasmonic enhancement are related to the material and geometric parameters of metallic nanostructures: the size,[106,107] pattern,[108,109] as well as the permittivity of the metal and surrounding material.[105,113] While size and geometric patterns can be extracted with modern imaging and surface profilometric tools, the complex permittivity is typically obtained from the reflection and transmission measurements of the metallic film with analysis using Drude model parametrization and Kramers−Krönig relations.[114–116] However, it is now clear that these approaches, which were developed for bulk (i.e., continuous) metals films, are less suitable for ultrathin metals[117] or aggregates of metal nanoparticles (NPs) with individual sizes of the order 50 nm and less.[118] In the latter case, it was suggested to modify the collision frequency of free electrons by an approximate, empirically determined size-dependent parameter in the Drude model.[119] However, it remains difficult to measure the complex permittivity of such nanostructures directly without resorting to various theoretical models for the electromagnetic mixing formulas between the NPs and their surroundings.[120] It is therefore desirable to investigate the permittivity or complex refractive index of ultrathin metal film by several different optical methods. Spectroscopic ellipsometry can obtain the complex permittivity of homogeneous thin films, with decreasing accuracy and resolution as the films get thinner.[83,121] Furthermore, when the films are not isotropic this method becomes significantly less reliable and requires nontrivial computations to extract the five parameters of interest (real and imaginary part of the refractive index for TE and TM light, plus the thickness if it is not determined by other means). For metallic thin films, i.e., for films with a negative real part of the permittivity, the surface plasmon resonance (SPR) technique can also be used to investigate the complex refractive index and thickness of metal films,[49,122] but only for TM polarized light and only for a limited range of dielectric refractive indices for the substrate and cover materials of the thin films. In particular, SPR cannot be excited at an ultrathin metallic film−air interface at infrared wavelengths because the anomalous size-dependent permittivity of the metal becomes positive.[123] In a more recent advance, interferometric picometrology was developed to measure the in-plane component of the complex refractive index of ultrathin gold films with a single wavelength, linearly polarized light beam at normal incidence.[124] In this new method, the complex index can be calculated from the amplitude change and phase shift of the complex reflection coefficient. A common issue with all thin film measurement techniques is that they are not well suited for curved surfaces. For the geometry considered here, a conformal coating on the surface of a 125 μm diameter optical fiber, the curvature represents a challenging measurement problem. Finally, for very thin films that perturb electromagnetic waves very little, the best way to increase the measurement accuracy is to use frequency-domain techniques, such as measuring the frequency shifts of resonator structures that have high quality (Q) factors.[125–127] It is this type of approach that is presented here with a resonating waveguide structure that is especially easy to fabricate and to measure.[41] The objective of the present study is twofold: the development of a technique to measure the permittivity of thin metal films on optical fibers and the characterization of such coatings made by a relatively high-temperature chemical vapor deposition (CVD) process.

The polarized evanescent electric field of resonantly excited optical fiber modes is used to probe ultrathin gold films deposited on the surface of the fiber in order to investigate the effective medium properties (refractive index and absorption) of the films at infrared wavelengths. Conformal gold films with different thicknesses were deposited on the cladding of optical fibers by CVD, as previously reported for other metals or gas precursors.[52–54] The excitation of the probing optical modes was carried out with a tilted fiber Bragg grating (TFBG) inscribed in the core of the optical fiber over lengths of several millimeters. The grating couples up to 90% of the light launched in the core to individual cladding guided modes and thus provides high signal-to-noise ratio resonances with Q-factors (resonance wavelength/resonance width) of ~10000.[41] Changes in the spectral properties of these cladding mode resonances were used to measure the complex permittivity of the coatings averaged over the full length of the gratings (5 mm in these experiments). The method further allows the separate measurement of film properties for TE (in-plane electric field) and TM light (out-of-plane) and as a result can detect the presence of birefringence or dichroism. These polarization-resolved measurements were obtained by exciting separately cladding-guided modes with azimuthally or radially polarized electric field at the cladding-coating boundary, which correspond respectively to TE and TM light polarizations for the electric field penetrating into the deposited coatings.[61] In parallel, scanning electron microscope (SEM) and atomic force microscope (AFM) images of the films that were deposited were used to provide shape, roughness, and average thickness data, thus completing the characterization of the films. An effective medium model for the deposited thin films was used to determine their complex permittivity in terms of the measured parameters. Finally, we extracted the metal volume filling fraction and aspect ratio of the NPs in the films by modeling the measured effective medium permittivity of the films composed of ellipsoidal gold NPs under TE and TM polarizations with a modified Clausius−Mossotti equation.

## 5.3               Experimental Section

### TFBG Probe Fabrication

TFBGs were written in hydrogen-loaded CORNING SMF-28 fiber with a pulsed KrF excimer laser using the phase-mask method.[41] This fiber is a standard telecommunication grade fiber that guides only one core mode at wavelengths between 1300 and 1620 nm. The grating period (555.2 nm) is chosen to reflect the core guided light at the "Bragg" wavelength around 1610 nm. At shorter wavelengths between 1520 and 1600 the core-guided light is coupled to the 125 μm diameter cladding of the fiber, which acts as a highly multitude waveguide. Figure 5.2 shows the reflected transmission spectrum of a TFBG in air prior to coating. Each of the resonances corresponds to a wavelength at which core-guided light is coupled to a specific group of polarized cladding modes. The length of the TFBG was chosen to be only 5 mm long in order to minimize the impact of an eventual thickness gradient on the measured quantities if the deposition was not uniform. A tilt angle of 10° was chosen in order to excite a large number of strong, high-order cladding mode resonances across the measurement range of our instrumentation (centered around 1550 nm). While TFBGs are normally measured in transmission, we implemented a reflective sensing configuration in order to have a single entry port in the deposition chamber and to avoid bending the fiber (Figure 5.3). This configuration was implemented by cleaving the fiber 1.5 cm downstream from the TFBG and coating the flat end with a sputtered gold mirror. While the outside face of the end mirror also gets coated during CVD, this has no influence on the measured properties as the mirror is thick enough to be impervious to additional coatings. A consequence of the reflective configuration is that the core-guided light goes through the grating twice and therefore that the grating power transfer function is squared.

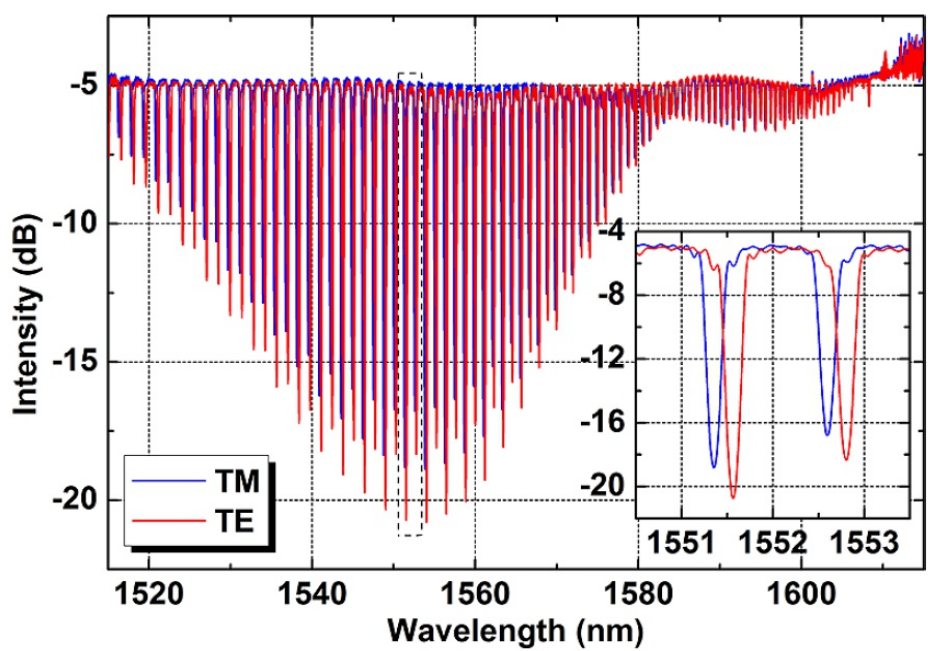

**Figure 5.2:** Typically reflected transmission spectrum of bare TFBG under TM and TE polarizations.

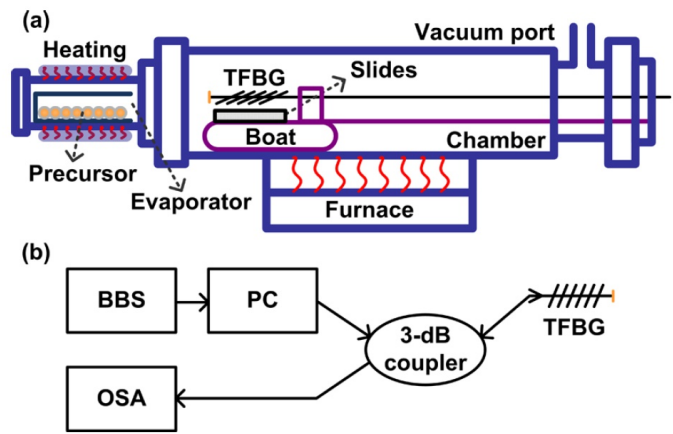

**Figure 5.3:** Schematic diagram of experimental process: (a) gold CVD setup for optical fiber; (b) TFBG spectral measurement system. A source (BBS) followed by a polarization controller generates a broad spectrum of light that is linearly polarized either parallel (TM) or perpendicular (TE) to the tilt plane of the grating. The core-guided light going through the grating is reflected by a gold mirror, goes through the grating a second time, and is detected by an optical spectrum analyzer (OSA).

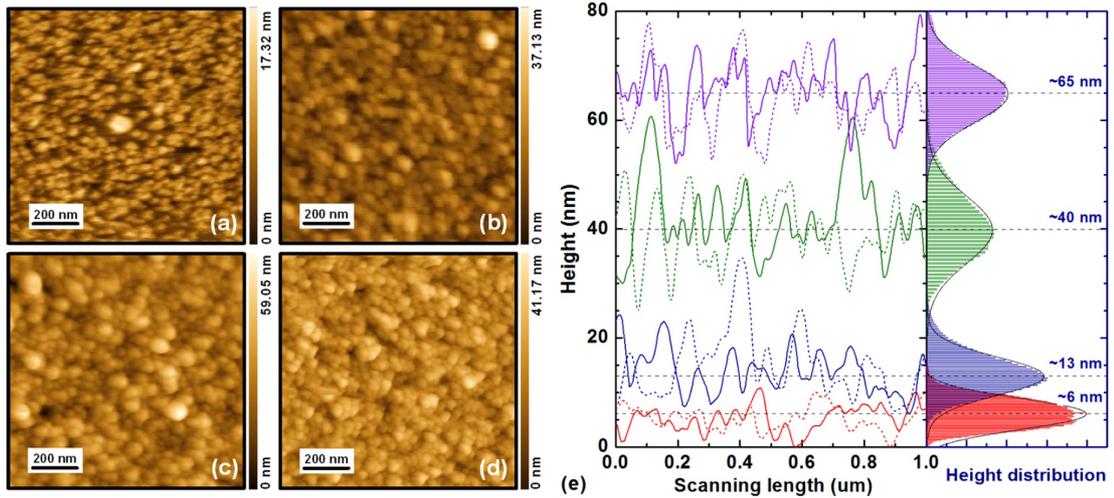

**Figure 5.4:** (a−d) AFM images of gold CVD films deposited on planar slides with effective thickness of 6, 13, 40, and 65 nm. (e) Two sample height profiles as well as height histograms of the AFM images shown in (a−d). The centers of the effective height distributions are displaced by the average step height of the film for each sample.

## Sample Preparation

Figure 5.3a illustrates the experimental setup of the gold CVD system. The fiber was located in a stainless steel deposition chamber, which itself was mounted within a furnace. An evaporator containing a vial loaded (under $N_2$ ) with a gold(I) iminopyrrolidinate ([Au(Me$_2$ -$^t$Bu-ip)]$_2$ ) CVD precursor was situated at the end of the chamber and surrounded by heating tape for external resistive heating by a Variac. At the beginning of the experiment the chamber was evacuated to a base pressure of 5 mTorr, and then the furnace temperature (substrate temperature) was set to 350 ° C. After the substrate achieved constant temperature, the evaporator was heated to a temperature of ~ 160 ° C to initiate volatilization of the precursor. Onset of deposition occurs when the evaporator temperature reaches ~130 ° C, at which point the vaporous precursor saturates the chamber, effectively depositing on all exposed surfaces including the cladding of the TFBG. The deposited precursor then thermally

decomposes into pure gold by a reaction that is still under investigation but that was confirmed to occur at temperatures above 300 °C.[27]    High-resolution X-ray photoelectron spectroscopy (XPS) analysis performed on witness slides coated with this gold material showed no significant or detectable impurities. Since the gold precursor is thermally stable between 150 and 300 °C, the thickness of the gold film can be controlled by adjusting the amount of precursor, the temperature of the deposition chamber, and the location of the samples to be coated relative to the inlet of the evaporator into the deposition chamber. A small piece of bare optical fiber and witness slides were coated simultaneously for *ex situ* microscopic imaging of the surface morphology of the gold films. These additional samples were located 3− 5 mm away from the TFBG to ensure a similar coating environment.

## Optical Measurements

The measurement system is shown in Figure 5.3b The fiber-coupled broadband source (JDSU BBS1560) launches light from 1520 to 1620 nm in the same type of single mode telecommunication fiber used for the TFBG. Even though the fibers are not of the polarization- maintaining type, a polarization controller (JDSU PR2000) can be used to generate a state of polarization that will evolve to a linear state either parallel (p-polarized, or TM) or perpendicular (s-polarized or TE) at the grating. As long as the fiber paths are not moved during the experiments, polarization remains stable. As indicated earlier, when the polarization state of incident light at the TFBG contains only TE- or TM-polarized light, the electrical field of the excited high-order cladding modes are predominantly azimuthal or radial at the cladding boundary, respectively, and the associated TFBG resonances occur at different wavelengths.[41,61] The light transmitted through the grating twice returns toward the light source in the same fiber, and a 50:50 fiber coupler picks off part of the reflected light to be measured by an ANDO AQ6317B optical spectrum analyzer with a nominal resolution of 0.01 nm.

By using the TFBG spectral measurement system shown in Figure 5.3b, the polarization-dependent permittivity of the thin gold film can be interrogated by in-plane (TE) and out-of-plane (TM) light and measured very easily by the polarization-dependent spectra of the TFBG. The system is completely fiber-coupled, and therefore no optical alignment is required.

While TFBGs can usually be used to monitor coatings formation (and *in situ* process temperature) during deposition,[52,54] it was not possible here. The thermal decomposition temperatures exceed the range over which FBGs are stable and therefore induce irreversible spectral changes upon return to room temperature. Therefore, all results shown here compare a coated and bare TFBG, with the bare spectrum measured after removing the coating (instead of prior to deposition), in order to remove the effect of the thermally induced spectral changes. Separate TFBGs were used for each coating thickness, and different thicknesses were obtained by changing the amount of gold precursor in the evaporator. Coatings were removed by etching in Aqua Regia (a fresh mix of nitric acid and hydrochloric acid with volume ratio of 1:3). This methodology yields accurate comparative spectral information about the wavelength shift and amplitude attenuation of the TFBG resonances induced by the gold coatings.

**Microscopy**

Scanning electron microscopy (SEM) and atomic force microscopy (AFM) were performed on Tescan and Agilent (Molecular Imaging) PicoSPM II systems, respectively. Imaging of fiber samples and coated witness slides was carried out by SEM for grain size analysis and AFM for surface morphology and thickness. In view of the roughness of the deposited coatings, the thickness data was calculated from the average step height at the boundary of films interrupted by abrasion of a small section.

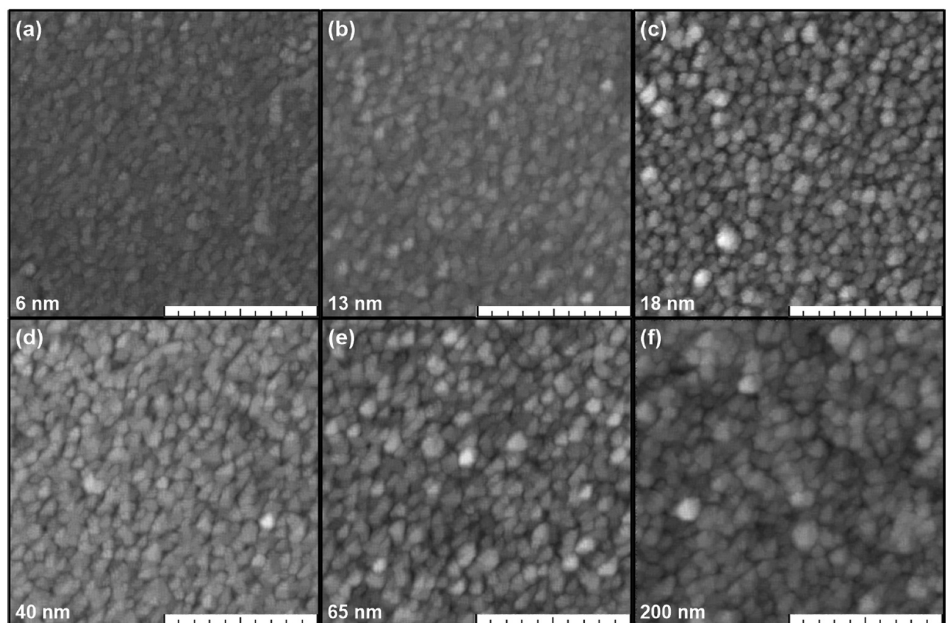

**Figure 5.5:** (a−f) SEM images of gold CVD films coated on optical fibers at various effective thicknesses (indicated at bottom-left of each image). The scale bar is 500 nm.

**Table 5.1:** Average thickness, thickness distribution (standard deviation and full width at half-maximum (fwhm) of the gaussian fitting curve of the height histograms), and average lateral size of gold NPs (± 5 nm) extracted from AFM and SEM images of CVD gold samples

| average thickness/nm | 6 | 10 | 13 | 18 | 25 | 40 | 65 | 200 |
|---|---|---|---|---|---|---|---|---|
| standard deviation/nm | 2.8 | 2.2 | 3.5 | 5.1 | 3.6 | 6.2 | 5.1 | 5.5 |
| fwhm/nm | 6.7 | 5.3 | 8.2 | 11.9 | 8.4 | 14.5 | 12.0 | 13.0 |
| lateral size/nm | 36.2 | 40.1 | 42.6 | 54.7 | n.a. | 50.6 | 61.5 | 77.6 |

## 5.4 Results

### Surface Characterization of Gold CVD Films

Figure 5.4 shows AFM images (a− d) and effective height profiles (e) of the gold films with various effective thicknesses. The effective height profiles were obtained by adding the average step height from scratched samples to the middle of the AFM surface height distributions of undisturbed coatings. As evidenced on the AFM images, the gold films are all composed of roughly spherical NPs regardless of average thickness. The thinner film (6 nm) appears to be made up of individual NPs (because the height distribution extends over ± 6 nm around the mean), while the AFM profiles of thicker films do not penetrate down to the substrates. In fact, the size distribution is nearly identical (± 15 nm) for the 40 and the 65 nm thick films. Furthermore, the lateral size of the NPs on the AFM images does not differ much (apart from the thinnest film). In order to eliminate possible measurement artifact due to the convolution of the AFM tip size with the profiles of the NPs, SEM analysis was made on the flat witness slides for each thickness, as shown in Figure 5.5. It is important to note that the apparently very high curvature of the fiber surface (1/(62500 nm)) should have no effect on the deposition rate for the effective thicknesses used here (6−200 nm). ImageJ software was used to measure the average lateral sizes of representative samples of the gold NPs at each thickness. The average lateral sizes (approximate diameters) are tabulated in Table 5.1 along with the thickness parameters extracted from the AFM measurements. The major size differences occur for the thinnest and thickest films measured (6 and 200 nm, respectively), with somewhat more modest changes (lateral sizes between 42.6 and 61.5 nm) for thicknesses between 13 and 65 nm. It is very clear that thin gold films deposited by this kind of CVD precursor grow by NP nucleation and do not become more uniform, even for the 200 nm thick gold film shown in Figure 5.5f.

## TFBG Spectral Responses

Figure 5.6 shows the experimental wavelength shifts and amplitude attenuations of a pair of cladding modes (one TE and one TM) with effective indices near 1.378. For each resonance, the peak-to-peak amplitude ($A_{p-p}$) is defined as the difference between the minimum transmission of a resonance and the nearest maximum in the spectrum, while the wavelength is determined as the midpoint of the spectral dip. These measured values are used to calculate changes in the effective index of the mode through eqs 1 and 2. The effective index is a measure of the mode phase velocity along the axis of the fiber, and it depends on the wavelength, the material indices, and the waveguide geometry. A mode effective index of 1.378, combined with the refractive index of the fiber cladding (1.4440), corresponds to light from this mode striking the cladding− coating boundary at an angle of incidence of 72º. The addition of a coating on the fiber leads to a change in the real and imaginary parts of the mode effective indices and as a result to a shift of the transmission spectrum resonances (in wavelength and amplitudes). Part (a) of Figure 5.6 shows clear spectral shifts, different for TE and TM modes, even for films thicknesses as small as 6 nm. The TE mode resonance shifts by 1.5 nm as the film thickness increases to 65 nm, while the TM resonance shifts by only 0.4 nm. This information is not sufficient however because the gold coating also introduces optical loss. In order to measure the imaginary part of the complex permittivity of the film, the increase in mode propagation loss must be measured. This is done by measuring the resonance amplitude, which depends on the grating length, coupling coefficient, and mode loss. The first two parameters do not change when a thin coating is applied (the coupling coefficient could change slightly with loss, but this is a second-order effect, negligible for the shifts observed here), so the amplitude shifts can be directly linked to the addition of the lossy coatings. Since different gratings were used for each thickness, the starting values of the TFBG resonance corresponding to the desired cladding mode are different (histograms in Figure 5.4b). As a result, the amplitude changes of the resonances appear irregular (line-symbol curves in Figure 5.4b. Fortunately, this is not a problem as the algorithm to extract mode loss from the amplitude change is independent of the initial values of the resonance amplitude.[128]

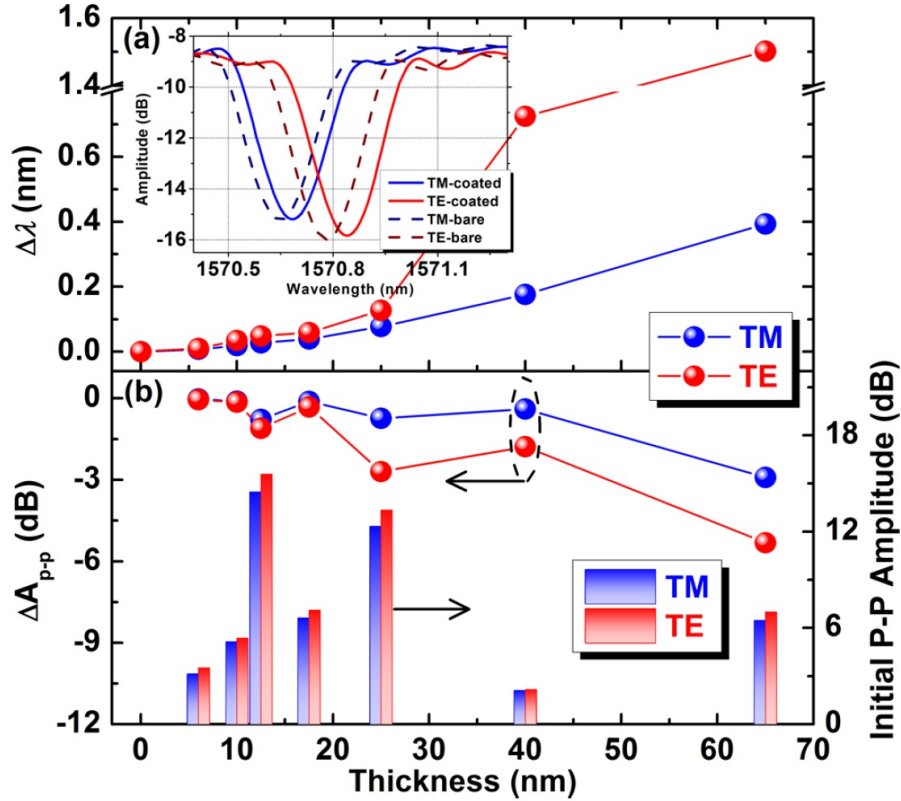

**Figure 5.6:** Relative wavelength shift (a) and peak-to-peak amplitude attenuation (b) of TE- and TM-polarized cladding mode resonances versus the effective thickness of gold film. The initial peak-to-peak amplitudes of the bare gratings are shown as a histogram in (b). The inset shows the measured spectral changes of the pair of cladding modes used for the 18 nm thick film.

**Theoretical Model for Complex Refractive Index of Gold CVD Films**

In order to calculate the complex refractive index ($n_f - ik_f$) of the gold NP films from the TFBG data, an effective medium approach is used, whereas an equivalent uniform film is used for the NP coating (note that we use the convention of a positive $k_f$ for optical loss in choosing the sign of the imaginary part in the expression for the complex refractive index). The thickness of the equivalent film is chosen to be equal to the average step height of the corresponding actual coatings. Figure 5.7 shows this correspondence schematically. Once this is done, the real and imaginary parts of the propagation constant of the optical modes guided by the structure of Figure 5.7b can be calculated and compared to the experimental data. The mode propagation constant is equal to $2\pi N_{eff}/\lambda$, where $\lambda$ is the wavelength of the light and $N_{eff}$ the mode "effective index" $\left( N_{eff} = N_{eff}^r - i N_{eff}^i \right)$, i.e., the quantity that is monitored here. The aim of the simulation is to find the complex refractive index of the gold NPs film $(n_f - ik_f)$ that

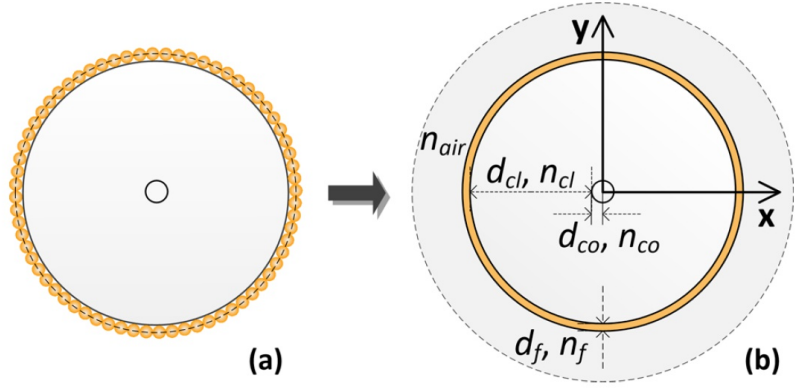

**Figure 5.7:** Cross section of multilayer model of optical fiber with gold NP film (dash line indicates the average thickness of the gold film) (a) and uniform gold film with equivalent thickness (b).

reproduces the spectral shifts shown in Figure 5.6, for each film thickness and for each mode polarization. Separate calculations for TE and TM modes allow the determination of any anisotropy (birefringence in $n_f$ or dichroism in $k_f$) in the deposited films. The data extraction algorithm proceeds as follows:

**1.** Find the effective indices (real in this lossless case) of the pair of modes of the bare fiber that were used in the experiments, using the following formula for the real part of the effective index $N_{eff}^r$ [41]

$$N_{eff}^r = \frac{\lambda_{cl} \cos \theta}{\Lambda - n_{co}} \qquad [1]$$

where $\lambda_{cl}$ is the wavelength of the cladding mode resonance, $\theta$ is the tilt angle of the grating planes, $\Lambda$ is the grating period, $n_{co}$ is the effective index of the core mode.

**2.** Using the grating length (using L = 10 mm because the light goes through the 5 mm long grating twice in this reflective configuration) as a fixed parameter, find the coupling constant $\kappa$ of the mode from the initial peak-to-peak resonance transmission amplitude T (corrected from the dB scale to linear units) using the following formula

$$\kappa = \frac{\tanh^{-1}(1-T)}{L} \qquad [2]$$

**3.** Use a " mode solver" to calculate the modes of the structure shown in Figure 5.7b for the $d_f = 0$ case and find the pair of modes with the value of $N_{eff}^r$ found in part 1. A commercial complex mode solver based on vectorial finite differences was used for this purpose (FIMMWAVE, by Photon Design). The fiber parameters, which remain fixed throughout the process, are the following: the thickness (radius) of core, $d_{co} = 4.15 \ \mu m$; the thickness of cladding, $d_{cl} = 58.35 \ \mu m$; the thickness of surrounding air layer is set as 17.5 μm; and the corresponding refractive indices of every layer at the wavelength of 1570 nm are $n_{co}$=1.45033, $n_{cl}$=1.44402, and $n_{air} = 1$.

**4.** For each thickness and wavelength shift value reported in Figure 5.6a, find the measured shift in $N_{eff}^r$ by eq 1.

**5.** Similarly, from the change in T , determine the imaginary part of the effective index of the mode of the coated fiber at each thickness. The formula relating $N_{eff}^i$ to T is a little more complicated, and we use a commercial grating solver for this purpose (Optigrating, by Optiwave).

**6.** The calculated complex effective indices of the modes at each thickness are reported in Figure 5.8. The final step is run the mode solver again, with the added equivalent layer, and to look for values of $n_f$ and $k_f$ that reproduce the shifts in the complex effective index of the modes found in steps 4 and 5. The final results are plotted in Figure 5.9.

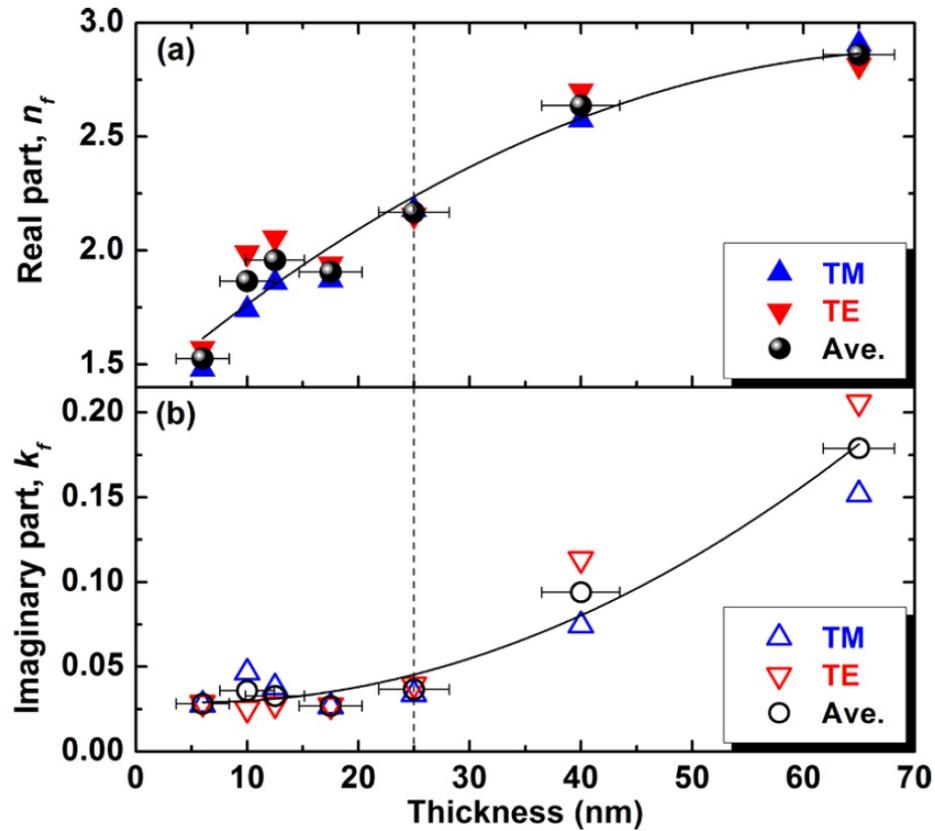

**Figure 5.9:** Calculated real (a) and imaginary (b) parts of complex refractive indices of gold NP film in TE- and TM-polarizations, and corresponding average values versus effective thickness. The dashed line at 25 nm represents the mean free path of electrons in gold. The horizontal error bar for each thickness is obtained from the standard deviation of the Gaussian fitting curve of the height histograms shown in Table 5.1.

**5.5      Discussion**

It is clear from Figure 5.8 that both the TE- and TM-polarized resonances shift to longer wavelengths for these thin gold films, but with a different shift magnitude, a difference that increases with thickness over the range shown here. If the coating refractive index was that of bulk gold (0.5− i10.9 at these wavelengths),[114] the wavelength shift should have been to shorter wavelengths because of the decrease in the refractive index at the cladding boundary. The overall mode loss also increases, which is expected from the addition of a lossy coating, by different values depending on polarization. However, these measurements can be somewhat misleading as they not only depend on the complex refractive index of the added film but also on its thickness, as well as on the polarization dependence of the fiber wave-guiding properties. Figure 5.9 reveals instead that the complex refractive index values for the film as probed by the TE and TM modes are essentially identical and therefore that the films are optically isotropic, apart from a weak increasing dichroism for thicknesses above 25 nm. The line in Figure 5.9 follows the average complex refractive index calculated from the TE and TM mode probe results. On the basis of the statistically determined measurement uncertainty of 1 pm in the central wavelength and of 5e−3 dB in the peak-to-peak amplitude of TFBG cladding mode resonances, the worst case error in the complex refractive index calculation was ± (0.03− i0.005), obtained for the case of the 6 nm thick gold film. This final uncertainty is smaller than the size of the plotted points in Figure 5.9. The most striking feature of Figure 5.9 is that the complex refractive index measured here differs markedly from the bulk values for gold at these wavelengths. In particular, and this is a feature that has important practical consequences, the high value of the real part of the film index relative to the imaginary part means that the real part of the film's permittivity $(\varepsilon_f^r = n_f^2 - k_f^2)$ is positive and therefore that these films cannot support the propagation of surface plasmon polaritons (SPP) (unlike gold films deposited by other methods including evaporation, sputtering, and electroless plating).[129] Also, the films deposited by CVD are 100 times less lossy than bulk gold for thicknesses below 20 nm, meaning that very little power is transferred from the fiber to the coatings.

These results also differ from those obtained for gold films of similar thicknesses deposited by other means. In particular, it has been shown that evaporated gold films have optical properties at 1550 nm wavelengths that were essentially identical to those of bulk gold, down to 20 nm thicknesses.[130] For thinner films down to 15 nm the imaginary part of the index decreases rapidly, and the real part increases from 0.5 to 1.0 (but the real part of the permittivity remains negative). The deposition rate for these evaporated films was 3 nm/min. Another recent paper studied even thinner evaporated films (deposited at room temperature with a deposition rate of 12 nm/min) and showed that the transition from discontinuous (low loss, high real index) to continuous (bulk) behavior occurred for thicknesses between 2 and 10 nm (results obtained at visible wavelengths, however).[124] There is a large body of literature, dating back to at least 1950, which indicates that these differences can be attributed to the morphology of the gold films deposited by the various methods [124,131] relative to that of the CVD gold films shown in Figure 5.5. It is clear that the CVD films do not have a tendency to morph from isolated NPs into continuous clusters of increasing size as the deposition progresses, as observed generally. The slow deposition rate (estimated to be near 1 nm/min) and high substrate temperature of 350 °C in this gold CVD process[31] yield a lower optical loss and larger (real) refractive index, due to the highly fragmented nature of the films which continue to grow via nucleation instead of aggregation. The behavior shown here was observed previously only for films deposited at very slow rates[132] or with high substrate temperatures.[132–134] It is interesting to note also the effective complex refractive index of films made up of very sparsely distributed gold NPs (10% volume fractions for 10 nm diameter particles distributed randomly on a flat surface) was measured to lie near $(1.7 - i0.1)$ at visible wavelengths,[83] almost identical to what was obtained here for the much denser 11 nm thick films, albeit at infrared wavelengths. Finally, Figure 5.9 shows the beginning of a saturation of the growth in the real index, an increase in loss, and the beginning of dichroism at a thickness of 25 nm (indicated by the vertical dashed line in the figure) which correspond to a generally accepted mean free path of electrons in gold.[124] It thus appears that when the nanoparticulate films reach this thickness, some paths form for electrons to transit between NPs, but very few.

In order to further investigate the electromagnetic properties of the films produced, we now proceed with an evaluation of the inherent optical properties of the NPs by modeling the films using the Clausius−Mossotti model. For the most general case of anisotropic complex permittivity, it is assumed that the CVD gold film is composed of gold NPs with an oblate spheroid shape (according to the information provided in Table 5.1). Such NPs have different polarizabilities in the major and minor axes. Figure 5.10 illustrates how the evanescent field of TE and TM cladding modes propagates through these spheroids on the surface of the fiber. The major (a) and minor (b) axes of the oblate gold spheroids are aligned with the TE and TM electric fields, respectively. The generalized Clausius−Mossotti equation for a layer of metal spheroids embedded in an insulator material is[135]

$$\frac{\varepsilon_{eff} - \varepsilon_i}{L_m^{TM,TE}\varepsilon_{eff} + (1 - L_m^{TM,TE})\varepsilon_i} = \phi \frac{\varepsilon_m - \varepsilon_i}{L_m^{TM,TE}\varepsilon_m + (1 - L_m^{TM,TE})\varepsilon_i} \qquad [3]$$

where $\varepsilon_{eff}$ is the effective permittivity of the mixture, $\varepsilon_m$ and $\varepsilon_i$ are the permittivities of metal and insulator, respectively, $L_m$ is the depolarization factor for TE or TM polarizations, and $\phi$ is the volume fraction of metal material. The depolarization factors in the major $L_m^{TE}$ (TE-polarization) and minor $L_m^{TM}$ (TM-polarization) axes can be described as a function of the aspect ratio of the oblate metal spheroid[136]

$$L_m^{TM} = \left(1 + 1.6\,x + 0.4\,x^2\right)^{-1}$$ [4a]

$$L_m^{TE} = \frac{\left(1 - L_m^{TM}\right)}{2}$$ [4b]

where $x$ is the aspect ratio (b/a ) of the oblate spheroid. Eq. 3 and 4 can be used to relate the measured effective permittivity (refractive index) of the gold films for each effective thickness to the properties of the metal inclusions by substituting suitable parameters for the aspect ratio and volume fraction of the gold NPs. For the permittivity of the NPs, the bulk values for gold at this wavelength were used. Figure 5.11 shows the real and imaginary parts of the refractive index for the two polarization states calculated by fitting eq. 3 to the real part of the index only (i.e., Figure 5.9a) with the best values for the metal fraction and aspect ratio. The Clausius−Mossotti model obviously fits very well, without having to use a modification of the bulk permittivity to account for the particles with diameters under 25 nm.[119] The metal volume fractions and aspect ratio used in the fit range from 30.5% to 68.3%, and from 0.8 to 1.0, respectively. The aspect ratios are consistent with those extracted from Table 5.1, while the volume fractions confirm that substantial air gaps remain at all thicknesses. Note that the volume fraction for closely packed spheres of any diameter is 74%.[137]

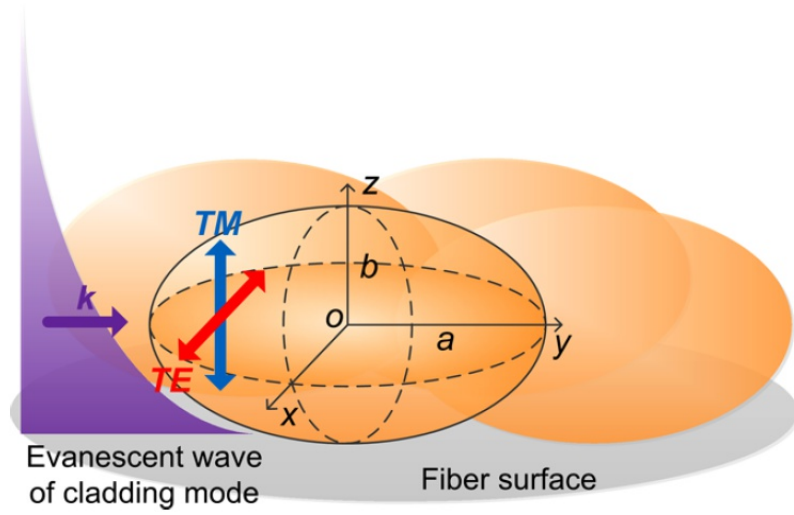

**Figure 5.10:** Schematic diagram of spheroidal gold nanoparticles deposited on fiber surface (assuming that the fiber surface of small area (μm²) is located in the xy plane), interacting with evanescent field of polarized cladding mode. The gold spheroids are oblate so that the major axis a and minor axis (polar axis) b are parallel and perpendicular to the xy plane, respectively.

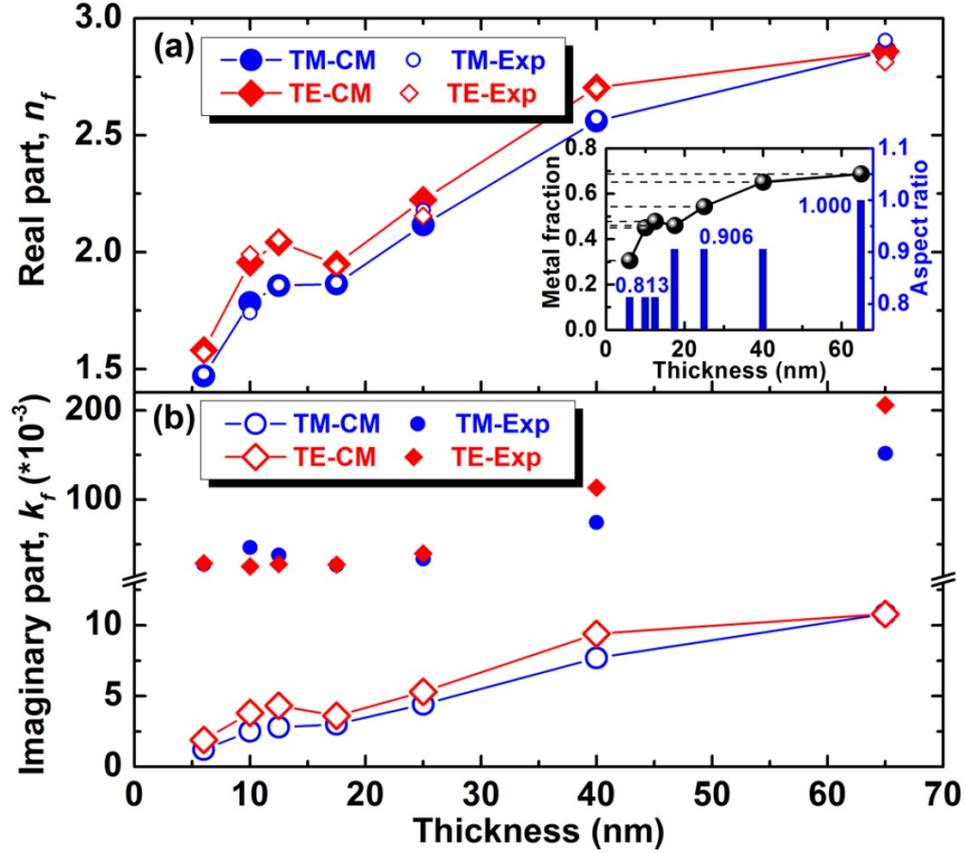

**Figure 5.11:** Real (a) and imaginary (b) parts of gold nanoparticle films calculated by the modified Clausius−Mossotti (CM) equation versus the measured effective thickness. The inset indicates the metal fraction and aspect ratio used in the CM model. The smaller hollow and solid scattered points (TM/TE-Exp) reproduce (from Figure 5.9) the real and imaginary part of the film index extracted from the measurements.

However, fitting the model to the real part of the permittivity results in a 1

order of magnitude underestimate of the imaginary part and fails to reproduce the observed increased dichroism at thicknesses larger than 25 nm. This is due to the fact that the imaginary part of the effective permittivity in the Clausius−Mossotti model comes solely from the imaginary part of the bulk permittivity of the metal inclusions, while the measurement of mode loss, from which the experimental effective medium properties are derived, include a strong scattering component and is therefore larger than modeled.[54] Since the fit indicates that the inclusions are nearly spherical, the imaginary part of the model fit is not expected to be anisotropic. Scattering, however, from NP aggregates that are very thin in the out-of-plane direction but almost continuously connected in the plane, especially at increasing thicknesses, should be stronger (more lossy) for TE light than for TM light. This is what is observed. Another point of comparison is a more recent theoretical model for thin (7 nm and less) but continuous stand-alone gold films based on density functional theory (DFT).[138,139] This DFT model predicts a sign change of the real part of the permittivity (from negative to positive) as well as a significant anisotropy between in-plane and out-of-plane components, when the interband transitions of the gold electrons are added to the intraband transitions. However, the DFT model also does not reproduce our results quantitatively as it does not apply to NP aggregates in its theoretical framework.

## 5.6     Conclusion

In conclusion, the cladding modes of an optical fiber were used to investigate the complex refractive index of CVD gold coatings on the fiber cladding surface. The films were prepared from a precursor with a high decomposition temperature above 300 ° C and slow deposition rates near 1 nm/min. The AFM and SEM images of the gold CVD films with different thicknesses indicate that the films remain composed of nanoparticle aggregates with lateral sizes not exceeding 72 nm, even for average film thicknesses up to 200 nm. This kind of growth is of the Vollmer−Weber type, i.e,. island growth without coalescence into uniform layers.[88] The real and imaginary parts of the refractive index (and permittivity) of the gold nanoparticle films differ markedly from the bulk values at infrared wavelengths, as was often observed for films with thicknesses below 10 nm. However, in the case reported here, the anomalous values remain for thicknesses up to 65 nm and beyond, as in the case of evaporated gold films formed at very slow rates. The in-plane and out-of-plane components of the film permittivity were measured separately and shown to be essentially equal, apart from a small increasing dichroism at thicknesses larger than 25 nm. The polarization-dependent refractive index of the metal−insulator mixture calculated by Clausius−Mosotti theory indicates that the gold CVD film consists of ellipsoidal particles with aspects ratios from 0.8 to 1.0 and air gaps corresponding to metal filling fractions increasing from 30% to 68% over the average thickness range from 6 to 65 nm. A notable consequence of these results is that this CVD gold process cannot be used to coat optical fibers for the generation of surface plasmon polaritons[48,94] because the films produced have a positive real permittivity. This is unfortunate because CVD produces inherently uniform and controllable film thicknesses around the fiber circumference, which helps to realize narrow and strong surface plasmon resonance devices. It remains to be determined whether post-processing of the films could increase the connectivity of the gold NPs to bring the permittivity of the resulting film closer to bulk values.[132] Finally, it was demonstrated that the TFBG device can be used for the characterization at near-infrared wavelengths of ultrathin metal films deposited by gas phase processes, down to thicknesses below 10 nm. The small TFBG probe can be inserted easily into deposition chambers to ensure that the film deposited on the probe is identical to that on adjacent substrates. It should be noted that for process temperatures below 200 °C the TFBG can be interrogated in real time during deposition and therefore provide film measurements as they grow.[54]

# Chapter 6

## Anisotropic Effective Permittivity of Ultrathin PVD-grown Gold Coatings on Optical Fiber in Air, Water and Saline Solutions

Zhou, W.*[2]; **Mandia, D.J.**[1]; Griffiths, M.B.E.[1]; Barry, S.T.[1]; Albert, J.*[2] "Anisotropic Effective Permittivity of Ultrathin PVD-Grown Gold Coatings on Optical Fiber in Air, Water and Saline Solutions" *Optics Express.* **2014**; *22*(26); pp. 31665-31676.

[1] Department of Chemistry, Carleton University, 1125 Colonel By Drive, Ottawa, Ontario, Canada, K1S 5B6

[2] Department of Electronics, Carleton University, 1125 Colonel By Drive, Ottawa, Ontario,

Canada, K1S 5B6

*Corresponding author

## 6.1        Abstract

The optical properties of an ultrathin discontinuous gold film in different dielectric surroundings are investigated experimentally by measuring the polarization-dependent wavelength shifts and amplitudes of the cladding mode resonances of a tilted fiber Bragg grating. The gold film was prepared by electron-beam evaporation and had an average thickness of 5.5 nm (± 1nm). Scanning electron imaging was used to determine that the film is actually formed of individual particles with average lateral dimensions of 28 nm (±8 nm). The complex refractive indices of the equivalent uniform film in air at a wavelength of 1570 nm were calculated from the measurements to be $4.84-i0.74$ and $3.97-i0.85$ for TM and TE polarizations respectively (compared to the value for bulk gold: $0.54-i10.9$).

Additionally, changes in the birefringence and dichrois of the films were measured as a function of the surrounding medium, in air, water and a saturated NaCl (salt) solution. These results show that the film has stronger dielectric behavior for TM light than for TE, a trend that increases with increasing surrounding index. Finally, the experimental results are compared to predictions from two widely used effective medium approximations, the generalized Maxwell-Garnett and Bruggeman theories for gold particles in a surrounding matrix. It is found that both of these methods fail to predict the observed behavior for the film considered.

## 6.2 Introduction

Ultrathin gold films with thicknesses under 10 nm show size- and geometry-dependent features in their optical and plasmonic properties that are quite different from those of bulk material.[118] With decreasing thickness the thin gold films evolve morphologically from continuous to granular, during which a metal-to-insulator transition occurs.[117,140] The ultrathin gold aggregates can be represented by a single layer of oblate nanoparticles (NPs), and such layers have important applications as selective light absorbers[141,142] and many kinds of plasmonic devices.[143,144] Using the effective medium approximation (EMA) theories of the early 1900s,[120] the complex effective permittivity of the aggregated gold films embedded in a homogeneous matrix (surrounding material) can be simulated. In the EMA models, the gold aggregates are represented by point dipoles embedded in the homogenous background, and the resulting effective permittivity of the mixture is dependent on the complex permittivities of the two constituents (gold and surrounding material), and on the volume fraction of gold. For experimental studies of thin gold films, spectroscopic ellipsometry is one of the most common measurement tools as it is highly sensitive to the optical properties of thin films.[140,145–147] However, most of the previous work in this area was focused on measuring the thickness-dependent complex permittivity calculated from the ellipsometric angles to investigate the effect of various thin film growth mechanisms, including the impact of the shape and distribution of the NPs making up the ultrathin films. Comparatively few studies exist on the impact of the surrounding medium permittivity of these films,[148,149] and in such cases the gold NPs films are always embedded in solid dielectric material, making it impossible to study the same film under different surroundings. Moreover, the anisotropic optical properties of the ultrathin metal films caused by the nonspherical shape of their NP constituents limit the measurement accuracy of the conventional measurement methods, such as spectral reflectometry,[49,140] ellipsometry, [145,146] and interferometric picometrology[124] This is one of the reasons for which previous works often neglected the optical anisotropy in measurements of gold films with thicknesses spanning the insulator-to-metal transition. The purpose of the present paper is to investigate the impact of the surrounding medium on the anisotropic optical properties of highly disordered, morphologically complex gold NP aggregates with an average thickness of 5.5 nm. This experimental study is carried out by probing the ultrathin film, deposited on the surface of the cladding of an optical fiber by twostep evaporation, using the polarized evanescent fields of the fiber cladding modes that are resonantly excited by a tilted fiber Bragg grating (TFBG).[41,61] The anisotropic properties of the thin film are inferred from the changes in the amplitudes and wavelengths of the various cladding modes as the film coated fiber is exposed to various surroundings. Those results are then compared with EMA calculations based on measured NP shapes and bulk values of their permittivities, and the discrepancies observed are discussed in relation with the assumptions made by both the Maxwell-Garnett (M-G) and Bruggeman EMA models.

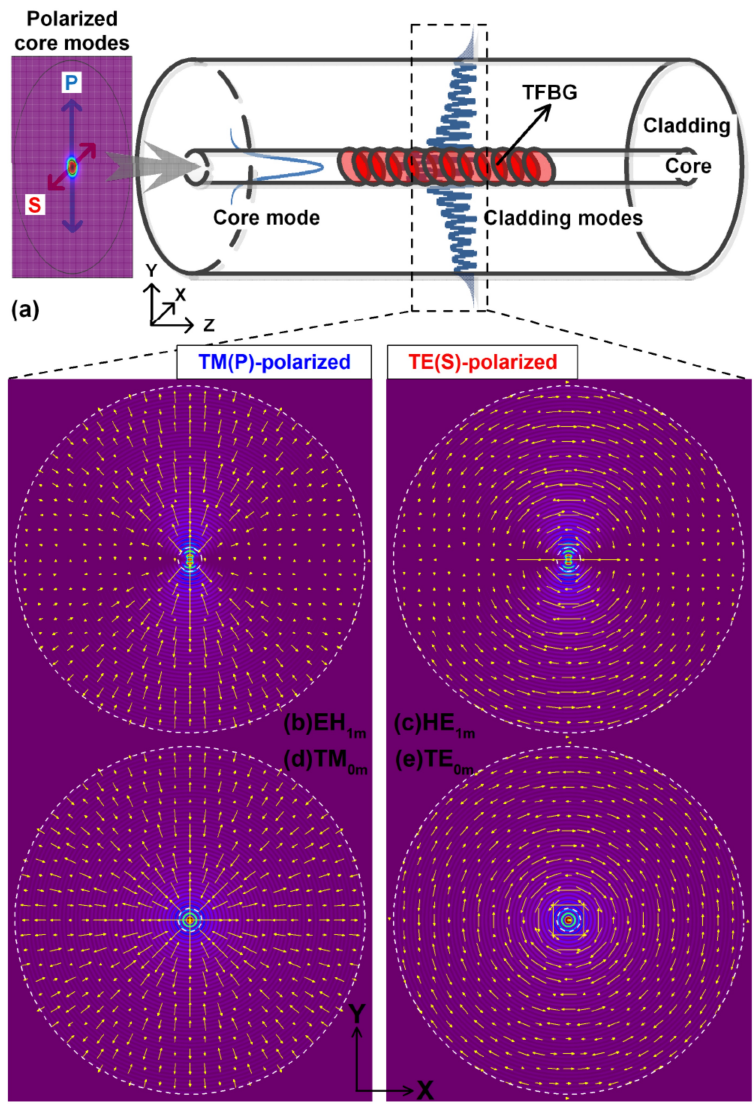

**Figure 6.1:** (a) Schematic diagram of TFBG under X-(S-) and Y-(P-) polarized core modes input. Electric field distributions of four typically vectorial cladding modes ((b) $EH_{1m}$, (c) $HE_{1m}$, (d) $TM_{0m}$, and (e) $TE_{0m}$) coupled from the X-(S-) and Y-(P-) polarized core modes by TFBG, respectively. (Note that the grating planes are parallel to X axis and tilted away from the Y axis). The white broken curves indicate the core and cladding edges of optical fiber.

In more detail, due to the inclination of the grating planes of a TFBG along a specific direction, two families of polarized cladding modes with radial or azimuthal electric fields at the fiber cladding boundary can be selectively excited by probing with linearly polarized input core-guided light (as shown in Fig. 1). [41,61] Figures 6.1(b)-(e) illustrate the simulated electric fields of four typical polarized cladding modes with the same radial order (number of zeros in field amplitude as a function of radial coordinate) in standard single mode fiber for telecommunications. $TM_{0m}$ and $EH_{nm}$ modes are radially polarized and result from P-polarized core mode input while the $TE_{0m}$ and $HE_{nm}$ modes are azimuthally polarized and result from S-polarized input. P- and S-polarizations refer to the orientation of the input electric field in the plane of incidence and out of the plane of incidence on the tilted grating fringes, respectively. In the following these mode families will be referred to as TM and TE to shorten the text. Due to the orthogonal electric fields at the interface of fiber cladding and surroundings, the TM- and TE-polarized cladding modes show distinct responses to surrounding refractive index (SRI) changes, but more strongly so when metal coatings or particles are present. [54,62,150] Therefore, the polarization-dependent properties of the cladding modes excited by TFBG can be used for surface plasmon resonance sensors [48,94,151] and measuring the anisotropic optical properties of thin metal coatings. [52,63] In this paper, the SRI-dependent spectral responses of the two kinds of polarized cladding modes are used to investigate the effect of the surrounding permittivity on the anisotropic effective optical properties of an ultrathin gold NP film with a mean thickness of 5.5 nm. Since the coated TFBG is both the substrate for the coating and the probe of its properties, successive measurements of a given coating in many surrounding environments can be very accurate as they are inherently independent of changes in either the film or the sensing probes, which remain unchanged. The complex effective permittivities of the coating are measured and compared with theoretical predictions based on both the generalized M-G and Bruggeman EMAs, wherein the input parameters are particle sizes and filling factors, extracted from scanning electron microscope (SEM) and atomic force microscope (AFM) image data, as well as literature values for the permittivity of bulk gold and of the surrounding media tested.

## 6.3        Sample Fabrication and Geometry

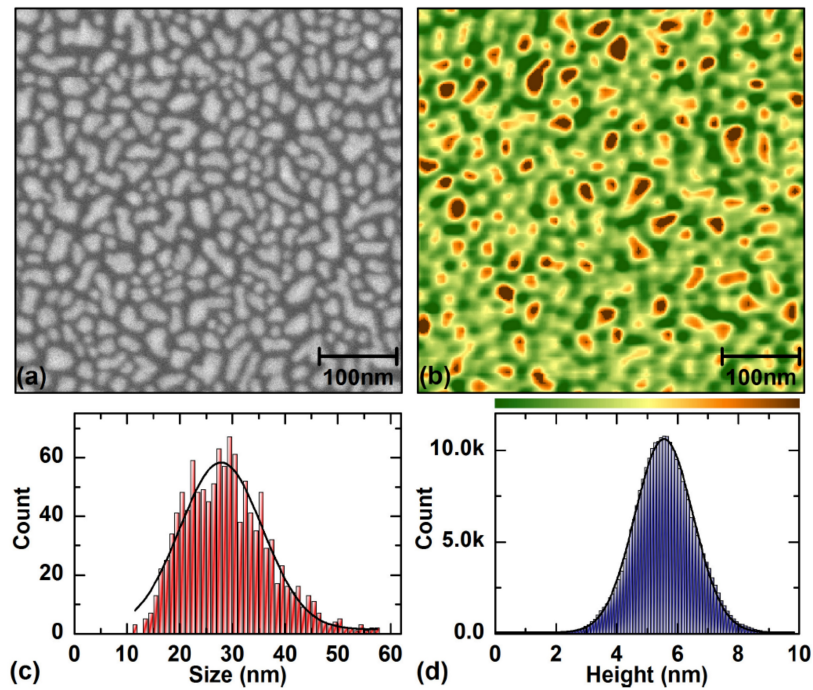

**Figure 6.2:** SEM (a) and AFM (b) images of the gold NPs film over an area of 0.5 x 0.5 μm² (with scale bar of 100 nm). (c) Histogram of gold NP lateral sizes based on 1172 particles distributed over an area of approximately 1.3 x 1.0 μm². (d) Height histogram of gold NPs film based on ~2.62 x 10⁵ extracted points from 1 x 1 μm² AFM image.

The TFBG used in this work was written in hydrogen-loaded CORNING SMF-28 fiber with a pulsed KrF excimer laser using the phase-mask method.[41] The hydrogenation process of the optical fibers is as follows: a pressure of 15.2 MPa, a temperature of 20 ℃, and a duration of 14 days. The length and the tilt angle were chosen at 10 mm and 10º to excite a large number of high-order cladding modes with strong evanescent fields for thin coating measurements. The Bragg wavelength is around 1613 nm, resulting from a phase mask period of 1114.8 nm. After cleaning the TFBG cladding surface with piranha solution (a mixture of sulfuric acid and hydrogen peroxide), it was placed in an electron-beam physical vapor deposition system at room temperature and under vacuum ($10^{-6}$ Torr). In order to have a relatively uniform gold coating around the fiber circumference, two gold deposition steps were conducted consecutively with the fiber being rotated by exactly 180º between the two deposition runs using a custom designed fiber holding fixture. A relatively uniform gold film with a thickness of ~5 nm was obtained under a deposition rate of ~6 nm/min. The estimated thickness was obtained by measuring a witness sample (a 2-inch silicon wafer with a 30 nm $SiO_2$ thermal oxide buffer layer) that was placed next to the TFBG for one deposition run. We now proceed to describe physical measurements of the actual coating morphology (thickness and NP shapes and filling factor) on the witness sample.

Figure 6.2 shows SEM (a) and AFM (b) images which show that the $SiO_2$ surface is completely covered by an irregular distribution of gold NPs. The evidence of small-scale coalescence of the aggregated gold NPs is already appearing for this thickness, resulting from the touching and merging of adjacent NPs. The quantitative analysis of the gold NP sizes and heights was carried out on the SEM and AFM images by using image processing software tools ImageJ and WSxM,[100] respectively. The distributions of the lateral size (diameter) and height of the gold NPs film are shown in Figs. 6.2(c) and 6.2(d), respectively. Based on the Gaussian fitting curves, the mean lateral size $d_{NP}$ and height $h_{NP}$ of the gold NPs are 28 nm with a standard deviation of 8 nm and 5.5 nm with a standard deviation of 1 nm, respectively. The relatively large standard deviations in lateral size and height reflect the randomness of the cluster formation at these thicknesses, as can be clearly observed in Figs. 6.2(a) and 6.2(b). It is very clear that this 5.5 nm thick gold evaporated film can be regarded as a mono-layer of oblate gold nanospheroids. Also, the volume factor f of the gold NPs is 68% ($\pm$ 5%), as obtained from the SEM image using the ImageJ software.

## 6.4        Experimental Results

Based on the experimental setup shown in Fig. 6.3(a) that includes a broadband source (BBS) (JDSU BBS1560), a polarization controller (PC) (JDSU PR2000), and an optical spectrum analyzer (OSA) (ANDO AQ6317B), the polarized spectra of TFBG under various SRIs can be measured.[63] Even though the nominal measurement resolution of this OSA is only 10 pm, the resonance line shapes were fitted with an inverted Gaussian function over several measurement points and the statistically significant uncertainty with which an individual resonance can be measured is $\pm$ 2 pm for the wavelength and $\pm$ 0.05 dB for the amplitude.[41] In this experiment, three different SRIs of 1, $1.315 - i1.21$ x $10^{-4}$,[152] and $1.360 - i3.24$ x $10^{-5}$ (at 1570 nm)[153] provided by air, deionized water, and a saturated NaCl-water solution were applied on the TFBG sample. In order to investigate the coating with the same cladding guided mode (resonance) in the three different SRIs, a resonance near 1566.5 nm (measured in air) was chosen, as the cut-off wavelength for cladding mode guidance in the largest SRI is around 1565 nm. The normalized spectra of the polarized cladding modes of the gold NP-coated TFBG under the three surroundings is shown in Fig. 6.3(b). The spectra were adjusted to compensate for any temperature dependence by shifting them so that the core mode resonances for all the measurements (located near 1613 nm) overlap perfectly.[41] All the spectra shown in Fig. 6.3 were obtained with the same TFBG, to eliminate any possible perturbation of the data by different underlying grating properties (it is not possible to fabricate exactly identical TFBGs by photosensitive processes). For comparison with the gold coated results, corresponding spectra of the same TFBG sample after etching of the gold coating are shown in Fig.6. 3(c). In Figs. 6.3(d) and 6.3(e), the floating columns indicate the extracted wavelength separations ($\lambda_{TE}-\lambda_{TM}$) and peak-to-peak (P-P) amplitude differences ($\Delta A_{TE}-\Delta A_{TM}$) between the TE- and TM-polarized cladding mode resonances under the three SRIs for both bare and coated TFBG samples.

It is obvious that the TE and TM cladding mode resonances both shift to longer wavelength under increasing SRI, as expected from standard waveguide theory, and roughly at the same average rate under these conditions (Fig. 6.3(d)). Also as expected, the wavelength difference between TE and TM resonances of the bare grating decreases from 0.173 nm to 0.013 nm with increasing SRI, indicating that the modes chosen approach their cut-off at the maximum SRI tested (since weakly guided and radiation modes no longer have polarization splitting)[150]. For the gold coated TFBG on the other hand the SRI dependence of the wavelength shifts completely opposite: the wavelength separation for the same pair of resonances is larger in air (0.210 vs 0.173 nm) and it further increases to 0.335 nm under the same changes of SRI as for the bare grating. So instead of reducing the polarization splitting, increasing the SRI makes the splitting larger when the TFBG is coated with the thin gold NP film.

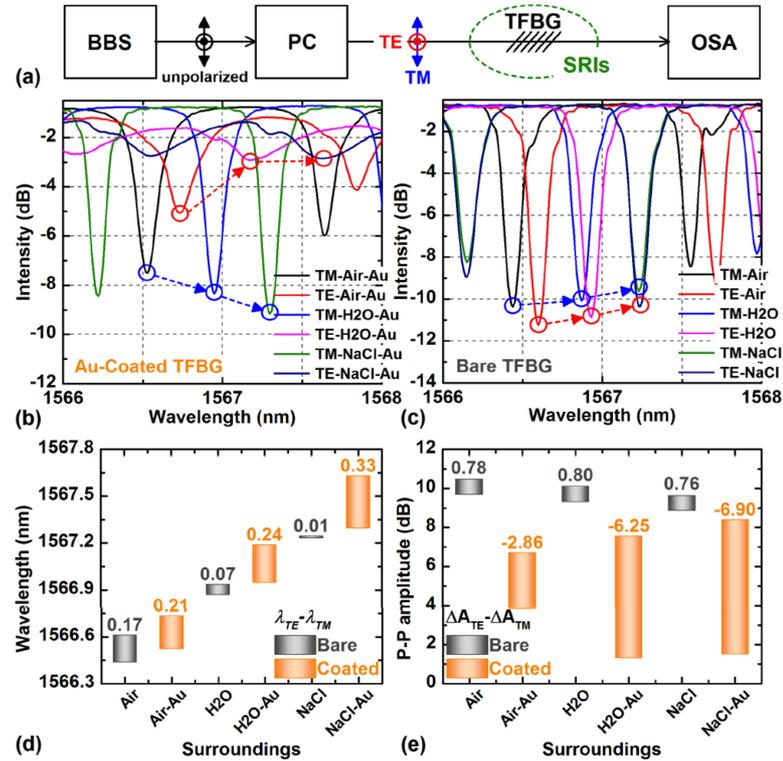

**Figure 6.3:** (a) Schematic diagram of experimental setup. Temperature-calibrated spectra of the gold NPs-coated (b) and the gold NPs-etched (bare) (c) TFBGs for TE- and TM-polarizations in air, DI water ($H_2O$), and saturated NaCl-water solution (NaCl) surroundings. (The red and blue circles are marked at the positions of the TE- and TM-polarized cladding mode resonances, respectively.) Wavelength separations (d) and peak-to-peak amplitude differences (e) between the pair of polarized cladding modes for the both of coated and bare TFBGs under the three SRIs. Note that the signs of the P-P amplitude differences under the cases of bare and coated TFBGs are opposite.

With regards to the amplitudes of the resonances as the SRI rises, they decrease slowly for the bare TFBG as the modes approach cut off, and the amplitudes of the TM modes are slightly lower, due to the weaker confinement of those modes relative to the TE ones. Again, the results for the grating with the gold coating are strikingly different. For the coated grating in air, the P-P amplitudes of all the modes are smaller than those of the bare grating, reflecting higher mode loss due to either absorption or scattering by the gold NPs. Upon increasing the SRI, however, the amplitudes of the TM modes recover (they become larger, indicating lower mode loss) while those of TE modes continue to decrease. In fact, the overall spectral responses of the coated TE and TM modes indicate clearly that for such gold coatings, TE modes (with their electric fields polarized parallel to the film) are much more perturbed and are also more sensitive to further SRI changes.

**6.5        Discussions**

In order to investigate the anisotropic optical properties of the gold NPs coating under the various SRIs, the effective indices of the polarized cladding modes have to be extracted from the spectral data. Then, those mode effective indices (along with the average coating thickness obtained by AFM) will be used to calculate the effective medium properties of the equivalent homogeneous layer that is added to the fiber waveguiding structure by the coating process. Figure 6.4 illustrates the evanescent fields of the cladding modes with TE- or TM-polarization along the gold NPs-coated TFBG surface (a) and the corresponding waveguide (b) where the NPs have been replaced by an effective medium layer with homogeneous properties. Since the added layer has a complex permittivity, its effect on the effective indices of the TE and TM cladding modes is widely different.[48,54,63,94,154]

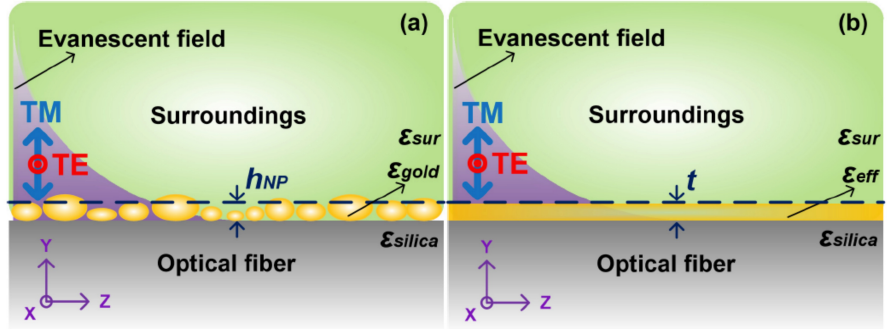

**Figure 6.4:** 2D schematic diagrams of a mono-layer of oblate gold NPs (a) and effective medium layer that consists of gold and surrounding materials (b) coated on optical fiber surface under evanescent fields of TE- and TM-polarized cladding modes propagating along the z-axis. The thickness t of the effective medium layer is equal to the average height $h_{NP}$ of the gold NP film measured from its AFM image. Note that the relative profiles of evanescent field and gold film shown above are not to scale, since the penetration depth of the cladding mode with effective index of ~1.366 in air is about 130 nm.

The real part of the effective index of a cladding mode $N_{eff}^{clad}$ can be obtained from the phase matching condition of the TFBG,

$$N_{eff}^{clad} = \frac{\lambda_{clad}\cos\theta}{\Lambda - N_{eff}^{core}} \qquad [1]$$

where $\lambda_{cl}$ is the wavelength of the cladding mode resonance, $\theta$ is the tilt angle of the grating planes, $\Lambda$ is the grating period, and $N_{eff}^{core}$ is the effective index of the core mode (~1.447). The formula relating the imaginary part $K_{eff}^{clad}$ (the mode extinction coefficient) to the amplitude of the cladding mode attenuation in the transmission spectrum requires the computation of the coupling coefficient κ between the incident core mode and the cladding mode of interest, given the known grating length and period, and the base fiber properties. The fundamental relationship between the coupling coefficient κ and the P-P transmission amplitude ΔA (converted from dB to linear scale) of the cladding mode can be written as,

$$\kappa = \frac{\tan h^{-1}\sqrt{1-\Delta A}}{L} \qquad [2]$$

where L is the grating length (10 mm). This relationship is correct for the lossless case (no gold film). Once loss is introduced (the film), the coupling coefficient is unchanged to first order and the observed decrease in the P-P amplitude is due to the appearance of an imaginary part (i.e. absorption) in the cladding mode propagation constant. This imaginary part is found by matching experimental results to simulations of TFBGs with lossy modes, carried out with a numerical tool (Optigrating, by Optiwave).[52,63] Figures 6.5(a) and 6.5(b) show the real and imaginary parts of complex effective indices for TE and TM modes of the gold coated fiber extracted from the experimental mode resonances when the coated fiber was immersed in the three SRIs (Fig. 6.3(b)). The real part of the effective indices increases with SRI and the difference (TE-TM) increases from 0.00038 to 0.0006. On the other hand, the calculation of the imaginary part of the effective index reveals that the extinction coefficient of the TE mode is enhanced from $1.51 \times 10^{-4}$ to $5.06 \times 10^{-4}$ in going from air to water, but decreases to $4.12 \times 10^{-4}$ in the salt-water solution. For the TM mode, extinction coefficient drops from $4.43 \times 10^{-5}$ in air to $7 \times 10^{-6}$ in the salt-water solution. It is important to note that the extinction coefficient for both TE and TM modes of the bare TFBG in the salt water solution is about $1 \times 10^{-6}$, which indicates absorption in the surrounding medium (due to its bulk extinction coefficient of 3.24 x $10^{-5}$) contributes little to the extinction coefficient of the modes of the coated fiber.

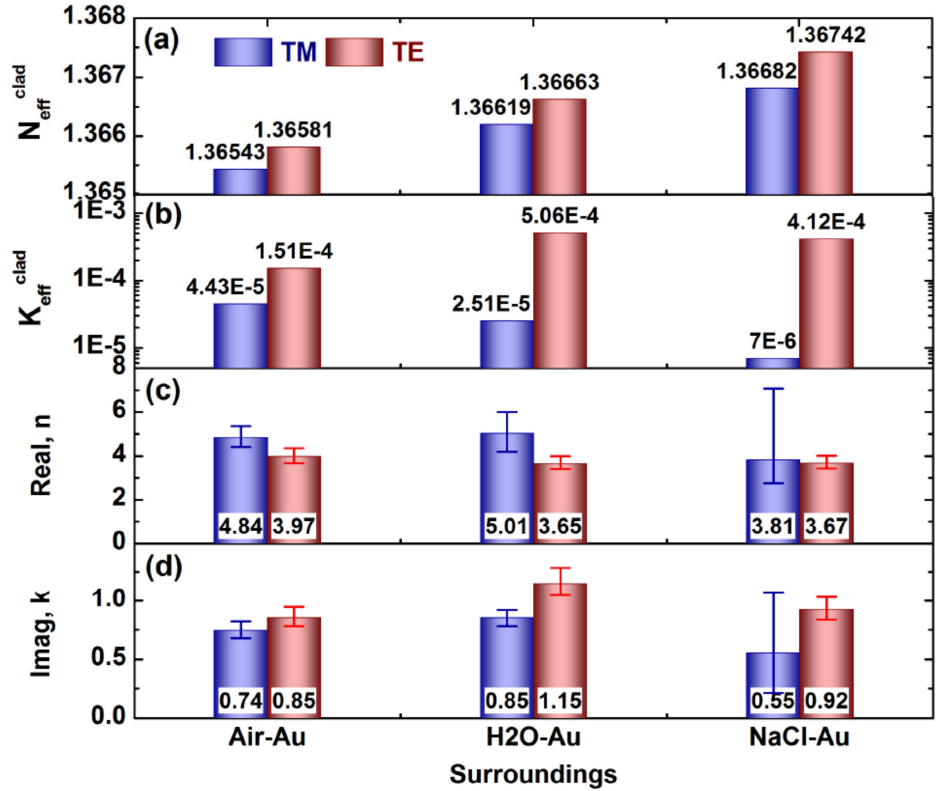

**Figure 6.5:** Real (a) and imaginary (b) parts of the complex effective indices of the TE and TM cladding modes $\left( N_{eff}^{clad} - iK_{eff}^{clad} \right)$ under the three SRIs; Real (c) and imaginary (d) parts of complex refractive indices $(n - ik)$ of the 5.5 nm thick effective homogeneous gold NPs film for TE and TM modes. The positive and negative errors of the complex refractive indices are evaluated from the thickness uncertainty of $\pm 1$ nm in FIMMWAVE simulations.

Now using the same optical fiber parameters (core and cladding) as those used in our previous paper,[63] a thickness of 5.5 nm for the effective medium layer and the different SRIs as the outer layer, a four-layer optical fiber model solved with a vectorial finite differences method (FDM) complex mode solver (FIMMWAVE, by Photon Design) can be used to find the complex refractive index of the gold NP film $(n-ik)$ that reproduces the measured mode effective indices.[52,63] While TE and TM modes always have different effective indices even for homogeneous, isotropic media,[155] in the present case it was not possible to find a unique value of the complex film refractive index to yield the measured TE and TM modes: a film anisotropy had to be introduced to account for in-plane (TE) and out-of-plane (TM) electric field polarizations. The calculated effective medium complex refractive indices of the gold NPs film under the three SRIs at the cladding mode wavelength around 1570 nm are shown in Figs. 6.5(c) and 6.5(d). The different real and imaginary parts of the refractive indices for the TE and TM polarizations under the various surroundings indicate that the 5.5 nm gold NP film is optically anisotropic at this near infrared wavelength. Furthermore, the SRI dependence of complex refractive indices demonstrate that the gold NPs coating on TFBG surface can be regarded as a composite material layer consisting of gold NPs and a given surrounding medium. Compared with the complex refractive index of bulk gold $(0.54−i10.9)$,[114] the calculated complex refractive indices of the gold NP layer has a 10 times higher real part and a 10 times lower imaginary part. In other words, the layer behaves much less as a metal than as a dielectric. This is consistent with the size-dependent optical properties of thin gold films investigated in other papers,[124,145–147] where the metallic character of gold is replaced with dielectric behavior with decreasing thickness, due to the confinement of free electrons caused by the interruption of percolation paths inside ultrathin metal films. Due to the electron mean free path of 25 nm,[124] electrons are always confined in the gold NPs with an average thickness of 5.5 nm in the out-of-plane direction, suggesting dielectric-like optical properties for the TM-polarization. But the small-scale coalescence of the aggregated gold NPs with average lateral size of 28 nm weakens the localization effect of electrons in the in-plane direction, which results in the more metallic complex refractive index of the gold NPs film for the TE-polarization (lower real part and higher imaginary part). For quantitative comparisons, the measured real parts of the complex permittivities ($n^2$-$k^2$) of the 5.3 nm and 6.2 nm thick gold evaporated films were around 25 and 10 at 1570 nm (0.79 eV) in air,[140] respectively. From Figs. 6.5(c) and 6.5(d), the corresponding real parts of the anisotropic complex permittivities of the 5.5 nm gold film in air are 22.8 and 15 for the TM and TE polarizations, respectively, which are included within the above reported values. Moreover, the estimated complex refractive indices of the gold evaporated films with "nominal" thicknesses of 4 nm and 5 nm were found to be about 5-i1.5 and 4.5−i2.5 at 1570 nm in air,[146] i.e. a similar real part as the one obtained here, but a much larger imaginary part. The origin of this discrepancy is that the gold films reported in ref [146] had larger-scale coalescence of the aggregated gold NPs than ours, which allowed more absorptive loss by Joule heating. The more connected gold films with lower thicknesses in ref [146] are probably caused by the difference between the mass-equivalent thickness and the effective thickness of the discontinuous gold films, which are correlated with each other by the filling factor (i.e. the mass-equivalent thickness of our gold NP film could be ~3.7 nm (5.5 x 68%).[140,156] Moreover, previous work on thin gold films (4-8 nm) used a quasi-homogeneous isotropic effective medium material for investigating their optical properties in visible and infrared wavelength regions.[124,146] Here, we provide further observations of the anisotropy of these films and demonstrate that directional depolarization effect in oblate gold NPs[63,147,148] and the resulting anisotropic permittivity of ultrathin gold material,[138,157] yields a measurable optical birefringence and dichroism.

For further investigating the anisotropic optical properties of the gold NPs coating under various SRIs, we now compare our measured results with theoretical calculations of the complex permittivity of an effective medium layer composed of gold NPs (using the bulk complex permittivity of gold at these wavelengths) and surrounding media. Two effective medium approximations (EMA) will be used, one based on the generalized M-G formula[158] and the other on the generalized Bruggeman formula.[159] From the SEM and AFM images shown in Fig. 6.2, the gold NPs can be represented as oblate gold spheroids embedded in the surrounding medium, where the minor axes of all oblate spheroids are perpendicular to the optical fiber surface. Thus, the generalized M-G and Bruggeman formulas can be respectively expressed as,[158,159]

$$\frac{\varepsilon_{eff}-\varepsilon_e}{\varepsilon_e+N_{TM,TE}\left(\varepsilon_{eff}-\varepsilon_e\right)}=f\,\frac{\varepsilon_i-\varepsilon_e}{\varepsilon_e+N_{TM,TE}\left(\varepsilon_i-\varepsilon_e\right)} \qquad [3]$$

$$\left(1-f\right)\frac{\varepsilon_{eff}-\varepsilon_e}{\varepsilon_{eff}+N_{TM,TE}\left(\varepsilon_e-\varepsilon_{eff}\right)}=f\,\frac{\varepsilon_i-\varepsilon_{eff}}{\varepsilon_{eff}+N_{TM,TE}\left(\varepsilon_i-\varepsilon_{eff}\right)} \qquad [4]$$

where $\varepsilon_{eff}$ is the effective permittivity of the effective medium layer, $\varepsilon_i$ and $\varepsilon_e$ are the permittivities of inclusion (gold) and surroundings, respectively, $N_{TM,TE}$ is the depolarization factor for TM or TE-polarization, and $f$ is the volume factor of the gold material. For the oblate gold spheroids, the depolarization factors of the TM- and TE-polarizations can be defined as a function of the aspect ratio $x$ of the oblate spheroid,[158]

$$N_{TM}=\left(1+1.6\,x+0.4\,x^2\right)^{-1} \qquad [5]$$

$$N_{TE} = \frac{(1 - N_{TM})}{2} \qquad\qquad [6]$$

where $N_{TM}$ and $N_{TE}$ are 0.752 and 0.124 for our structure, based on the following expression for $x = h_{NP} / d_{NP}$. With Eqs. (3) and (4), the calculated complex refractive indices of the effective medium coating under the three SRIs for the out-of-plane (TM) and in-plane (TE) directions are shown in Fig. 6, where the corresponding results obtained from the experimental data are also indicated.

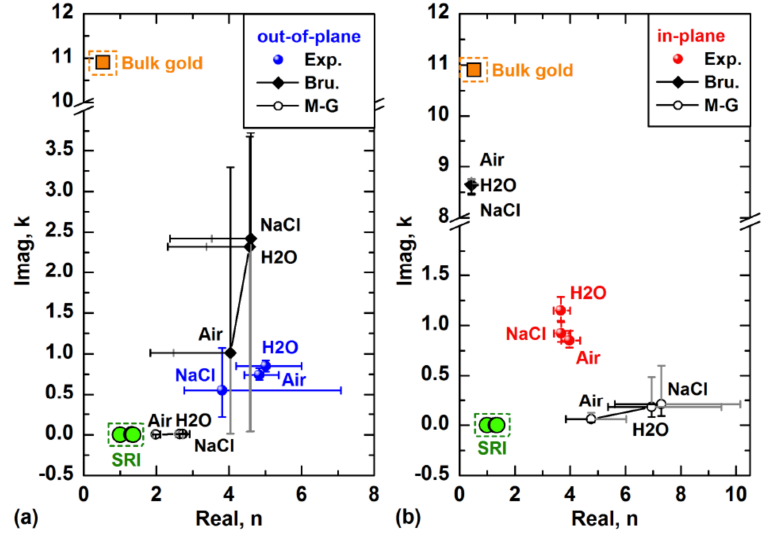

**Figure 6.6:** Complex refractive indices of the effective medium coating of the three surroundings for (a) out-of-plane (TM) and (b) in-plane (TE) directions calculated by M-G and Bruggeman EMAs, compared with the experimentally obtained results and the bulk permittivities of the medium constituents (gold and SRI). The gray and black error bars indicate the deviations of complex refractive indices caused by the minimum aspect ratio x of 4.5/36 and the maximum x of 6.5/20 in EMA models, respectively, based on the average height of 5.5 nm ( ± 1 nm) and the average lateral dimensions of 28 nm ( ± 8 nm) of the gold NPs. Note that the both of gray and black error bars show in negative direction for the real parts of Bruggeman EMA in Fig. 6.6(a).

Based on these comparisons of the three groups of complex refractive indices, some general trends can be observed. As expected from the EMA of metal particles in dielectric media, the large increase in the real part of the index of the film relative to that of its constituents is relatively well modeled, with the notable exception of the Bruggeman model for the in-plane (TE) polarization. In this latter case the model predicts a complex index that is very close to that of bulk gold (regardless of SRI), while the experimental data retains a strong dielectric behavior. It is as if the EMA layer was acting as a continuous metal barrier for light in the cladding and shielding it completely from the surroundings, which is not the case according to the experimental data but which does occur for thicker films.[62] The other observation is that the M-G EMA systematically underestimates the coating loss (k), as expected in this case,[160] while the Bruggeman model does the opposite (overestimates). An underestimated EMA value for loss is expected because the experimental values for loss come from two contributions in our experiments: absorption by gold NPs and scattering of the evanescent fields of cladding guided modes. Since the EMA approach does not include the effect of scattering, it was expected that those models would underestimate k relative to the experiments. It is clearly not the case for the Bruggeman EMA which again reflects its tendency to overestimate the loss due to absorption (i.e. the metallic character of the ultrathin coating with such aspect ratio of its constituent NPs). In fact, the applicability of the various EMAs to the different steps of growing gold NPs film deposited on glass that was described in [161] indicates that the M-G formula is quite suitable for sparsely distributed gold NP films while the Bruggeman formula is applicable when there is large-scale coalescence in the gold film growth. It looks from the results presented here that for intermediate levels of coalescence of ultrathin gold NP films, both EMA formulas fail in reproducing the experimental data. Some similar failures of the Bruggeman EMA were also observed in fitting the measured complex conductivities (directly related to complex permittivity) and optical transmittance at far-infrared frequency for gold films around the insulator-to-metal transition.[162,163] Our results for the anisotropic complex refractive index show that this gold NP film at a thickness of 5.5 nm is thicker than the critical value for the maximum dielectric constant, in agreement with measured critical thicknesses between 5.3 nm and 6.2 nm for light at a wavelength of 1570 nm (0.79 eV).[140] The slightly weaker insulating properties of the gold NPs film for the in-plane polarization (TE) are understandable since the dipole coupling between the gold NPs should occur first for in-plane polarization and help the gold film tend toward metallic material as the film grows.[147] However, the significant differences in the birefringence and absolute values of the measured effective medium results as a function of SRI are unexpected and may reveal new and important effects that occur near the insulator-to-metal transition of gold at these thicknesses and wavelengths.

## 6.6      Conclusion

In conclusion, the SRI-dependent effective refractive indices of the anisotropic 5.5 nm gold NPs film deposited by gold evaporation method were investigated with the polarized cladding modes excited by a TFBG. The complex refractive indices for modes polarized in the in-plane (TE) and out-of-plane (TM) directions, and the corresponding effective thin coating parameters were extracted from the measured polarization-dependent wavelength shifts and amplitude changes of the cladding modes around 1570 nm with help from simulations carried out with an accurate four-layer complex mode solver. For the extracted complex refractive indices of the gold NPs film embedded in different media, the results show that the real parts are 10 times larger than for bulk gold while the imaginary parts of the film are 10 times smaller. Such ultrathin gold films with only small scale coalescence have therefore a strong dielectric nature. It is worth noting that for the gold NP sizes obtained here, the localized surface plasmon resonance (LSPR) wavelength is near 750 nm, i.e. well outside of our measurement spectral range. While some of the NPs have larger sizes and hence longer LSPR wavelengths, the randomness of the sizes would smear out any significant impact. Therefore no plasmonic effect is expected to come into play. The slightly weaker insulating optical properties of the gold NPs film for the TE-polarization than those for TM-polarization suggest the proximity of an insulator-to-metal transition and a critical thickness lower than 5.5 nm at 1570 nm, for the deposition conditions used. Moreover, comparing the experimental results with predictions from the generalized M-G and Bruggeman EMAs, it was revealed that these effective medium models fail to correctly account for the properties of the NP films at such thickness for near infrared wavelengths. The EMA models also do not correctly account for the effect of augmenting the surrounding medium permittivity on the birefringence and dichroism of the gold NPs film. In view of the increasing importance of thin metal coatings in plasmon-assisted sensing and in photovoltaic devices, the polarization-resolved, high sensitivity measurements reported here show that ultrathin, partially connected metal films have anisotropic effective complex permittivities that are not well modeled by currently accepted EMA theories that are based on bulk metal properties with shape and density factors alone. It was also demonstrated clearly that the evolution of the equivalent film properties with SRI presents some anomalous features that are not consistent with our understanding of the effect of thin perturbations on guided modes, especially in the case of saline solutions.

# Chapter 7

Chemical Vapor Deposition of Anisotropic
Ultrathin Gold Films on Optical Fibers: Real-time
Sensing by Tilted Fiber Bragg Gratings and Use of
a Dielectric Pre-coating

**Mandia, D.J.\*[1]**; Zhou, W.[2]; Ward, Matthew J.[3]; Joress, H.[3]; Giorgi, J.B.[4]; Gordon,
P.G.[1]; Albert, J.[2]; Barry, S.T.[1] " Chemical Vapor Deposition of Anisotropic Ultrathin
Gold Films on Optical Fibers: Real-time Sensing by Tilted Fiber Bragg Gratings and
Use of a Dielectric Pre-coating" *Proc. SPIE.* **2014**; *9288*; pp. 92880M-1-92880M-10

[1] Department of Chemistry, Carleton University, 1125 Colonel By Drive, Ottawa,
Ontario, Canada, K1S 5B6

[2] Department of Electronics, Carleton University, 1125 Colonel By Drive, Ottawa,
Ontario,
Canada, K1S 5B6

[3] Cornell High Energy Synchrotron Source, Cornell University, 161 Synchrotron
Drive, Ithaca, New York, USA, 14853

[4] Department of Chemistry, University of Ottawa, 10 Marie Curie Pvt., Ottawa,
Ontario, Canada, K1N 6N5

\*Corresponding author

*For preliminary raw XAFS and post-processed, first-shell data, see figure 7.5*

## 7.1        Abstract

Tilted fiber Bragg gratings (TFBGs) are refractometry-based sensor platforms that have been employed herein as devices for the real-time monitoring of chemical vapour deposition (CVD) in the near-infrared range (NIR). The core-guided light launched within the TFBG core is back-reflected off a gold mirror sputtered onto the fiber-end and is scattered out into the cladding where it can interact with a nucleating thin film. Evanescent fields of the growing gold nanostructures behave differently depending on the polarization state of the core-guided light interrogating the growing film, therefore the resulting spectral profile is typically decomposed into two separate peak families for the orthogonal S- and P-polarizations. Wavelength shifts and attenuation profiles generated from gold films in the thickness regime of 5-100 nm are typically degenerate for deposition directly onto the TFBG. However, a polarization-dependence can be imposed by adding a thin dielectric pre-coating onto the TFBG prior to using the device for CVD monitoring of the ultrathin gold films. It is found that addition of the pre-coating enhances the sensitivity of the P-polarized peak family to the deposition of ultrathin gold films and renders the films optically anisotropic. It is shown herein that addition of the metal oxide coating can increase the peak-to-peak wavelength separation between orthogonal polarization modes as well as allow for easy resonance tracking during deposition. This is also the first reporting of anisotropic gold films generated from this particular single-source gold (I) iminopyrrolidinate and CVD process. Using an ensemble of X-ray techniques, the local fine structure of the gold films deposited directly on the TFBG is compared to gold films of similar thicknesses deposited on the $Al_2O_3$ pre-coated TFBG and witness slides.

## 7.2 Introduction

Largely due to their utility in a variety of capacities such as gas sensing,[67] solution-based protein sensing,[79] and refractometry,[73] ultrathin noble metal nanostructured films have been a highly attractive material under the umbrella of fiber optic-based sensing. The use of a TFBG device has recently been employed for the real-time monitoring of the growth and nucleation of gold films from a single-source gold (I) iminopyrrolidinate precursor from CVD. While the use of spectroscopic ellipsometry[83] or quartz crystal microbalance (QCM) data[93] are the conventional *in-situ* tools for monitoring deposition processes, the TFBG is an emerging technology that can also provide highly useful thin film metrics instantaneously.

Use of the TFBG's transmission amplitude spectrum allows for single-resonance tracking and aids in demodulation of the S- and P-polarized attenuation profiles. In this particular study, we explore the use of an alumina ($Al_2O_3$) pre-coating in generating a well-defined polarization-dependence property that has been observed previously for other copper and gold films.[52,63] We demonstrate that the peak-to-peak separation of the cladding modes for the orthogonal S-and P-polarization states can be increased with thickness of the $Al_2O_3$ dielectric layer. By inducing the higher sensitivity with respect to both resonance families, the polarization-dependent property responsible for anisotropic thin films can be obtained for these particular gold films, which have thus far been exclusively isotropic in nature.[63] The dielectric pre-coating has many promising implications, particularly towards further optimization of the TFBG as an *in-situ* deposition monitoring device for vapour-phase deposition processes such as CVD and development of gold-coated SPR-based sensor devices with extreme sensitivities.

## 7.3    Experiments

### TFBG Fabrication and *In-situ* CVD Monitoring Experiments

Figure 7.1 depicts the experimental design of the CVD experiments, including the collection of the optical data from the OSA. The custom hot-walled CVD reactor is composed of a stainless steel Barnstead tube furnace that is fitted with an in-line k-type thermocouple and proportional-integral-derivative (PID) relay circuit that finely controls the heating rate of the furnace (i.e. substrates such as the TFBG). The gold-containing guanidinate and iminopyrrolidinate CVD precursors, (1) [Au(N$^i$Pr)$_2$CN(Me)$_2$]$_2$ and (2) [Au(Me$_2$-$^t$Bu-ip)]$_2$, respectively, were synthesized according to protocols reported elsewhere.[30,31] The TFBG is fed through a groove that was milled and epoxied into a KF-25 flange cap and sits directly above the single-source precursor boat in the case of (1) and in a separately heated bubbler in the case of (2). In all cases, the precursor is typically loaded into a bubbler within a drybox under N$_2$ atmosphere for safe transfer to the CVD apparatus prior to each experiment. Base pressures of 30-45 mTorr are typically achieved and the substrates are heated to the target temperature of 200 ºC (for "Low T." samples), 300 ºC, or 350 ºC ("High T." samples) prior to deposition. Target temperatures of the precursors (i.e. bubbler temperatures) are typically selected based on the onset-of-volatilization temperature established by thermogravimetric analysis studies for the particular precursor. ~200 ºC and ~130 ºC are the typical volatilization temperatures used for (1) and (2), respectively. Upon volatilization of the precursors, the CVD chamber saturates to pressures of ~60-80 mTorr and this involves all exposed surfaces within the chamber, including the TFBG, to be coated by the target metal film. The custom-built fiber guide also contains an $Al_2O_3$-passivated metal boat to hold $Si(100)$ and glass witness slides adjacent to the TFBG portion of the optical fiber during depositions. The *ex-situ* data (AFM, SEM, XANES, XAFS, GID) for the gold films presented herein was primarily performed on the coated witness slides. The TFBG-polarized transmission amplitude spectrum was used to monitor the depositions on an optical spectrum analyzer (OSA). Downstream of the OSA is a polarization controller and coupler that provide a broad spectral "comb" made up of individual cladding modes that are either S -(parallel to grating plane) polarized or P-(perpendicular to grating plane) polarized. A TFBG grating 4 mm (instead of 1 cm to minimize thickness gradients across the TFBG) in length was inscribed within the fiber core of Corning SMF-28 optical fibers by the phase-mask method.[41] Grating tilt angle of 10º was selected and a grating-to-grating spacing (period) of 555.2 nm was generated using a pulsed KrF laser source. Due to the annealing effect of standard FBGs at elevated temperatures (T > 200 ºC), the attenuation of the Bragg mode (~1620 nm) is used as an internal standard during spectral monitoring and is subtracted out during post-processing calculations of the wavelength shift-induced refractive indices of the TFBG from the attenuation profiles.

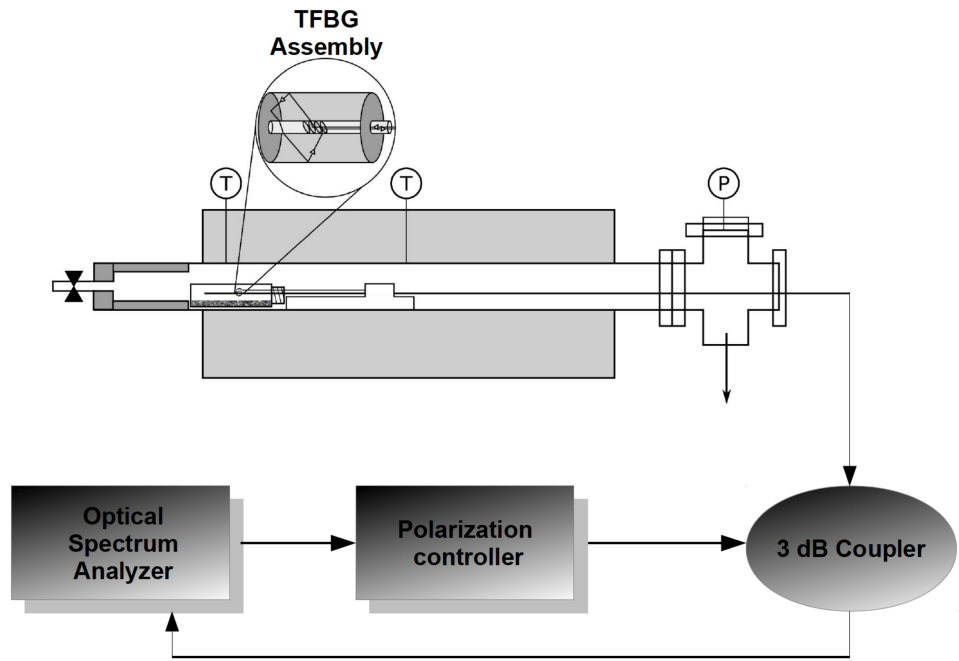

**Figure 7.1:** Schematic of the CVD system with the *in-situ* TFBG sensor assembly magnified for clarity. Thermocouples ("T") are typically located upstream and downstream from the precursor vial. In the case of gold (I) iminopyrrolidinates, a separate bubbler extension is used.

Low-voltage field-emission scanning electron microscopy (FE-SEM) and non-contact atomic force microscopy (NCAFM) were performed on JSM-7500F (JEOL) and PicoPlus II (Molecular Imaging) microscopes, respectively. Scratched zones in the films were generated mechanically and height profile statistics is based on differences between the maximum feature height on the substrate and the maximum height feature on the film surface. These are also correlated with histograms generated from the entire AFM images. Low-voltage SEM was performed on the gold coated Si(100) witness slides as well as the gold-coated alumina pre-coated Si(100) witness slides at beam voltages of 5 kV. After tuning to the resonant frequency of the tip-cantilever system (Nanosensors, PPP-NCL), the frequency is selected such that it falls within the non-contact or deflection region of the tip-sample force curve, effectively making it true non-contact mode AFM. NC-AFM height profiles and roughness (average and root-mean-square) were calculated via WSxM 5.0 (develop 6.3) SPM software suite.[100] X-ray photoelectron spectroscopy (XPS) of gold films from gold (I) iminopyrrolidinate is reported elsewhere,[31] but the compositional analysis collected herein for gold films from gold (I) guanidinates was performed similarly (on the Specs Gmbh analysis chamber system at base pressures of 1 x 10 [-10] Torr), with Ar[+]-sputtering cycles (2-min./cycle) at 0.7 keV. Sputtering was performed until a base composition value of the C 1s was obtained. Atomic layer deposition (ALD) was employed prior to CVD experiments to put a uniform coating of $Al_2O_3$ on the TFBG and Si(100) witness slides. 50 nm and 100 nm pre-coatings of $Al_2O_3$ were obtained using a Picosun R-150 series thermal ALD tool. 516 and 1032 cycles (ALD cycle is alternating pulses of $(CH_3)_3Al$ and $H_2O$) were used to generate the 50 nm and 100 nm $Al_2O_3$ films, respectively.

## Synchrotron radiation characterization: XANES and GID

X-ray absorption near-edge spectroscopy (XANES) measurements were carried out at the bending magnet-based F3 beamline of the Cornell High Energy Synchrotron Source (CHESS) at Cornell University, Ithaca, NY. CHESS is a 5.3 GeV hard X-ray light source which operates in top-up mode (positrons) at a ring current of 200 mA. A silicon (220) double-crystal monochromator (DCM) with an energy resolution (of ~$10^{-4}$ was used to scan X-ray energy across the gold $L_3$-edge (11.919 keV for Au[0]). Thin-film samples were mounted on a goniometer sample stage and were set to an angle just greater than the critical angle (calculated using the AFM-extracted sample thickness and roughness data then experimentally refined by scanning the sample angle across the calculated critical angle) to ensure the entire thickness of the gold films were probed. The detection mode was fluorescent X-ray yield recorded using a Hitachi 4-element silicon vortex detector with an XIA DXP XMAP processor. The detector count rate was kept below 140 000 counts to insure a linear dead time response. All spectra were normalized to dead-time corrected $I_0$, measured using an ion chamber upstream of the sample filled with 100% $N_{2(g)}$. A gold reference foil standard used for energy calibration was measured in transmission mode downstream of the sample between two ion chambers filled with 100% $N_{2(g)}$. All XANES data was calibrated, normalized, and processed using the Athena XAS software package by Demeter. Grazing incidence diffraction (GID) studies were carried out on G2-line at CHESS on a 6-circle diffractometer. Energies between 8 and 16 keV are available for use[164] and in this study, only the sample angle ($\mu$) was rotated during analysis of the gold diffraction rings at indices (200) and (220). Data was post-processed on Matlab using an X-ray utilities package and plotted in reciprocal space. (Figure 7.4b).

## 7.4       Results and Discussion

TFBG-polarized spectral monitoring of gold CVD from gold (I) guanidinates and gold (I) iminopyrrolidinates has resulted in the deposition of anisotropic[54] and isotropic[63] films (with slight dichroism only at thicknesses of > 25 nm) thus far. The use of particular cladding mode resonances from the TFBG-polarized transmission amplitude spectra has been employed to monitor the optical behaviour of the gold films during CVD. Gold films from both precursors result in rough nanocrystalline films, with RMS roughnesses often > 10% of the film thickness. The gold films from gold (I) guanidinates showed spectral attenuation during growth, but only on the order of 10 minutes, whereas gold films from the more recent gold (I) iminopyrrolidinate precursor often induced spectral attenuation for > 1 h. By tracking resonances around 1559 nm, the gold films from the gold (I) guanidinates showed similar attenuation profiles but a strong wavelength shift in the S-polarized peak family of -0.5 nm and no change observed for the P-polarized modes.[54] The peculiar wavelength shift differences from the gold films under both polarization modes and inability to support surface plasmon resonance (SPR) was attributed to the highly roughened and granular nature of the films.[54,63] Moreover the growth rate of 37 nm/min, compared to only 1.1 nm/min for the ultrathin gold films from gold (I) iminopyrrolidinates, was far too high to be able to generate conductive and uniform thin films with high purity. In fact, XPS was later employed to confirm that the gold (I) guanidinate precursor typically decomposes after chemisorption to the substrate (TFBG and witness slides) and this was evidenced by a 13 at.% base value obtained for the C 1s peak envelope after 3 low-energy (0.7 keV) sputtering cycles with $Ar^+$. This was rather high when compared to the 3.2 at.% carbon content found in the gold films obtained from gold(I) iminopyrrolidinates. The latter films were subject to further study, particularly due to their isotropic optical behaviour and therefore lack of polarization dependence. There was precedence in the literature[65,78] to employ a thin dielectric layer ($Al_2O_3$) over the TFBG prior to deposition of the gold over-layer in order to allow a larger wavelength separation of the TFBG resonance modes and enhance the polarization-dependence in the P-polarized peak family by increasing the SRI sensitivity of the TFBG. This has also been done for other fiber configurations (e.g. long-period gratings)[70] and would ideally generate films that are applicable to SPR-based TFBG sensor applications. Figure 7.2 is the transmission amplitude spectrum in the NIR region of the TFBG sensor before pre-coating of $Al_2O_3$, after pre-coating, and after CVD of ultrathin gold films onto the sensitized, pre-coated TFBG. Interestingly, an observable wavelength shift on the order of 500 pm occurs for the 100 nm thickness (Figure 7.2, top) of $Al_2O_3$ pre-coating only for the P-polarized modes, leaving the S-polarized completely unshifted. Moreover, the P-polarized resonance amplitudes increased an average of 2 dB after addition of the $Al_2O_3$ whereas they attenuated by roughly the same amount in the S-polarized peak family. An overall (peak-to-peak difference between gold coating and bare TFBG resonances) wavelength shift of 0.4 nm was observed in the P-polarized mode for the 100 nm $Al_2O_3$ pre-coated sample whereas only 0.15 nm wavelength shift was observed overall in the 50 nm $Al_2O_3$ pre-coated samples. Additionally, peak families under both polarizations attenuated after addition of the 50 nm $Al_2O_3$ pre-coating and an isotropic response to the addition of the pre-coating and gold coating from CVD was observed. That is, a small wavelength shift (< 0.15 nm ) occurs for both of light modes and this directly corroborates our previous findings for the gold coatings on the TFBG with similar thicknesses (see Table 7.1 "Low T." Au-Si(100))[63].

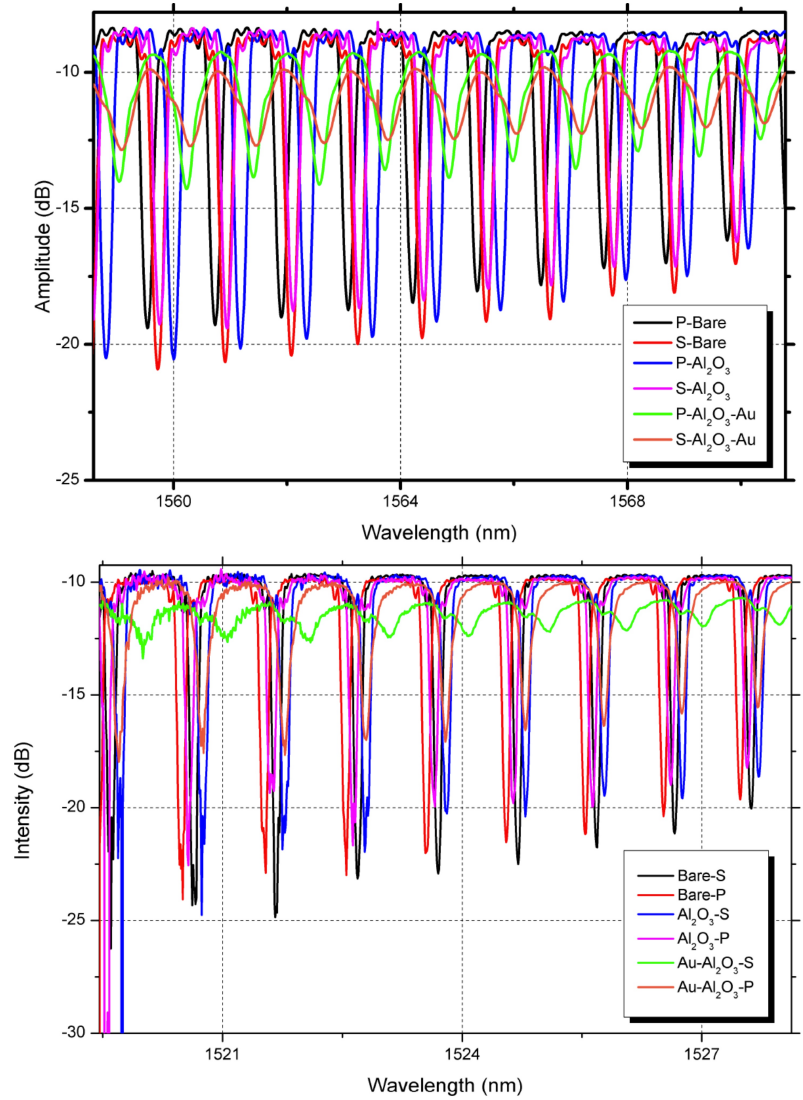

**Figure 7.2:** Transmission amplitude spectra of the bare TFBGs, Al₂O₃ pre-coating as deposited by ALD at thicknesses of 50 nm (bottom) and 100 nm (top), and spectral evolution of the TFBG during deposition of the ultrathin gold films by CVD onto the Al 2O3 pre-coated TFBGs.

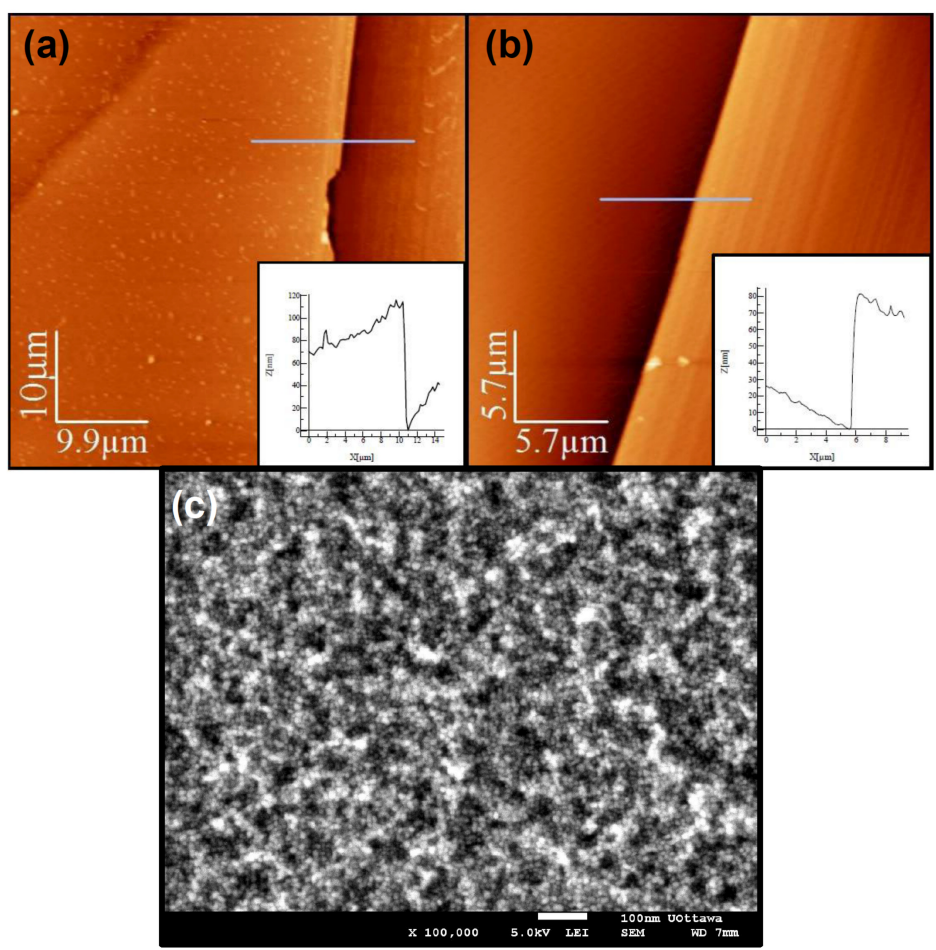

**Figure 7.3:** (a) 10 x 10 µm NC-AFM topographic image of the 100 nm $Al_2O_3$ pre-coating and (b) the 50 nm $Al_2O_3$ on a Si(100) witness slide. Insets are the height profiles for the corresponding thin films (note: statistics are generated from the profiles and correlated to the histograms generated for the same image). (c) SEM image of the 100 nm Au film deposited on the 100 nm $Al_2O_3$ pre-coated witness slide shown in (a). note: grey lines in the AFM images correspond to the height profiles in the insets.

XANES is a powerful local spectroscopic technique that employs the hard X-ray radiation generated by a synchrotron to provide information of the local electronic structure of a thin film. It has been used herein to correlate with the XPS data of the thin films on the Si(100) witness slides, particularly to elucidate the local coordinating environment of the gold absorbers (atoms) in the thin films. The gold films as deposited by CVD directly on the TFBG from the gold (I) iminopyrrolidinates are typically quite lossy, but the island-like growth mode of the ellipsoidal nanoparticles prevents coalescence into a continuous uniform film that has a negative real component of the effective permittivity (i.e. supports SPR).[63] Use of XANES and GID was required to explore the effect a thin dielectric pre-coating may have on the overall film morphology and the spectral evolution of the same gold films generated from CVD. XANES also provides conclusive information about the exact oxidation state of the gold in the film and the nature of the base carbon signal provided by the XPS/sputter data. In this case, it was found that the Au $L_3$ edge features are largely dominated by the bulk $Au^0$ species, with minor Au-C and Au-O contributions. Unlikely $2p \rightarrow 5d$ transitions that can occur with a concurrent increase in edge energy with oxidation state were not observed. The major edge features at 11.945 keV and 11.967 keV correspond to metallic $Au^0$ and occur for particle sizes down to 1 nm.[165] Because of the minor variation that is expected for beam-line intensity with respect to the absorption edge, the small differences in absorption edge energies cannot be accurately resolved.

**Table 7.1:** Summary of NC-AFM and XANES data for selected gold coatings on the TFBG and on the $Al_2O_3$ pre-coated TFBG

| Sample | Thickness (nm) | $R_{RMS}$ (nm) | Au-$L_3$ Edge Energy (keV) |
|---|---|---|---|
| [a]Au-Si(100)(High T.) | 45 | 13 | 11.9226 |
| [b]Au-Si(100)(Low T.) | 50 | 14 | 11.9234 |
| Au-$Al_2O_3$(50 nm)-Si(100) | 51 | 9 | N/A |
| Au-$Al_2O_3$(100 nm)-Si(100) | 100 | 12 | 11.9223 |
| [c]Au-$SiO_2$ (TFBG) | 50 | N/A | 11.9196 |

[a] "High T." corresponds to the substrate temperature, which was T=350 ºC
[b] "Low T." corresponds to the substrate temperature, which was T=200 ºC
[c] Thickness determined from AFM of a coated Si(100) witness slide adjacent to the fiber during CVD. Roughness parameters of the films on the fiber have not been determined.

However, we speculate that because grain sizes of the gold nanoparticles on the pre-coated slides and TFBG are only 30 nm versus an average of 50 nm for the gold nanoparticles directly on the Si(100) substrates,[63] the relatively smaller particles resolve at slightly higher $L_3$-edge absorption energies. GID data collected for the gold films directly on Si(100) witness slides (Figure 7.4b, right) and on the pre-coated witness slides (Figure 7.4b, left) along the two major diffraction peaks ((111), (200)) indicates that the thin films deposited on the pre-coated substrates are a lot more powder-textured and low stress. This is likely a result of the smoothing effect induced by the Al$_2$O$_3$ pre-coating on the deposited gold films. In fact, the root-mean-square roughnesses of gold films (Table 7.1) directly on the Si(100) witness slides is on average ~25% of the film thickness itself, whereas the roughnesses of the gold films for similar thicknesses deposited on the dielectric pre-coating are ~10-14% of the film thickness. These roughness parameters in both cases are far from the atomically smooth roughnesses observed in films generated by other deposition methods such as sputtering or ALD.

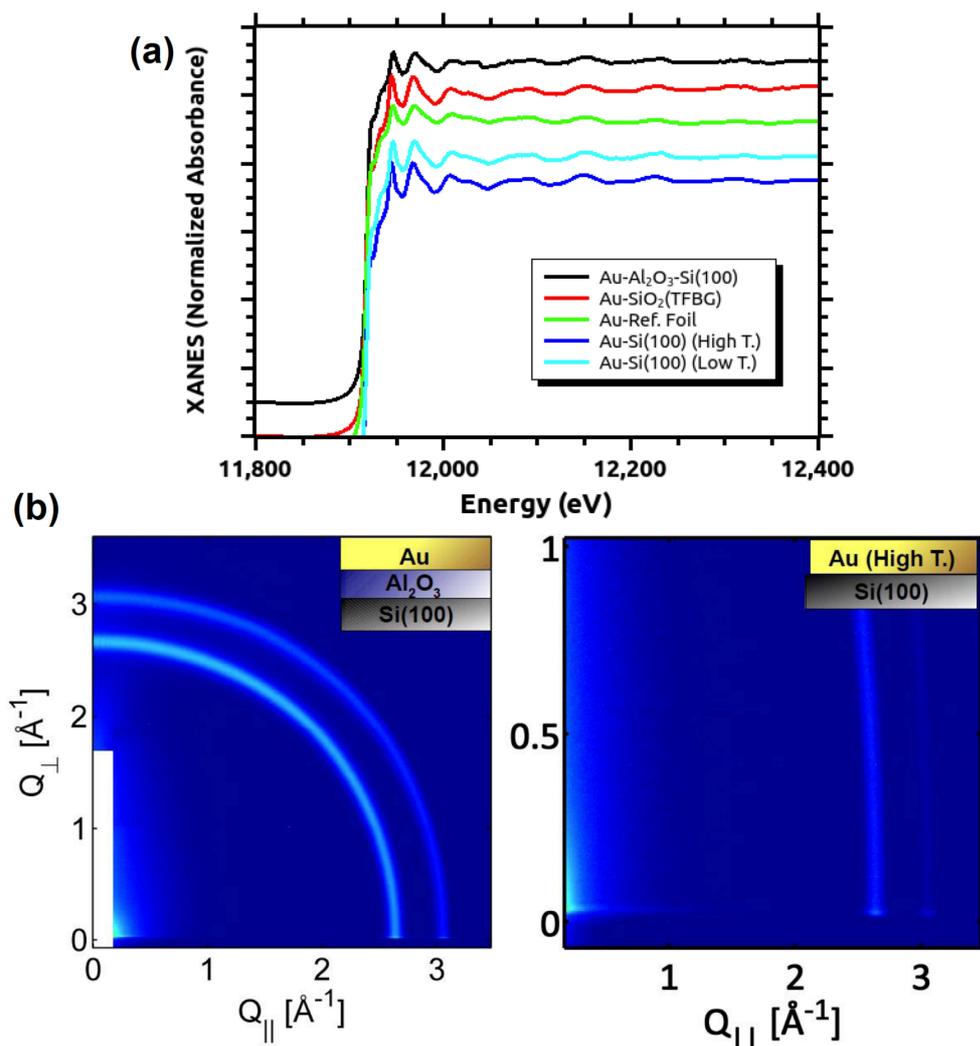

**Figure 7.4:** (a) XANES and (b) GID data collected at CHESS and plotted in reciprocal space for the selected films in Table 7.1. Note that the XANES spectra are normalized but vertically stacked for clarity around the Au L$_3$ edge. Left GID pattern is for the 100 nm Au film on the 100 nm Al$_2$O$_3$ pre-coated witness slide.

The anisotropic property seen in previously reported gold films from gold (I) guanidinates from CVD has been retained for the gold films deposited from a single-source gold (I) iminopyrrolidinate precursor simply by the addition of a thin dielectric pre-coating. By effectively pre-depositing a thin layer of $Al_2O_3$ by ALD onto the TFBGs, the wavelength shift induced on the P-polarized resonances (i.e. resonances responsible for the propagation of the surface plasmon polaritons) is enhanced by 0.3-0.4 nm. Moreover, the isotropic response of the S- and P-polarization modes is eliminated and the polarization-dependent attenuation (birefringence) of the TFBG-polarized transmission amplitude spectra is optimized.

Further studies are required to study the effect of the dielectric layer at various thicknesses and to quantify the exact SRI sensitivity enhancement provided by the $Al_2O_3$ pre-coating.

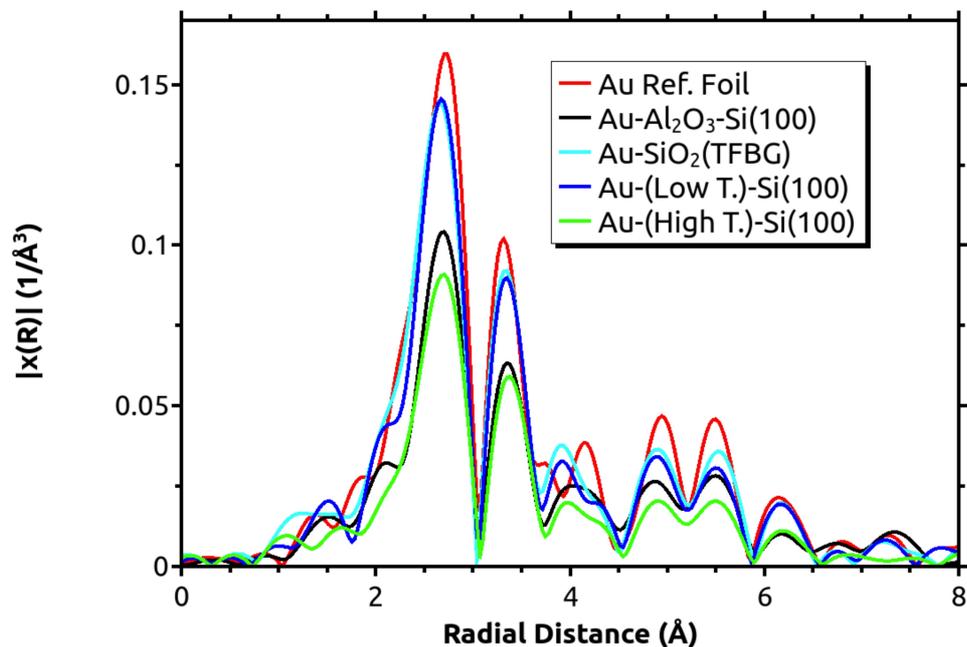

**Figure 7.5:** Fourier transforms of the $k^3$-weighted fine structure fitted data from raw X-ray absorption fine structure (XAFS) data obtained at CHESS of the gold films directly on the $Al_2O_3$-pre-coated TFBG and on uncoated Si witness slidse. A k-range of ~4.50 to 11.53 Å$^{-1}$ was selected, with Fourier transforms perform with a Hanning window applied.

# Chapter 8

## The Effect of ALD-grown $Al_2O_3$ on the Refractive Index Sensitivity of CVD Gold-coated Optical Fiber Sensors

**Mandia, D.J.\*[1]**; Zhou, W.[2]; Ward, Matthew J.[3]; Joress, H.[3]; Sims, J.J.[4]; Giorgi, J.B.[4]; Albert, J.[2]; Barry, S.T.[1] "The Effect of ALD-grown $Al_2O_3$ on the Refractive Index Sensitivity of CVD Gold-coated Optical Fiber Sensors" *Nanotechnology.* **2015**; *26*(43) 43002; pp. 1-12

[1] Department of Chemistry, Carleton University, 1125 Colonel By Drive, Ottawa, Ontario, Canada, K1S 5B6

[2] Department of Electronics, Carleton University, 1125 Colonel By Drive, Ottawa, Ontario,

Canada, K1S 5B6

[3] Cornell High Energy Synchrotron Source, Cornell University, 161 Synchrotron Drive, Ithaca, New York, USA, 14853

[4] Department of Chemistry, University of Ottawa, 10 Marie Curie Pvt., Ottawa, Ontario, Canada, K1N 6N5

\*Corresponding author

## 8.1        Abstract

The combined effect of nanoscale dielectric and metallic layers prepared by atomic layer deposition (ALD) and chemical vapor deposition (CVD) on the refractometric properties of tilted optical fiber Bragg gratings (TFBG) is studied. A high index intermediate layer made up of either 50 nm or 100 nm layers of $Al_2O_3$ (refractive index near 1.62) was deposited by ALD and followed by thin gold layers (30–65 nm) deposited from a known single-source gold (I) iminopyrrolidinate CVD precursor. The fabricated devices were immersed in different surrounding refractive indices (SRI) and the spectral transmission response of the TFBGs was measured. Preliminary results indicate that the addition of the dielectric $Al_2O_3$ pre-coating enhances the SRI sensitivity by up to 75% but this enhancement is highly dependent on the polarization and dielectric thickness. In fact, the sensitivity decreases by up to 50% for certain cases. These effects are discussed with support from TFBG simulations and models, by quantifying the penetration of the evanescently coupled light out of the fiber through the various coating layers. Additional characterization studies have been carried out on these samples to further correlate the optical behaviour of the coated TFBGs with the physical properties of the gold and $Al_2O_3$ layers, using atomic force microscopy, X-ray photoelectron spectroscopy, and an ensemble of other optical and X-ray absorption spectroscopy techniques. The purity, roughness, and morphology of gold thin films deposited by CVD onto the dielectric-TFBG surface are also provided.

## 8.2 Introduction

Gold and dielectric materials such as alumina ($Al_2O_3$) are widely used in a variety of applications such as optoelectronics,[165–168] light-harvesting for solar cell applications,[169,170] and plasmonics.[171] One of the current (and benchmark) high-performance methods of generating highly-dense, uniform, and conformal $Al_2O_3$ thin films is by atomic layer deposition (ALD).[172,173] In particular, the use of alternating exposures of trimethylaluminum (TMA) and $H_2O$ is the most well-established ALD process for $Al_2O_3$ that exists in the literature.[15] Due to the excellent film thickness control offered by the cyclic and self-limiting surface chemistry of this process, alumina's optical properties are highly controllable. In fact, many applications have benefited from $Al_2O_3$ as the dielectric material because of its stable refractive index (RI) (typically 1.62) and remarkable compatibility with Si substrates such as Si(100).[174] Because of its ability to coat geometrically complex surfaces (e.g. high aspect ratio structures), ALD is also employed in many applications[175] that require uniform coverage of a dielectric and metallic material with minimal to no defects.

Recently, many reports regarding the sub-wavelength optical properties of nanostructures such as Au nanoparticles (AuNPs) have highlighted the importance of incorporating a dielectric material to improve the performance of waveguiding structures that require high electric field localization (such as those utilizing surface plasmon resonance, SPR). Recent studies involving hybrid metal-dielectric waveguides,[176] AuNP assemblies,[105,108] and ultrathin AuNP clusters[124] have all emphasized the importance of a dielectric component in enhancing the plasmonic performance (i.e. allowing the films to support surface plasmon polaritons, SPPs) of the nanostructured materials. Also, the inclusion of AuNPs directly into a dielectric matrix such as an optical fiber has shown improvements to AuNP-based sensing experiments in the UV to near infrared range (NIR).[20] Recently, photonic crystal fiber LPGs were employed to detect humidity ingression in the relative humidity range of 20%–40% with remarkable sensitivity, making these point sensors promising candidates for corrosion detection in reinforced concrete structures and other infrastructure applications.[20,177,178] Numerous other examples have recently demonstrated the use of optical fiber technology as a method for refractive index sensing, with a particular focus on fiber-optic configurations such as long period fiber gratings,[179] D-shaped plastic optical fiber sensors,[180] exposed core fiber Bragg gratings (FBGs),[181] and a variety of other FBGs such as tilted FGBs (TFBGs).[45,53,54,182] TFBGs are the fiber-optic platforms employed in the present work.

TFBGs are ideally suited for these studies because of their capability to excite a number of cladding modes whose properties can be probed very precisely by measuring the wideband spectral response of the grating transmission.[41] Moreover, by referencing to the attenuation and wavelength shift of the core mode back-reflection of the TFBG transmission spectrum, the cross-sensitivity issues with high temperatures or strain during *in-sit*u chemical vapor deposition (CVD) monitoring are removed; this is one of the main advantages that TFBGs offer over other fiber-optic configurations.[41] Additionally, due to breakup of the circular symmetry caused by the tilted grating planes, the polarization state of the incident core mode controls the orientation and polarization of the excited cladding modes at the cladding boundary.[41] This is of utmost importance for metal coatings as the boundary conditions for metal-dielectric interfaces depend very strongly on the polarization state of the light. For example, cladding modes with radial polarization (corresponding to transverse magnetic, or TM polarization at the interface) at the cladding boundary can be excited selectively and be used to couple light into surface plasmon waves on metal-coated fibers. A gold-coated TFBG refractometer based on SPR was demonstrated using this technique,[48] and shown to yield a great enhancement in the minimum surrounding refractive index (SRI) detection level.[50] As well, it was recently demonstrated that by depositing oriented silver nanowires on optical fibers, a localized SPR at communication wavelengths can be excited by cladding modes with azimuthally polarized (corresponding to transverse electric, or TE) evanescent electric fields that are parallel to the nanowires.[154] Even without plasmonic enhancement, an ultrathin gold coating with an anomalous complex permittivity (i.e. positive real part) can also enhance the SRI sensitivity of TFBGs, especially for the TE-polarized cladding modes.[182]

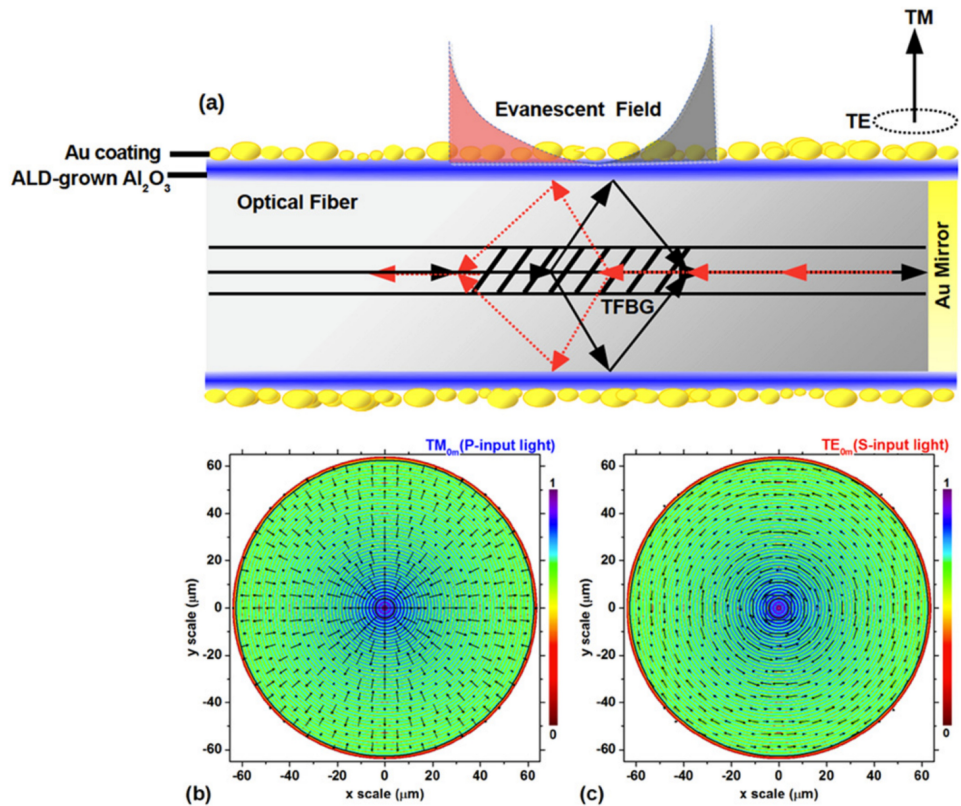

**Figure 8.1:** (a) Side-view cross-section schematic of the TFBG with the dielectric, ALD-grown Al$_2$O$_3$ and CVD-grown Au coating, with a sputtered gold film on the fiber end to act as a mirror. The bottom two panels represent electric field distributions of the vectorial cladding modes corresponding to TM-polarized mode (b) and TE-polarized mode (c).

In this work, the SRI sensitivity enhancement for TFBGs coated with a hybrid film including an inner dielectric layer (Al$_2$O$_3$ ALD film) and an outer metal layer (Au CVD film) are investigated. The expected enhancement arising from the additional dielectric layer comes from the fact that the RI of the coating is higher than that of the fiber cladding glass and thus 'pulls' the electromagnetic fields out of the fiber and into the overlaying gold, making them more sensitive to small SRI changes. Figure 8.1 depicts a TFBG structure coated by such a hybrid film. Due to the inclination of the grating planes of a TFBG along a specific direction, two families of polarized cladding modes with radial (TM) or azimuthal (TE) electric fields at the fiber cladding boundary can be selectively excited by linearly polarized input core-guided light aligned parallel (P) or perpendicular (S) to the grating planes, respectively (as shown in Figure 8.1). The SRI sensitivities of the two families of the polarized cladding modes are experimentally investigated for the TFBGs coated by the hybrid films with different thicknesses, and compared with the cases of the Au-only CVD coatings. Finally, the power density profiles of the evanescent fields of the polarized cladding modes in structures with 50 nm and 100 nm thick $Al_2O_3$ film-coated optical fibers are simulated under various SRIs.

## 8.3　　　　Methods

### TFBG Sensing Principle

The TFBGs were written by the phase-mask method [41] in hydrogen-loaded CORNING SMF-28 ($GeO_2$-doped $SiO_2$ core) fibers using a KrF excimer laser. Based on the phase matching condition, the forward propagating core mode can be coupled into a backward propagating core mode and a number of cladding modes by the tilted grating planes. The resonant wavelengths of the excited core mode $\lambda_{co}$ and cladding modes $\lambda_{cl}$ can be expressed as,

$$\lambda_{co} = 2 n_{eff}^{co} \Lambda \qquad\qquad [1]$$

$$\lambda_{cl} = (n_{eff}^{co} + n_{eff}^{cl}) \Lambda \qquad\qquad [2]$$

where $n_{eff}^{core}$ and $n_{eff}^{cl}$ represent the effective indices of the core mode and cladding mode resonances, respectively, with $\Lambda$ being the grating period. Due to the evanescent fields around the optical fiber surface, the effective indices of the cladding modes ($n_{eff}^{cl}$) depend on the SRI and change when the SRI is perturbed in any way, resulting in measurable resonance wavelength shifts. Thus, the SRI dependence of the cladding modes can be enhanced by coating dielectric and/or metal films on the TFBG surface, where the enhancement amplitude is mainly dependent on the optical properties (permittivity) and thickness of the coatings.

## Film Depositions on Optical Fibers

A full and detailed description of the gold CVD process on TFBG substrates and synthesis of the gold(I) iminopyrrolidinate CVD precursor are reported elsewhere.[31,54,63] It should be noted that the sputtered gold mirror on the fiber end is thick enough (~300 nm) that the reflected transmission spectrum of the TFBG is not affected by layers deposited thereafter or by the high temperatures of the CVD process. The only purpose of the end mirror is to be able to measure the TFBG transmission by capturing the reflected light (as discussed below in the section on spectral measurements) instead of having to loop the fiber back out of the deposition chamber. Deposition of the 50 nm and 100 nm $Al_2O_3$ dielectric layers was performed by ALD on a Picosun R150 Thermal ALD tool.  $Al_2O_3$ was deposited by using alternating exposures of trimethyl aluminium (TMA, 98%, Strem) and deionized water at a substrate temperature of 150 ºC in $N_2$ carrier gas. Precise thickness control of the $Al_2O_3$ layers was obtained by using 516 cycles and 1032 cycles of TMA and $H_2O$ for the 50 nm layer and 100 nm layer, respectively.[183] CVD of the Au coatings could be deposited with thickness control to the nearest 1 nm. Si (100) with native oxide witness slides were used for ex situ measurements. This substrate was chosen to allow comparison of the deposited films with literature examples, which all use this substrate.

### X-ray Reflectivity (XRR)

High-energy (hard X-ray) synchrotron XRR was performed to characterize the Au coatings on the Si(100) and $Al_2O_3$-coated witness slides. This was carried out at the G2 beamline of the Cornell High Energy Synchrotron Source (CHESS) on a 6-stage diffractometer equipped with a Be(0001) monochromator. A comprehensive review of the G2 beamline diffractometer and its six-circle κ goniometer system is available elsewhere.[164] The X-ray beam energy for the XRR experiments was 10.04 keV and data was collected over the angular range of $0.25° \leqslant \theta \leqslant 4.00°$ increments. Analysis of the reflectivity data was performed using the Motofit package within an Igor Pro 6.36 environment. Motofit employs a genetic algorithm-based, slab-model fitting procedure that allows the momentum transfer vector, otherwise known as Parrat's recursion formula,[184,185] $Q=4\pi(\sin\theta)/\lambda$ to be calculated for a multi-layer stack based on the scattering layer densities and constituent layer roughness parameters. This was used to directly estimate film thickness based on the periodicity of the Kiessig fringes.

### X-ray Photoelectron Spectroscopy (XPS) and $Ar^+$ Etching

High-resolution XPS (14.27 keV Al Kα source) and $Ar^+$-sputtering were performed in a multi-technique ultra-high vacuum system (Specs, Gmbh). A witness slide (Au-$Al_2O_3$(100 nm)-Si(100)) was mounted to a copper sample holder and transferred through a load-lock system to the main manipulator arm of the analysis chamber. To prevent potential surface charging effects during XPS, the sample was mounted such that it could be grounded through the manipulator arm and out to a reference potential. Base pressure of the analysis chamber during XPS acquisitions ranged between 8 x $10^{-10}$ and 1 x $10^{-9}$ Torr. Typical $Ar^+$-sputtering cycles lasted 2 min each at an acceleration voltage of 1.71 keV in a background pressure of 1 x $10^{-5}$ Torr. Due to Ar-incorporation into the films, $Ar^+$-sputtering is typically halted after a maximum of ~40 min. A Phoibos hemi-spherical analyzer system and detector were used for analysis, and both CasaXPS and Igor Pro 6.36 software suites were used for XPS peak plotting and deconvolution. All fitting procedures performed used the relative sensitivity factors from the Scofield library, which is built into the CasaXPS software.

**Atomic Force Microscopy (AFM)**

Thickness measurements by AFM were performed in intermittent contact mode on a Molecular Imaging (Agilent) PicoSPM II system with a custom isolation chamber. A diamond-tip pen was used to put a small scratch in the sample witness slides prior to mounting them on the AFM stage. To direct the AFM tip into the scratched zone of the sample, an optical microscope fitted with a CCD camera was used to get the cantilever directly over one side of the scratch. AFM images were collected on 4 buffer systems (2 topographic, 1 amplitude, 1 phase) for each sample in Pico Scan 5.0 software using Nanosensors (uncoated) PPP-NCL AFM tips (RF ~164 kHz). Image analysis to obtain height profiles and root-mean-square (RMS) roughness parameters of the samples was performed in WSxM 5.0 Develop 7.0 software.[100] Due to tip-sample convolution effects on the lateral (xy) resolution of AFM images, ImageJ software was used for particle size analysis on the microscopic (AFM and SEM) images.

**Spectral Measurements**

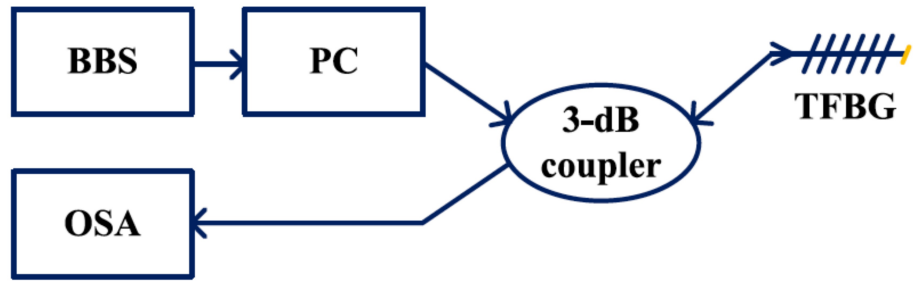

**Figure 8.2:** Schematic diagram of the experimental setup for reflected transmission measurements of TFBG probes.

By cleaving the fiber downstream from the TFBG area and depositing a sputtered gold mirror thicker than 300 nm on the cleaved fiber end, the reflective transmission measurement can be implemented on the TFBG probe. Figure 8.2 shows the experimental setup of the TFBG spectral measurement, where a broadband source (BBS) (JDSU BBS1560) launches light from 1520 to 1620 nm into the fiber core, a polarization controller (JDSU PR2000) is used to generate a TE- or TM-polarized core mode at the grating, and an optical spectrum analyser (OSA) (ANDO AQ6317B) measures the reflected transmission spectrum from the TFBG probe with a wavelength resolution of 0.01 nm. The light transmitted through the TFBG twice returns towards the BBS in the same fiber and a 3 dB fiber coupler picks off part of the reflected light to be measured by the OSA. The measured light intensity levels are relative and measured in dB. The spectral measurements of the coated TFBGs under different surrounding refractive indices (SRIs) were performed in a variety of NaCl (saline) solutions with salt mass concentrations ranging from 0% (DI water) to 24%. The three kinds of TFBGs used in this study (Bare-TFBG, Au-coated TFBG, Au–Al$_2$O$_3$-coated TFBG) were affixed to a fiber holder and subsequently immersed in the solutions by pipetting the solution onto the fiber such that a meniscus coats only the area of the fiber containing the TFBG. Since the penetration depths of the fiber evanescent waves do not exceed a few micrometers, this method is equivalent to full immersion of the fibers in the saline solutions. Therefore, the *ex-situ* spectral measurements of the coated TFBGs can be obtained under highly accurate polarization states, and temperature effects removed from the measurement by measuring resonance positions relative to that of the core mode, since all resonances have the same temperature dependence while the core mode is inherently insensitive to the SRI.

It should also be noted that standard telecommunication optical fiber was used due to its outstanding waveguiding properties between 1500 and 1620 nm, and because fiber-based instrumentation for these wavelengths is readily available. The transmission spectra of the two TFBGs used for the $Al_2O_3$ pre-coating experiments were identical with the exception of the Bragg resonance wavelength, which was used to normalize the spectra and remove any cross-sensitivity effects due to temperature or strain.

# 8.4    Results and Discussions

## Surface Characterization of Au Films

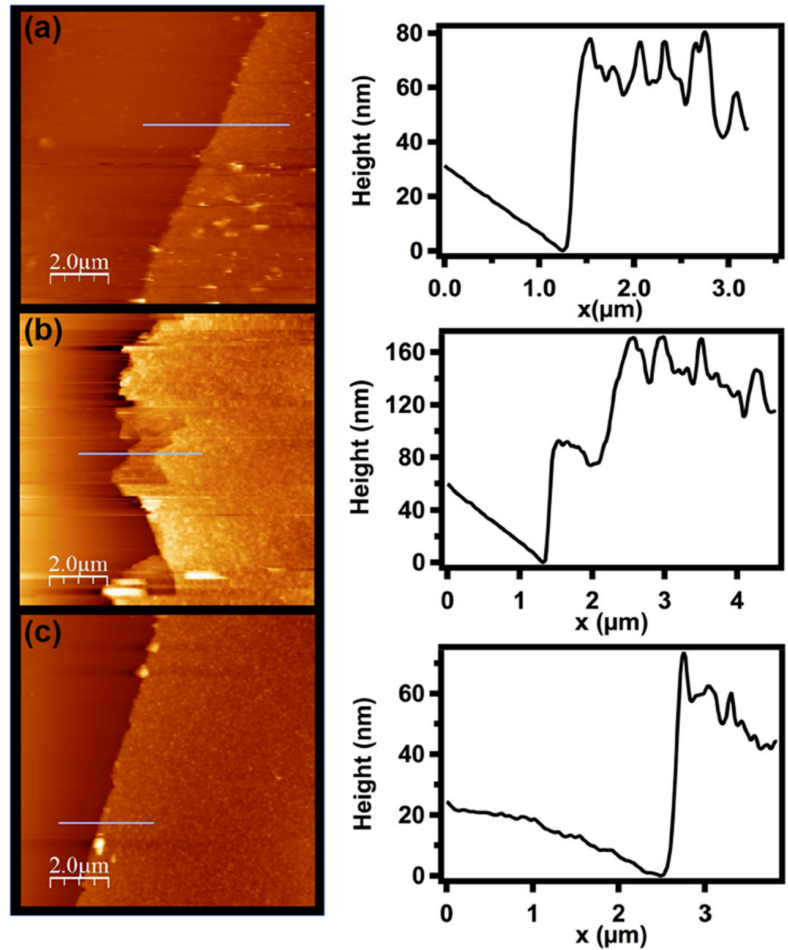

**Figure 8.3:** 8 μm x 8 μm AFM topographic images of film-scratched regions for (a) Au coating on 50 nm Al₂O₃ pre-coated Si(100) witness slide, (b) Au coating on 100 nm Al₂O₃ pre-coated Si(100) witness slide, (c) Au coating on Si(100) witness slide. Height profiles for the selected cross-section (gray lines in the AFM images) of each AFM topographic image are shown on the right.

For further characterization of the CVD process, Au coatings were also deposited on Si(100) and Al$_2$O$_3$-coated Si(100) witness slides at the same time as on the fibers. Figures 8.3(a)– (c) depicts the intermittent contact mode AFM images of

**Table 8.1:** Thickness and roughness parameters for the Au coatings as-deposited on Si(100) or Al$_2$O$_3$ pre-coated Si(100) witness slides.

| Sample | AFM Thickness (nm) | XRR Thickness (nm) | *Particle Size (nm) | RMS Roughness (nm) |
|---|---|---|---|---|
| Au-Si(100) | 51.1 | N/A | 14.27 | 18.43 |
| Au-50 nm Al$_2$O$_3$-Si(100) | 52.1 | 45.53 | 11.21 | 5.40 |
| Au-100 nm Al$_2$O$_3$-Si(100) | 99.1 | N/A | 9.39 | 15.4 |

*Particle size determination was verified by post-processing of AFM (Figure 3a-c) and SEM images (not shown) using ImageJ.

these coatings along with sample height profiles of the film scratches to reveal constituent layer thicknesses. The thickness and roughness parameters derived from the AFM measurements are summarized in table 8.1. The rms roughness of the Au coating as deposited directly on Si(100) is ~36% of the thickness of the film itself, which is quite characteristic of a roughened and unstressed granular film. However, with the addition of a 50 nm Al$_2$O$_3$ coating interface between the Au coating and Si substrate, an Au coating of very similar thickness (52.1 nm) has an rms roughness equivalent to only

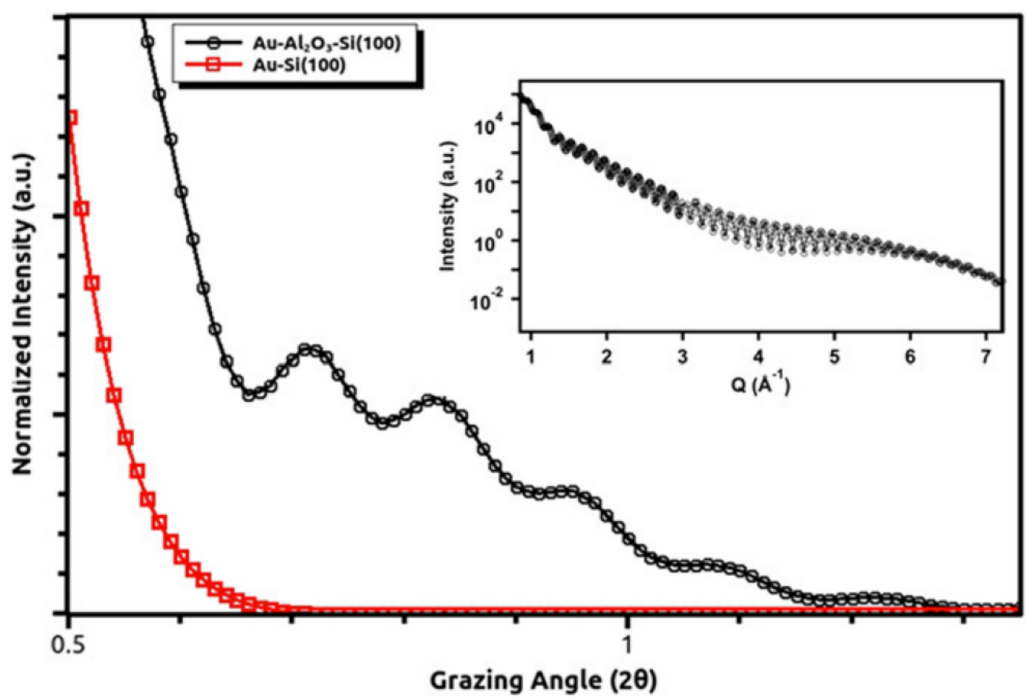

**Figure 8.4:** X-ray reflectivity profile of the Au coating on 50 nm $Al_2O_3$ pre-coated Si(100) witness slide (black trace) and Au coating of similar thickness (table 8.1) on Si(100) witness slide (red trace). Note: critical angle for the reflectivity experiments was set to 0.075°. Au film thickness (from Kiessig fringes of black trace) was extracted from the reflectivity plot in the inset.

~10% of that thickness. Furthermore, it was found that the average particle size decreases from 14.27 nm to 9.39 nm for the Au coatings deposited directly on Si(100) and ALD $Al_2O_3$ pre-coated Si(100) witness slide, respectively. The AFM height profile for the Au-100 nm $Al_2O_3$-Si(100) sample in figure 8.3(b) shows that there is an intermediate plateau in the height that roughly equals the height of the interfacial $Al_2O_3$ layer followed by another plateau ~150 nm, which corresponds to the Au coating deposited as the top layer (i.e. Au over-coating). XPS and $Ar^+$-sputter depth profiling verified that the $Al_2O_3$ layer is relatively inert despite causing better wetting of the Au coating from CVD when compared to deposition directly onto Si(100) (vide infra).

This de-roughening of the Au coating induced by the $Al_2O_3$ interfacial layer was further investigated by XRR. Mobility or inter-diffusion of the constituent layers was evident by the presence of Kiessig fringe patterns (figure 8.4) in the Au-50 nm $Al_2O_3$-Si(100) sample, whereas a mirror-like reflectivity that is characteristic of thicker and/or highly roughened films was found for the Au-Si(100) sample (Au layer thickness same as found in table 8.1). The Kiessig fringes can be used to directly estimate film thicknesses, even in a multi-layer stack.[186] By obtaining the absolute reflectivity (inset of figure 8.4) of the film in Q space, the periodicity of the fringes was used to directly calculate the top constituent layer (Au coating) thickness, which was found to be 45.53 nm (table 8.1). Due to the high interfacial roughness of the Au-Si (100) film (red trace, figure 8.4), a fringe pattern necessary for thickness, density, or any other thin film metrics could not be generated. Similarly, the reflectivity from the Au-100 nm $Al_2O_3$-Si(100) sample (not shown) was overshadowed by the constituent layer thickness and roughness parameters of the sample and therefore a Kiessig fringe pattern was not obtained. The deviation between AFM and XRR generated thicknesses (table 8.1) is due to the different length scales on which they operate; where XRR covers more lateral range compared to the local and profilometric AFM method.[187] Nevertheless, the surface de-roughening effect of the interfacial dielectric ALD $Al_2O_3$ layers was apparent from both the AFM and XRR data, and this effect induces better uniformity of the Au over-coatings, which has strong optoelectronic implications that are discussed later in this article.

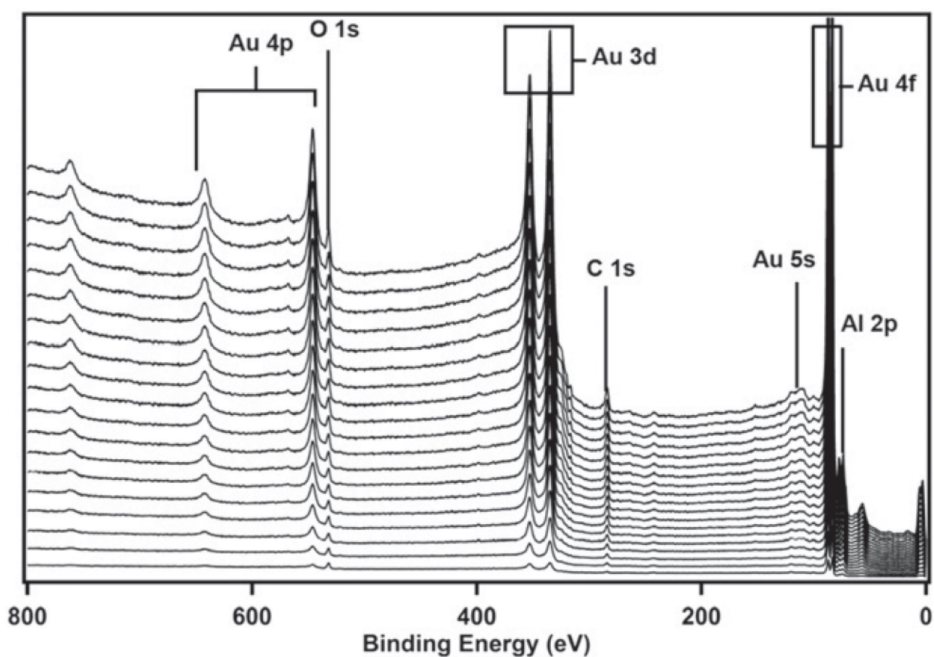

**Figure 8.5:** XPS survey scan of the Au coating on the 100 nm $Al_2O_3$ pre-coated Si(100) witness slide. The regions denoting the relevant surface species were also subject to high-resolution scans (figure 8.6). Note: the survey spectra have been vertically stacked for clarity.

To further probe the Au–$Al_2O_3$ interface, the Au-100 nm $Al_2O_3$–Si(100) system was studied by XPS and $Ar^+$-sputter depth profiling. Overall, the XPS data suggests that the $Al_2O_3$ interfacial layer partially intermixed with the Au layer. This likely occurs during the CVD process, where the substrate temperature was 350 ºC–400 ºC. This intermixing caused features in the Al 2p spectra resulting from Au–Al and Au– Al–$AlO_x$ (sub- and super-oxide) species, but approached stoichiometric $Al_2O_3$ after > 35 min of Ar$^+$-sputtering.

Figures 8.5 and 8.6 show the high-resolution survey scans, as well as the peak envelope data for the Au 4f (after 5 Ar$^+$-sputter cycles) and Al 2p regions (as-deposited and after 19 Ar$^+$-sputter cycles). Also shown in figure 8.6(d) is the Au:Al ratio as a function of Ar+-sputtering time. The as-deposited films had a composition of mostly Au (78.24 at.%), C (4.85 at.%) and Al$_x$O$_y$ (11.42 at.% Al, 4.30 at.% O) and impurity-level concentrations of Si that ranged from 0.19–1.19 at.% throughout the analysis. The high-resolution Al 2p regions in figures 8.6(a) and 8.6(b) reveal a very complex distribution of Al species that appear to be intermixed with the Au layer presumably from the CVD process. Interestingly, in addition to contributions from the expected oxidized Al species (figure 8.6(a), as-deposited sample) which appear at 74.79 eV and 75.19 eV for the spin–orbit split Al 2p$_{3/2}$ and Al 2p$_{1/2}$ species, respectively, there are also contributions for metallic Al species at 73.47 eV and 73.87 eV, which suggests a possible phase separation.[188] Evidence of intermixing of the Al$_2$O$_3$ layer with the Au over-coating is more evident from the presence of the Au 5p$_{1/2}$ peak found at 71.8 eV[189] and also another Al 2p contribution at 77.3 eV that is likely due to the presence of Al/AlO$_x$ (superoxidic) or Al/Au/AlO$_x$ within the 99.1 nm Au coating [174,190] It is evident from figure 8.6(c) that the Au 4f region (after 15 min of Ar⁺-sputtering) remains quite constant, with dominant peaks at 83.70 eV and 87.34 eV corresponding to the Au $4f_{7/2}$ and Au $4f_{5/2}$ regions, respectively, that are characteristic of the spin–orbit splitting components of bulk Au⁰.[191] The slight asymmetry of the peak envelopes is due to the partial incorporation of Al in the Au coating and the relatively stable Au: Al ratio (figure 8.6(d)) throughout Ar⁺-sputtering.

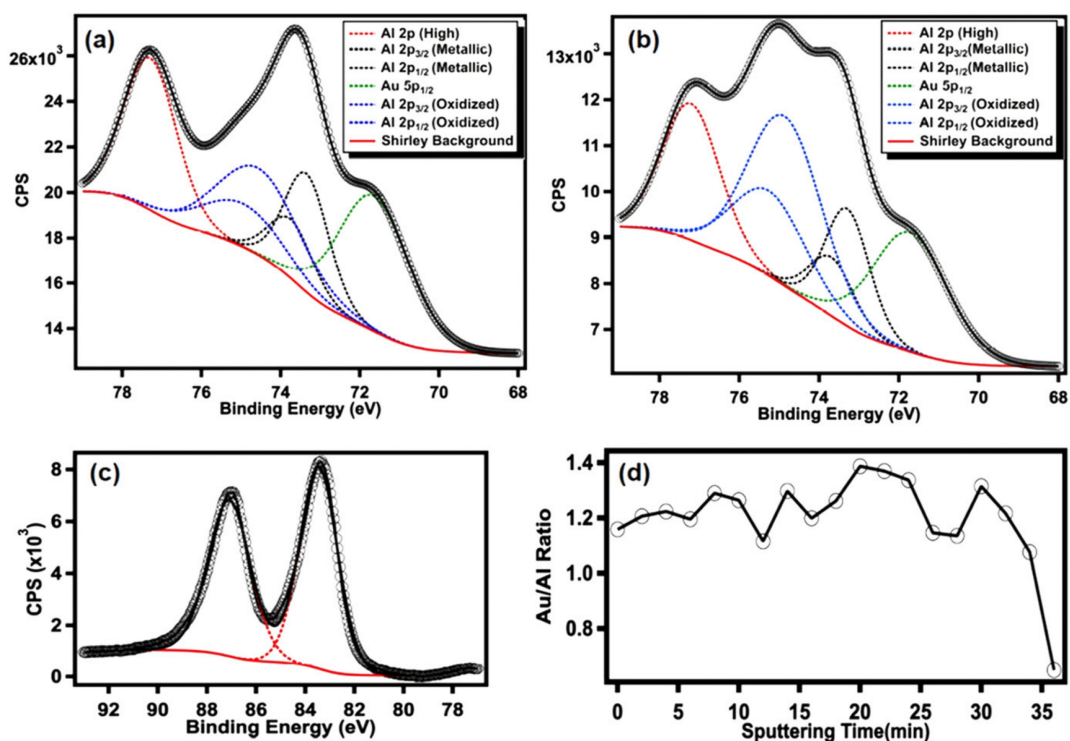

**Figure 8.6:** High-resolution XPS peak envelopes of the Au coating on 100 nm $Al_2O_3$ pre-coated Si(100) witness slide (a) Al 2p region (asdeposited), (b) Al 2p region after ~20 Ar⁺-sputter cycles (~36 min in (d)), (c) Au 4f region and (d) the Au/Al ratio throughout Ar⁺-sputter depth profiling. Note: typical Ar⁺-sputter cycle is ~1.71 keV/2 min bombardment. All peak regions were fit to a Shirley background and mixed Gaussian–Lorentzian functions were used for peak fittings.

## Spectral Response of the Coated TFBGs

Two TFBG sensors were fabricated with Bragg wavelengths at 1562.04 nm and 1611.45 nm. In order to compare similar cladding modes in the two samples, and due to the different Bragg wavelengths, two pairs of polarized cladding modes (with a wavelength separation of 0.13 nm between the TE and TM mode of each pair) located at the same spectral distance from their respective Bragg wavelength were used, i.e. around 1529 nm and 1574 nm. As a result, the effective indices of the two pairs of cladding modes were around 1.384 with a chromatic dispersion-caused difference of 0.006. Thus, the selected cladding modes of each fiber had similar SRI dependences.

**Table 8.2:** Bragg wavelengths, grating period, and wavelengths of selected cladding mode resonances of two bare TFBG samples.

| Sample | $\lambda_{co}$ / nm | $\Lambda$ / nm | $\lambda_{cl}$-TM / nm | $\lambda_{cl}$-TE / nm |
|--------|--------|--------|--------|--------|
| TFBG-1 | 1562.04 | 539.4 | 1529.37 | 1529.50 |
| TFBG-2 | 1611.45 | 556.4 | 1573.96 | 1574.09 |

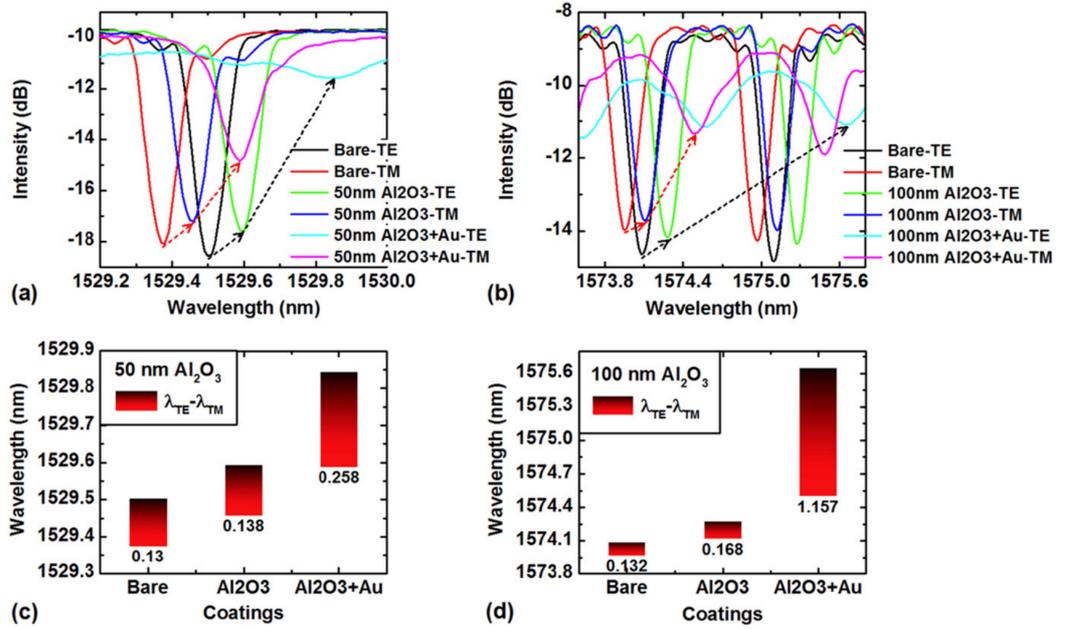

**Figure 8.7:** Polarized spectra of TFBG-1 (a) and TFBG-2 (b) before and after coating $Al_2O_3$ and Au films (all spectra are normalized with Bragg resonances). (c) and (d) show the extracted wavelength separation between two polarized cladding mode resonances for each coating from the two TFBG samples, respectively. The different widths of the cladding mode resonances are caused by the different grating lengths of 10 mm and 4 mm for TFBG-1 and TFBG-2, respectively.

The resonance wavelengths of the core mode and selected cladding modes of the two bare TFBG samples are shown in table 8.2. Figures 8.7(a) and 8.7(b) shows the measured spectra of the two TFBGs around the selected polarized cladding mode resonances for bare, $Al_2O_3$-coated, and $Al_2O_3$-Au-coated cases, respectively. The tracked wavelength shifts are marked by arrows. It can be seen that the wavelength separation increased from 0.130 nm to 0.138 nm for the 50 nm $Al_2O_3$ coating, while it increased from 0.132 nm to 0.168 nm for the 100 nm $Al_2O_3$ coating case. It is expected that the increased thickness of the $Al_2O_3$ coating with a refractive index higher than that of the fiber cladding material (i.e. 1.44 of $SiO_2$ versus 1.65 for $Al_2O_3$) can enhance the wavelength separation between the two polarized cladding modes.[65] Also, both polarized cladding mode resonances had a similar amplitude attenuation of ~1 dB, which could be caused by the absorption coefficient of $Al_2O_3$ material (or by a change in the cladding mode profile that would impact the coupling between the core mode and the selected cladding modes). On the other hand, the spectral responses of the two TFBG samples show very large polarization dependence after coating with the gold CVD films, where the TE-polarized cladding modes had much larger wavelength shifts and amplitude attenuations than the TM-polarized ones, especially for the TFBG-2 sample shown in figure 8.7(b). This is consistent with previous work,[63] but in the present study we also find that the wavelength position of the TE-polarized cladding mode shifts (~1.7 nm) by more than the inter-resonance spacing for the 100 nm $Al_2O_3$–Au coating (~0.6 nm, figure 8.7(b)). Based on the spectral responses of the two TFBG samples induced by the CVD Au films, the thicknesses of the CVD Au coatings were about 30 nm and 65 nm estimated from the polarized wavelength separations for TFBG-1 and TFBG-2, respectively, as determined from the calibrations provided in ref [63].

### SRI Sensitivity

Figure 8.8 shows the spectra of the polarized cladding mode resonances that are used for SRI monitoring, where the SRIs of the saline solutions with different mass concentrations are calibrated from our previous work.[45] The spectral responses of two additional TFBG samples with the Bragg wavelengths at ~1611 nm and with only ~30 nm and ~65 nm thick CVD Au coatings (i.e. no $Al_2O_3$) were also tested with the same saline solutions for the SRI sensitivity comparisons. It can be seen that the two polarized cladding mode resonances shifted to longer wavelengths with increased SRI for each coating case. However, the two TFBG samples with ~65 nm Au coatings (regardless of the $Al_2O_3$ coating) had the stronger polarization-dependent spectral responses induced by the different SRIs, where the wavelength shifts of the TE-polarized cladding mode resonances were much smaller than that of the TM-polarized ones (as shown in figures 8.8(b) and 8.8(d)). Figure 8.9 shows the extracted wavelength shifts of the polarized cladding mode resonances with the normalized point at 1.315 (SRI of DI water at those wavelengths) from figure 8.8. By using a linear fitting function, the SRI sensitivities of the polarized cladding modes of the TFBG samples coated with different films can be obtained from the linearizations (as shown on the right of figure 8.9).

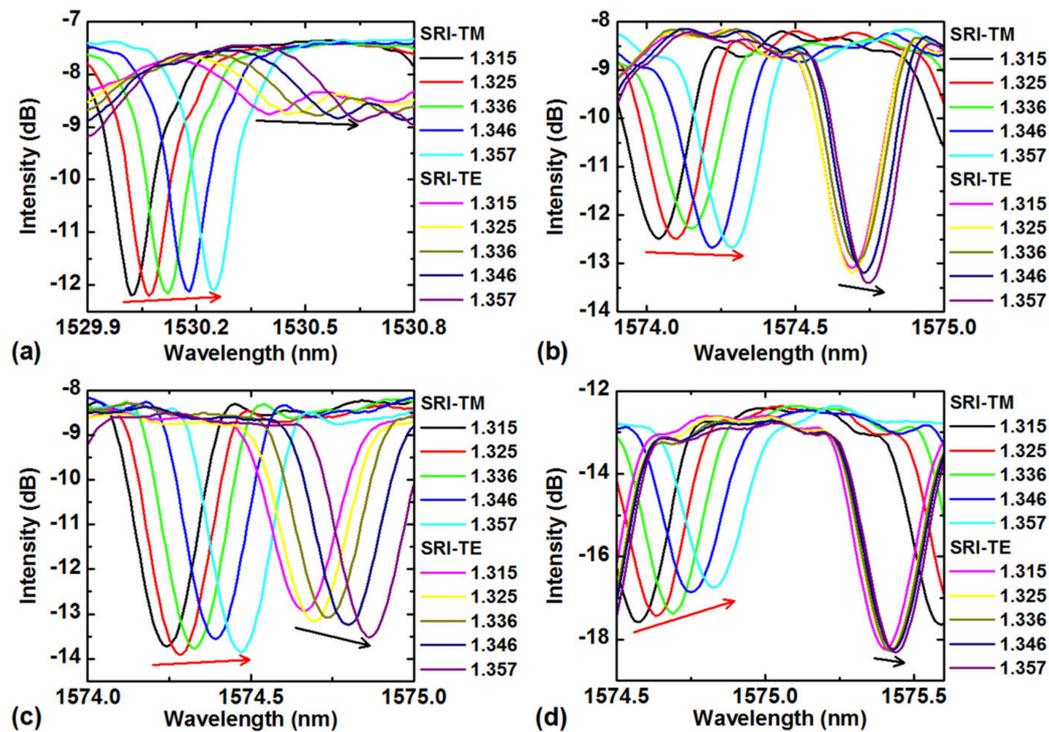

**Figure 8.8:** Polarized spectra of coated TFBG-1 (50 nm $Al_2O_3$ + Au) (a) and coated TFBG-2 (100 nm $Al_2O_3$ + Au) (b) measured under various SRIs (NaCl solutions). The corresponding polarized spectra of TFBGs only coated by the CVD Au films with similar thicknesses of ~30 nm and ~65 nm measured under various SRIs are also shown in (c) and (d), respectively. The arrows indicate the wavelength shift directions of the polarized cladding mode resonances.

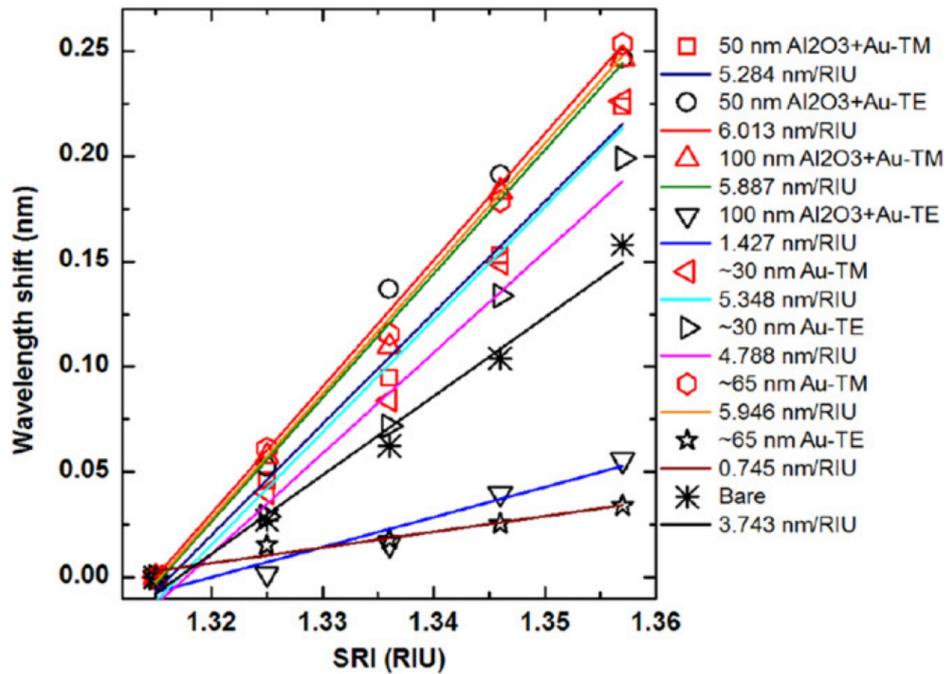

**Figure 8.9:** Wavelength shifts versus SRIs for TFBGs coated by $Al_2O_3$ + Au films and only Au films. The SRI sensitivity for each TFBG sample is extracted by the linear fitting slope. The reference SRI sensitivity of a bare TFBG sample measured under unpolarized light is also shown.

Based on the extracted polarization-dependent sensitivities for each coating case, it is found that the TE- and TM-polarized cladding modes have the quite similar SRI sensitivities (with a maximum difference of ~0.73 nm/RIU) for the ~30 nm Au film-coated TFBGs (both with 50 nm $Al_2O_3$- 30 nm Au and single 30 nm Au coatings), while the TM-polarized cladding modes have the much higher SRI sensitivities (~5.2 nm/RIU at maximum) for the ~65 nm Au film-coated TFBGs (both with 100 nm $Al_2O_3$- 65 nm Au and single

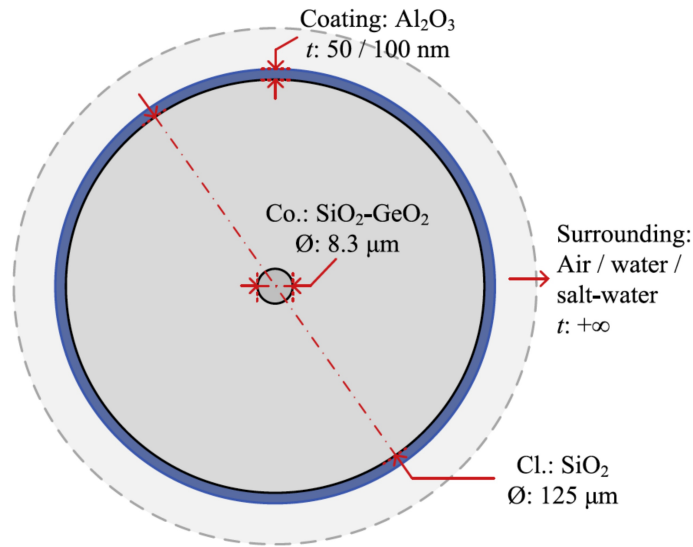

**Figure 8.10:** Four-layer model parameters used for the FIMMWAVE simulation of the power density profiles (figure 8.11) for the uniform (ALD-like) 50 nm and 100 nm $Al_2O_3$ coatings. TE and TM modes with effective indices of 1.392 934 and 1.392 725 (calculated from bare optical fiber in air), respectively, were selected for simulation of the power density profiles under different coatings and surroundings (SRIs: air, water, NaCl (sat.)). Note: 'Co' is core and 'Cl' is cladding.

65 nm Au coatings). For the 30 nm Au CVD film, the evanescent fields of the both TM- and TE-polarized cladding modes can penetrate through the Au aggregate layer,[63] and because of the relatively small thickness and aggregated surface of such CVD Au films, the effective permittivity of the Au film is highly dependent on the SRIs for either polarization, similar to ultrathin Au films obtained by PVD.[182] Thus, with the 30 nm Au coating, the wavelength shifts of the TE- and TM-polarized cladding modes are mainly caused by the SRI-induced effective permittivity change of the CVD Au coating. These can be even larger than the SRI change magnitude since the SRI induced wavelength shifts of both the TE- and TM-polarized modes for the TFBGs with 30 nm of Au are larger than that of the bare TFBG. Moreover, at this Au thickness the 50 nm $Al_2O_3$ layer enhances the SRI sensitivity by 25.6% for the TE-

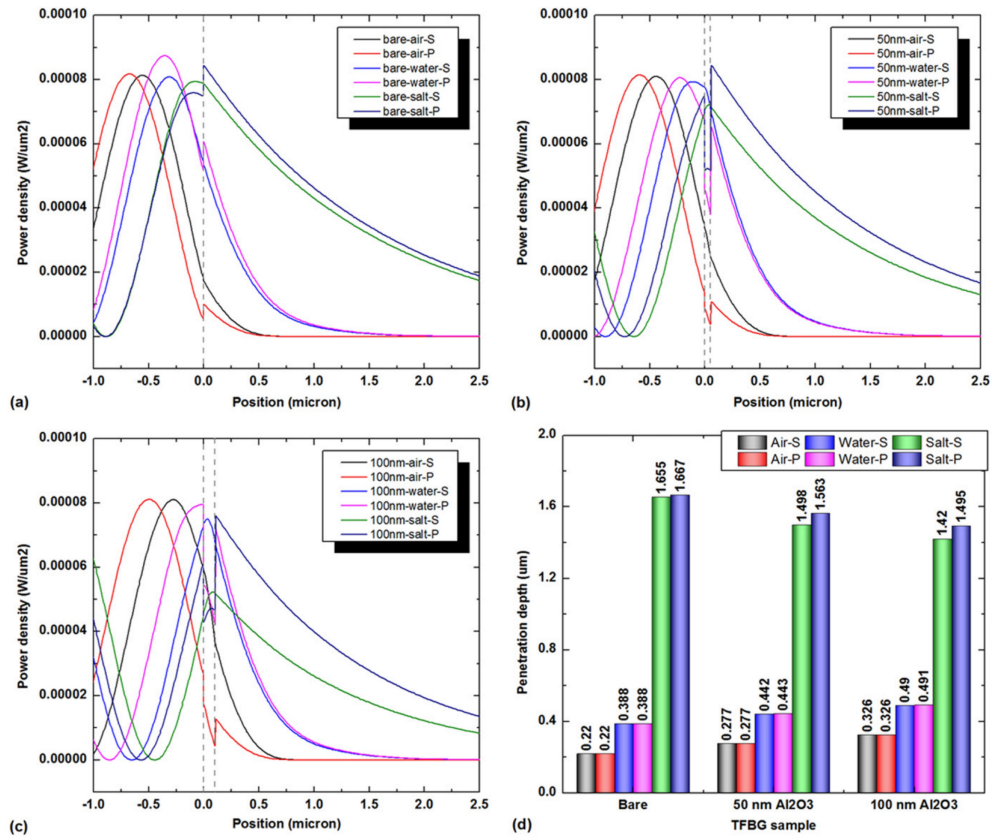

**Figure 8.11:** Power density profiles generated from the four-layer model simulation (figure 10) for the S (TE) and P (TM)-polarized cladding modes of (a) bare TFBG, (b) 50 nm Al$_2$O$_3$-TFBG, (c) 100 nm Al$_2$O$_3$-TFBG in various SRI conditions. (d) Evanescent field penetration depths extracted from the power density profiles in (a)–(c) for the S (TE) and P (TM)-polarized modes in various SRI conditions (air, water, NaCl(sat.))

For the 65 nm Au coating case (both the 100 nm Al$_2$O$_3$- 65 nm Au and single 65 nm Au coatings), the evanescent field of the TE-polarized cladding mode is mostly localized within the Au aggregates of the CVD film and does not 'feel' the SRI changes. However, the TM-polarized cladding mode can still penetrate through the Au layer to detect the SRI changes, giving it as much as 8 times more SRI sensitivity than the TE-polarized mode. It is also interesting to note that the 100 nm Al$_2$O$_3$ layer enhanced the (weak) SRI sensitivity of the TE-polarized cladding mode by 91.5%, but not at all for the TM one.

## Simulations of the Evanescent Field Penetration in the SRI

To elucidate the effect of the ALD-grown $Al_2O_3$ dielectric layer on the SRI properties of the CVD-grown Au coatings, four-layer optical fiber simulations using FIMMWAVE (by Photon Design) were carried out. Figure 8.11 shows the power density profiles of the TE- and TM-polarized evanescent fields as a function of radial position within the TFBG cross-section. It is clear in all of the power density profiles (figures 8.11(a)–(c)) that the tangential electric fields of the TE-polarized cladding modes tended to be continuous at the cladding-TFBG boundary layer whereas the electric fields of the TM-polarized cladding were discontinuous across permittivity interfaces. The mode power density profiles were used to calculate the penetration depths of the evanescent fields for the polarized cladding modes of the bare, 50 nm $Al_2O_3$-coated, and 100 nm $Al_2O_3$-coated TFBGs. The optical and geometric parameters of the optical fiber are depicted in figure 8.10 with the refractive indices of the core, cladding, $Al_2O_3$ coating, and surroundings set as 1.4495, 1.4437, 1.6211 and SRI (1 for air, 1.315 for water, and 1.360 for sat. NaCl solution [182]), respectively.

Looking at figure 8.11(d), it is apparent that the addition of an $Al_2O_3$ dielectric layer drastically enhanced the penetration depths of the evanescent field for both orthogonally polarized cladding modes in air or water surroundings, but decreases them in the saturated NaCl solution surroundings. The overall decrease in evanescent field depth for both polarized modes in the saturated NaCl solution was not, however, uniform as the $Al_2O_3$ layer thickness increases (i.e. the TE-polarized mode evanescent field depths decreased by ~0.16 nm and ~0.24 nm for the 50 nm and 100 nm Al2O3 coatings, respectively; similarly, the TM-polarized mode evanescent field depths decreased by ~0.1 nm and ~0.17 nm for the 50 and 100 nm $Al_2O_3$ coatings, respectively). These $Al_2O_3$ precoating thicknesses were selected to exploit the polarization-dependence of the TFBG. If the thickness of the $Al_2O_3$ plus the Au metal layer exceeds 150 nm, then the out-of-plane (TM) polarization mode shows excellent SRI sensitivity, while the in-plane (TE) polarization component no longer has sensitivity to the SRI. Thus, the two thicknesses show the benefit of this enhancement (with 100 nm $Al_2O_3$) compared to the absence of this benefit, but while still having $Al_2O_3$ (with 50 nm $Al_2O_3$). In general, the anisotropic (polarization-dependent) property of the TFBG is exacerbated as the dielectric layer thickness increases and it is an important result that will prompt further investigation into how the SRI sensitivity of the TFBG sensor could be enhanced.

## 8.5     Conclusions

In this study, uniform 50 nm and 100 nm conformal ALD-grown $Al_2O_3$ coatings were successfully used to enhance the RI sensitivity of a TFBG-inscribed optical fiber sensor for in situ monitoring of Au thin film growth by CVD, as well as for solution-based refractometric measurements in various environmental SRI conditions (e.g. air, water, and NaCl solutions). It was shown by XPS that some intermixing of the $Al_2O_3$ interfacial layer with the Au coatings does occur from the CVD process, but the overall enhancement of RI sensitivity and evanescent field depth for both the orthogonally polarized cladding modes (TE and TM) is still observed. By controlling the interfacial ALD-grown $Al_2O_3$ layer thickness, the sensitivity of the TFBG was greatly enhanced and this effectively makes it an even more versatile and promising material with many optical applications.

# Chapter 9

## Atomic Layer Deposition of Gold Metal

Griffiths, M.B.E.[1]; Pallister, P.J.[1]; **Mandia, D.J.[1]**; Barry, S.T.*[1] "Atomic Layer Deposition of Gold Metal" *Chem. Mater.* **2016**; *28*(1); pp. 44-46.

[1] Department of Chemistry, Carleton University, 1125 Colonel By Drive, Ottawa, Ontario, Canada, K1S 5B6
*Corresponding author

## 9.1    Introduction

Nanoparticulate gold is a useful material being used in applications like gas-sensing,[192] heterogeneous catalysis,[193] and by virtue of its surface plasmon resonance, photonics.[194] Currently, solution-phase reduction of gold salts in the presence of facet-specific surfactants and capping agents is the method of choice for producing gold nanoparticles. Solution-phase synthesis can be a drawback if the nanoparticles are needed for an application where the surfactant might interfere. One straightforward method of controlling the nucleation and growth of metallic nanoparticles is atomic layer deposition (ALD). ALD is a layer-by-layer, gas-phase deposition technique that has been used to deposit a wide array of metallic, semiconducting, and insulating films and features.[195] Gold nanoparticles have been deposited by several vapor-based methods including physical vapor deposition (PVD) and chemical vapor deposition (CVD).[196] Materials that are deposited by ALD are grown from a gas-phase precursor chemical that, by virtue of surface chemical reactions, react on a surface to form the target material. It is surface chemistry, not surfactants, that controls the size and shape of deposited nanofeatures. The versatility of this technique is reflected in the number of available processes that exist to deposit a variety of different materials.[16] A significant amount of research has been undertaken to study the deposition of metal films (including copper),[197] but the other group 11 metals have proved more challenging.[198–200] Reported here for the first time is gold metal deposition by ALD.

**9.2                Experimental**

**Precursor Synthesis**

All manipulations were performed with rigorous exclusion of air using an Mbraun Labmaster 130 glovebox, Schlenk techniques, and $N_2$ ($\geq$99.998) gas. NMR analysis was performed using a 300 MHz Bruker Avance spectrometer. $HAuCl_4 \cdot xH_2O$ (49.9 weight % Au) was purchased from Strem Chemicals and used as received. Trimethylphosphine was also purchased from Strem Chemicals as 5 g in an ampoule, and was diluted to 1.31 M solution in toluene and then stored in an $N_2$-atmosphere glovebox in a teflon screwcap bottle. MeLi (1.6 M in diethyl ether) was purchased from Sigma-Aldrich as 4 x 25 mL bottles packaged under nitrogen in SureSeal septa and used as received. Anhydrous dichloromethane, tetrahydrothiophene, and methyl iodide were purchased from Sigma-Aldrich and was used as received. Diethyl ether was purified using an Mbraun Solvent Purification System.

**Tetrahydrothiophenegold(I) chloride:** $HAuCl_4.xH_2O$ (49.9 weight % Au, 20.45mmol) was dissolved in 120mL of 5:1 EtOH : $H_2O$ in a 400 mL beaker. 4.2 mL (47.64 mmol) of tetrahydrothiophene (THT) was added dropwise causing the precipitation of a flocculent white solid. This was stirred for 1 hr and then filtered, washed sequentially with 3 x 20 mL of ethanol, 3 x 20 mL of diethyl ether, and then dried under high vacuum for 12 hours. The yield was 90 – 98%, based on multiple trials (typical mass $\approx$ 6.5g).

**Trimethylphosphinogold(I) chloride:** In a glove box, dry tetrahydrothiophenegold(I) chloride (6.55 g, 20.4 mmol) was dissolved in a minimum volume of dichloromethane in a 250 mL Schlenk flask. Dilute PMe₃ (15.5 mL, 1.31 M in toluene, 20.4 mmol) was added drop-wise, and the reaction was stirred for 1 hour. In a fume hood, this solution was filtered through a medium frit, washing with dichloromethane to pass all of the product. The resulting solution was left in a fume hood overnight to evaporate, resulting in the formation of a bright white crystalline solid.

Yield = 6.20 g, 99 %. [1]H-NMR (300 MHz, CDCl₃): δ 1.62 (d, (PMe₃)), [2]$J_{H-P}$ = 11.4Hz.

**Trimethylphosphinotrimethylgold(III):** 5.533 g (18 mmol) of (PMe₃)AuCl was placed in a 500 mL Schlenk flask along with a stir bar and was then dried under high vacuum for 2 hours. Meanwhile, a 100 mL dropping funnel was flame dried under high vacuum, then cooled to room temperature under vacuum. Both Schlenk flask and dropping funnel were then backfilled with $N_2$. The (PMe₃)AuCl was suspended in 300 mL of dry $Et_2O$, and the dropping funnel was then attached under a stream of $N_2$ from both flask and funnel. The top of the dropping funnel was then fitted with a dry rubber septum, flowing nitrogen from the Schlenk arm of the 500 mL flask. With the apparatus assembled, the suspension was cooled to -78 ºC in a dry ice / isopropanol bath. One bottle of MeLi (25 mL, 40 mmol, 1.6 M in $Et_2O$) was transferred to the dropping funnel via cannula using a slight vacuum assist from the Schlenk arm. The Schlenk arm was only slightly cracked to vacuum, and then closed again. This was repeated until all the MeLi has been transferred. Care was taken not to keep the system under active vacuum since this was found to cause MeLi to precipitate in the dropping funnel. An anti-backflow bubbler was used on the nitrogen line to ensure that no air could enter the system during the vacuum-assist. Once transferred, the cannula was removed, and MeLi solution was added very slowly drop-wise at a rate of 1 drop every 6-8 seconds. A very slow addition was key to a high yield since the intermediate formed was very thermally unstable. A total addition time of 50-60 min was ideal. Once all the MeLi was added, the funnel was rinsed with 20 mL of dry diethyl ether, which was then added dropwise to the stirring solution. By this point, the suspension often appeared yellow to green in colour. This suspension was stirred for 2 hours after completion of MeLi addition at − 78 ºC. MeI (1.91 mL, 30.6 mmol) was then added very slowly dropwise over 10 min, and the reaction was stirred for a further 2 hours at -78 ºC. The colour of the suspension usually returned to white at this point, but yellow or grey was also observed. The reaction was allowed to warm to room temperature over 1 hour, and was then cooled in an ice bath. 40 mL of distilled water was added to the funnel, and this was then added dropwise to the stirring suspension at an initial rate of one drop every 10 seconds. Colour change or consistency change of the reaction warranted stopping addition and allowing for the reaction to subside before continuing again. When addition was done too quickly, the flask would turn dark purple. If done correctly, a slight purple to clear ethereal solution formed above an aqueous solution, generally about 30 minutes of drop-wise addition. Gradual darkening of the solution was normal. The ethereal layer was separated using a separatory funnel without shaking, and dried over excess MgSO$_4$ which served as both a drying agent and a nanoparticle sequestering agent. The solution was filtered and dried on a rotary evaporator yielding a colourless liquid. If this liquid was contaminated by (PMe$_3$AuMe), then an additional purification step was required before   proceeding. (PMe$_3$AuMe): [1]H-NMR (300 MHz, CDCl$_3$): δ 1.43 (d, 9H, (PMe$_3$), [2]J$_{H-P}$ = 8.8 Hz), δ 0.33 (d, 3H, (Au(CH$_3$)), [3]J$_{H-P}$ = 8.4 Hz).

**Purification Step:** If (PMe$_3$)AuCH$_3$ was observed by NMR dissolved in the product, the mixture was then dissolved in 3 volume equivalents of neat methyl iodide and stirred in a sealed flask for 6 hours. The MeI was then evaporated, leaving a mixture of the product, (PMe$_3$)AuMe$_3$, and (PMe)AuI, the latter of which is non-volatile. This liquid (or purified/recycled material) was purified by simple vacuum distillation (80 ℃, 100 mTorr) using a cold water condenser and an ice-cooled receiving flask. The product was dislodged from the condenser periodically by gentle heating with a heat gun or Bunsen burner until it liquified and flowed again into the receiving flask. Yield = 5.30 g, (93 %). m.p. = 22 ℃ . $^1$H-NMR (300 MHz, CDCl$_3$): δ 1.49 (d, 9H, (P(CH3)3), $^2$J$_{H-P}$ = 9.6 Hz), δ 0.81 (d, 3H, (Au(CH$_3$)$_2$(CH$_3$)), $^3$J$_{H-P\ (trans)}$ = 9.6 Hz), δ 0.09 (d, 6H, (Au(CH$_3$)$_2$(CH$_3$)), 3JH-P (cis) = 7.6 Hz). 13C- {1H}-NMR (75 MHz, CDCl$_3$): δ 12.02 (d, P(CH$_3$)$_3$, $^1$J$_{C-P}$= 28.9 Hz), δ 10.87 (d, (Au(CH$_3$)$_2$(CH$_3$)), $^2$J$_{C-P}$ (trans) = 133.0 Hz), δ 6.46 (d, (Au(CH$_3$)$_2$(CH$_3$)), $^2$J$_{C-P}$ (cis) = 28.9 Hz). 31P- {$^1$H}-NMR (121 MHz, CDCl$_3$): δ -9.75 (P(CH$_3$)).

## Thermogravimetric Analysis (TGA)

In a typical experiment, ~20 mg of **1** was loaded in a platinum pan which was hung on the microbalance arm of a TA Instruments Q500 TGA, which is housed in an inert atmosphere MBraun glovebox. All experiments were purged by nitrogen gas. In the ramp experiment (Figure 9.4a), a ramp rate of 10 ℃/min was employed from room temperature to 500 ℃. For the isothermal experiment (Figure 9.4b) the temperature was ramped automatically to a set-point of 65 ℃ and held at this temperature for 175 minutes.

## Vapor Pressure Determination

To calculate a Langmuir vapor pressure curve for **1**, ~50 mg of **1** was loaded in a platinum pan which was hung on the microbalance arm of a TA Instruments Q500 TGA, which is housed in an inert atmosphere MBraun glovebox. A temperature program was run such that the temperature was ramped by 10 ºC and then held isothermally for 10 minutes. This was repeated from room temperature to 400 ºC. The slope of the linear weight loss per unit time at each isothermal step was calculated and expressed with respect to the area of the pan. This was graphed on a logarithmic scale against 1/temperature to extract a relationship for mass delivery by temperature (Figure 9.5). A known method[201] was used to calculate the Langmuir expression for vapor pressure.

## Plasma-enhanced Atomic Layer Deposition

All thin films were deposited using a Picosun R200 ALD reactor with a Picoplasma remote plasma source system (plasma RF power of 2800 W). Oxygen gas (≥99.999%) was mixed with Argon gas (≥99.999%) at flow rates of 110 and 185 sccm respectively during plasma ignition. A charged screen between the plasma source and the substrate blocked charged species from participating in surface chemistry. The reactor pressure and temperature were 5 hPa and 120 ºC, respectively. Silicon (100) with native oxide and soda lime microscope slides were used as substrates, and were pretreated with 10 plasma pulses (14 s $O_2$ plasma | 10 s $N_2$) before beginning ALD experiments. Milli-Q deionized water was used in the water bubbler, and was cooled to 19 ºC using a Pelletier cooler.

A fresh sample of **1** was loaded into a cleaned glass vial and then inserted into the Picosolid booster crucible before each experiment. The headspace of the bubbler was purged of air and replaced with nitrogen before heating to 85 ºC In a typical experiment, 1000 cycles with a duration of 4 s volatilized approximately 550 mg of **1**, and 1250 10 s cycles used between 1 and 1.2 g of **1**. The typical pulse sequence was 4 s Au | 14 s O2 plasma | 0.1 s water each separated by a 10 s nitrogen gas (≥99.999%) purge. The purge times were chosen to be reliable, rather than optimal. The flow rates of $N_2$ carrier gas for the gold precursor and water were 120 sccm and 100 sccm, respectively. Leftover precursor from each experiment was stored in a separate vial, and the booster vial wiped clean with ethanol. Accumulated spent precursor was repurified by vacuum distillation as described above in the synthetic procedure and used without noticeable loss of film quality.

**Thin Film Characterization**

Typically, the gold films were deposited using 100 cycles using 4s pulses of **1**, 14 s pulses of $O_2$ plasma, and 0.1 s pulses of water with 10s nitrogen gases purges between each step. These films were imaged using a Tescan Vega-II XMU Variable Pressure Scanning Electron Microscope with 20 keV potential (Figure 9.6). The films appeared well connected and uniform. Attempts at characterizing the resistivity were unsuccessful due to poor adhesion of the gold metal film on glass and silicon/$SiO_2$ substrates. XPS spectra were collected in the analysis chamber of a Specs/RHK multi-technique ultrahigh vacuum system using a Phoibos 100 SCD power supply, hemispherical analyzer, and detector. Whenever possible, the sample being analyzed was electrically grounded. An XR-50 X-ray source containing an Al anode (400 W) was used in this study (14.26 keV Al Kα source), and analyses were performed within a base pressure range of 7 x 10^-10 Torr to 1 x 10^-9 Torr. Survey (Figure 9.7) and high-resolution scans of the as-deposited gold films were collected using SpecsLab2 software, and post-processed and deconvoluted using CasaXPS. High-resolution spectra were fitted to a Shirley background and fitted using the appropriate Scofield-based factors for photoelectron cross-section. The film appears to have a nucleation delay of 96 cycles according to the growth per cycle fit (Figure 9.8), and this is presently under scrutiny in our lab.

### 9.3    Results and Discussion

We synthesized trimethylphosphinotrimethylgold(III) (**1**) from a literature preparation as the chemical precursor; a compound that has previously been employed in chemical vapor deposition.[202,203] This compound is air and water stable, is a liquid at room temperature, and has a vapor pressure that follows this Langmuir equation:[201]

$$\ln(p) = 0.059T - 1.65 \qquad [1]$$

where p is pressure in Pascals, and T is temperature in Celsius.

Heating **1** to 85 ℃ was sufficient to deliver the precursor into the deposition furnace. To reduce this compound at the surface to metallic gold, we found that both hydrogen gas and hydrogen plasma were not effective: at temperatures lower than 130 ℃, no deposition was observed, and at temperatures over 130 ℃ , **1** begins to decompose to gold metal without a reducing co-reagent (i.e., by a chemical vapor deposition mechanism).

Noble metal deposition by ALD can employ molecular oxygen as a co-reagent: the strategy is to oxidize the organic ligand system, leaving the noble metal behind.[204] We found that molecular oxygen did not react with **1** under 130 ℃. Other oxygen-containing co-reactants (e.g., $H_2O$ and $O_3$) were similarly unreactive with **1** under thermal conditions. However, using a plasma-assisted process with **1** showed that oxygen plasma produced a film of gold metal by ALD at 120 ℃.

Using just oxygen plasma, the films were discolored and nonmetallic in appearance: typically they were brown or dark purple. These films discolored and delaminated over time in ambient conditions and developed a foul odor. Energy-dispersive X-ray spectroscopy (EDX) and X-ray photoelectron spectroscopy (XPS) (Figure 9.1, Table 9.1) showed that gold metal was certainly present, but so too were oxygen and phosphorus.

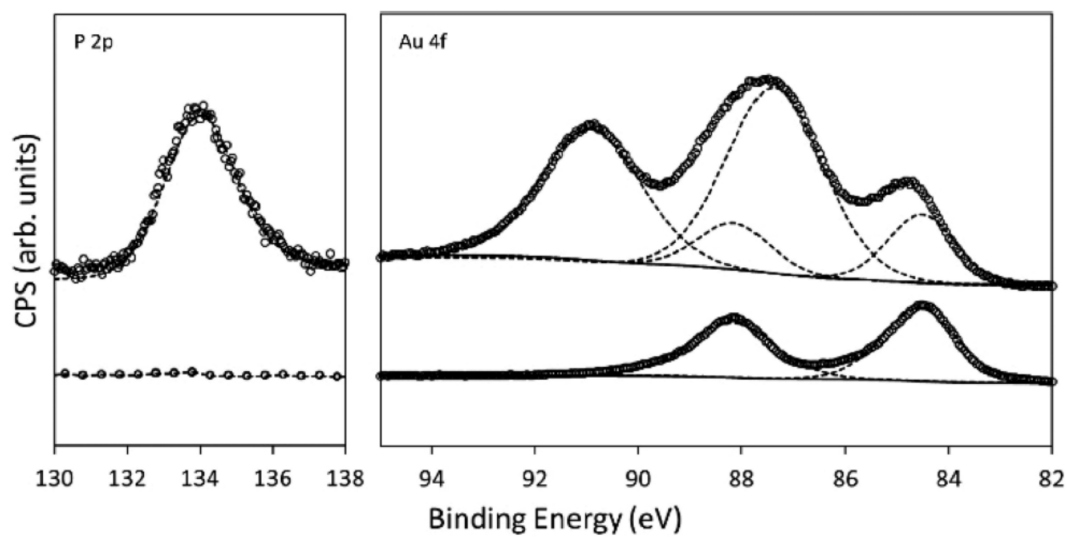

**Figure 9.1:** XPS spectra of the phosphorus 2p and gold 4f regions for the film deposited without a water pulse (top), and with a water pulse (bottom).

**Table 9.1:** Compositional analysis by XPS of the film formed without water and the film formed with water[a]

| Element | Without Water (at. %) | With Water (at. %) |
|---------|-----------------------|--------------------|
| Au | 76.16 | 91.52 |
| C | 5.61 | 6.65 |
| N | 1.43 | 0[b] |
| P | 6.26 | 0[b] |
| O | 10.54 | 1.83 |

[a] These data represent the surface measured "as is", and likely contains carbon and oxygen from atmospheric sources.
[b] These species were below the detection limit of XPS.

In the XPS spectrum, a peak was observed centered at 134.1 eV, which is in the range of the P(V) oxidation state. There was also evidence of Au(0), with peaks at 84.5 and 88.2 eV, and Au(III) at 87.5 and 90.9 eV. This suggests that some of the phosphine ligand from the precursor remained with the gold adatom at the surface, and was converted into a gold−phosphine containing film when reacted with oxygen plasma.

Although the nature of this film is not presently understood, further characterization has shed some insight. Trimethylphosphine is known to react with oxygen to produce $P_2O_5$ and other phosphorus(V) oxides. Because treatment of $P_2O_5$ with water yields phosphoric acid, the impure film was sonicated in $D_2O$, and $D_3PO_4$ was observed as a major product by $^{31}P\{1H\}$ NMR spectroscopy (1.12 ppm rel. to $H_3PO_4$), suggesting the presence of phosphorus(V) oxides in the film. Interestingly, rinsing these films with water was enough to seemingly dissolve them, even though they were well-adhered to the silicon substrate.

The simplest chemistry that rationalizes these facts is that the oxygen plasma combusts the precursor, forming oxides of both gold and phosphine in the process:

$$Me_3P\text{-}Au\text{-}Me_3 \ + \ [O^*] \quad \rightarrow \quad Au_2O_3 \ + \ P_2O_5 \ + \ 6CO_2 \ + \ 9H_2O$$

Gold metal is then slowly formed from the gold oxide, which agrees with the presence of both oxidation states in the XPS. This conversion occurs in the bulk at 160 ºC,[205] but the higher chemical potential of a deposited monolayer could allow this conversion to happen at lower temperature:

$$Au_2O_3 \quad \rightarrow \quad 2Au^0 \ + \ 3/2O_2$$

In an attempt to eliminate the phosphorus impurity in the film, water was chosen as a ternary reactant with the purpose of hydrolyzing the phosphorus impurity to phosphoric acid, which would then be volatilized away. Using an ABC-type pulse sequence of 4 s of compound 1, 14 s oxygen plasma, and 0.1 s water produced very pure gold metal films at 120 ºC. XPS analysis of these films showed very clearly that there was only metallic gold(0) (84.5 and 88.2 eV) with no phosphorus signals observed (Figure 9.1, Table 9.1). EDX also confirmed the absence of phosphorus in the film. Because gold oxide converts to gold metal faster in the presence of water,[205] this explains the formation of the metallic film. Likewise, the phosphorus oxide is converted into volatile hydrogen phosphate, and is removed during the purge step:

$$P_2O_5 + 3H_2O \rightarrow 2H_3PO_4$$

Using this process, gold thin films were deposited using a range of process parameters with high fidelity on both silicon and borosilicate substrates. Thickness measurements were calculated from the *k*-ratios measured by EDX using a program called GMRFilm,[206] and the resulting growth rates demonstrated ALD-characteristic self-saturating growth after 2 s pulse lengths of **1** (Figure 9.2a). A growth rate of 0.5 Å/cycle was observed, and is typical for metal films.[196] A graph of growth rate per cycle shows that there is an induction period (Figure 9.8). This is a point of ongoing research in our lab. The films themselves were highly reflective metallic gold mirrors (Figure 9.2b).

Because our desire is to exploit this technique to form metallic nanoparticles, we were very interested in studying the effect of a small number of cycles of the process. Deposition of gold metal on polymer-coated copper transmission electron microscopy (TEM) grids was done with 2 and 5 cycles. With two ALD cycles, EDX showed the presence of gold, but features were difficult to differentiate from the background carbon of the polymer film. With five ALD cycles, obvious gold

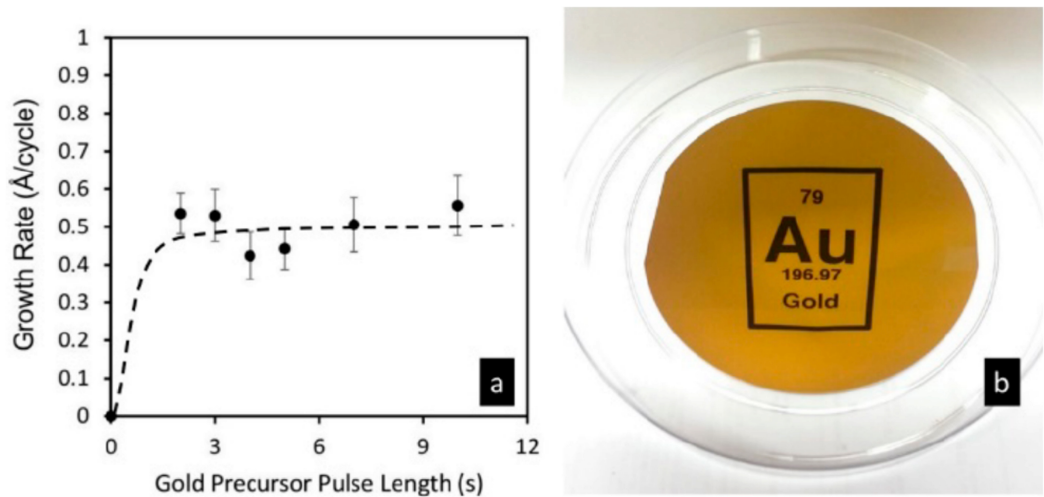

**Figure 9.2:** Gold metal deposition from **1**. Panel a shows the saturation curve as the exposure of 1 is altered. Panel b shows a photo of the as-deposited film (using a 4 s pulse) with the chemical symbol for "Au" reflected from the surface of a 100 mm wafer.

nanoparticles were observed by transmission electron microscopy (TEM) and EDX

(Figure 9.3). The lattice spacing of atoms in the TEM was 0.23 nm, which is indicative of the Au(111) crystal face.

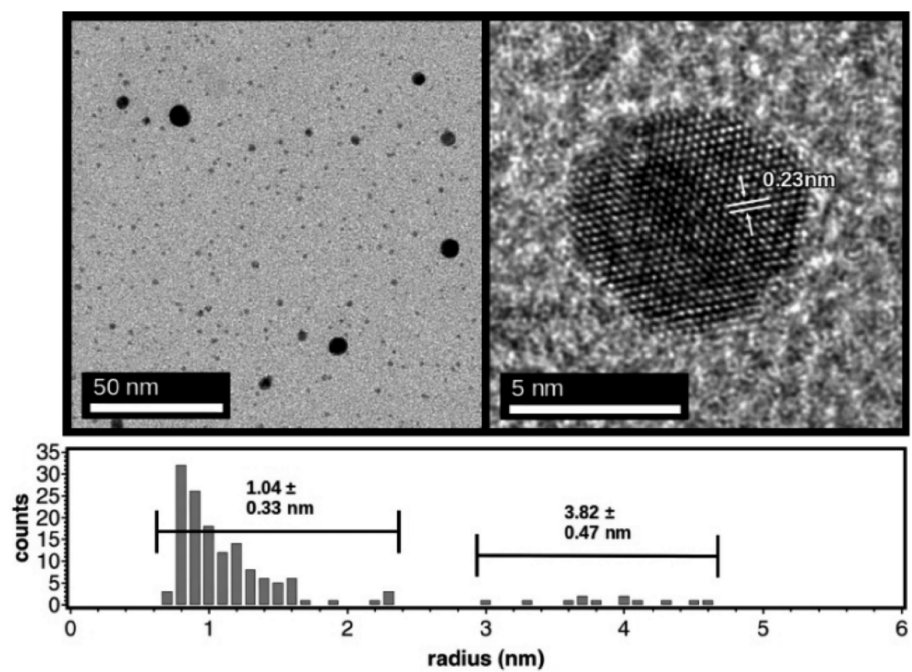

**Figure 9.3** TEM of gold metal nanoparticles deposited on a TEM grid after five ALD cycles using **1** with oxygen plasma and water as co-reactants. The histogram shows the distribution of particle sizes.

The deposited gold metal particles showed some polydispersity, which we attribute to mobility of the reduced gold adatoms formed during the plasma step. At five ALD cycles, the particles had a bimodal dispersity, clustered in small and large particles. The large particles were few in number and had an average radius of 3.8 (±0.5) nm whereas the smaller particles were much more numerous and had an average diameter of 1.0 (±0.3) nm with the most common radius being 0.8 nm.

This shows that ALD-deposited gold can produce nanoparticles on a substrate without the need for solution-phase chemistry and capping agents (other than the inherent ligand system). The overall growth during ALD cycles as well the movement of adatoms on the surface to various nucleation points suggests a Vollmer−Weber type growth. This is expected due to the energetics of the plasma step and the propensity for precursor nucleation onto metallic gold rather than the carbon mesh of the TEM grid.

The compound trimethylphosphinetrimethylgold(III) has excellent precursor characteristics: it is a liquid at room temperature, and has great volatility. It is tolerant of ambient water and oxygen under standard conditions, which makes it simple to handle in laboratory conditions. The process is robust and repeatable, and allows Ångström-level control over the deposition of gold metal nanostructures. This newly discovered gold process is a vast improvement over typical chemical vapor deposition processes.

## 9.4    Supporting Information

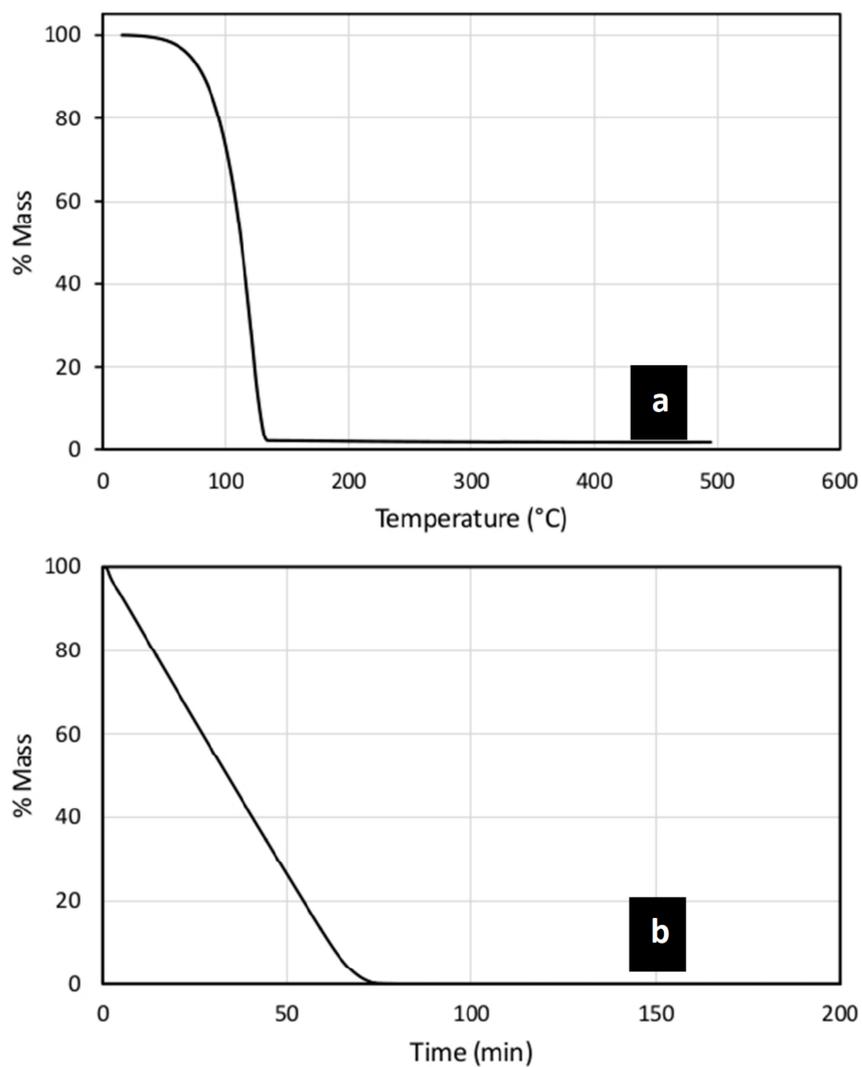

**Figure 9.4:** The thermogravimetric (TG) traces for 1. a) The standard ramp TG with a ramp rate of 10 ºC/min; b) the isothermal TG holding 1 at 65 ºC.

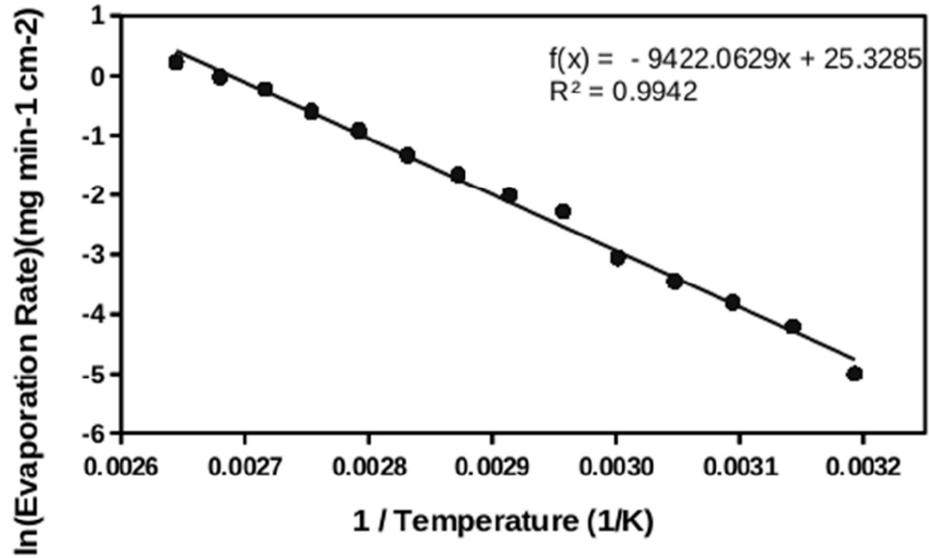

**Figure 9.5:** The evaporation rate for 1. The inset is the fitted linear regression and coefficient of determination for the fit.

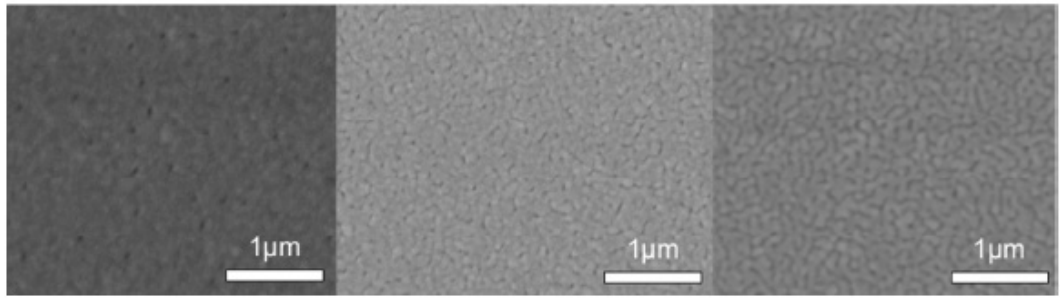

**Figure 9.6:** Scanning electron microscopy images of gold metal films deposited by **1**. Each of these independent depositions were comprised of 1000 atomic layer deposition cycles using 4s pulses of 1, 14 s pulses of $O_2$ plasma, and 0.1 s pulses of water with 10s nitrogen gas purges between each step.

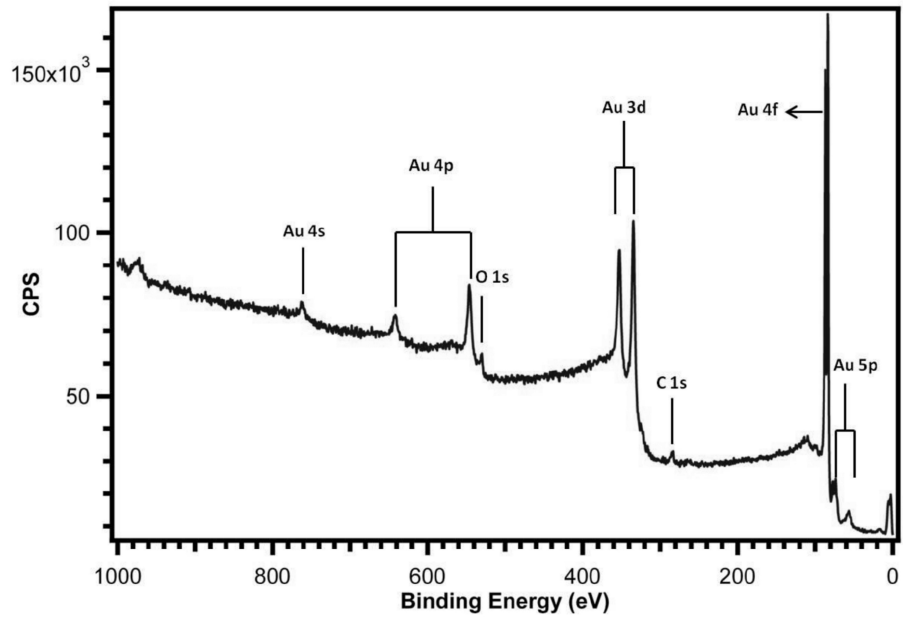

**Figure 9.7:** An XPS survey spectrum of a gold film deposited on a Si(100) substrate using 4s pulses of **1**, 14 s pulses of $O_2$ plasma, and 0.1 s pulses of water with 10s nitrogen gases purges between each step.

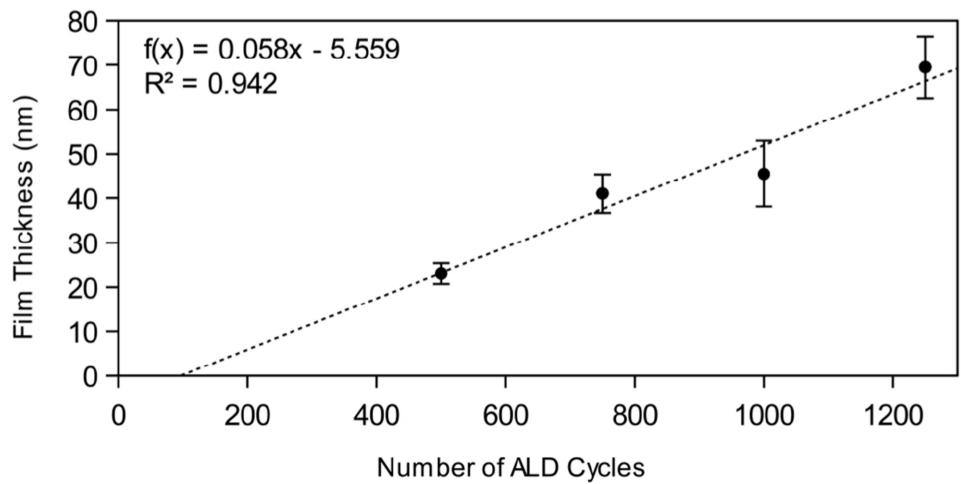

**Figure 9.8:** A graph of the growth rate as a function of cycle number.

# Chapter 10

## *In-Situ* Optical Monitoring of Gold and Alumina Atomic Layer Deposition: Development of an Ultra-high Sensitivity Plasmonic Optical Fiber-based Sensor

**Mandia, D.J.**[*1] ; Zhou, W[2].; Pallister, P.J.[1]; Albert, J.[2]; Barry. S.T.[1]
*Unpublished.* **2016**

[1] Department of Chemistry, Carleton University, 1125 Colonel By Drive, Ottawa, Ontario, Canada, K1S 5B6

[2] Department of Electronics, Carleton University, 1125 Colonel By Drive, Ottawa, Ontario, Canada, K1S 5B6

*Corresponding author

## 10.1 Abstract

TFBGs are an emergent optical fiber-based technology that have shown a variety of applications, particularly with respect to *in-situ* optical sensing.[41] Linearly polarized light that is launched through the core of an optical fiber can be out-coupled through the cladding layer of the TFBG and attenuate based on local refractive index (RI) changes occurring in the surrounding medium (e.g., during the nucleation and growth of a thin film on the fiber surface).

In the present work, a TFBG was successfully used as the sensing platform for *in-situ* and real-time monitoring of ALD, for both gold metal and alumina. The TFBG-polarized transmission data of gold metal plasma-enhanced ALD (PEALD) indicates that the insulator-to-metal transition for gold occurs between 84 and 110 ALD cycles, which, based on a reported growth rate of 0.5 Å/cycle,[17] corresponds to an effective thickness range of 4.2 to 5.5 nm, respectively. Moreover, the ALD gold-coated TFBGs (after 900 cycles, with an effective thickness of ~45 nm) support surface plasmon resonance (SPR) when subjected to various media with different surrounding refractive indices (SRIs), yielding a remarkable and unprecedented RI sensitivity of ~550 nm/RIU, which is orders of magnitude larger than the highest SRI sensitivity for CVD-grown gold nanoparticle films (6.013 nm/RIU, chapter 8) on the 50 nm $Al_2O_3$-pre-coated TFBG device. To compare, a TFBG coated with gold by PVD with the same thickness had a bulk RI sensitivity of 506 nm/RIU. This work confirms that the optimum RI sensitivity (550 nm/RIU) for gold coated TFBG sensors is achievable by ALD. Additionally, the feasibility of a TFBG as an *in-situ* optical and spectroscopic sensor for ALD is demonstrated by this result.

## 10.2    Preliminary results: *in-situ* optical fiber monitoring of ALD for alumina and gold

Thus far in the ALD literature, there are countless examples of *in-situ* characterization techniques such as spectroscopic ellipsometry (SE)[36], quartz crystal microbalance (QCM)[34], Fourier-transform infrared (FT-IR) spectroscopy,[38] and various X-ray absorption spectroscopic (XAS) methods such as XPS for characterizing ALD processes.[207] However, many of the mentioned techniques lack sensitivity in the low ALD cycle regime (<20 cycles) and, for SE in particular, the extracted reflection or transmission data typically is mathematically treated using the Drude model, which assumes the permittivity constants of bulk metals (i.e. not ultra-thin, nanoparticulate assemblies).[124] This preliminary work with *in-situ* diagnostics using an optical fiber-based ellipsometer describes, for the first time, the use of optical fibers photo-inscribed with TFBGs to monitor ALD processes. By exciting separately polarized cladding-guided modes in the TFBG (azimuthally in TE or radially in TM) the growth of alumina films by the highly-robust and benchmark $(CH_3)_3Al$ and $H_2O$ thermal ALD process,[15] as well as gold metal films by a recently reported gold metal ALD process[17] (also described in Chapter 9) were monitored in real-time. In this particular work, a Picosun R200 Plasma ALD tool was adapted to accommodate the TFBG sensing apparatus (OSA, polarization controllers, broadband source, and coupler if needed) by running multiple optical fibers through a custom-built optical fiber sample holder (Figure 10.1). Based on this design, multiple optical fibers (up to 10) can be channeled through "v" grooves milled into the aluminum sample plate leaving only the TFBG portion of the optical fibers (i.e. the sensing platform) exposed directly in the deposition zone during ALD. This would ensure self-consistency or homogeneity between TFBG samples while also allowing for enough statistics to determine how effectively and uniformly the ALD coatings can decorate multiple TFBGs. The only difference from the *in-situ* optical measurement methodology already presented throughout the thesis for the CVD processes is that optical measurements during ALD were performed in transmission mode (closed-loop) instead of the typical reflection mode where a gold mirror is sputtered (300 nm) ~2 cm downstream of the TFBG. This results in lower leakage of the core-guided light due to diffuse scattering and improves the depth (by multiple dB) and sensitivity of the cladding mode resonances.[41] Figure 10.2 depicts the direct optical measurement of alumina thin film growth during ALD for the TE-polarized (panel a) and TM-polarized (panel b) modes. The isotropic wavelength shift of the TE and TM modes begins at as early as 5 ALD cycles with an overall wavelength red-shift of 0.4 nm from 0 to 1000 cycles. In fact, as shown previously with 50 nm and 100 nm alumina coatings by ALD on TFBGs,[208] the wavelength separation ($\lambda_{TE}$-$\lambda_{TM}$) increases from ~0.13 nm to nearly 0.2 nm as the film thickness increases from 50 nm (~500 cycles) to 100 nm (~1000 cycles). This enhanced wavelength separation arises from the addition of a more "lossy" (higher refractive index) alumina layer (RI ~ 1.65) on the $SiO_2$ cladding (RI ~1.44: a similar enhancement effect has been shown for ZnO films, as well.[65] However, it has been found that the enhancement of sensitivity to the SRI of the orthogonally polarized cladding modes of a TFBG is only observed for SRI < 1.315 (the SRI of water), after which the evanescent field depth decays exponentially.[208] In any case, the use of the workhorse alumina ALD process as a calibration experiment to test our *in-situ* optical

ALD monitoring technique revealed that, similar to SE, the TE and TM cladding mode resonances offer cycle-for-cycle resolution of thin film growth in real-time based on local RI changes of the medium in which the fiber is located (TM polarized modes) or RI changes in the metal or metal oxide layer itself (TE polarized modes).

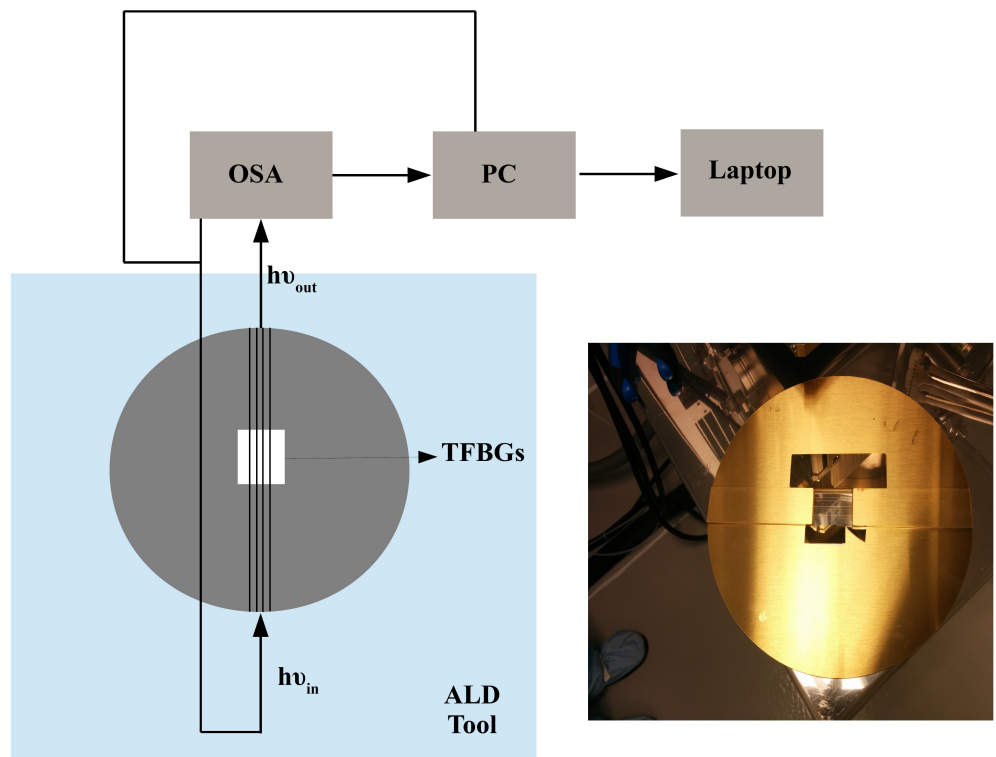

**Figure 10.1:** Schematic of the *in-situ* ALD sample holder with actual picture (right) of the sample holder after 900 cycles of gold ALD. In this closed-loop (transmission mode) system, linearly-polarized input light (hν$_{in}$) interrogates the surrounding TFBG surface through the cladding during ALD and returns a perturbed light output (hν$_{out}$) that can be further analyzed. (note: OSA is the optical spectrum analyzer, PC is the polarization controller).

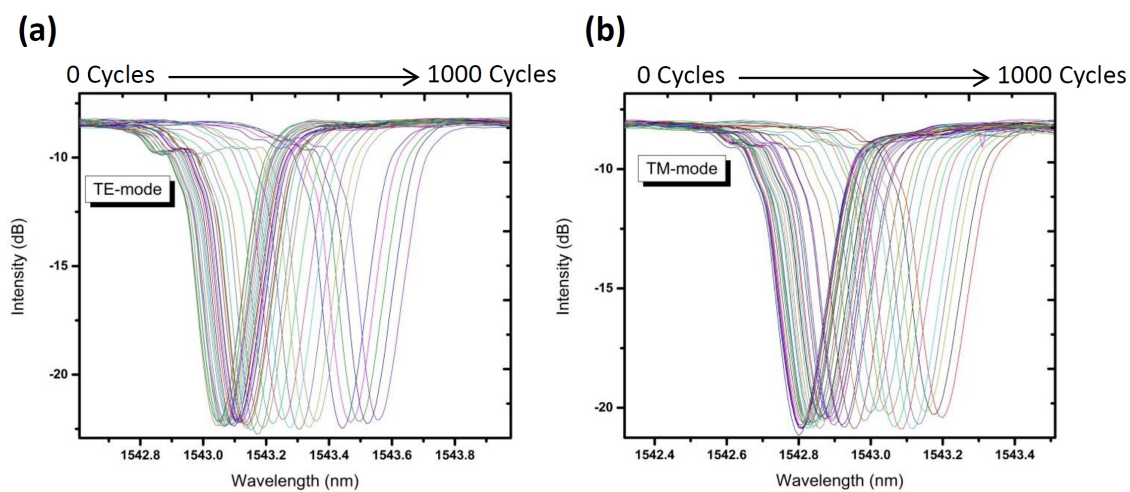

**Figure 10.2:** (a) *in-situ* TE-polarized transmission spectrum of alumina thermal ALD from 0 to 1000 cycles, (b) *in-situ* TM-polarized transmission spectrum of alumina thermal ALD from 0 to 1000 cycles. (note: 1 cycle is TMA + $H_2O$)

Follow-up experiments for *in-situ* optical monitoring of gold ALD were carried out as per the ALD pulse sequence described in Chapter 9 with the following modified ABC-pulse sequence: 6s $(CH_3)_3AuP(CH_3)_3$ | 20s $O_2$ plasma | 0.5s $H_2O$ with intermittent 10s $N_2$ (in Ar carrier gas) purge cycles to remove precursor byproducts. Because it is a plasma ALD process, the optical fiber polymer jacket was stripped prior to inserting the fiber into the ALD tool. This avoids contamination of the resultant films that would occur due to reaction of the jacket and $O_2$ plasma. A 50-cycle $O_2$ plasma pretreatment was used to clean the substrate (Si or glass witness slides, TFBGs, bare fiber) surfaces prior to any ALD experiments. In all *in-situ* gold ALD monitoring experiments, up to 6 fibers (3 TFBGs, 3 bare fibers) were used to ensure that identically uniform, continuous, and "ALD-like" gold coatings were obtained each time with as close to identical effective optical properties as possible. Optical monitoring in transmission mode was undertaken in similar fashion to what has been described previously.

SPR from metal-clad (particularly gold) optical fibers has been the focal point of many theoretical[47,50] and experimental studies,[48,52,64,79,94] and this is due to the fact that metal layers which support surface plasmon polaritons (SPPs) impose a sensing interface that is extremely sensitive to the surrounding chemical environment. Grating-assisted, selective excitation of SPR in the near-infrared (NIR) range can only occur if a specific cladding mode couples to an SPR, resulting in loss of its own amplitude and possibly neighbouring modes depending on the thickness of metal layer. The location of the resulting perturbed cladding mode(s) (SPR "notch") is a direct measure of the cladding mode effective index (as described by the phase-matching rule described earlier), making the metal-clad TFBG a bulk refractive index sensor. Figure 10.3 features the polarization-dependent transmission spectra of a gold-coated TFBG after 900 gold ALD cycles (effective thickness of $45.3 \pm 4.5$ nm confirmed by AFM thickness measurements on Si witness slides and shown in Figures 10.6d 10.6f) after submersion in di-$H_2O$. Detection of SRI changes, however, is only achievable through the TM-mode (radially excited electric field) since the evanescent field of the resulting gold layer or nanostructure extends far into the cladding and into the surrounding medium but is forbidden within the gold film itself.[54] This is immediately obvious in Figure 10.3 since the excitation of SP resonances occurs in the TM-polarized cladding modes but after 90º rotation of the linearly-polarized input light to the TE-polarization state, the SPR signature disappears.

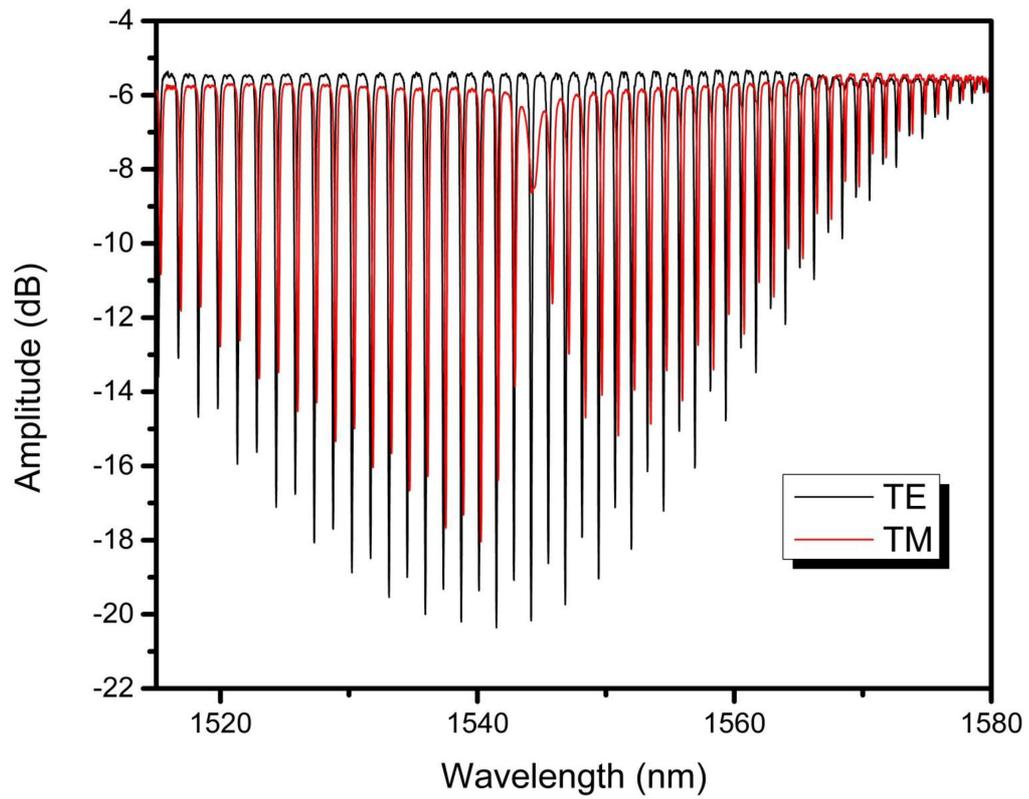

**Figure 10.3:** TFBG-polarized transmission spectra of a gold-coated (by ALD, 900 cycles) TFBG immersed in diH₂O. The lossy TM resonance at 1545 nm (notch) is characteristic of gold films that support SPR.

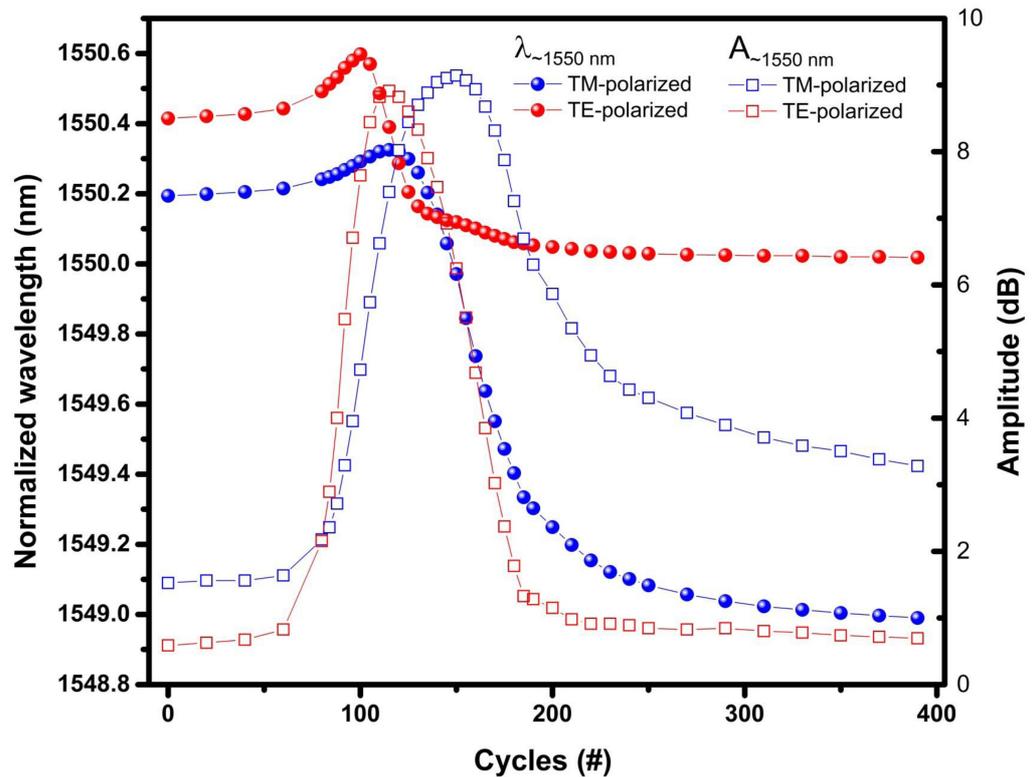

**Figure 10.4:** *In-situ* spectral evolution (wavelength shift at resonance located at 1550 nm and associated attenuation) of the gold plasma ALD process during the first 400 cycles. Solid circles correspond to the wavelength shift data and hollow squares correspond to the attenuation of the resonance at 1550 nm at various cycle numbers.

Figure 10.4 depicts the spectral evolution of the *in-situ* optical monitoring of the gold ALD process in the form of time-resolved wavelength shift as well as the attenuation profiles of the TM- and TE-polarized cladding modes. As the first few gold layers deposit, the azimuthally-excited TE modes, as expected, undergo a wavelength shift and attenuation just prior to the radially-excited TM modes. Perhaps the most striking feature, which was never observed in any of the CVD monitoring experiments, was the full regeneration of the cladding mode amplitudes for both polarization states. On the basis of this data and the associated wavelength shift of the orthogonally-polarized cladding modes occurring between 84-110 cycles, it was concluded that film closure and the insulator-to-metal transition of the gold films from the ALD process occurs between effective thicknesses (recalling the growth rate of ~0.5 Å/cycle) 4.2 to 5.5 nm. The percolation threshold for evaporated gold films (on Si/SiO$_2$) is typically reported to be ~6.5 nm,[140] however our results indicate the gold films produced from ALD are more optically continuous than this. Unlike the previously described gold CVD processes wherein the granular gold films maintain their non-metallic properties at thicknesses as high as 65 nm,[63] we directly observed (in real-time) this insulator-to-metal transition of the gold films during ALD and thereby generated films supporting SPP propagation (negative effective real permittivity with a complex effective permittivity that is orders of magnitude higher). For comparison to the gold-coated TFBG from ALD (900 cycles, 45.3 ± 4.5 nm), a TFBG was coated with gold by evaporation (PVD) with a thickness of 45.5 ± 5.1 nm, as measured by AFM on a gold-coated Si witness slide (Figures 10.6c, 10.6e). Cross-sectional SEM (Figure 10.5a-b) revealed that the ALD-grown gold coating is extremely uniform around the circumference of the optical fiber, whereas the gold coating grown by PVD was oblate. This was attributed to the fact that the optical fiber sample must be rotated 180º halfway through the sputtering process to ensure total coverage. The calculated root-mean-square roughness (R$_{RMS}$) of the PVD gold coatings (Figure 10.6a) was 1.28 nm and relatively lower than the ALD-grown gold coating which was 5.23 nm (Figure 10.6b), however, the gold coatings from ALD were much more continuous at such a thickness compared to the PVD gold coatings and show more coalescensce of the gold nanoparticle aggregates.

As mentioned, the SPP propagation that lends itself to SPR can easily be observed by finding the cladding mode with the maximum loss. That is, at the gold-SRI boundary, the electric field distribution is such that SPR is either masked (TE-polarized resonances) or exhibited (TM-polarized resonances). To exploit the evolution of the anisotropic effective permittivity of the gold films after 900 ALD cycles, polarization-dependent loss (PDL) analysis was employed in order track the SPR resonance during small changes in the SRI. By sweeping across the same wavelengths used for the transmission mode (1520-1620 nm) *in-situ* ALD process monitoring, the gold-coated TFBG devices (from ALD and PVD) exhibit a PDL (dichroism) at wavelengths within 5 nm of the lossy cladding mode that occurs in the amplitude spectrum. This is due to the variation of peak-to-peak amplitude ($A_{TE}$-$A_{TM}$) and wavelengths ($\lambda_{TE}$- $\lambda_{TM}$) as the cut-off mode is approached. The demodulation technique used to decompose the polarization-dependent transmission spectra of a TFBG into the PDL is thoroughly described in more detail elsewhere.[50] Figures 10.7 and 10.8 contain the PDL data for the PVD-grown and ALD-grown gold-coated TFBG sensors, respectively, as a function of increasing SRI (0% to 26% NaCl solutions). It is worth noting that prior to the PDL measurements, the TFBG sensors were calibrated for absolute refractometric sensitivity in the NIR range by calculating the effective RI of the cutoff cladding mode for each mass concentration (%) of NaCl solution.[45]

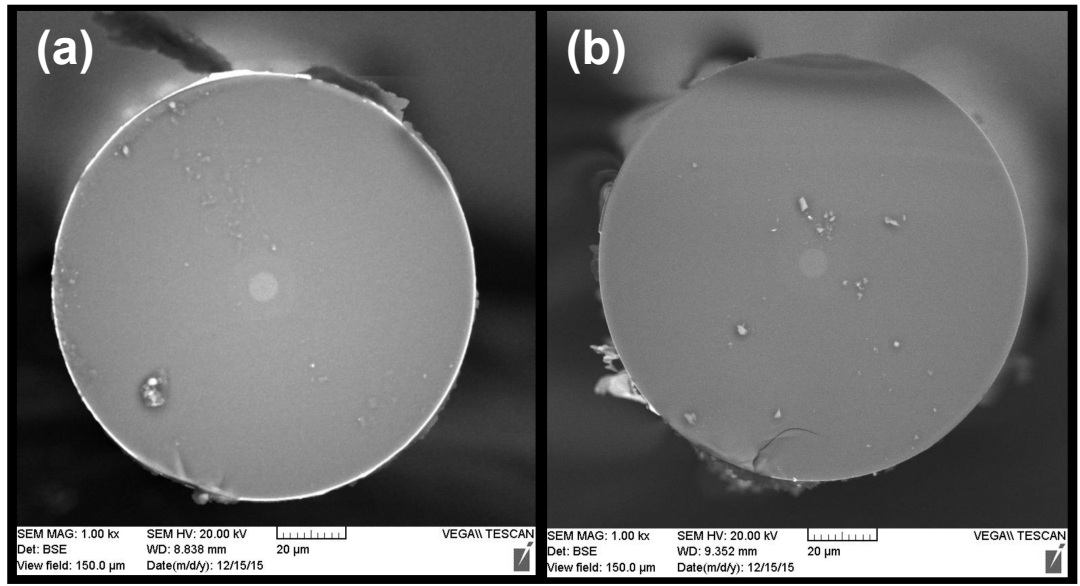

**Figure 10.5:** Cross-sectional (end-on) SEM image of a gold-coated fibers after (a) 900 cycles of gold ALD and (b) after sputtering (evaporation of gold). Note that the small grey circle in the middle is the optical fiber core (diameter of ~8.2-8.4 µm).

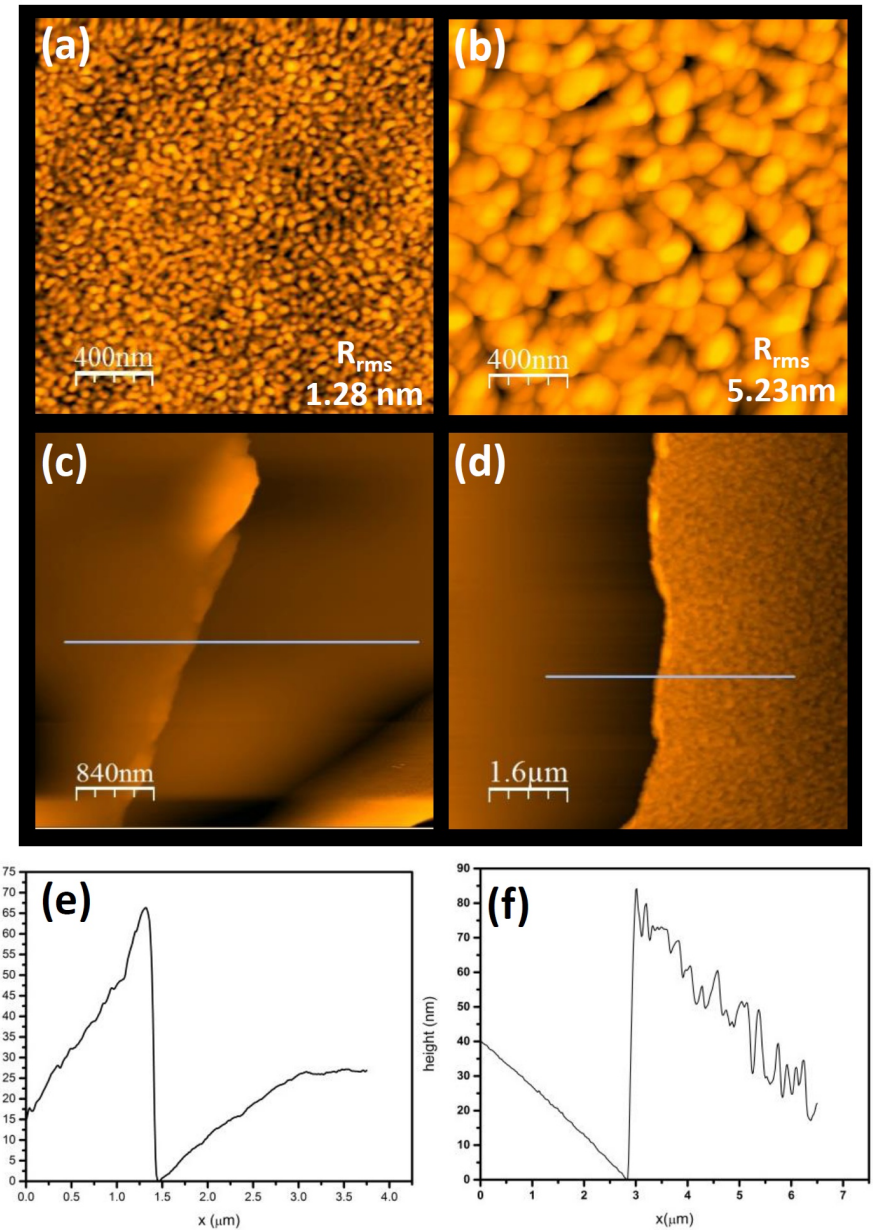

**Figure 10.6:** (a) 2 x 2 μm topographic AFM image of the PVD-grown gold film on Si(100) witness slide with root-mean-square roughness ($R_{rms}$) indicated. (b) 2 x 2 μm topographic AFM image of the ALD-grown gold film on Si(100) witness slide with corresponding $R_{rms}$ indicated. (c) and (d) are topographic AFM images of film-scratched zones of the witness slide for the PVD-grown and ALD-grown gold films, respectively. Grey lines correspond to the height profiles in (e) and (f) for the PVD-grown and ALD-grown films on Si(100), respectively.

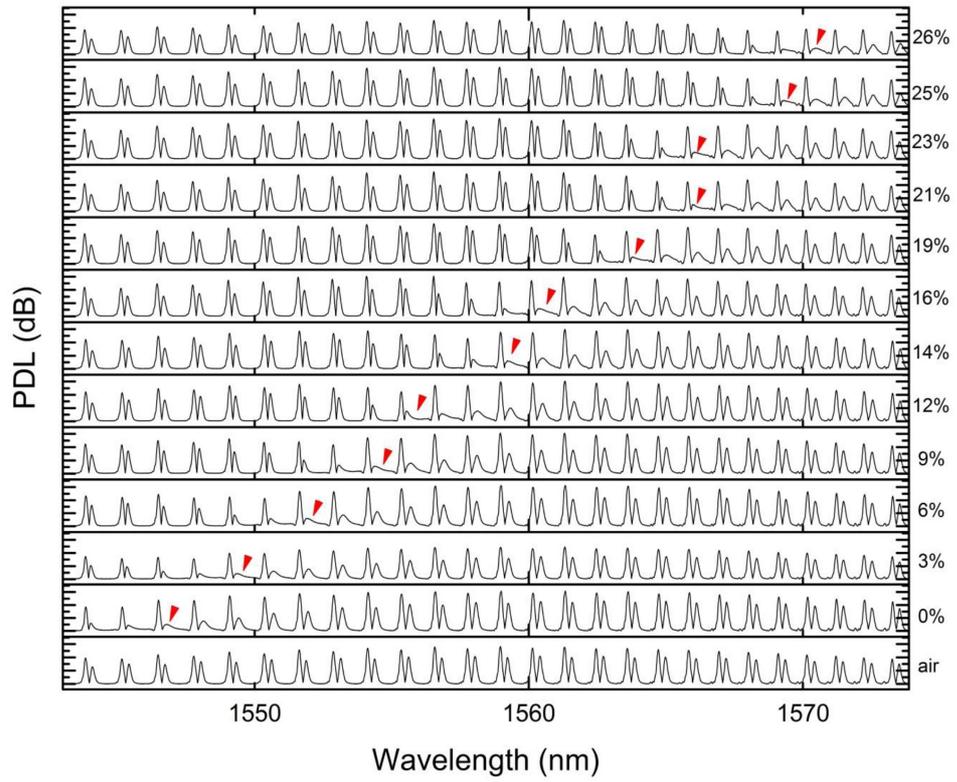

**Figure 10.7:** Stacked TM-polarized PDL spectra of the gold-coated (by PVD) TFBG under various SRIs ranging from air (SRI=1) to saturated (26% by mass) NaCl solution (SRI=1.361). Red arrows indicated the location of the SPR (lossy) resonance.

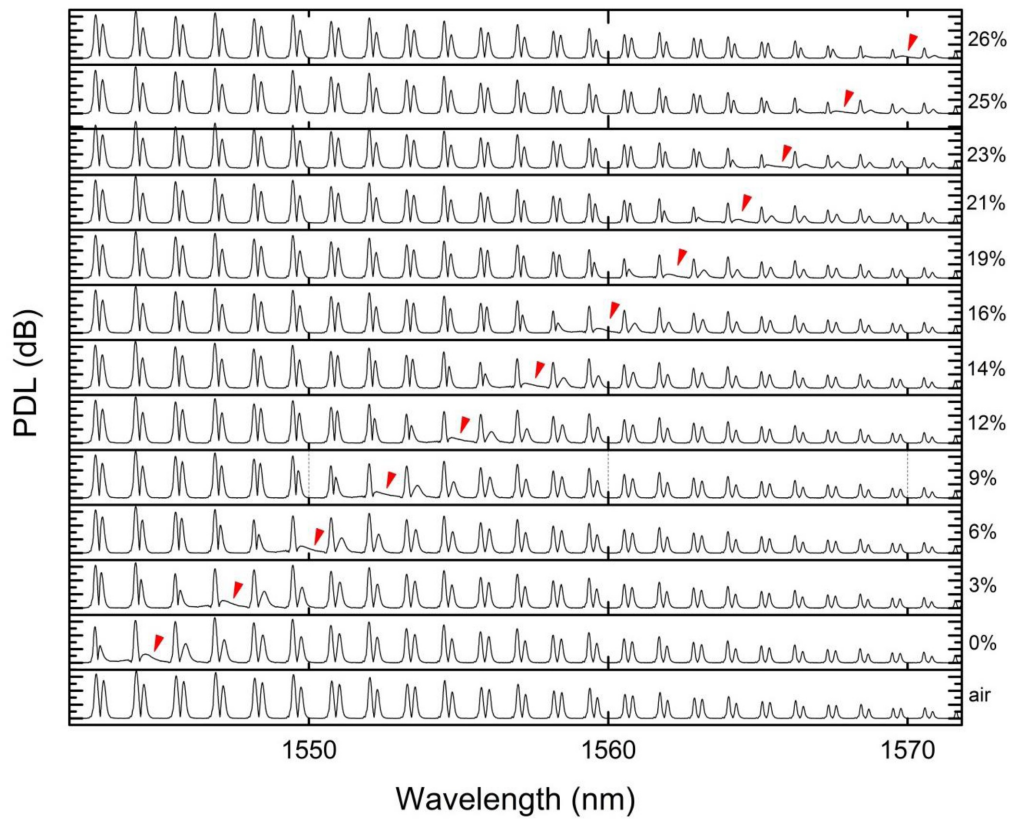

**Figure 10.8:** Stacked TM-polarized PDL spectra of the gold-coated (by 900 cycles of gold ALD) TFBG under various SRIs ranging from air (SRI=1) to saturated (26% by mass) NaCl solution (SRI=1.361). Red arrows indicated the location of the SPR (lossy) resonance.

It appears that at SRIs near saturation, the PVD-grown gold-coated TFBG no longer generates significant mode power coupling between the core and cladding (surrounding medium). After selection of the most lossy cladding mode (SPR wavelength) wavelength at each SRI, the refractive index sensitivities of the ALD-grown and PVD-grown gold-coated TFBG sensors were calculated from the slope of the linearizations shown in Figure 10.9. Remarkably, the refractometric sensitivity of the gold-coated SPR sensor from the ALD process was at the theoretical limit of ~550 nm/RIU (RIU=RI units) whereas the TFBG sensor with the PVD gold coating had only 506 nm/RIU. Additionally, when propagating the statistical error on precision of 3 different TFBG sensors used in the same 900-cycle ALD process (Figure 10.9), it is 7% less precise when compared to that of the single SPR-TFBG sensor tested with the PVD-grown gold coating. On the basis of this statistical analysis, the variance across the 3 different SPR-TFBG devices (ALD-grown gold case) is less than the variance obtained from the single SPR-TFBG device for the PVD-grown gold case. The aforementioned theoretical limit (i.e. maximum) arises from the fact that the core effective index ($n_{eff,\ core}$) and grating period ($\Lambda$) are physical properties of the TFBG unaffected by the surrounding medium, which sets up the following boundary condition:[47]

$$\lambda_{SPR} = \left(n_{eff,SPR} + n_{eff,core}\right)\Lambda \qquad [1]*$$

$$\frac{\delta\lambda_{SPR}}{\delta n_{ext}} = \frac{\delta n_{eff,SPR}}{\delta n_{ext}}\Lambda \qquad [2]$$

$$\frac{\delta\lambda_{SPR}}{\delta n_{ext}} \leqslant \Lambda \qquad [3]$$

*Phase-matching condition also defined in earlier chapters

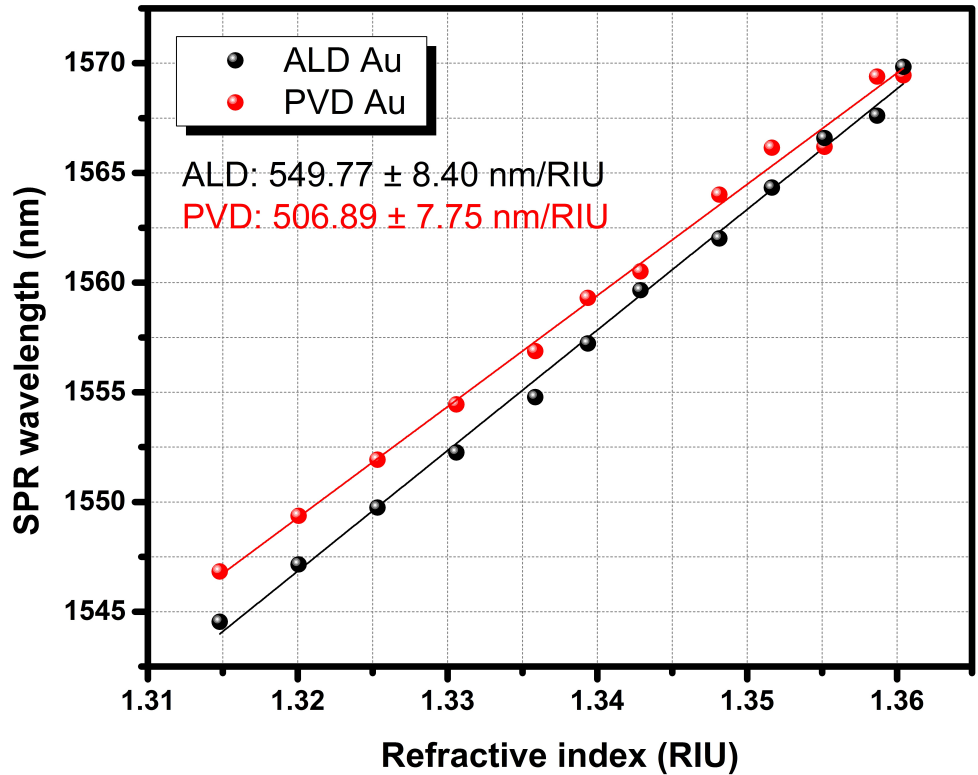

**Figure 10.9:** Refractometric (SRI) sensitivity of the ALD-grown gold-coated (black circles) TFBG and the PVD-grown gold-coated TFBG (red circles) based on the location of the attenuated resonances from in the associated PDL spectra (Figures 10.7-8). Error calculation for the ALD-grown gold-coated sample is based on the precision between 3 different TFBG sensors used during the same deposition experiment. The associated error for the PVD-grown gold coating case is the standard deviation for measurements on 1 TFBG sensor.

where $\delta n_{ext}$ and $\delta\lambda_{SPR}$ indicate the changes in SRI and SPR wavelength, respectively. Since the grating period after fabrication of the 10º TFBGs in this work is typically in the 550-555 nm range, this means the total effective index shift (SPR) and, consequently, refractive index sensitivity can never be greater than 550 nm/RIU. Overall, the extracted spectral data from the ALD-grown gold-coated TFBG sensor from Figures 10.3, 10.7-10.9 demonstrates that the gold coatings from the ALD process enhance the sensitivity of the SPR mode as the cladding diameter of the TFBG decreases. Interestingly, the SPR mode sensitivity for our ALD-grown gold coating (~45 nm) has an ultimate refractometric sensitivity identical to a 70 nm gold coating on a TFBG from simulations.[47] This further confirms that the optical anisotropy of the ALD gold thin films begins even sooner in the deposition than what has been observed by SE on flat substrates during gold evaporation (PVD). Moreover, the gold films are not only reproducibly uniform, but they are also optically conductive (negative real component of the permittivity) and the SPR technique described herein can be used to probe the complex refractive index of highly continuous noble metal coatings such as gold.

## 10.3    Summary and General Outlook

In this chapter, we have demonstrated a new *in-situ* method to optically monitor thermal or plasma-enhanced ALD processes using TFBG technology. We firstly use the well-established AB-type thermal ALD process for alumina thin films to calibrate the spectral measurement setup and then proceed to the ABC-type process for gold metal PEALD already described (Chapter 9). In this work we provide *in-situ* optical characterization of gold metal thin films as they nucleate and grow during ALD and show, through a direct measurement of the insulator-to-metal transition for gold metal (on glass), that ALD-grown gold thin films are even more uniform, continuous, and optically conductive than their PVD counterparts. Since these anisotropic gold metal films support SPP propagation, their optical properties were further exploited to develop a robust SPR-based optical sensor with a sensitivity that is both ~10% higher than one fabricated with a PVD-grown gold coating and at the theoretical sensitivity limit for a gold-clad SPR sensor operating in the NIR range. For comparison, the SRI sensitivities of the CVD-grown gold coatings presented throughout this thesis were all at least 100 times smaller for the same thicknesses. In conclusion, the modulated index of a TFBG is perfectly suited for the direct measurement of CVD and ALD processes and the use of gold metal ALD has led to the development of an SPR-based optical sensor with ultimate refractometric sensitivity.

# Chapter 11

Conclusion

To summarize, the tilted grating planes induced by the inscription of a TFBG structure in the optical fiber's core allow for the circularly polarized core-guided light to be decomposed into two unique and orthogonal channels of linearly polarized light (TE and TM). Initially, the nucleation and growth of gold films via CVD was monitored using 10º-tilted TFBGs and showed excellent sensor performance at thicknesses ranging from 5 to 200 nm. *In-situ* CVD monitoring measurements were performed in TFBG reflection mode, which had the advantage of a single entry point for the TFBG into CVD reaction chamber, limiting spectral loss. The spectral evolution of the orthogonally-polarized cladding modes excited by the TFBG provided very distinct optical profiles that could be related to the absorption or scattering behaviour of the cladding-guided light being coupled into the "lossy" gold nanostructures during thin film growth. In the case of the more thermally stable gold(I) iminopyrrolidinate, this TFBG-based method of evanescent field absorption spectroscopy was useful for individual resonance tracking of both the TE- and TM-polarization states in the NIR range as the electric field extending into the nucleating gold nanoparticles was slowly perturbed. In the case of the more thermally unstable gold(I) guanidinate precursor, reaction quenching using $N_2$ carrier gas at various points during the attenuation allowed for careful *ex-situ* analysis by SEM and AFM on witness slide samples to corroborate the experimentally obtained growth rates (222 nm/min during attenuation and 37 nm/min overall) by the TFBG witness probe. In the case of the gold nanoparticle films grown by CVD from gold(I) iminopyrrolidinate, polarization-dependent absorption occurs at thicknesses higher than 25 nm, implying that the films go from discontinuous (isotropic with respect to real and complex permittivities) to relatively continuous (anisotropic), but still far from conductive (even at 200 nm). Using the Maxwell-Garnett EMA and experimental evidence from AFM image analysis, it was confirmed that the high concentration of air-gaps in the CVD-grown gold nanoparticle films is due to the relatively low aspect ratio of the ellipsoidal gold nanoparticles and their lack of coalescence, even at the high CVD process temperatures (300-350 ºC), into continuous films that support SPR. The evolution of the anisotropy in the gold nanoparticle films at thicknesses higher than 25 nm was evident by a birefringence in $n_f$ (real part of the effective permittivity; scattering component) and concurrent dichroism in $k_f$ (imaginary part of the effective permittivity; absorption or "dielectric" component that governs SPR). Interestingly, we observed a similarly anomalous effective permittivity for PVD-grown gold near the insulating regime (5.5 nm) of the insulator-to-metal transition for gold. Enhancements to the SRI sensitivity of the "dielectric" gold coatings from the CVD processes improved from 2.49 nm/RIU to 6.02 nm/RIU with the addition of an ALD-grown $Al_2O_3$ interfacial layer (50 nm) prior to gold CVD. The addition of the dielectric pre-coating improved the peak-to-peak amplitude sensitivity of the cladding modes and also increased the wavelength separation between the orthogonally polarized cladding modes, which consequently improved single resonance tracking during deposition experiments. Lastly, we successfully incorporated the TFBG "ellipsometer" directly into a Picosun R-200 plasma ALD tool and obtained cycle-resolved wavelength shift and attenuation data of thermal $Al_2O_3$ ALD (for calibration), which, expectedly, yielded isotropic and uniform $Al_2O_3$ coatings for up to 1000 cycles. After development of a plasma-enhanced gold ALD process, we monitored gold ALD from nucleation delay (< 10 cycles) through the insulator-to-metal transition (4.2.-5.5 nm; 84-120 cycles) and up to 900 cycles (~45 nm) to generate an ALD-grown gold-coated SPR-TFBG sensor with a refractive index sensitivity 2 orders of magnitude higher than the dielectric gold nanoparticle coatings from CVD (6-200 nm), the ultrathin (5.5 nm) PVD gold-coated TFBG, the $Al_2O_3$-pre-coated TFBG devices with CVD grown gold over-coatings and 43 nm/RIU more sensitive than an SPR-TFBG sensor with a of PVD-grown gold coating of similar thickness (~45). With the development of these SPR-TFBG sensors using ALD-grown gold, we have demonstrated the robustness and versatility of TFBGs as both an *in-situ* diagnostic tool for any CVD or ALD process, but also as an optical fiber sensor with unprecedented refractometric sensitivity.

239.